\documentclass[12pt, A4,twoside]{report}

\usepackage{fancyhdr}
\usepackage{graphicx,psfrag,amsmath}
\usepackage{nccmath} 
\usepackage{chapterbib}
\usepackage{cite}
\usepackage[usenames,dvipsnames]{color}
\usepackage{setspace}
\usepackage[margin=16pt,font={small,stretch=1.1}]{caption}
\usepackage{enumitem}
 \fancyhfoffset[LE]{.5 cm}
\clearpage{\pagestyle{empty}\cleardoublepage} 

\renewcommand{\chaptermark}[1]{\markboth{\chaptername\
\thechapter}{}}

\newcommand{\e}{\eta}

\newcommand{\subs}[1]{_\mathrm{#1}}

 \newcommand{\mpl}{M\subs{p}}
\newcommand\sigb{\bar\sigma}
\newcommand\ytt{\tilde {y}}

\newcommand\mzt{{\tilde{m}}_0}
\newcommand\Azt{\tilde{A_0}}
\newcommand{\s}{\Sigma}
\newcommand{\Sigb}{{\overline\Sigma}}
\newcommand{\oot}{\overline {126}}

\newcommand{\ovl}{\overline}
\newcommand{\nnu}{\nonumber\\}

 \def\be{\begin{equation}}
\def\ee{\end{equation}} \def\bea{\begin{eqnarray}}
\def\eea{\end{eqnarray}} 

\begin{document}
\newpage  \pagenumbering{roman}\thispagestyle{empty}
{\Large\baselineskip 27pt
\begin{titlepage}
\begin{center}
%\vspace*{1in}
 {\bf{\textsf{NEW MINIMAL SUPERSYMMETRIC SO(10) GUT PHENOMENOLOGY AND ITS COSMOLOGICAL IMPLICATIONS }}} \par
\vspace{0.7in} {\bf{\textsf{A THESIS}}}\\ \textsf{ {Submitted to the\\
FACULTY OF SCIENCE\\ PANJAB UNIVERSITY, CHANDIGARH\\ for the
degree of} }\\\par \vspace{0.1in}\textbf{\textsf{DOCTOR OF
PHILOSOPHY}}\\\par \vspace{0.8in}\textsf{\textbf{2014}} \par
\vspace{0.3in} \par \vspace{0.8in} \textbf{\textsf{ILA GARG}}
\par \vspace{0.4in}
 \par
\vspace{0.38in}%\begin{figure}[h!]
%\centering
 %   \includegraphics[scale=0.5]{PU.eps}\\
 %\end{figure}
{\textsf{DEPARTMENT OF PHYSICS\\CENTRE OF ADVANCED STUDY IN
PHYSICS\\PANJAB UNIVERSITY\\CHANDIGARH, INDIA}}
\end{center}
\end{titlepage}
}

\thispagestyle{empty}

 \mbox{}\newpage \thispagestyle{empty} \vspace*{4.in}
\begin{center}\emph{{\huge\textbf{Dedicated To My Parents}}}
 \end{center}
\newpage\thispagestyle{empty}
\mbox{}\newpage \addcontentsline{toc}{chapter}{Acknowledgements}
\thispagestyle{plain} \begin{center}{\Large \bf
\emph{Acknowledgements}}\end{center} \medskip I express my deep
gratitude to my thesis supervisor Prof. C. S. Aulakh for his
constant help and support during my Ph.D. His knowledge, vision
and understanding in particle physics is invincible. He helped me
to improve my personality as a scientist. I am also thankful to
him for providing a healthy and open environment for research.

 I am very thankful to my family members for their constant moral
support, love, faith and prays for me. Without their support I
won't be able to make it.

 I would like to say thanks to my
co-scholar Charanjit Kaur whom I know from my M.Sc. days. She is a
part of this journey from the
very beginning. Healthy discussions with her always helped me a pile to tackle even difficult problems.

 I would like to recognize my friends Tanvi Vashisht, Jagdish Kaur, Rajni Jain, Simanpreet Kaur, Amandeep
 Chahal, Pankaj Verma, Rohit Sharma, Sukhjit Kaur, Sonika Goyal
  for having delightful moments during tough times and making life easy away from home.
   A special thanks to  Anubha Jindal, Pranati Rath, Deepika Makkar, Aman Jindal for
   their constant support and encouragement.

 I sincerely pay my regards to the Chairperson of our Department for
the research facilities provided. I would like to say thank to all the members of this department for providing me
 all kind of help. I am also thankful to council of scientific and industrial research (CSIR) of India for
providing me the generous funding during the course of my research
work.

 Above all I am grateful to God for providing me the strength
and good health to accomplish this task. Last but not least I
thank all who were directly or indirectly involved in the
fulfillment of this project.
\medskip\bigskip\medskip\medskip\\
{\bf
Date:01/04/2015~~~~~~~~~~~~~~~~~~~~~~~~~~~~~~~~~~~~~~~~~~~~~~~~~~~~~~~~~~~~~~~~~~~~Ila
Garg}

\newpage\thispagestyle{empty}
\mbox{}\newpage
\addcontentsline{toc}{chapter}{Abstract}
\thispagestyle{plain} \begin{center}{\Large \bf
Abstract}\end{center} Supersymmetric GUTs based on SO(10) gauge
group are leading contenders to describe particle physics beyond
the  Standard Model. Among these the ``New minimal supersymmetric
SO(10) grand unified theory" (NMSGUT) based on Higgs system
10+120+210+126+$\overline{126}$ has been developing since 1982.
 It now successfully fits the whole standard Model gauge coupling,
  symmetry breaking and
 fermion mass-mixing data  as well  as the neutrino mass and
 mixing data in terms of  NMSGUT
 parameters and just 6 soft supersymmetry breaking parameters defined at
  the GUT scale.
  In this thesis we study  the phenomenology of NMSGUT,
its implications for  inflationary and Cold Dark matter cosmology
and develop Renormalization group(RG) equations for the flow of
NMSGUT couplings  in the extreme ultraviolet.

 We carry out a long
calculation to show that threshold corrections due to heavy
 particle spectra  to the matching
condition between matter Yukawa couplings of the effective Minimal
Supersymmetric Standard Model (MSSM) and the NMSGUT Yukawa vertex
provide a generic mechanism to cure the long standing problem of
fast Baryon decay rate due to dimension five operators.
 Incorporating these corrections we are able to find
regions of the parameter space of NMSGUT superpotential where wave
function renormalization of the effective MSSM Higgs doublets is
close to the dissolution value ($Z_{H,\overline{H}}=0$). The
SO(10) Yukawas required to match the MSSM fermion data are then
lowered after the rescaling required  to achieve canonical kinetic
terms in the effective MSSM. It is these lowered SO(10) Yukawas
will also determine the dimension 5 Baryon number  violation
operator coefficients. Thus the associated rates can be suppressed
to levels which are compatible with current experimental bounds.

Within the context of a supersymmetric model based on an extension
of the MSSM gauge group we show that superpotential corresponding
to supersymmetric Type-I seesaw
  for neutrino mass can provide inflection point inflation
  along a flat direction associated with a gauge invariant
combination of the Higgs, slepton and right handed sneutrino
superfields. The Majorana mass of a right handed neutrino sets the
scale of inflation. In this scenario the inflation parameters and
fine-tuning required for flat inflaton  potential is controlled by
superpotential parameters rather than soft susy breaking terms
  which is in contrast to Dirac neutrino-inflation connection
  and therefore  more stable under radiative corrections.
  Reheating after   inflation is necessary for the success of
  big Bang nucleosynthesis. Reheating  in context of
   supersymmetric seesaw inflation (SSI) via `instant preheating'
  mechanism is presented. The high reheat temperature
  $\sim$ $10^{11}$-$10^{13}$ GeV
  in this scenario requires large gravitino mass greater than $  50 $ TeV to
   avoid a gravitino problem,
  pointing towards a high Susy breaking scale.
 The embedding of SSI in NMSGUT, which accommodates
   seesaw mechanism to explain neutrino masses
  and also predicts Susy breaking scale of O(10-100 TeV), is carried out.
  We discuss the obstacles that arose while embedding SSI in the NMSGUT
  and revisit the same in
  the light of new results of large value of  tensor to scalar
  component ratio ($r$) of cosmic microwave background fluctuations
   by BICEP2 experiment. We find an improvement in the
  number of e-folds by 4-orders of magnitude but are still short by
   a factor of  about 50.

  In NMSGUT R-parity conservation down to weak scale facilitates
  Neutralino as lightest stable supersymmetric particle suitable for cold dark matter
  candidate. We evaluate the  relic density of dark matter(DM) on NMSGUT
  parameter space (using   the  public DARKSUSY code)
  for   the realistic parameter sets found using GUT scale threshold
   corrections. We find that  solutions with light smuons can
    yield acceptable DM relic densities.

  In addition we also calculated  the 2-loop renormalization group
  equations (RGE) for the
  parameters of NMSGUT. We show that using RGEs of NMSGUT parameters
   in the energy range from Planck to  GUT scale one can generate negative soft
 MSSM   Higgs mass squared parameters : which
 were found to be essential   for the explanation the distinctive normal hierarchy
   sparticle spectra  found  in the NMSGUT
   at the large tan$\beta$ values so far investigated.

\newpage\thispagestyle{empty}
\addcontentsline{toc}{chapter}{List of Publications}
\thispagestyle{plain}  \begin{center}{\Large \bf List of
publications}\end{center}
\begin{enumerate} \item C. S. Aulakh and
{\bf I. Garg}, ``Supersymmetric Seesaw Inflation", Phys. Rev. D.
{\bf 86}, 065001 (2012) [arXiv:hep/ph-1201.0519v4 ] \item
C.~S.~Aulakh, {\bf I. Garg} and C.~K.~Khosa,
  ``Baryon stability on the Higgs dissolution edge: threshold corrections and suppression of baryon violation in the NMSGUT,''
  Nucl.\ Phys.\ B {\bf 882}, 397 (2014)
  [arXiv:1311.6100 [hep-ph]].
 \end{enumerate}

\newpage\thispagestyle{empty}
\mbox{}\newpage \tableofcontents
 \mbox{}\newpage
\addcontentsline{toc}{chapter}{List of Figures}
\begin{spacing}{1.15}
 \listoffigures
 \addcontentsline{toc}{chapter}{List of Tables}
  \listoftables
 \end{spacing}

\addtocontents{toc}{\protect\rule{13.9 cm}{.1pt}\par}
\mbox{}\newpage\pagenumbering{arabic}
\chapter{Introduction}
 With the advancement of technology and precision of experimental measurements, the last few decades
have added a lot to our understanding of elementary particles and their interactions.
The theory of strong, weak and electromagnetic interactions, i.e. the standard model (SM) of particle physics has been
 developed and established during this time.
  All the particles of SM (three complete families of elementary leptons and quarks)
   except Higgs boson (responsible for mass of massive particles in SM via Higgs mechanism)
  were known by the end of 20th century.
In July, 2012, ATLAS and CMS, the two major experimental set-ups
at Large Hadron Collider, jointly announced the discovery of a particle at 125
GeV \cite{Atlas,CMS} which has properties of SM Higgs boson. This discovery completed
the SM in terms of its particle content. The Standard
Model now explains low energy phenomena to a precision of better than
1$\%$ \cite{pdg}. But then the question arises, is it the ultimate
theory or just a limit of some new theory at high scale? This
question is all the more acute as SM is not able to answer some
important questions which arise naturally. For example, in the
 SM without ad-hoc dimension 5 operators neutrinos are massless. However from neutrino
oscillations \cite{oscillation1,oscillation2} it is well
established that neutrinos are massive. SM has no explanation for
hierarchial structure of fermion masses and number of generations
restricted to three (Flavor problem). The structural instability
is another issue i.e. electroweak masses are unstable to loop
corrections from heavy particles in the region between weak scale
and Planck scale. It has problems in cosmological sector as well.
SM doesn't contain a suitable candidate for dark matter and has no
explanation for accelerated expansion of universe i.e. concept of
dark energy.

 All these
limitations of SM led particle theorists to develop Grand Unified
theories. The main insight of such theories is that the observed
phenomena coded in the peculiar structure of the SM are the low
energy manifestation of a fundamental unified theory with a larger
gauge symmetry which like the SM symmetry is spontaneously broken
(but at a high scale) resulting in the observed SM as the
effective theory to an excellent approximation. Grand unification
was proposed by J. Pati and Abdus Salam \cite{ps} in 1974 by
considering a gauge group $SU(4)_c \times SU(2)_L \times SU(2)_R$
as symmetry of nature. The SU(4) group unified the color SU(3)
group with $U(1)_{B-L}$ so that Lepton number becomes ``fourth
color'' and the $SU(2)_R$ group parallels the chiral action of
$SU(2)_L$ on left chiral fermions thus restoring parity.
Thereafter H. Georgi and S. Glashow \cite{gg} gave a model which
embeds the SM in a SU(5) gauge theory with a SM fermion family in
$\overline{\textbf{5}}+\textbf{10}$ of SU(5). The other minimal
possibility is SO(10)\cite{hp}. More refined models are
Supersymmetric GUT Models. Susy GUTs are strongly motivated by
near exact unification of running gauge couplings at energy scale
$\sim$ $10^{16.33}$ GeV in the minimal supersymmetric standard
model (MSSM). Supersymmetry is crucial to maintain the large
separation
between Electroweak and GUT physics scales.

 Supersymmetric Grand unified theories
based on the SO(10) gauge group have many attractive features:
\begin{itemize}
\item  The fundamental {\bf16} dimensional spinor representation
of SO(10) contains one family of 15 SM fermions and right handed
neutrino necessary for neutrino mass (via Seesaw mechanism
\cite{seesaw}).
 \item Third generation Yukawa unification in large tan$\beta$
 $\simeq$ 50 region \cite{Carena,hallrathempf}.
  \item Pati Salam group $SU(4) \times SU(2)_{L} \times
SU(2)_R$ is subgroup of SO(10) so it automatically embeds the
minimal supersymmetric Left-Right models
\cite{MSLRM1,MSLRM2,MSLRM3}. \item B-L even vevs result in
R-parity preservation down to the weak scale \cite{rpar3} and
consequently a stable Lightest Supersymmetric Particle(LSP) which
is the most suitable Weakly Interacting Massive Particle(WIMP)
dark matter candidate in modern cosmology. \item A specific SO(10)
model \cite{aulmoh,ckn,babmoh,abmsv,bmsv,bmsv1,bsv} based on the
{\textbf{10+210+126+}}$\overline{\textbf{126}}$ Higgs GUT system
has the further remarkable virtues of:
    \end{itemize}
    \begin{enumerate}
    \item Solution of symmetry breaking at the GUT scale \cite{abmsv}.
\item An explicit calculations of superheavy spectra are available
\cite{ag1,bmsv,ag2,fuku04}.\item Such theories have intriguing
gauge unification threshold effects which can raise the
unification scale close to the Planck scale \cite{ag2}. \item An
old conjecture regarding likely futility \cite{dixitsher} of Susy
SO(10) due to uncontrollable threshold corrections from GUT scale
fields has been disproved \cite{ag2}. Threshold corrections can
emerge in a well controlled way and even lead to interesting
mechanisms and phenomena which are characteristics of MSGUTs.
\end{enumerate}
 The so called New Minimal Supersymmetric SO(10) GUT (NMSGUT)
 \cite{nmsgut,nmsgutIII} based on the {\textbf{10+210+120+126+}}$\overline{\textbf{126}}$ Higgs GUT system
 has all the above mentioned features. The main objective of this thesis is to
contribute to making the NMSGUT a possibly viable and complete
theory of particle physics and cosmology with the special emphasis
on cosmology. The idea is to develop the theory on the same lines
and as explicitly as the Standard Model. In order to do this,
besides improvements of the computer codes already developed
\cite{nmsgut,nmsgutIII} for calculating the effects of the NMSGUT
at intermediate  and low energies  we need to confront the theory
to various open problems of particle physics and cosmology. The
specific objectives of this thesis are as follows:
\begin{enumerate}
\item Study of threshold effects of heavy fields appearing due to
GUT scale symmetry breaking and consequently lowering of the
strength of d=5, $\Delta$ B $\neq$ 0 operators resulting in
natural suppression of proton decay rate.
 \item The
study of supersymmetric seesaw inflation (SSI). \item The
reheating mechanism in context of SSI. \item BICEP2 revolution and
NMSGUT inflation. \item The testing of NMSGUT spectra in dark
matter searches for neutralino as a dark matter candidate. \item
The study of NMSGUT RGEs in the region between Planck scale and
GUT scale.
\end{enumerate}

\section{Outline of thesis}
The outline of thesis is as follows:

 In Chapter {\bf2} we start by
discussing the NMSGUT Higgs system, its superpotential and
symmetry breaking scheme. We then discuss the crucial superheavy
spectrum of NMSGUT and the corrections due to these superheavy
fields to the unification scale $M_{X}$, the QCD coupling at $M_Z$
($\alpha_3(M_Z)$) and the gauge coupling at unification scale
($\alpha_G$): at one loop order. We demonstrate the use of the
fermion mass formulae and SM fermion fitting criteria in NMSGUT.
Then we present formula for dimension 5 (``d=5'') baryon number
violation operators and discuss the problem of fast proton decay
in NMSGUT.

 Having the tools described in Chapter 2 in hand
  we present the GUT threshold effects to SO(10) Yukawas in Chapter {\bf3}. The
  extremely lengthy calculations involving the combination of
  several thousand terms were done separately by three
  collaborators: Myself, another Ph.D student C. K. Kaur and our
  thesis advisor Prof. C. S. Aulakh. This work is published as
  \cite{BStabHedge}. Such collaboration was essential to weed out
  errors over an iterative checking process lasting over one year.
  We have divided the presentation of results between the two
  theses since this work is the basis of both and is shared
  equally. We present the one loop formulas for threshold corrections and
give illustrative examples to underline the
 significance of the GUT scale threshold effects and the need to include
 them. We then discuss about lowering of strength of baryon violation d=5 operator
 due to inclusion of threshold corrections. We discuss various aspects of
 fitting criteria together with threshold effects and give an example solution of NMSGUT parameters
 which fit fermion mass-mixing data and are compatible with B decay limits.
 We also present full expression for correction factors
 to fermion(anti-fermion) line while those for Higgs line can be found in \cite{BStabHedge,charanthesis}.

In Chapter {\bf 4} we review early universe cosmology. Starting with the Big
 Bang and the puzzles of modern cosmology, we discuss the idea of inflation
 to solve these puzzles. We present the ingredients to have an inflationary epoch and
 constraints from observations on inflaton potential. We discuss in detail the inflation models based on MSSM
 flat directions (inflection point inflation) and the fine tuning
 problem in such models. Then we discuss an example of Dirac-neutrino-inflation
  connection in this context.

The nature of neutrino mass may be Dirac or Majorana with the
latter carrying a special attraction because it allows for a neat
explanation for the smallness of neutrino masses, we present the
Majorana-neutrino-inflation connection in Chapter {\bf 5}. We
start with discussing about a generic renormalizable inflection
point inflation (GRIPI) model followed by supersymmetric seesaw
inflation (SSI) and present the conditions on inflationary
potential parameters to achieve slow roll inflation conditions. We
also discuss about the advantages of our scenario of inflation
over Dirac-neutrino-inflation scenario. This work is published as
\cite{SSI}.

 In Chapter {\bf 6} we first discuss the necessity of
the reheating after inflation. Then we present
 one of the mechanism of reheating called ``instant preheating''.
 We present basic steps of how the inflaton energy is dumped into the
 MSSM degrees of freedom after the end of inflation in
 context of SSI. Then we present the evolution of densities
 of different $\chi $ modes and estimates of reheating temperature.

In Chapter {\bf 7}  we discuss embedding of supersymmetric seesaw
inflection point inflation in context of NMSGUT. Then in the light
of BICEP2 results we present an analysis of renormalizable
inflation potential to achieve slow roll inflation without any
fine tuning. Then we revisit Susy Seesaw inflation in
NMSGUT. The obstacles and possible solutions
 to achieve inflation in NMSGUT are discussed.

In Chapter {\bf 8} we first review the main issues concerning Dark
matter. Then we discuss about possibility of neutralino as
candidate for cold dark matter. The relic density calculations for
NMSGUT low energy MSSM spectrum using DARKSUSY package are
presented. The results and constraints from relic density
calculation on NMSGUT parameter space are discussed.

 In Chapter
{\bf 9} we present two loop renormalization group equations for
soft parameters for NMSGUT superpotential. This calculation was
also very lengthy and done in collaboration with C. K. Khosa and
Prof. C. S. Aulakh and hence we present here half of the very
extensive formulas.
 The exploitation of these RGEs has not been done fully but a qualitative view of
 how soft parameters will run in the region between Planck scale and NMSGUT scale is
 presented along with the explanation of why this running may play
 a crucial role in completing the NMSGUT.

 In Chapter {\bf 10} we present a summary and direction of future
 research emerging from this thesis.
\newpage

\chapter{ New Minimal Supersymmetric SO(10) GUT}
\section{Introduction}
The New Minimal Supersymmetric Grand Unified theory (NMSGUT)
\cite{nmsgut,nmsgutIII} is based on $\textbf{210} \oplus
\textbf{10} \oplus \textbf{120} \oplus \textbf{126} \oplus
\overline{\textbf{126}}$ Higgs system of SO(10) gauge group. The
representations $\textbf{10}(H_{i}),\textbf{120}(\Theta_{ijk})$
are real 1 and 3 index (totally antisymmetric (TAS)),
$\textbf{126}(\Sigma_{ijklm})$ is complex (5 index, TAS, self
dual)  and $\textbf{210}(\Phi_{ijkl})$ is 4 index antisymmetric
tensor. Here i,j,k,l,m =1,2...10 run over the vector
representation of SO(10). The fundamental spinor representation is
\textbf{16} dimensional.
 The tensor product $\mathbf{16 \times 16}$ = \textbf{10+120+126}, so
there are 3 possible fermion mass generating (FM) Higgs:
$\textbf{10},\textbf{120},\overline{\textbf{126}}$. The
$\overline{\textbf{126}}$ plays a dual role i.e. it also acts as
AM (adjoint multiplet type) which breaks the SO(10) gauge symmetry
to MSSM together with $\textbf{210}$ plet.

 The original MSGUT (Minimal Susy GUT) \cite{aulmoh,ckn,babmoh,abmsv,bmsv,bmsv1,bsv} was the
renormalizable globally supersymmetric theory (minimal in terms of
parameter counting) in which $\textbf{120}$ plet Higgs was not
considered and $\textbf{10}$ and $\textbf{126}$ were used to fit
fermion data. But MSGUT failed to fit the neutrino mass in the
viable parameter space with the available seesaw mechanism (Type I
and Type II). In MSGUT Type I is dominant over Type II and Type I
seesaw masses are one order of magnitude short \cite{blmdm,gmblm}
of those required by atmospheric neutrino oscillations. To
overcome this difficulty NMSGUT with $\textbf{120}$ Higgs
multiplet (which is quite naturally eligible since it is one of
the three multiplets
$\textbf{10},\textbf{120},\overline{\textbf{126}}$ that have
renormalizable SO(10) invariant couplings to the SO(10) fermion
matter ($\textbf{16}$) bilinears) was proposed \cite{nmsgut}. In
NMSGUT the CKM structure and charged fermion hierarchy is
generated by $\textbf{120}$ along with $\textbf{10}$. The
$\overline{\textbf{126}}$ representation is eased from its duty to
fit charged fermion masses. But it performs an important task to
enhance the Type I seesaw neutrino masses via its ultra weak
couplings (because Type I is inversely proportional to
$\overline{\textbf{126}}$ coupling ). Type-II seesaw contribution
is directly proportional to its coupling so gets further
suppressed. It also lowers the right handed neutrino masses to
$10^9-10^{13}GeV << M_X$ compatible with leptogenesis
\cite{leptogen}.

\section{NMSGUT superpotential and symmetry breaking}
 The full superpotential of NMSGUT containing the mass terms and
 trilinear couplings is given schematically as

 \bea  W_{NMSGUT}&=& m 210^2 +  M_{H}
 10^{2}+
 \lambda 210^3 + M 126\cdot \oot
 + \eta  210 \cdot  126 \cdot \oot \nnu &+&
 10\cdot  210 (\gamma 126 +\bar\gamma  \oot)
 +{M_{\Theta}} 120\cdot 120  +  {k} 10\cdot 120\cdot
210\nonumber\\  &+&  {\rho}   120\cdot 120\cdot 210  +  { \zeta}
  120\cdot 126\cdot 210 + {
\bar{\zeta}}  120\cdot  { {\oot}} \cdot 210  \nonumber\\& +&h_{AB}
16_A \cdot 16_B.10 + f_{AB} 16_A \cdot 16_B.\bar{126}
 + g_{AB}16_A \cdot
16_B.120 \eea Details of tensor contractions can be found in
\cite{ag2,nmsgut}.
 Here  $h_{AB},f_{AB}$ are symmetric matrices of
Yukawa couplings of the $\mathbf{10,\oot}$ Higgs multiplets to the
$\mathbf{16_A .16_B} $ matter bilinear and $g_{AB}$ is
antisymmetric matrix  for the coupling of newly added Higgs
representation ${\bf{120}}$ to
 $\mathbf{16_A .16_B} $. One of the complex symmetric matrices can be
made real and diagonal by a choice of SO(10) flavor basis. The
complex Yukawas contain 3 real and 9 complex i.e 21 real
parameters.

   The symmetry breaking scheme of NMSGUT is same as MSGUT. If one looks at the Pati- Salam decomposition of
   ${\bf 120}$ given below, it doesn't contain any SM singlet.  So the 120-plet doesn't take part in GUT scale symmetry breaking
   where $SU(2)_L \times U(1)_Y$ is unbroken.
    \bea \Theta_{ijk}(120)= \Theta_{\mu\nu}^{(s)}(10,1,1)+  \Theta^{\mu\nu}_{(s)}(\overline{10},1,1)+
    \Theta_{\nu \alpha \dot{\alpha}}^{\mu}(15,2,2)\\\nonumber
     \Theta_{\mu\nu\dot{\alpha}\dot{\beta}}^{(a)}(6,1,3)+ \Theta_{\mu\nu\alpha\beta}^{(a)}(6,3,1)+
     \Theta_{\alpha\dot{\alpha}}(1,2,2)\eea
The GUT scale vevs that break
     the SO(10) gauge symmetry (one step symmetry breaking) down to SM symmetry are \cite{aulmoh,ckn,abmrs}
     \bea <(15,1,1)>_{210} \,\, : \,\, <\phi_{abcd}> &=& \frac{a}{2}
      \epsilon_{abcdef} \epsilon_{ef}\\
      <(15,1,3)>_{210} \,\, : \,\, <\phi_{ab \tilde{\alpha}\tilde{\beta}}>
      &=&
      \omega \epsilon_{ab} \epsilon_{\tilde{\alpha}\tilde{\beta}}\\
      <(1,1,1)>_{210} \,\, : \,\, <\phi_{\tilde{\alpha}\tilde{\beta}\tilde{\gamma}\tilde{\delta}}>
      &=&
      p  \epsilon_{\tilde{\alpha}\tilde{\beta}\tilde{\gamma}\tilde{\delta}}\\
<(10,1,3)>_{\overline{126}} \,\, : \,\,
<\overline{\Sigma}_{\hat{1}\hat{3}\hat{5}\hat{8}\hat{0}}>
&=& \overline{\sigma}\\
<(\overline{10},1,3)>_{126} \,\, : \,\,
<\Sigma_{\hat{2}\hat{4}\hat{6}\hat{7}\hat{9}}> &=& \sigma
      \eea
 For preservation of Susy, vanishing of D-terms gives the condition
      $|\sigma|=|\overline{\sigma}|$.

An important simplification of this theory is that the GUT scale
vevs and therefore the mass spectrum are all expressible in terms
of a single complex parameter $x$ which is a solution of the cubic
equation \cite{abmsv}. \be 8 x^3 - 15 x^2 + 14 x -3 = -\xi (1-x)^2
\label{cubic} \ee where  $\xi ={{ \lambda M}\over {\eta m}} $.
Then the dimensionless vevs (denoted by tilde) in units of
($\frac{m}{\lambda}$) are $\tilde{\omega}$ = -$x$ and \be
\tilde{a}= \frac{x^2+2 x -1}{1-x}\,\,;\,\, \tilde{p}=\frac{x(5
x^2-1)}{(1-x)^2}\,\,;\,\,
\tilde{\sigma}\tilde{\overline{\sigma}}=\frac{2 \lambda x (1-3 x
)(1+x^2)}{\eta (1-x)^2} \ee
Once one has all the superheavy vevs
in terms of this parameter $x$ the superheavy mass spectrum which
results due to the SO(10) symmetry breaking and responsible for
all the GUT threshold effects can be calculated
\cite{bmsv,ag1,ag2}.

\section{NMSGUT spectrum}
 The technique \cite{ag1,ag2} to find the heavy GUT spectrum
 is to decompose the SO(10) invariants of NMSGUT superpotential to
 Pati-Salam invariants. Then substituting the GUT
 vevs, one gets the superpotential in the MSSM vacuum in terms of
 MSSM invariants. The results for MSGUT are given in
 \cite{ag1,ag2} and for those extra terms corresponding to ${\bf
 120}$ can be found in \cite{nmsgut}. These heavy fields fall into
  26 MSSM irreducible representation (irreps) types which are named
 after 26 letters of alphabet \cite{ag2,nmsgut}.

{\bf Gauge/gaugino masses}: The massive gauge boson together with
its longitudinal mode (Goldstone scalar)
 and real scalar partner of longitudinal mode as four
 bosonic degrees of freedom forms a massive supermultiplet. Its gaugino and the chiral fermion
 super-partner of Goldstone scalar pair form one Dirac fermion superpartner (with four degrees of freedom).
 These together form the so called massive vector gauge supermultiplet. Total 33 massive gauge/gauginos will
 appear due to symmetry breaking to MSSM. These gauge/gauginos lie in the PS representation (6,2,2)$\oplus$(1,1,3)
 and in addition triplets and anti-triplets in (15,1,1). These obtain mass by pairing with the AM Higgs fermion and their
  mass can be obtained by putting vevs of AM Higgs in the PS decomposition of gaugino Yukawa term given as:
  \be g\sqrt{2}[\frac{1}{3!} <\tilde{\Phi}^*_{ijkl}> \lambda_{im}\Phi^{F}_{mjkl}
  +\frac{1}{2.
  4!}(<\tilde{\Sigma}^*_{ijkln}>\lambda_{im}\Sigma^{F}_{mjkln}
  +<\tilde{\overline\Sigma}^*_{ijkln}>
  \lambda_{im}\overline\Sigma^{F}_{mjkln})]\ee
For example the $\overline{E}[\overline{3},2,-\frac{1}{3}]$
$\oplus$ $E[3,2,\frac{1}{3}]$ gauge field with mass $
m_{\lambda_E}$= $g
\sqrt{(4|a-\omega|^2+2|\omega-p|^2+2|\sigma|^2)}$. The complete
mass matrix for E is $6\times 6$. The expression for full matrix
is given in Eqn.(\ref{Ematrix}).

{\bf AM chiral masses}: The masses of chiral supermultiplets are
calculated by putting AM Higgs vevs in the PS decomposition of the
quadratic and trilinear terms in $W_{AM}$ \cite{ag1,ag2,nmsgut}.
There
 are three types of mass terms involving fermions from chiral supermultiplets:
\begin{itemize}
\item {\bf Unmixed chiral}: They don't mix with other massive chiral or gauge/gaugino fields.
 If the conjugate of chiral field exists then they form a Dirac field else the fermions have Majorana masses.
 In NMSGUT there are 11 ($A[1,1,4]$, $B[6,2,\frac{5}{3}]$, $I[3,1,\frac{10}{3}]$, $M[6,1,\frac{8}{3}]$,
$N[6,1,\frac{-4}{3}]$, $O[1,3,-2]$, $U[3,3,\frac{4}{3}]$,
  $V[1,2,-3]$, $W[6,3,\frac{2}{3}]$, $Y[6,2,\frac{-1}{3}]$, $Z[8,1,2]$)
  Dirac supermultiplets and 2 Majorana supermultiplets
  (S[1,3,0],Q[8,3,0]). Here square brackets contain $SU(3)_C \times SU(2)_L \times U(1)_{Y}$
  quantum number for respective fields, for example
    \bea O[1,3,-2] = \frac{\overrightarrow{\Sigma}_{44(L)}}{\sqrt{2}}
    \quad\quad \bar{O}[1,3,2]= \frac{\overrightarrow{\bar{\Sigma}}_{(L)}^{44}}{\sqrt{2}}\eea
    form a Dirac supermultiplet with mass 2$(M+\eta(3a-p))$.

 \item {\bf Mixed pure chiral}: They mix with each other but not with massive gauge/gaugino fields.
 There are 8 ($R[8,1,0]$, $h[1,2,1]$, $t[3,1,\frac{2}{3}]$, $C[8,2,1]$, $D[3,2,\frac{7}{3}]$, $K[3,1,\frac{8}{3}]$, $L[6,1,
 \frac{2}{3}]$, $P[3,3,\frac{2}{3}]$) fields of this type. Here we give the
example of most important pure chiral field i.e NMSGUT Higgs field,

\bea
[1,2,-1](\bar{h}_1,\bar{h}_2,\bar{h}_3,\bar{h}_4,\bar{h}_5,\bar{h}_6)\oplus
[1,2,1](h_1,h_2,h_3,h_4,h_5,h_6)&\equiv&(H^{\alpha }_{\dot{2}},
\bar\Sigma_{\dot{2}}^{(15)\alpha},\nonumber\\\Sigma_{\dot{2}}^{(15)\alpha},
\frac{\Phi_{44}^{\dot{2}\alpha}}{\sqrt{2}},O^{\alpha}_{\dot{2}},
O_{\dot{2}}^{(15)\alpha}\hspace{2mm}) \oplus (H_{\alpha
\dot{1}},\bar\Sigma_{\alpha \dot{1}}^{(15)},\Sigma_{\alpha
\dot{1}}^{(15)},\frac{\Phi_{\alpha}^{44\dot{1}}}{\sqrt{2}},
O_{\alpha\dot{1}},O_{\alpha\dot{1}}^{(15)}) \eea In NMSGUT the
Higgs mass matrix ${\cal H}$ is $6\times 6 $, two additional rows
and columns (relative to the MSGUT) come from the 120-plet.

\be {\scriptsize\begin{pmatrix} -M_H &
\bar{\gamma}\sqrt{3}(\omega-a) & -\gamma\sqrt{3}(\omega + a)&
-\bar{\gamma}\bar{\sigma}&kp & -\sqrt{3}ik\omega \cr
 -\bar{\gamma}\sqrt{3}(\omega+ a)& 0 & -(2M + 4\eta(a+ \omega))&0 &
 -\sqrt{3}\bar{\zeta}\omega & i(p+2\omega)\bar{\zeta}\cr
\gamma\sqrt{3}(\omega-a) & -(2M + 4\eta(a- \omega))&0 & -2\eta
\bar{\sigma}\sqrt{3}& \sqrt{3}\zeta\omega& -i(p-2\omega)\zeta\cr
-\sigma\gamma & -2\eta\sigma\sqrt{3}&0 & -2m + 6\lambda(\omega-a)&
\zeta\sigma & \sqrt{3}i\zeta\sigma\cr
pk& \sqrt{3}\bar{\zeta}\omega& -\sqrt{3}\omega\zeta&
\bar{\zeta}\bar{\sigma}& -m_{\Theta}&
\frac{\rho}{\sqrt{3}}i\omega\cr
\sqrt{3}ik\omega&i(p-2\omega)\bar{\zeta}&
 -i(p+2\omega)\zeta& -\sqrt{3}i\bar{\zeta}\bar{\sigma}&
  -\frac{\rho}{\sqrt{3}}i\omega& -m_{\Theta} - \frac{2\rho}{3}a \end{pmatrix}}
  \label{hmatrix}\ee
To get the masses, the above matrix is diagonalized after imposing the fine
tuning condition $Det {\cal H} =0$ (details in section 2.5).\\

 \item {\bf Mixed chiral-gauge}: These mix with gauge fields as well as among themselves.
 There are 5 such types of fields $G[1,1,0]$ and $X[3,2,\frac{4}{3}]$, $E[3,2,\frac{1}{3}], F[1,1,
2], J[3,1,\frac{4}{3}]$. For example
\end{itemize}

\bea [\bar 3,2,-{1\over 3}](\bar E_1, \bar E_2,\bar E_3,\bar
E_4,\bar E_5,\bar E_6) \oplus [3,2,{1\over
3}](E_1,E_2,E_3,E_4,E_5,E_6)&\equiv& (\Sigma_{4 \dot
1}^{\bar\mu\alpha}, \Sigb_{4\dot 1}^{\bar\mu \alpha},\nonumber\\
\phi^{\bar\mu 4\alpha}_{(s)\dot 2} , \phi^{(a) \bar\mu
4\alpha}_{\dot 2},\lambda^{\bar\mu 4\alpha}_{\dot 2},\Theta_{4
\dot{1}}^{\bar\sigma\alpha}) \oplus  (\bar\Sigma_{\bar\mu \alpha
\dot{2}}^{4}\s_{\bar\mu\alpha\dot 2}^4,\phi_{\bar\mu 4\alpha\dot
1}^{(s)}, \phi_{\bar\mu 4\alpha\dot
1}^{(a)},\lambda_{\bar\mu\alpha\dot
1},\Theta_{\bar\sigma\alpha}^{4 \dot{1}}) \eea
 The 6$\times$ 6 mass matrix ${\cal E}$ is
\be {\scriptsize
\begin{pmatrix}-2(M+\e(a-\omega))&0&0&0&0&(i \omega -ip + 2ia )\zeta\cr 0&-2(M+\e(a-3\omega))&
-2\sqrt{2} i\eta\sigma&2i\eta\sigma&ig\sqrt{2}\bar\sigma^*& (-3 i\omega +ip + 2ia )\bar\zeta\cr
0&2i\sqrt{2}\eta\bar\sigma&-2(m+\lambda(a-\omega))&-2\sqrt{2}\lambda\omega&2g(a^*-\omega^*)& -\sqrt{2}\bar{\zeta}\bar{\sigma}\cr
0&-2i\eta\bar\sigma&-2\sqrt{2}\lambda\omega&-2(m-\lambda\omega)&\sqrt{2} g(\omega^*-p^*)& \bar{\sigma}\bar{\zeta}\cr
0&-ig\sqrt{2}\sigma^*&2g(a^*-\omega^*)&g\sqrt{2}(\omega^*-p^*)&0&0\cr
( -i\omega +ip - 2ia )\bar\zeta&(3i\omega -ip - 2ia )\zeta&
-\sqrt{2} \zeta \sigma & \sigma\zeta&0& -(m_{\Theta}
+\frac{\rho}{3}a - \frac{2}{3}\rho \omega)\cr
\end{pmatrix}}\label{Ematrix}\ee

The mass matrix $\cal E$
  has the usual super-Higgs structure : complex conjugates of
 the 5th row and column after omitting the diagonal entry,
 provides left and right null eigenvectors
of the  chiral $5 \times 5 $ submatrix ${\bf E}$ obtained by
omitting the fifth row and column.
 ${\cal E}$ has non zero determinant although the determinant of ${\bf{E}}$ vanishes.

\section{Importance of NMSGUT spectrum}
   Due to these super heavy
 fields present in the theory there are corrections at one loop to
 unification scale $M_{X}$, the QCD
coupling at $M_Z$ ($\alpha_3(M_Z)$) and the gauge coupling at
unification scale ($\alpha_G$) as well as to fermion Yukawa
couplings.
 We discuss here about the corrections to $\alpha_{G},\alpha_{3}(M_{Z}),Log_{10}(M_{X})$ and
 corrections to fermion Yukawa couplings in the next chapter. Here $M_X$ is taken to be the mass of
  lightest gauge multiplet which mediates the
 proton decay and not a scale ($M_X^0 = 10^{16.33} GeV$) where the three MSSM gauge couplings meet(since in general
 they do not at a point: due to threshold effects etc).

 The threshold corrections due to heavy spectra on the matching relations
 of gauge couplings of low scale ($M_Z$=91.1876 GeV) and GUT scale
 ($M_X$) are done using the technique of Weinberg \cite{weinberg} and Hall \cite{hall}.
 The relation between two scale gauge couplings is given as:
 \bea \frac{1}{\hat{\alpha}_i(M_Z)}=\frac{1}{\alpha_G
 (M_X)}+8\pi b_i ln\frac{M_X}{M_Z}+4\pi
 \sum_{j}\frac{b_{ij}}{b_j}ln X_j-4\pi\lambda_i(M_X)+...
\eea Here $\hat{\alpha}_i=g_i^2/4\pi $
 are MSSM gauge coupling in SU(5)GUT normalization. Second and third term corresponds to one and two loop gauge
 running with $X_j=1+8\pi b_j \alpha_G(M_X^0) ln(M_{X}^0/M_S)$. $b_i$ and $b_{ij}$ are coefficients of one and two loop
 gauge beta functions of MSSM \cite{martinvaughn}. $\lambda_i$ is
 the leading one loop contribution of superheavy thresholds given
 as \cite{weinberg, hall} in $\overline{\text{MS}}$ scheme:
 \bea \lambda_i(\mu) =
 -\frac{2}{21}(b_{iV}+b_{iGB})+2(b_{iV}+b_{iGB})ln\frac{M_V}{\mu}+2
 b_{iS}ln\frac{M_V}{\mu}+2 b_{iF}ln\frac{M_F}{\mu}\eea
 Where V, GB, S, F corresponds to vectors, Goldstone boson, scalar and fermions respectively.
 The formulas for threshold corrections are given \cite{ag2,gmblm,langpol}
as: \begin{align} \Delta^{(th)}(ln{M_X}) &={{\lambda_1(M_X) -
\lambda_2(M_X) } \over{2(b_1 - b_2)}} \nonumber \\
 \Delta_X &\equiv\Delta^{(th)}(Log_{10}{M_X}) = .0222 +0.4873({{\bar b}'}_1 -{{\bar b}'}_2 )
 Log_{10}{{M'}\over  {M_X}} \nonumber \\
 \Delta_3&\equiv\Delta^{(th)} (\alpha_3 (M_Z))\nonumber\\&=
 {{100 \pi (b_1-b_2)\alpha(M_Z)^2}\over{[(5b_1+3b_2-8b_3)sin^2\theta_w(M_Z)-3(b_2-b_3)]^2}}
 \sum_{ijk}\epsilon_{ijk}(b_i-b_j)\lambda_k(M_X)\nonumber\\&
= .000311667 \sum_{M'} (5 {{\bar b}'}_1
 -12{{\bar b}'}_2 +7{{\bar b}'}_3) Log_{10}{{M'}\over{M_X
 }} \nonumber \end{align}
 \bea \Delta_G\equiv\Delta^{(th)}(\alpha_G^{-1}(M_X)) &=& \frac{4 \pi(b_1
\lambda_2(M_X)-b_2 \lambda_1(M_X))}{b_1-b_2}\nonumber \\
&=&-1.27+  0.1786\sum_{M'}( 6.6 {{\bar b}'}_2 - {{\bar b}'}_1)
Log_{10}{{M'}\over {M_X }}
   \eea
Where ${\bar b'}_i = 16\pi^2 b_i'  $ are  1-loop $\beta$ function
coefficients for different superheavy multiplets having mass $M'$.

  To satisfy the perturbativity conditions the following limits
were imposed on corrections: \bea
-20.0\leq \Delta_G &\equiv&  \Delta  (\alpha_G^{-1}(M_X))  \leq 25 \nonumber \\
3.0 \geq  \Delta_X &\equiv &\Delta (Log_{10}{M_X}) \geq - 0.03\nonumber \\
-.017< \Delta_{3} &\equiv & \Delta\alpha_3(M_Z)  <
-.004\label{criteria} \eea
An important point to mention is that
the
 corrections depend only on ratios of masses and are independent of overall
 scale parameter which is chosen to be $m$ (mass of {\bf 210} plet).
 The unification scale is defined as $ M_X = M_X^0 10^{\Delta_X}$ and $M_X$=
 $m_{\lambda_{X}}$=$|m/\lambda|$ $g_{10}$ $\sqrt{4|\tilde{a}+\tilde{\omega}|^2+2
|\tilde{p}+\tilde{\omega}|^2}$.  The overall scale $|m|$ is then
calculated by the RG analysis and given by \bea
|m_{RG}|&=&10^{16.33+\Delta_
X}\frac{|\lambda|}{g_{10}\sqrt{4|\tilde{a}+\tilde{\omega}|^2+2
|\tilde{p}+\tilde{\omega}|^2}} GeV \label{uniscale}\eea Here
$g_{10}$ is the SO(10) gauge coupling (threshold corrected).

\section{Fermion mass formulae}
The $6 \times 6 $ mass matrix \cal{H} for MSSM Higgs type doublets is most
important for fixing the tree level formulae for Yukawa couplings. The  fine tuning condition
$Det {\cal H} =0$  is needed to keep a pair of Higgs doublets
$H_{(1)}, {\bar H}_{(1)}$ (defined via left and right  null
eigenstates of the mass matrix ${\cal H}$) light. The composition
of these null eigenstates in terms of the GUT scale doublets is
specified by the so called ``Higgs fractions''
$\alpha_i,{\bar\alpha}_i $. The ``Higgs fractions'' are important
as they determine how much different original GUT scale doublets
will contribute to the EW symmetry breaking. They may be derived
from the bi-unitary transformation ($h_i= U_{ij} H_j$, $\bar{h}_i= \bar{U}_{ij} \bar{H}_j$)
 that diagonalizes $\cal H$:
$\overline{U}^T {\cal H} U $ = $\Lambda_h$ then $U_{i1}$ =
$\alpha_i$, $\overline{U}_{i1}$=$\overline{\alpha}_i$.

To get the Dirac masses of fermion replace $<h_i>\rightarrow
\alpha_i v_u, <\bar h_i>\rightarrow \bar\alpha_i v_d$ in Dirac
mass matrices, where $v_{u,d}$ are the vevs of the light MSSM
doublets $H_{1},\bar H_{1}$. The fermion Dirac masses can be
calculated from the
 decomposition of $\bf{16\cdot16\cdot (10 \oplus 120\oplus \oot)}$ given
 in \cite{ag1,ag2,blmdm}  and this yields \cite{nmsgut}:
\bea { y}^u &=&  ( {\hat  h} + {\hat  f} + {\hat  g} )\quad ;\quad
{ y}^d = {({\hat {r}}_1} {\hat  h} + { {\hat {r}}_2} {\hat  f}  +
{\hat {r}}_6 {\hat  g})\\\nonumber { y}^{\nu}&=& ({\hat  h} -3
{\hat  f}  + ({\hat {r}}_5 -3) {\hat {g}})\quad ;\quad { y}^l ={ (
{\hat {r}}_1} {\hat  h} - 3 {  {\hat {r}}_2} {\hat  f} +
   ( {\hat {\bar{r}}_5} -
   3{\hat {r}}_6){\hat  g})\\\nonumber
{\hat {r}}_1&=& \frac{ \bar\alpha_1}{\alpha_1};\quad {\hat
{r}}_2= \frac{ \bar\alpha_2}{\alpha_2} ;\quad
   {{\hat {r}}_5}= \frac{4 i \sqrt{3}{\alpha_5}}{\alpha_6+ i
   \sqrt{3}\alpha_5}  \nnu
 {\hat {r}}_6 &=&
\frac{{{\bar{\alpha}}_6}+ i \sqrt{3}{{\bar{\alpha}}_5}}{\alpha_6+
i \sqrt{3}\alpha_5}  ;\quad {\hat {\bar{r}}_5}=
\frac{4 i \sqrt{3}{{\bar{\alpha}}_5}}{\alpha_6+ i
   \sqrt{3}\alpha_5}\\\nonumber
{\hat  g} &=&2i g {\sqrt{\frac{2}{3}}}(\alpha_6 + i\sqrt \alpha_5)
\quad;\quad \hat  h = 2 {\sqrt{2}} h \alpha_1 \quad;\quad\hat  f =
-4 {\sqrt{\frac{2}{3}}} i f\alpha_2
 \eea
The Yukawa couplings of matter fields with \textbf{120} Higgs
field included give no additional contribution to the  Majorana
mass matrix
 of the superheavy neutrinos $\bar\nu_A$  so it remains same as MSGUT.
 \bea M^{\bar\nu}_{AB}=  8 {\sqrt{2}} f_{AB} {\bar\sigma}\eea
 The Type I contribution is obtained by eliminating $\bar\nu_A$

\bea W &=& {\frac{1}{2}}  M^{\bar\nu}_{AB} \bar\nu_A\bar\nu_B  +
\bar\nu_A m^{\nu}_{AB} \nu_B  + .....\rightarrow {\frac{1}{2}}
M^{\nu (I)}_{AB} \nu_A\nu_B  + ....\nnu M^{\nu(I)}_{AB} &=&
-((m^{\nu})^T (M^{\bar\nu})^{-1} m^{\nu})_{AB}  \eea
From the Type
II seesaw  (the ${\bf{120}}$ plet  contributes new terms)  one
obtains \cite{ag2,nmsgut} contribution to the light neutrino
Majorana mass :

\be M_{\nu}^{II}=16 i f_{AB}<{\bar{O}}_- > = 16 i f_{AB}({
 i{\frac{\gamma}{\sqrt{2}}}}   \alpha_1 + {i\sqrt{6}}\eta \alpha_2 - {\sqrt{3}}\zeta \alpha_6 + i
\zeta \alpha_5) \alpha_4 ({{v_u^2}\over {M_{\Theta}}}) \ee
where
$M_{\Theta}= 2 (M + \eta (3a-p))$. NMSGUT considers Type I
  and Type II mechanisms together. Then one does not
  have any freedom to switch one of them off at will.

\section{SM fermion fitting criteria in NMSGUT and its features}
Fitting to SM fermion data is done by $\chi^2$ analysis. Fitting
is done at two scales, $M_{X}^0$ and $M_{Z}$. The two scale
fitting process is looped via 2-loop MSSM RGEs
\cite{martinvaughn}.

 {\bf $M_{X}^0$ scale fitting}:  This requires us to find a set of superpotential parameters of
 NMSGUT such that the fermion Yukawas calculated from these parameters match the experimental data evolved to $M_{X}^0$.
 The fitting at GUT scale is done in terms of 38 free superpotential parameters of NMSGUT.
 It includes 21 real parameters from the fermion Yukawa sector ($h_{AA}(3), f_{AB}(12), g_{AB}(6)$).
 One of the symmetric matrices is made real diagonal using the freedom to make U(3) rotations on the 16-plet kinetic terms
  (here $h_{AB}$ is chosen to be real diagonal).
 24 parameters come from scalar trilinear and quadratic terms($\eta$, $\rho$, $\kappa$, $\gamma$, $\bar\gamma$, $\zeta$,
 $\bar\zeta$, $\lambda$, m, M, $\tilde {m_{\Theta}}$, $M_H$). Using phase redefinitions of 5 Higgs fields the number of
 parameters reduces to 24-5=19. $M_H$ is kept fixed by fine tuning condition. m is also fixed by equation (\ref{uniscale}).
 So there are basically 37 free parameters. These parameters will determine the MSSM Yukawa and neutrino masses at $M_{X}$
  scale with the formula given in section 4.
  Then 2-loop RGEs extrapolate central values of SM data (it includes fermion Yukawas, CKM mixing angles, neutrino mass
   splitting and mixing data, total 18 parameters) to MSSM one loop corrected unification scale $M_{X}^0=10^{16.33}$ GeV
   (corrections from Right handed neutrino and Susy particles are not considered). The $\chi^2$ function at $M_{X}^0$
   is given as
 \bea  \chi^2= \sum_{i=1}^{18}\frac{(O_i-\bar{O_i})^2}{\delta O_i^2}\eea
 Here $O_i$ are calculated from NMSGUT parameters, $\bar{O_i}$ are the target values and
 $\delta O_i$ are the uncertainties in the target values based on extrapolating the fermion data uncertainties from the weak scale
  to GUT scale \cite{antuschspinrath}. The downhill simplex method (AMOEBA subroutine)
  given by Nelder and Mead \cite{pressteukolsky} is used to do the $\chi^2$ fit. The 37 parameters are thrown at random to
  start, then the subroutine AMOEBA search in 37 dimensional parameter space to find best fit.

Features of GUT scale fitting:
\begin{itemize}
\item Regions of NMSGUT parameter space exist where a fit for 18
SM parameters with $\chi < .1$ can be achieved.
  \item  It successfully fits the Neutrino masses and large mixing angles.
  \item  The tiny values of the $\oot$ couplings results in three righthanded neutrino masses
  to be much lighter than $M_X$ and having the range $\sim 10^8-10^{13} GeV$, required for Leptogenesis.
\item  The most striking feature of NMSGUT fitting at $M_X$ is in
fitting to down and strange quark. NMSGUT like other GUTs require
large tan$\beta$ $\approx$ 50 for t-b-$\tau$ Yukawa unification.
The price to achieve that is that the down and strange quark
Yukawa couplings come out to be 4-5 times
 smaller than the SM required central values at $M_{Z}$.
 However instead of taking this as a no-go \cite{grimuslavoura} it was reinterpreted
 \cite{nmsgut,nmsgutIII} as necessity of Susy threshold corrections to d,s quark Yukawa
couplings in large tan$\beta$ region which are to be just what is
required to allow the matching.
\end{itemize}

{\bf $M_{Z}$ scale fitting}: The best fit found at $M_X$ scale
along with five SUGRY-NUHM (Supergravity-Non universal Higgs mass)
type soft Susy breaking parameters ($m_0$ (universal scalar mass),
$m_{\frac{1}{2}}$ (universal gaugino mass), $A_0$ (universal
trilinear coupling), $m_{H^2,\bar{H}^2}$) is run down to $M_Z$
scale using two loop MSSM RGEs \cite{martinvaughn}. The $\chi^2$
function at this scale is defined as: \bea \chi^2_{Susy}= \sum_{i}
(1-\frac{y'^{MSSM}_{i}}{y^{SM}_i})^2\eea Here
$y'^{MSSM}_i=y^{GUT}_i [Sin\beta({\bf cos\beta})(1+\eta_i)]$ is
the susy threshold corrected Yukawa for up({\bf down}) type
fermions. $y^{GUT}_i$ is the value of the Yukawa coupling from the
GUT fit run down to $M_{Z}$.
 $\eta_i$ is the correction factor. The
corrections include gluino, chargino \cite{piercebaggar} and bino
\cite{antuschspinrath} loop corrections. The corrections are
applied in the absence of generation mixing. The leading
contribution of Susy correction to top quark comes from the gluon
and squark-gluino loop. In case of down type quarks, bottom quark
Yukawa coupling receive corrections from gluino  due to large
tan$\beta$. However it is cancelled or even dominated by chargino
corrections due to large $A_{0} y_t$. However for down and strange
quark the gluino corrections (Yukawa is small for first two
generations) reduce the SM Yukawa required to match with run down
values of NMSGUT Yukawas. In case of charged leptons there are no
gluino corrections. The fitting is again done throwing randomly
the Susy breaking parameters at GUT scale (running them down to
$M_Z$ scale) using downhill simplex method. The s-particle spectra
is calculated at tree level using subroutines from SPHENO
\cite{spheno}. Then Susy threshold corrections are calculated and
applied to run down values of GUT fermion Yukawas to compare with
SM Yukawas.

Features of $M_Z$ scale fitting and Susy Spectra:
\begin{itemize}
\item The most striking feature of $M_Z$ scale fitting is
requirement of large $\mu, A_0$ parameters required to have large
Susy threshold corrections to fit down and strange quark.

 \item
Large and negative value of $m^2_{H,\bar H}$ ensures that the
third s-generation is heavier than first two independently from
the value of $A_0$. Thus NMSGUT predicts normal s-hierarchy
opposite to most of other GUTs which predicts inverted
s-hierarchy.
 \item Pure Bino is the lightest supersymmetric particle (LSP)
 which can be a suitable candidate for dark matter. Also in some cases NMSGUT
 provides light smuon close to the mass of Bino in some cases to
 provide co-annihilation channel as well as allow explanation of muon g-2 anomaly.
\item An interesting feature is that the ratio of gaugino masses
diverges significantly from the ratio 1:2:7 :: $M_1:M_2:M_3$
expected from the one loop running of the gauge couplings. \item
S-particles are in range 2 TeV-50 TeV, gauginos $\sim$ 0.1 -0.2
TeV, Higgsinos $\sim$ 100 TeV and in some cases smuon can be as
light as a few hundred GeV.
 \item The MSSM Higgs mass is
calculated at 1-loop level (corrections from stop and sbottom
quarks) and able to achieve mass value 124-126 GeV as dictated by
experiments. Also the large value of $A_0, \mu$ ( O(100 TeV))
parameter suggest large mass for light Higgs due to loop
corrections.  This feature of NMSGUT is known from 2008
\cite{nmsgut}, well before discovery of Higgs in 2012 made it a
commonly acceptable assumption. Also these large values are quite
natural for a decoupled/Mini Split type Susy s-spectra
\cite{Decoupled,minisplit}.
\end{itemize}

 \section{Dimension five operators for B,L violation}
Quarks and Leptons are grouped together in the same irreducible
representation in Grand Unified theories. This leads to conversion
of quarks into leptons e.g. proton decay via exchange of heavy
gauge or Higgs fields. However non observation of the decay of
proton put stringent constraints on Susy GUTs. The experiment at
Super Kamiokande water Cherenkov radiation detector has given
lower bound on the decay rate of proton to kaon and positron
channel as \cite{kamiokande} \be \Gamma_{p \rightarrow K^+
\bar{\nu}} > (3.3 \times 10^{33} \text{yrs})^{-1}; \,\,\,\,\,\,\,
\Gamma_{p \rightarrow e^+ \pi^0} > (1.0 \times 10^{34}
\text{yrs})^{-1}\ee In Non Susy GUTs the leading contribution to B
violation comes from d=6 operators. The decay rate calculated from
d=6 B violation operators is proportional to $\frac{1}{M_X^2}$,
here $M_X$ is mass of exchanged particle. In Non-Susy theories
$M_X$ is of O($10^{15}$ GeV)  and in Susy theories it is O(
$10^{16}$ GeV). So the life time calculated from the d=6 operators
comes out to be O($10^{36}$ yrs) or more.

 However in Susy GUTs the d=5 and 4, B violation operator can also
 contribute. The d=4 (renormalizable) operators only arise in R-Parity violating theories.
 Since NMSGUT preserves R-parity so they are not relevant here.
 The d=5 operators involve two fermions and two sfermions with
 exchange of a color triplet Higgsino.
   In NMSGUT the effective superpotential for $B+L$
violating processes (although B-L preserving as it is a part of
gauge symmetry) due to exchange of color triplet superheavy chiral
supermultiplets contained in the $\mathbf{10,\oot,120}$ Higgs
multiplets are given in \cite{ag1,ag2} for $\mathbf{10,\oot}$ and
\cite{nmsgut} for $\mathbf{120}$. The complete d=4 superpotential
in terms of MSSM decompositions is: \bea
  W_{FM}&=& 2\sqrt{2}h_{AB}[\bar{t}_1 (\epsilon{\bar{ u}}_A { d}_{B } +
Q_A { L}_{B }) + t_1(\frac{\epsilon}{2}Q_A { Q}_{ B } + \bar{ u}_A
\bar{
e}_{B}-\bar{d}_A\bar{\nu}_B)]\nonumber\\&&-2\sqrt{2}h_{AB}[\bar{h}_1
({\bar{ d}}_A { Q}_{B } + \bar{ e}_A { L}_{B }) + h_1(\bar{{ u}}_A
{ Q}_{ B } + \bar{ \nu}_A { L}_{B})]\nonumber\\&&
+4\sqrt{2}f_{AB}[t_2 (\frac{\epsilon}{2}Q_A { Q}_{ B }- \bar{ u}_A
\bar{e}_{B }+\bar{\nu}_A\bar{d}_B) + \bar{t}_2(Q_A { L}_{ B }
-\epsilon \bar{ u}_A
\bar{d}_{B})]\nonumber\\&&+4\sqrt{2}f_{AB}[\frac{i}{\sqrt{3}}\{\bar{h}_2
(\bar{d}_A { Q}_{ B }- 3\bar{e}_A L_{B }) + h_2(\bar{u}_A {Q}_{ B
} -3\bar{ \nu}_A L_{B})\}\nonumber\\&&+2(\bar{C}_1\bar{d}_A
Q_B-C_2\bar{u}_A Q_B)+2(E_1\bar{d}_A L_B-D_2\bar{u}_A
L_B)\nonumber\\&&+2(\bar{D}_2\bar{e}_A Q_B-\bar{E}_2 \bar{\nu}_A
Q_B)]\nonumber\\&&
+4f_{AB}[\bar{\Sigma}^{\dot{\alpha}\dot{\beta}}\bar{Q}_{A\dot{\alpha}}\bar{Q}_{B\dot{\beta}}+
2i(\bar{A}\bar{e}_A\bar{B}-G_5
\bar{\nu}_A\bar{\nu}_B)-2\sqrt{2}i\bar{F}_1\bar{e}_A\bar{\nu}_B\nonumber\\&&
+(\bar{W}Q_A Q_B+2\bar{P}Q_A L_B+\sqrt{2}\bar{O}L_A
L_B)\nonumber\\&&-2i
t_4(\bar{d}_A\bar{\nu}_B+\bar{u_A}\bar{e}_B)+2\sqrt{2}i(K
\bar{d}_A\bar{e}_B-J_1\bar{u}_A \bar{\nu}_B)]\nonumber\\
&&+2\sqrt{2}g_{AB}[\bar{h}_5 ({\bar{ d}}_A { Q}_{B } + \bar{ e}_A
{ L}_{B }) - h_5(\bar{{ u}}_A { Q}_{ B } + \bar{ \nu}_A { L}_{B})]
\nonumber\\ & &  -2\sqrt{2} g_{AB}[{\sqrt{2}} \bar{{ L}}_2 {Q}_A {
Q}_{B} + F_4 { L}_A { L}_{ B} + \sqrt{2}
\bar{t}_6 { Q}_A { L}_{ B} \nonumber\\
& &  + 2{\sqrt{2}}{ L}_2 \bar{{ u}}_A \bar{ d}_B +
\sqrt{2}t_6(\bar{{ u}}_A \bar{ e}_B - \bar{ d}_A \bar{ \nu}_B )+
2\bar{F}_4 \bar{ \nu}_A
\bar{ e}_B]\nonumber\\
 & & -2\sqrt{2}g_{AB}[2\bar{C}_3 \bar{ d}_A { Q}_B - 2 C_3 \bar{ u}_A { Q}_B +
\frac{i}{\sqrt{3}}\bar{h}_6( \bar{ d}_A { Q}_B- 3\bar{ e}_A { L}_B
) \nonumber\\&&- \frac{i}{\sqrt{3}}h_6 (\bar{ u}_A { Q}_B -3 \bar{
\nu}_A { L}_B)+2
\bar{D}_3 \bar{ e}_A { Q}_B - 2\bar{E}_6 \bar{ \nu}_A { Q}_B\nonumber\\
& & +2 E_6 \bar{ d}_A { L}_B - 2D_3 \bar{ u}_A { L}_B ]
-2i{\sqrt{2}} g_{AB}[\epsilon \bar{J}_5 \bar{ d}_A \bar{ d}_B
\nonumber\\ && + 2 K_2 \bar{ d}_A \bar{ e}_B - \epsilon \bar{K}_2
\bar{ u}_A \bar{ u}_B - 2 J_5 \bar{ u}_A \bar{ \nu}_B
\nonumber\eea \bea && -{\sqrt{2}} \epsilon  \bar{t}_7 \bar{ d}_A
\bar{ u}_B -{\sqrt{2}} t_7(\bar{ d}_A \bar{ \nu}_B - \bar{ e}_A
\bar{ u}_B)] - 2 g_{AB}[\epsilon P_2 { Q}_A { Q}_B+ 2 \bar{P}_2 {
Q}_A { L}_B ]
 \eea
The  exchange of a Higgsino that couples to matter with a given
$B+L$ will give rise to a $B+L$ violating effective $d=5$ operator
if it has a nonzero contraction with a conjugate (MSSM)
representation Higgsino which couples to a matter chiral bilinear
with a $B+L$ different from the conjugate of the first $B+L$
value. In NMSGUT in addition to familiar color triplets
 $[\bar 3,1,\pm {\frac{2}{3}}]\subset {\bf{120}}$  i.e  $\{ \bar
t_{(6)},\bar t_{(7)}\}[\bar 3,1,{\frac{2}{3}}]$ and
$\{t_{(6)},t_{(7)}\} $ but $P[3,3,\pm{\frac{2}{3}}]$ and
$K[3,1,\pm{\frac{8}{3}}]$ multiplet types also contribute to
baryon violation. Also due to mixing in P($P_1,P_2$), K($K_1,K_2$)
a number of fresh contributions appear which were not there in
MSGUT without ${\bf{120}}$.

After integrating out the heavy triplet Higgs supermultiplets the
effective $d=4$ Superpotential for Baryon Number violating
processes in the NMSGUT \cite{ag2,nmsgut} is given (to leading
order in $m_W/M_X$) by:

\bea  W_{eff}^{\Delta B\neq  0} = -{L}_{ABCD} ({1\over 2}\epsilon
{ Q}_A { Q}_B { Q}_C { L}_D) -{ R}_{ABCD} (\epsilon {\bar{ e}}_A
{\bar{u}}_B { \bar{ u}}_C {\bar{ d}}_D) \eea where the
coefficients are

and \bea R_{ABCD} &=&{\cal S}_1^{~1} {\tilde h}_{AB} {\tilde
h}_{CD}
 - {\cal S}_1^{~2}  {\tilde h}_{AB} {\tilde f}_{CD} -
 {\cal S}_2^{~1}  {\tilde f}_{AB} {\tilde h}_{CD} + {\cal S}_2^{~2}  {\tilde f}_{AB} {\tilde f}_{CD} \nonumber \\
 &-& i{\sqrt 2} {\cal S}_4 ^{~1} {\tilde f}_{AB} {\tilde h}_{CD}
+i {\sqrt 2} {\cal S}_4 ^{~2} {\tilde f}_{AB} {\tilde f}_{CD}
\nnu& +& {\cal S}_6 ^{~1} {\tilde g}_{AB} {\tilde h}_{CD} - i
{\cal S}_7 ^{~1} {\tilde g}_{AB} {\tilde h}_{CD} -  {\cal S}_6
^{~2} {\tilde g}_{AB} {\tilde {\tilde f}}_{CD}+ i   {\cal
S}_7^{~2} {\tilde g}_{AB} {\tilde {\tilde f}}_{CD}
\nonumber \\
&+&   i{\cal S}_1 ^{~7} {\tilde h}_{AB} {\tilde g}_{CD} -i  {\cal
S}_2 ^{~7} {\tilde {\tilde f}}_{AB} {\tilde g}_{CD}+ \sqrt{2}
{\cal S}_4 ^{~7} {\tilde {\tilde f}}_{AB} {\tilde g}_{CD}\nonumber
\\&+&  i {\cal S}_6 ^{~7} {\tilde g}_{AB} {\tilde g}_{CD}  +{\cal S}_7
^{~7} {\tilde g}_{AB} {\tilde g}_{CD}- \sqrt{2} ({\cal
K}^{-1})_1^{~2}
{\tilde {\tilde f}}_{AD}{\tilde g}_{BC}\nonumber\\
&-&  ({\cal K}^{-1})_2^{~2} {\tilde g}_{AD}{\tilde g}_{BC}
 \eea

 \bea  L_{ABCD} &=& {\cal S}_1^{~1} {\tilde h}_{AB} {\tilde h}_{CD}
+ {\cal S}_1^{~2} {\tilde h}_{AB} {\tilde f}_{CD} +
 {\cal S}_2^{~1}  {\tilde f}_{AB} {\tilde h}_{CD} + {\cal S}_2^{~2}  {\tilde f}_{AB} {\tilde
 f}_{CD}\nnu
&-&  {\cal S}_1^{~6}  {\tilde h}_{AB} {\tilde g}_{CD} -
 {\cal S}_2^{~6}  {\tilde f}_{AB} {\tilde g}_{CD}
 +  \sqrt{2}({\cal P}^{-1})_2^{~1} {\tilde g}_{AC}{\tilde f}_{BD}\nonumber\\
 &-&   ({\cal P}^{-1})_2^{~2} {\tilde g}_{AC}{\tilde g}_{BD}
 \eea
here ${\cal S}= {\cal T}^{-1} $ and ${\cal T} $ is the mass matrix
for $[3,1,\pm 2/3]$-sector  triplets :
 $W={\bar t}^i {\cal T}_i^j t_j
+...$, and

 \bea {\tilde h}_{AB} = 2 {\sqrt 2} h_{AB}  \qquad {\tilde f}_{AB} = 4
{\sqrt 2} f_{AB} \qquad  {\tilde g_{AB}} = 4 g_{AB} \eea Using
these formulas the Baryon decay rates are estimated in NMSGUT on
the basis of fits of fermion data. In the next section we will
discuss about fits found in NMSGUT with baryon decay rates $\sim
10^{-28} \text{yrs}^{-1}$ which are faster than experimental value
by six orders of magnitude.

\section{Problem of fast proton decay ($\tau \sim 10^{28}$ yrs )}
In NMSGUT, the raising of unification scale $M_{X}$ near to Planck
scale due to threshold effects leads to suppression of gauge
mediated d=6 Baryon violation operators. A suppression of 4-6
orders of magnitude to $10^{36}$ yrs can be achieved relative to
the non-Susy case. However the d=5 operators are a cause of
concern. The effective dimension=4 operators are dressed with
sparticles, which will eventually yield d=6 effective 4 fermion
operators for baryon decay via exchange of
 gaugino/Higgsino. These operators are proportional to $\frac{1}{M_H M_{Susy}}$ (where
 $M_{Susy}$ is the mass scale of s-particles) and give
 leading contribution to proton decay \cite{weinberg1,sakai}.
 So to calculate the decay rate the knowledge of sparticle
spectrum and their mixing is required.

 The formulas of
\cite{lucasraby,gotonihei} are used to calculate the baryon decay
rates. The (162 complex) operator coefficients ($L_{ABCD},
R_{ABCD}$) are run down from $M_{X}$ to $M_{Z}$ using RGEs and
then along with the Susy spectrum calculated at this scale the run
down values of these parameters are inserted into the formulas of
\cite{gotonihei}. The rates are calculated in terms of parameters
of NMSGUT superpotential required for fermion fitting at GUT scale
and 5 Susy breaking parameters required for fitting at $M_{Z}$
scale. The two example sets \cite{nmsgut} which account for viable
SM fermion fits but fail to satisfy currents limits on baryon
decay rates are given in Table \ref{BDECI}.

\begin{table}
 $$
 {\small\begin{array}{|c|c|c|}
 \hline
 {\rm parameter}& {\rm Soln. ~1}& {\rm Soln. ~2}\\ \hline
 \tau_p(M^+\nu)&
    8.1 \times 10^{ 28}&  1.7 \times 10^{ 28}\\
 \Gamma(p\rightarrow \pi^+\nu) &
    3.1 \times 10^{ -30} & 7.2 \times 10^{ -30}\\
  {\rm BR}( p\rightarrow
 \pi^+\nu_{e,\mu,\tau}) &\{2.6 \times 10^{ -5}, 0.09 ,0.91\}&
 \{ 3.04 \times 10^{ -5}, 0.01, 0.99 \}\\
 \Gamma(p\rightarrow K^+\nu) & 9.2\times 10^{ -30}&   5.2 \times 10^{ -29}\\
 {\rm BR}( p\rightarrow K^+\nu_{e,\mu,\tau}) &\{
   1.1 \times 10^{ -4} , 0.27, 0.73\}  &\{ 5.45 \times 10^{ -5}, 0.01, 0.99 \} \\

 \hline\end{array}
 }$$
\caption{\small{$d=5$ operator mediated proton lifetimes
$\tau_p$(yrs), decay rates
 $ \Gamma ( yr^{-1} )$ and Branching ratios in the dominant Meson${}^++\nu$ channels. }} \label{BDECI}\end{table}

\section{Conclusion}
 Despite many attractive structural features and successes of NMSGUT the solutions found for complete fermion fits
 are unable to fulfill constraints from proton decay. However \cite{nmsgutIII} proposed that in Minimal
 renormalizable Susy SO(10) theories, large wave function corrections
 are possible to light field propagators entering into the fermion Yukawa vertex due to large number of heavy fields.
 This suppresses the SO(10) Yukawa required to match with MSSM
 Yukawa. The suppressed SO(10) Yukawa will determine the d=5, $\Delta$ B
 $\neq$ 0 operators which will give the decay rate compatible with experimental limits.
   However in the work, while calculating the threshold
 corrections it was assumed that the MSSM Higgs is
 dominantly made from the $h_1$ i.e. $\alpha_1$ is dominating. In this thesis and associated work, the complete 1-loop
 calculation is presented. In next chapter we will present full threshold corrections to
 SO(10) Yukawa vertex and the consequent lowering of the d=5, $\Delta$ B
 $\neq$ 0 operators.

\newpage\thispagestyle{empty} \mbox{}\newpage

\chapter{Threshold Corrections and Suppression of d=5, $\Delta$ B $\neq$ 0 Operators}
\section{Introduction}
Threshold corrections at both $M_X$ (Superheavy threshold
corrections to $M_{X}$, $\alpha_3(M_Z)$, $\alpha_G $) and $M_{S}$
(Susy threshold corrections to fit down and strange quark Yukawas)
played a central role in earlier NMSGUT calculations and fits of
the fermion hierarchy \cite{nmsgut,nmsgutIII}. We discussed about
threshold corrections to $M_{X},\alpha_3(M_Z),\alpha_G, y_{d,b,s}$
in the previous chapter. In this chapter we will discuss about the
importance of threshold corrections from dressing (by superheavy
fields) of light Higgs, fermion and anti-fermion line leading into
matter-Higgs Yukawa vertex in the NMSGUT. Again the claim of
\cite{dixitsher} about futility of threshold corrections by
arguing that the large number of fields and their masses above or
below the GUT scale can give rise to un-controllable uncertainty
in low energy predictions is proved wrong. We argue that without
performing any actual calculation one can't come to any
conclusion. The most appealing feature of NMSGUT is that one has
the full spectrum \cite{ag1,ag2,nmsgut} of all the heavy fields
appearing due to symmetry breaking. So after having information
about interactions and superheavy mass spectrum in hand one can
calculate the threshold corrections. To start with, a preliminary
calculation (with some defects) of the wave function
renormalization of the Higgs line in the matter
fermion-antifermion-MSSM Higgs vertex was done \cite{nmsgutIII}.
It was assumed a good
  approximation to take the MSSM Higgs
 to be dominantly made up of the $\mathbf{10}$-plet so that one can ignore the admixture of the other 5 MSSM type Higgs
    doublets pairs present in the theory. However in practice it is found that viable
    parameter sets often imply sizable admixtures of Higgs multiplets other than $\textbf{10}$-plet.
    So to complete the threshold calculations
we considered the other five remaining doublets(from
$\textbf{120},\textbf{126},\overline{\textbf{126}},\textbf{210}$)
too and applied the threshold corrections to each doublet
\cite{BStabHedge}. The importance of these corrections can be
counted from the observation that for the tree level fits found in
\cite{nmsgut}, the wave function corrections (due to the large
number of fields) violates even basic constraints such as
positivity of wave function renormalization ($Z>0$) very badly.
While searching the parameter space for viable fermion fits,
positivity of the wave function
  renormalization factors $Z_{f,H}$ is crucial. We found that despite the
  large number of fields involved in the calculation, the correction
  factors $Z_{f,H}$ can be  reduced by a factor of several hundred keeping the positive sign undisturbed.
   Consequently it drastically
reduces the magnitude of the SO(10) couplings required to fit the
data at $M_{X}$ relative to those found in \cite{nmsgut}! The
Baryon decay suppression mechanism we use relies on this very
(large N facilitated) limiting value being approached  i.e.  $
Z_{H,\ovl H}\simeq 0$. The solutions found \cite{BStabHedge}
respecting $1>>Z\simeq 0$ for light field renormalization  also
improved the
  perturbative status of the couplings so determined.
  With these small couplings (much smaller than before \cite{nmsgut})
  we are able to achieve tree level fermion fits and gauge
  unification \cite{nmsgut}.
 The  higher loop
 corrections will also behave the same way unless the theory
  has a pathologically ill defined perturbation expansion.
 The results from fits \cite{BStabHedge} have shown that the complexity of the spectra
  effectively enhances the possibilities for finding arrangements of parameters for which the
  feared breakdown does not take place. Our results favor the view that
  there is a natural tendency for a ``Higgs dissolution edge''
  to form when implementing the unusual requirement of a fine
  tuned light MSSM Higgs pair to separate out of a surplus of
  superheavy MSSM doublets. We found realistic fits of the earlier
type \cite{nmsgut} but now fully viable inasmuch as the
$d=5,~\Delta B\neq 0$ lifetimes can be $ 10^{34}$ yrs or more.

\section{One loop formula for GUT thresholds}
The method of threshold corrections to gauge couplings is given by
Weinberg and Hall \cite{weinberg,hall}. Its generalization was
done by Wright \cite{wright} for calculating high scale threshold
corrections to Yukawa couplings. It has been available for long
time but not utilized fully, possibly due to the assumption that
such corrections have negligible effect on the underlying theory.
However in our opinion the GUT scale threshold corrections can be
utilized in a favorable way without disturbing the already present
features of theory under consideration. In this section we present
the basic formalism to calculate the one loop thresholds in SO(10)
NMSGUT.

 As we know that due to non-renormalization theorem in
supersymmetric theories the superpotential parameters are
renormalized only due to wave function corrections
\cite{WessZumino}. So we need to calculate the correction factors
for three legs (fermion, anti-fermion and Higgs) of SO(10) Yukawa
vertex which will eventually give us the corrected Yukawa coupling
(when we redefine the fields). However to calculate the
corrections for the large number of heavy fields which couple to
the light fermions and MSSM Higgs at SO(10) Yukawa and gauge
vertices, though straightforward, is very laborious and time
consuming. One needs to decompose the SO(10) superpotential terms
into Pati-Salam labels which is already available in
\cite{ag1,ag2,nmsgut} and then to MSSM group. For calculation of
fermion (anti-fermion) line correction factor we need to look at
the trilinear invariants of the matter fields inside 16-plet with
the Higgs (10,120,126,$\overline{126}$) and gauge field (45-plet).
The calculation of correction factor to Higgs line is much more
complicated than matter lines as it involves huge number of terms
($\sim$ 1100) due to mixing among different Higgs. To perform the
calculation we need to change the weak basis of heavy field
supermultiplet mass matrices to diagonal mass basis. For a given
complete set of mass matrices and trilinear coupling
decompositions \cite{ag1,ag2,nmsgut} one can easily compute this
mass basis.

\begin{figure}[htbp]
 \begin{center}
\includegraphics[scale=0.4]{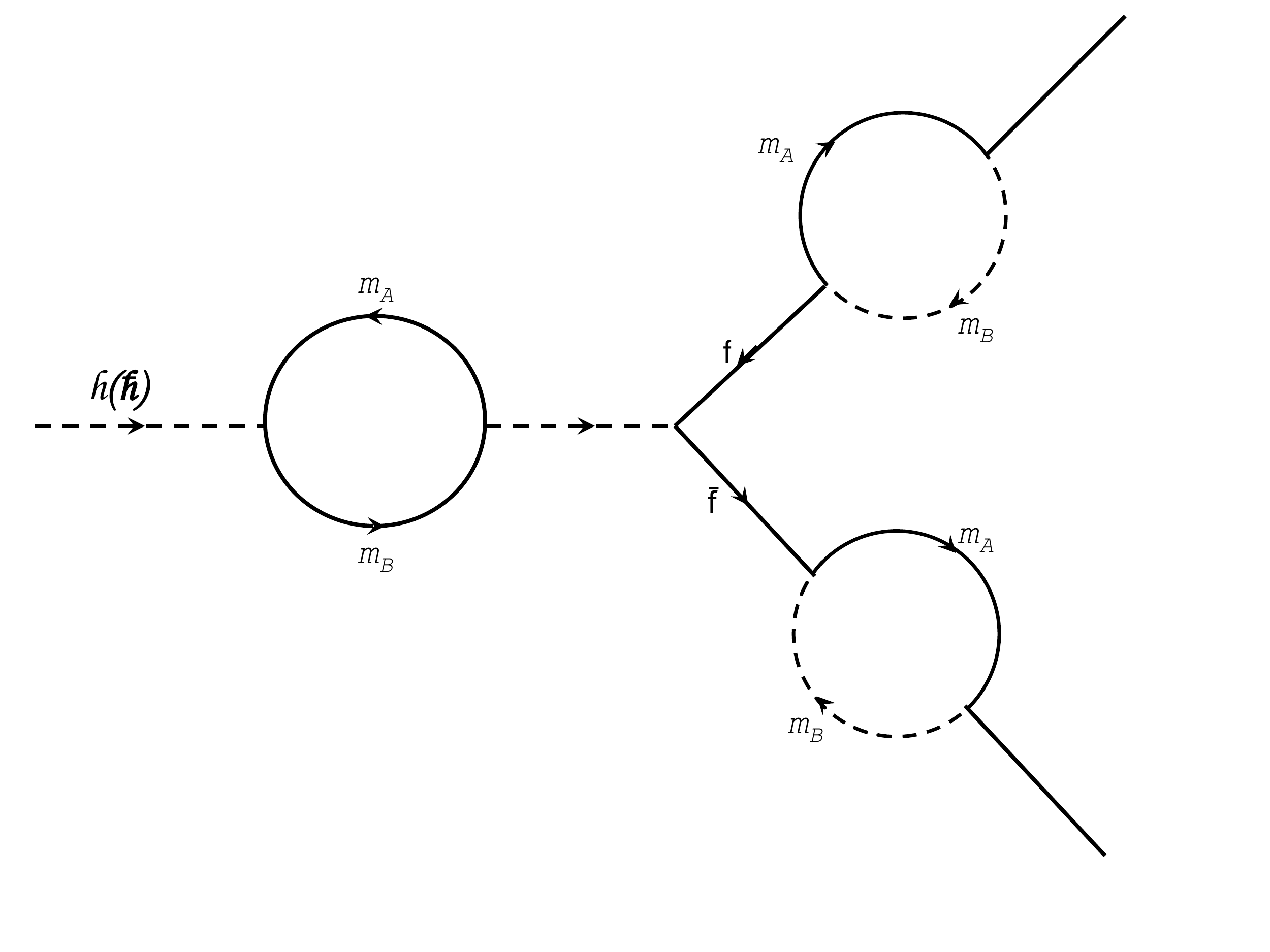}\caption{1-loop correction to SO(10) Yukawa vertex.}
\label{loop}\end{center} \end{figure}

 Let us first work out the expression for correction factor for generic scalar and fermion field. Then
 we present our explicit expressions (Appendix {\bf A}).
 For a generic heavy field type $\Phi$ (conjugate
$\overline \Phi$) the mass terms in the superpotential diagonalize
as :
 \bea{\overline{\Phi }}= U^{\Phi}{\overline{\Phi' }} \quad ;
 \qquad {\Phi } = V^{\Phi}\Phi'\quad \Rightarrow \quad
 {\overline{\Phi}}^T M \Phi ={\overline{\Phi'}}^T M_{Diag}
 \Phi'\eea
The circulation of heavy supermultiplets within the one loop
insertions on each of the 3 chiral superfield lines
 ($f_c=\bar f,f,H_f=H,\ovl{H}$) enters the matter Yukawa vertices (a pictorial
view of it is shown in figure \ref{loop}) : \bea {\cal L} = [f_c^T
Y_f f H_f]_F + H.C. +.... \label{treeyukawa}\eea

From \cite{wright} one finds that a finite wave function
renormalization in the Kinetic terms : \bea {\cal L}=[\sum_{A,B}(
{\bar f}_A^\dagger (Z_{\bar f})_A^B {\bar f}_B +{f}_A^\dagger
(Z_{f})_A^B {f}_B ) + H^\dagger Z_H H + {\ovl H}^\dagger Z_{\ovl
H} {\ovl H}]_D +..\eea Here $A,B=1,2,3$ run over fermion
generations and $H,\ovl H$ are the light Higgs doublets of the
MSSM.  As we discussed in the previous chapter that the light
Higgs superfields are linear combinations of 6 Higgs doublets
$h_i,\bar h_i, i=1...6$ which come from the GUT multiplets:\bea
H=\sum_i \alpha_i h_i \qquad ;\qquad
 \ovl H=\sum_i \bar \alpha_i \bar h_i \eea where  $ \alpha_i,\bar \alpha_i$ are the usual Higgs
fractions \cite{abmsv,ag1,ag2,nmsgut}.

Now we need the kinetic terms for the matter and Higgs field in
canonical form. For that we need to diagonalize the correction
factors $Z_{f,\bar f}$. Let $U_{Z_f},{\ovl U}_{Z_{\bar f}}$ be the
unitary matrices which diagonalize $Z_{f,\bar f}$ to positive
definite form $\Lambda_{Z_f,Z_{\bar f}}$ ($U^\dagger Z U=
\Lambda_Z$). We thus have a new basis after performing this
transformation as:\bea f &=& U_{Z_f} \Lambda^{-\frac{1}{2}}_{Z_f}
\tilde f =\tilde U_{Z_f} \tilde f\qquad ; \qquad \bar f=
U_{Z_{\bar f}} \Lambda^{-\frac{1}{2}}_{Z_{\bar f}} \tilde{\bar
f}=\tilde U_{Z_{\bar {f}}} \tilde {\bar f}\nnu
H&=&\frac{\widetilde H}{\sqrt{Z_H}}\qquad ; \qquad \ovl
H=\frac{\widetilde {\ovl H}}{\sqrt{Z_{\ovl
H}}}\label{f2ftilde}\eea After substituting these expressions in
eq. (\ref{f2ftilde}) and eq. (\ref{treeyukawa}) we get \bea {\cal
L}&=&[\sum_A( {\tilde{\bar f}}_A^\dagger \tilde{\bar f}_A +
{\tilde{f}}_A^\dagger \tilde{f}_A ) + \widetilde{H}^\dagger
\widetilde{H} + \widetilde{{\ovl H}}^{\dagger} \widetilde{{\ovl
H}}]_D + [\tilde{\bar f} ^T \tilde{Y}_{f} \tilde{f} \widetilde{H}_
{f}]_F + H.c. +..\nonumber \eea   \be\tilde Y_f= \Lambda_{Z_{\bar
f}}^{-\frac{1}{2}} U_{Z_{\bar f}}^T {\frac{Y_f}{\sqrt{Z_{H_f}}}}
U_{Z_f} \Lambda_{Z_f}^{-\frac{1}{2}} = \tilde{U}_{Z_{\bar f}}^T
{\frac{Y_f}{\sqrt{Z_{H_f}}}} \tilde{U}_{Z_f} \label{loopYukawa}\ee
Now it is $\tilde Y_f$ different from original $Y_f$ obtained
\cite{abmsv,ag1,ag2} from the SO(10) Yukawas,  should be compared
with MSSM Yukawa while fitting at matching scale.

For any light Chiral field $\Phi_i$ the generic form of
corrections is ($Z= 1 -{\cal K}$) :\bea{\cal K}_i^j=-
{\frac{g_{10}^2}{8 \pi^2}} \sum_\alpha {Q^\alpha_{ik}}^*
{Q^\alpha_{kj}} F(m_\alpha,m_k) +{\frac{1}{32 \pi^2}}\sum_{kl}
Y_{ikl} Y_{jkl}^* F(m_k,m_l) \eea
 Here first term represent corrections from gauge sector and
 second term corresponds to Yukawa sector. $g_{10}= g_5/\sqrt{2}$ and $g_5$ are  the SO(10) and $SU(5)$
gauge couplings. The example of the terms for gauge coupling and
Yukawa coupling has been given in previous chapter.

 $F$ is known as the symmetric 1-loop Passarino-Veltman function.
 When both the fields circulating inside the loop are heavy
 then $F(m_1,m_2)$ has form
\bea F_{12}(M_A,M_B,Q)={1\over {(M_A^2- M_B^2)}}( M_A^2\ln
{M_A^2\over Q^2} -M_B^2\ln    {M_B^2\over Q^2} )- 1 \eea When one
field is light (say $M_B\rightarrow 0$) then it reduces to just
\bea F_{11}(M_A,Q)=F_{12}(M_A,0,Q)= \ln {M_A^2\over Q^2} - 1 \eea
 One should keep in mind that if one of the heavy fields in the loop has MSSM doublet type
 $G_{321}$ quantum numbers $[1,2,\pm 1]$ then sum over light-light loops
must be avoided because that calculations are part of MSSM scale
radiative corrections.

  \section{Importance of threshold corrections}
The importance of the threshold corrections at $M_X$ for the
matter fermion Yukawas can be counted from what values one obtains
for $Z_{f,{\bar{f}},H,\ovl H}$ using parameters from the previous
fits found in \cite{nmsgut} (obtained without including GUT scale
threshold corrections) .

\begin{table}
 $$
 \begin{array}{|c|c|c|c|}
\hline
 \multicolumn{4}{|c|}{\mbox{SOLUTION 1}} \\
 \hline
 \mbox{Eigenvalues}(Z_{\bar u})& 0.928326& 0.930946& 1.031795\\
 \mbox{Eigenvalues}(Z_{\bar d})& 0.915317& 0.917464& 0.979132\\
 \mbox{Eigenvalues}(Z_{\bar \nu})& 0.870911& 0.873470& 0.975019\\
 \mbox{Eigenvalues}(Z_{\bar e})& 0.904179& 0.908973& 0.971322\\
 \mbox{Eigenvalues}(Z_{Q})& 0.942772& 0.946127& 1.027745\\
 \mbox{Eigenvalues}(Z_{L})& 0.911375& 0.916329& 0.997229\\
 Z_{\bar H},Z_{H}& -109.367 & -193.755 & \\
\hline \multicolumn{4}{|c|}{\mbox{SOLUTION 2}}\\
 \hline
 \mbox{Eigenvalues}(Z_{\bar u})& -7.526729& -7.416343& 1.192789\\
 \mbox{Eigenvalues}(Z_{\bar d})& -7.845885& -7.738424& 1.191023\\
 \mbox{Eigenvalues}(Z_{\bar \nu})& -8.830309& -8.681419& 1.234923\\
 \mbox{Eigenvalues}(Z_{\bar e})& -7.880892& -7.716853& 1.238144\\
 \mbox{Eigenvalues}(Z_{Q})& -9.203739& -9.109832& 1.171956\\
 \mbox{Eigenvalues}(Z_{L})& -9.797736& -9.698265& 1.217620\\
 Z_{\bar H},Z_{H}& -264.776 & -386.534 & \\
 \hline
 \end{array}
 $$
 \caption{\small{
 Examples showing the eigenvalues of the wavefunction
 renormalization matrices $Z_f$ for fermion lines and for
MSSM Higgs ($Z_{H,\overline H}$) for solutions found in
\cite{nmsgut}. In both solutions $Z_{H,{\ovl H}}$ are so large
that they imply the kinetic terms change sign and in Solution 2
even $Z_{f,\bar f}$ have large negative values!\label{table1} }}
 \end{table}

In the first solution of table (\ref{table1}) we see that the wave
function corrections to Higgs line have large negative values so
they can change the sign of kinetic terms thus violating unitarity
condition. Moreover in Solution 2 fermion line corrections also
turn out to be negative along with Higgs correction. So we can say
that the parameter sets (of superpotential) found previously in
\cite{nmsgut} as well as all previous GUTs that claim to account
for observed charged fermion and neutrino data are doubtful.
However we see that the problem suggests the solution as well:
When the correction factors are imposed and we put a constraint on
the correction factor to remain positive $Z>0$, then not only
finding the good fits like the previous ones \cite{nmsgut}, we are
able to find regions of the parameter space where matter Yukawa
couplings and other super potential parameters are lowered
significantly thus improving the status of the model w.r.t.
perturbativity conditions.
 \section{Suppression of d=5 operators for B,L violation}
The important thing to notice is that the coefficients
$L_{ABCD},R_{ABCD}$ of the $d=5 $ baryon decay operators are
determined in terms of SO(10) Yukawa couplings $(h,f,g)_{AB}$
which also determine the MSSM yukawa. The effective superpotential
can be obtained by integrating out the heavy chiral
supermultiplets that is responsible for baryon decay (details are
discussed in the previous chapter): \bea
 W_{eff}^{\Delta B\neq 0} = -{ L}_{ABCD} ({1\over 2}\epsilon
{ Q}_A { Q}_B { Q}_C { L}_D) -{ R}_{ABCD} (\epsilon {\bar{ e}}_A
{\bar{ u}}_B { \bar{ u}}_C {\bar{ d}}_D) \eea
Now once we get the
one loop corrected Yukawa $\tilde Y_f$ (see equation
(\ref{loopYukawa}), we need to  diagonalize it to mass basis ( let
us denote it by primes) using bi-Unitary transformation.  The
unitary matrices $(U_f^{L,R})$ which will do it are made up of the
left and right eigenvectors of $\tilde Y_f$. The phases must be
fixed to make the diagonal form $(U^L_f)^T {\tilde Y}_f U^R_f
=\Lambda_f$ positive definite : \bea W &=& (\bar f')^T \Lambda_f
f' {\tilde H_f}\nnu
 f&=& \tilde U_{Z_f} U^R_f f'= {\tilde U_{f}}' f'\nnu
 \bar{f}&=& \tilde U_{Z_{\bar{f}}} U^L_f \bar{f}'= {\tilde U_{{\bar{f}}}}' \bar{f}'\eea
Due to these transformations the $d=5, \Delta B =\pm 1$ operator
coefficients in terms of primed basis become \bea L_{ABCD}'
&&=\sum_{a,b,c,d} L_{abcd} (\tilde U_Q')_{aA} (\tilde
U_Q')_{bB}(\tilde U_Q')_{cC} (\tilde U_L')_{dD} \nnu R_{ABCD}'
&&=\sum_{a,b,c,d} R_{abcd} (\tilde U_{\bar e}')_{aA} (\tilde
U_{\bar u}')_{bB}(\tilde U_{\bar u}')_{cC} (\tilde U_{\bar
d}')_{dD} \eea
 Now while searching with the additional requirement
of suppression of $d=5, \Delta B =\pm 1$ operator coefficients
$L'_{ABCD}, R'_{ABCD}$ (so that they give proton lifetime proton
lifetime $\tau_p
> 10^{34}$ yrs), we find that the search engine inevitably target
those regions of SO(10) parameter space where $Z_{H,\ovl H} <<1$.
This results into lowering of SO(10) Yukawa couplings required to
match the MSSM. They are lowered to such a significant extent
\cite{BStabHedge} that they become much smaller than those
\cite{nmsgut} when threshold corrections were not included. Now
these suppressed Yukawa couplings will enter into
$L'_{ABCD},R'_{ABCD}$ to make them lower. An important point to
note is that $L'_{ABCD},R'_{ABCD}$ are
 not boosted by $Z_{H,\bar{H}}$ as they  don't have an external Higgs line
 and $Z_{f,\bar f}$ have very small effect as they are close to one.
The condition to achieve the required baryon decay rates during
search for fits is discussed in the next section along with the
other features of fitting criteria.

The effect of GUT scale threshold corrections on the MSSM and
other GUT parameters is also illuminating and worth mentioning.
The MSSM superpotential $\mu$ parameter is larger than that
derived from the GUT parameters by the factor $(Z_H Z_{\bar
H})^{-1/2}$. The factor also appears for the soft Susy breaking
parameter $B$.  The matter sfermion soft masses are affected by
$Z_f^{-1}$ which will be very close to 1 so as to make no
difference. However the soft Higgs masses will be enhanced by
$Z_{H/{\overline H}}^{-1}$.  The trilinear coupling parameter
$A_0$ does not change because the change of wave function
corrections are absorbed by the Yukawa coupling which defines it
($A=A_0 \tilde Y$). While searching it is these corrected
parameters that are found in our fits.

Now an important part of calculations at $M_X$ are the masses of
right handed neutrino $(M_{\bar{\nu}})_{AB}\sim f_{AB}<\bar
\sigma>$. However the vev $<\bar \sigma>$ is protected by the
non-renormalization theorem. It is fixed since it is defined in
terms of the parameters $m,\lambda,M,\eta$ and the
 corresponding field fluctuation is not a part of the low
 energy effective theory. The heavy loops  will
  redefine   $ f_{AB}\rightarrow
\tilde{f}_{AB} =({\widetilde U_{\bar \nu}}^T f{\widetilde U_{\bar
\nu}})_{AB}  $ along with  $Y^{\nu}_{AB}\rightarrow \tilde{Y}^{
\nu}_{AB}  $(eqn(\ref{loopYukawa})). So in Type I Seesaw when the
right handed neutrinos $\bar\nu$ are integrated out the factors
$\tilde{U}_{\bar\nu}$ actually cancel out and we are left with
$\tilde{U}_{\nu} Z_H^{-1}$ that changes the formula obtained
without threshold corrections. So we have considered the effect of
one-loop GUT threshold corrections to the gauge, Yukawa, Seesaw,
B-decay operators and GUT scale Susy breaking parameters.

\section{ Search strategy and conditions imposed on fits}
In this section we discuss the search strategy and conditions
imposed on the search engine while doing SM data fit. These
conditions includes the requirement of lowering of B-decay as well
as constraints from current experimental data \cite{pdg}.
\begin{enumerate}
 \item Strict positivity condition is imposed on the wave function
 renormalization factor:  \bea Z_{f,\bar f,H,\bar H}>0
\label{strictpert}\eea Without this the kinetic terms would become
ghost like.
 \item The suppression of $d=5,\Delta B\neq 0$
operators (which consists of LLLL and RRRR operators) is achieved
by putting a penalty on the maximum value of these operators as
$Max(\tilde{O}^{(4)})< 10^{-5 }$ ($\tilde{O}$ is the dimensionless
operator in units of $|m/\lambda|$ so in dimensionful terms $Max(
{O}^{(4)})< 10^{-22 } GeV^{-1}$). It results into proton lifetimes
  above $10^{34}$ yrs. When threshold corrections were not included the
   maximal absolute value of $O^{(4)}$ was of $O(10^{-17} GeV^{-1})$ which leads to decay rates
   $\sim 10^{-27}\,$ yr${}^{-1}$ \cite{nmsgut}.
 The interesting fact is that while looking for fits along with this condition we find those parameter regions where
  $Z_{H,\ovl H}$ $\approx$ 0. However $Z_{f,\bar f}$ undergo  minor changes since the $ {\bf16-}$plet Yukawas
 are suppressed.

 \item Along with the contribution of GUT threshold corrections, we incorporated as an
 improvement the contribution of Susy threshold
effects on gauge unification parameters
$\alpha_3(M_Z),M_X,\alpha(M_X)$. This is important is the sense that NMSGUT
generated Susy spectrum is highly spread out so then can contribute to a significant amount.  The correction factors are :
 \bea
\Delta^{Susy}_{\alpha_s} & \approx & \frac{-19\alpha_s^2}{28\pi} \ln\frac{M_{Susy}}{M_Z}\eea

Here
\be M_{Susy} =  \prod_i m_i^{-{\frac{
5}{38}} (4 b_i^1 -9.6 b_i^2 +5.6 b_i^3)} \label{msusy}\ee
 is the weighted sum over Susy particles as given in \cite{langpol}.

\be \Delta_X^{Susy} = \frac{1}{11.2\pi } \sum_i (b_1 -
b_2)Log_{10}\frac{m_i}{M_Z} \ee

\be \Delta_G^{Susy} = \frac{1}{11.2\pi}\sum_i (6.6\, b_2 - b_1)
\ln\frac{m_i}{M_Z}\ee

Here $b_1$, $b_2,b_3$ are the 1-loop $\beta$ function coefficients
of U(1), SU(2), SU(3) in the MSSM  respectively.
Among these $\Delta^{Susy}_{\alpha_s}$ can
 have significant  value to change the allowed range at GUT scale. So
  we limit our search program  for following range of $\Delta^{Susy}_{\alpha_s}$.
\be -.0146< \Delta^{Susy}_{\alpha_s}  <-0.0102  \ee

 \item All the Susy particle masses are calculated at tree level except Higgs mass.
The Higgs masses are calculated using the 1-loop
 corrected electroweak symmetry breaking
 conditions and  1-loop effective potential using a subroutine \cite{spheno}
 based on \cite{loophiggs}. The Higgs($h^0$) mass in range  $124\, GeV < m_{h^0} < 126 \, GeV$ are ensured via a penalty.

\item  The LHC Susy searches have put \cite{susy2013} lower limit
of $ 1200$ GeV for the Gluino mass which is quite model
independent. So a tolerable Gluino mass is achieved via penalty as
$M_{\tilde G} >1$ TeV.

\end{enumerate}

\section{Example fit with description}
In this section we present an example fermion fit after including
GUT threshold effects. We give a detailed description of the model
parameters. Some of the old features of NMSGUT are still there
along with some new features which we will discuss here. The
complete set of tables (\ref{table a}-\ref{table g}) of example
fit is given in Appendix B.

 In Table \ref{table a} the complete
set of NMSGUT parameters defined at the one loop unification scale
 $M_X^0=10^{16.33}\,$ GeV-which is the GUT-MSSM matching scale-
 along with the soft Susy breaking parameters ($m_0,m_{1/2},A_0,B,
  M^2_{H,\bar H}$) values and the superpotential parameter $\mu $ are given.
  The soft Susy breaking parameters are of N=1 Supergravity GUT type. However
 Non-universal soft Higgs masses (NUHM) for the MSSM Higgs are allowed as they
 originate from the mixture of different SO(10) irreps which have their own soft masses
  and renormalization of Supergravity induced masses by RG flows from $M_{Planck}$ to $M_{GUT}$.
Just like the previous fits found in \cite{nmsgut}, negative values for these soft masses are found. It can be possible
 if the soft masses of at least some of the SO(10) representations are negative themselves.
  From the values of SO(10) matter Yukawa couplings ($h_{AA},f_{AB},g_{AB}$) it is clear that they are suppressed
  and the superpotential couplings are reduced ($\eta$ undergoes a large reduction).
 The mass spectrum of superheavy fields is less spread in magnitude which improves the status of the
  NMSGUT on the scale of perturbation.
 The Type I and Type II neutrino seesaw masses along with right handed Majorana neutrino masses are also given.
 In the second block of the same table are the soft Susy breaking parameters at $M_{X}$ scale.
 The universal gaugino parameter $|m_{1/2}|$ is quite small (0-100 GeV) compared to other soft
 parameters $m_0,A_0, M_{H,\bar H}$ which are of O(100 TeV). In this solution it is approximately 0.
 The values of B and $\mu$ at $M_{X}$ scale are determined by RGE running from $M_{Z}$ to $M_{X}$ of
 values calculated from electroweak symmetry breaking conditions \cite{nmsgut,piercebaggar}.
 In the third block the changes($\Delta_{X,G,3}^{Susy}$) in
 gauge unification parameters from Susy breaking  scale threshold corrections are given.
 The benefit of imposing $1>>Z>0$ condition can be best appreciated if one compares the solutions
 found in \cite{nmsgut} with solution given here and in \cite{BStabHedge}. The effective
 Susy breaking scale defined in Eqn. (\ref{msusy}) is found to be $M_{Susy}$= 2.887 TeV.

In Table \ref{table b}  the values of the target fermion parameters i.e two
loop RGE extrapolated, Susy threshold corrected MSSM Yukawas,
mixing angles, neutrino mixing angles and their mass squared
differences are given. The uncertainties in their values are
calculated as in \cite{antuschspinrath} along with the achieved
values and pulls. It represents an excellent fit with most of the
fractional errors O(1.0\%).  Next is the eigenvalues of  the GUT
scale threshold correction factors for fermion, anti-fermion and
Higgs lines $Z_{f,\bar f, H,\ovl H}$ of Yukawa vertices. It is
clear from the table that amoeba search engine has found that
region where
  $Z_{H,\bar{H}}$ $\approx$ 0. As the magnitude of SO(10) matter Yukawas is reduced so the
  $Z_{f,\bar f}$ are close to one and almost same for the three generations.
   Then come ``Higgs fractions'' \cite{abmsv,ag2,nmsgut}
  $\alpha_i, {\bar \alpha}_i$ which play a crucial role in the
 fermion mass formulae \cite{ag2,blmdm,gmblm,nmsgut}.
 We note that the values of the $\alpha_1, {\bar
\alpha}_1$ given in the table are taken real by convention
although phases are used in threshold corrections. Because the
overall phase of the $\alpha, \bar{\alpha}$ does'nt affect the
other physical parameters of NMSGUT so we continue to take it real
as in \cite{nmsgutIII}.

In Table \ref{table c} the first column contains the running
values of the SM masses at $M_Z$ scale. Next column contains mass
values calculated from run down values of GUT scale threshold
corrected Yukawas from $M_{X}$ to $M_{Z}$ using two loop RGEs
\cite{martinvaughn}. Then large $\tan\beta$ driven Susy radiative
correction are applied to these masses (third column) which are
then compared with the SM values.

  Table \ref{table d} contains the values of the soft supersymmetry
 breaking parameters at $M_{Z}$ scale. The kind of soft parameter
 set the NMSGUT prefers is one of its distinct and remarkable features. The
survival of the NMSGUT  is linked to a
  rarely considered and different kind of soft Susy spectrum with large
  $\mu,A_0,B > 100$ TeV and third generation sfermion masses heavier than first two generations and in
 the 10-50 TeV range. However in some cases \emph{right chiral} sfermions especially the
smuon (see Solution 1 of \cite{BStabHedge}) emerges at O(100) GeV
which is close to experimental lower limits on sfermion masses.
Light smuon solutions are interesting in the sense that they allow
a significant supersymmetric contribution to the muon $g-2$
anomaly and they can also provide a coannihilation channel with
the LSP (pure Bino in this case) and so reduce the relic density
to levels required to validate it as a dark matter candidate.
 We will discuss this feature in Chapter 8 in detail. An important point worth
 mentioning is that despite the very small value of $M_{1/2}$ $\approx$ 0 at $M_{X}$ we
 are able to get the reasonable gaugino mass parameters because the large $A_0$ driven two loop RGEs
 are capable of generating acceptable gaugino masses of O(1 TeV). Also the correlation between $M_{1},M_{2},M_{3}$ deviates
 significantly from 1:2:7 just like previous fits \cite{nmsgut,nmsgutIII}.

 In Table \ref{table e}  the Susy particle masses
 determined using two loop RGEs parameters given in the previous table without generation mixing
and in Table \ref{table g} masses with generation mixing are
given. Since they are so similar they justify the use of diagonal
values to estimate the Susy threshold corrections.

The values of B-decay rates for our example
solution are given in Table \ref{BDEC}. The details of contribution from different
channels (gluino, neutralino and chargino) can be found in
\cite{charanthesis}.

\begin{table}[htb]
 $$
{\small \begin{array}{|c|c|c|c|c|c|}
 \hline
 \tau_p(M^+\nu) &\Gamma(p\rightarrow \pi^+\nu) & {\rm BR}( p\rightarrow
 \pi^+\nu_{e,\mu,\tau})&\Gamma(p\rightarrow K^+\nu) &{\rm BR}( p\rightarrow K^+\nu_{e,\mu,\tau})\\ \hline
    3.5 \times 10^{ 34}
 &
    2.1 \times 10^{ -36}
 &
 \{
   1.7 \times 10^{ -3}
 ,
   0.18
 ,
   0.81
 \}&
    2.6 \times 10^{ -35}
 &\{
   1.8 \times 10^{ -3}
 ,
   0.2
 ,
   0.8
 \} \\
 \hline\end{array}}
 $$
\caption{\small{$d=5$ operator mediated proton lifetimes
$\tau_p$(yrs), decay rates
 $ \Gamma ( yr^{-1} )$ and Branching ratios in the dominant Meson${}^++\nu$ channels. }} \label{BDEC}\end{table}
\section{Conclusions}
We showed that the superheavy threshold
corrections to Higgs (and matter) kinetic terms and thus to the
Yukawa couplings play a critical role due to the
large number of fields involved in dressing each line entering the
effective MSSM vertices. In fact
  care must be taken to maintain positivity of the kinetic terms
  after renormalization which is otherwise generically badly violated : in
particular by the   fits found earlier. As a result we find that
searches incorporating threshold corrected Yukawa couplings, and a
constraint to respect B-decay limits, naturally flow to region of
parameter space that has weak Yukawa couplings and $Z_{H,\ovl H}$
close to zero and hence imply strong lowering of the required
SO(10)  matter Yukawa couplings.  The mechanism that we
demonstrate is likely to work in any realistic GUT since the
features required are so generic and the necessity of
implementation of threshold corrections while maintaining
unitarity undeniable. Since its success depends on $Z_H$
approaching zero while remaining positive rather than fine tuning
to some specific parameter values our mechanism  is likely to be
robust against 2 and higher loop corrections.  Moreover the large
wave function renormalization driven threshold/matching effects
can also have notable influence on  soft supersymmetry breaking
parameters, enhancing $\mu,M^2_{H,\ovl H}$  relative to their GUT
scale values consistent with the patterns found in our fits here
and before.
 The central result of this work is that close attention must
be paid to the consequences of the fact that MSSM Higgs multiplets
derive from multiple GUT sources. Analyses of GUT models that
neglect the multiple GUT level origin of MSSM Higgs and the
resulting large threshold corrections to tree level effective MSSM
couplings should no longer be accepted uncritically.

\clearpage
\section{ Appendix A}
Here we present the explicit form of expressions for correction
factors to fermion, anti-fermion lines. As we mentioned that great
care must be taken while doing these calculations. so we did many
consistency checks to make sure that we have considered
contributions from all members of multiplets.

The correction factor to fermion (anti-fermion) line  is given as
$Z_{f,(\bar{f})}=1-{\cal K}^{f,(\bar f)}_\Phi$.  ${\cal K}^f_\Phi$
signifies the loop corrections on the matter fermion ($f$) line
with loop containing heavy multiplet $\Phi$. Using the formulae in
Section \textbf{2} gives:
   \bea  ({ {16\pi^2}}) {{\cal K}}^{\bar{u}}&=&
K^{\bar{u}}_{\bar{T}}+K^{\bar{u}}_{T}+2
K^{\bar{u}}_{H}+\frac{16}{3}K^{\bar{u}}_{C}
+2K^{\bar{u}}_{D}+K^{\bar{u}}_{J}+4K^{\bar{u}}_{L}+4K^{\bar{u}}_{\bar
K}+16 K^{\bar{u}}_{M}\nnu &&- 2 {g_{10}^2} (0.05
F_{11}(m_{\lambda_G},Q) + F_{11}(m_{\lambda_J},Q) +
F_{11}(m_{\lambda_F},Q) +4 F_{11}(m_{\lambda_X},Q) \nnu &&+2
F_{11}(m_{\lambda_E},Q))\\
K^{\bar{u}}_{\bar{T}}&=&2\sum_ {a =
1}^{\mbox{7}}\biggr(\bar{h}U^T_{1a}-2\bar{f}U^T_{2a}-\sqrt{2} i
\bar{g}U^T_{7a}\biggr)^*
\biggr(\bar{h}U^T_{1a}-2\bar{f}U^T_{2a}\nnu &&+\sqrt{2} i
\bar{g}U^T_{7a}\biggr) F_{11}(m^T_{a}, Q)\\
K^{\bar{u}}_{T}&=&\sum_ {a =
1}^{\mbox{7}}\biggr(\bar{h}V^T_{1a}-2\bar{f}V^T_{2a}-2\sqrt{2} i
\bar{f}V^T_{4a}-\sqrt{2}\bar{g}V^T_{6a}+\sqrt{2} i
\bar{g}V^T_{7a}\biggr)^*
\biggr(\bar{h}V^T_{1a}-2\bar{f}V^T_{2a}\nnu &&-2\sqrt{2} i
\bar{f}V^T_{4a}+\sqrt{2}\bar{g}V^T_{6a}-\sqrt{2} i
\bar{g}V^T_{7a}\biggr) F_{11}(m^T_{a}, Q)\\
K^{\bar{u}}_{H}&=& \sum_ {a =
2}^{\mbox{6}}\biggr(\bar{h}V^H_{1a}-\frac{2i}{\sqrt{3}}\bar{f}V^H_{2a}-
 \bar{g}V^H_{5a}+\frac{i}{\sqrt{3}} \bar{g}V^H_{6a}\biggr)^*
\biggr(\bar{h}V^H_{1a}-\frac{2i}{\sqrt{3}}\bar{f}V^H_{2a}+
 \bar{g}V^H_{5a}\nnu &&-\frac{i}{\sqrt{3}} \bar{g}V^H_{6a}\biggr)
F_{11}(m^H_{a}, Q)\eea
\be K^{\bar{u}}_{C}=\sum_ {a =
1}^{\mbox{3}}\biggr(-4\bar{f}V^C_{2a}+
 2\bar{g}V^C_{3a}\biggr)^*\biggr(-4\bar{f}V^C_{2a}-
2\bar{g}V^C_{3a}\biggr) F_{11}(m^C_{a}, Q)\ee \be
K^{\bar{u}}_{D}=\sum_ {a = 1}^{\mbox{3}}\biggr(-4\bar{f}V^D_{1a}+
 2\bar{g}V^D_{3a}\biggr)^*\biggr(-4\bar{f}V^D_{1a}-
2\bar{g}V^D_{3a}\biggr) F_{11}(m^D_{a},Q)\ee \bea
K^{\bar{u}}_{J}&=&\sum_ {a = 1,a \neq
4}^{\mbox{5}}\biggr(-4i\bar{f}V^J_{1a}+
 2i\bar{g}V^J_{5a}\biggr)^*\biggr(-4i\bar{f}V^J_{1a}-
 2i\bar{g}V^J_{5a}\biggr) F_{11}(m^J_{a}, Q)\\
 K^{\bar{u}}_{L}&=&\sum_ {a =
1}^{\mbox{2}}\biggr(2\sqrt{2}i\bar{f}V^L_{1a}+
 \sqrt{2}\bar{g}V^L_{2a}\biggr)^*\biggr(2\sqrt{2}i\bar{f}V^L_{1a}-
 \sqrt{2}\bar{g}V^L_{2a}\biggr) F_{11}(m^L_{a}, Q)\\
 K^{\bar{u}}_{\bar{K}}&=&2\sum_ {a = 1}^{\mbox{2}}
(-i\bar{g})^*(i\bar{g}) |U^K_{2a}|^2F_{11}(m^K_{a}, Q)\\
K^{\bar{u}}_{M} &=&(2i\bar{f})^*(2i\bar{f}) F_{11}(m^M, Q) \eea

\bea({ {16\pi^2}}) {\cal K}^{\bar{d}}&=&
K^{\bar{d}}_{\bar{T}}+K^{\bar{d}}_{T}+2K^{\bar{d}}_{\bar
H}+\frac{16}{3}K^{\bar{d}}_{\bar C}+2K^{\bar{d}}_{E}+K^{\bar{d}}_{
K}+4K^{\bar{d}}_{L}+4K^{\bar{d}}_{\bar J}+16K^{\bar{d}}_{N}\nnu
&&-2 {g_{10}^2} (0.45 F_{11}(m_{\lambda_G},Q) +
F_{11}(m_{\lambda_J},Q) + F_{11}(m_{\lambda_F},Q) \nnu && +
2F_{11}(m_{\lambda_X},Q)+4F_{11}(m_{\lambda_E},Q)) \eea \bea
K^{\bar{d}}_{\bar{T}}&=&2\sum_ {a =
1}^{\mbox{7}}\biggr(\bar{h}U^T_{1a}-2\bar{f}U^T_{2a}+\sqrt{2} i
\bar{g}U^T_{7a}\biggr)^*
\biggr(\bar{h}U^T_{1a}-2\bar{f}U^T_{2a}\nnu &&-\sqrt{2} i
\bar{g}U^T_{7a}\biggr) F_{11}(m^T_{a}, Q)\\
K^{\bar{d}}_{T}&=&\sum_ {a =
1}^{\mbox{7}}\biggr(-\bar{h}V^T_{1a}+2\bar{f}V^T_{2a}-2\sqrt{2} i
\bar{f}V^T_{4a}+\sqrt{2}\bar{g}V^T_{6a}+\sqrt{2} i
\bar{g}V^T_{7a}\biggr)^* \biggr(-\bar{h}V^T_{1a}\nnu
&&+2\bar{f}V^T_{2a}-2\sqrt{2} i
\bar{f}V^T_{4a}-\sqrt{2}\bar{g}V^T_{6a}-\sqrt{2} i
\bar{g}V^T_{7a}\biggr) F_{11}(m^T_{a}, Q)\\
K^{\bar{d}}_{\bar{H}}&=&\sum_ {a =
2}^{\mbox{6}}\biggr(-\bar{h}U^H_{1a}+\frac{2i}{\sqrt{3}}\bar{f}U^H_{2a}+
 \bar{g}U^H_{5a}-\frac{i}{\sqrt{3}} \bar{g}U^H_{6a}\biggr)^*
\biggr(-\bar{h}U^H_{1a}+\frac{2i}{\sqrt{3}}\bar{f}U^H_{2a}\nnu &&-
 \bar{g}U^H_{5a}+\frac{i}{\sqrt{3}} \bar{g}U^H_{6a}\biggr)
F_{11}(m^H_{a}, Q)\\ K^{\bar{d}}_{\bar{C}}&=&\sum_ {a =
1}^{\mbox{3}}\biggr(4\bar{f}U^C_{1a}-
 2\bar{g}U^C_{3a}\biggr)^*\biggr(4\bar{f}U^C_{1a}+
2\bar{g}U^C_{3a}\biggr) F_{11}(m^C_{a}, Q)\\
K^{\bar{d}}_{E}&=&\sum_ {a = 1,a \neq
5}^{\mbox{6}}\biggr(4\bar{f}V^E_{1a}-
 2\bar{g}V^E_{6a}\biggr)^*\biggr(4\bar{f}V^E_{1a}+
 2\bar{g}V^E_{6a}\biggr) F_{11}(m^E_{a},Q)\\
K^{\bar{d}}_{K}&=&\sum_ {a =
1}^{\mbox{2}}\biggr(4i\bar{f}V^K_{1a}-
 2i\bar{g}V^K_{2a}\biggr)^*\biggr(4i\bar{f}V^K_{1a}+
 2i\bar{g}V^K_{2a}\biggr) F_{11}(m^K_{a}, Q)\eea
\bea K^{\bar{d}}_{L}&=&\sum_ {a =
1}^{\mbox{2}}\biggr(2\sqrt{2}i\bar{f}V^L_{1a}-
 \sqrt{2}\bar{g}V^L_{2a}\biggr)^*\biggr(2\sqrt{2}i\bar{f}V^L_{1a}+
 \sqrt{2}\bar{g}V^L_{2a}\biggr) F_{11}(m^L_{a}, Q)\\
 K^{\bar{d}}_{\bar{J}}&=& 2\sum_ {a = 1,a \neq 4}^{\mbox{5}}
(i\bar{g})^*(-i\bar{g}) |U^J_{5a}|^2F_{11}(m^J_{a}, Q)\\
K^{\bar{d}}_{N} &=& (2 i\bar{f})^*(2i\bar{f}) F_{11}(m^N, Q) \eea

\bea ({ {16\pi^2}}) {\cal K}^{\bar e}&=& 3
K^{\bar{e}}_{T}+2K^{\bar{e}}_{\bar H}+K^{\bar{e}}_{\bar
F}+6K^{\bar{e}}_{\bar D}+3K^{\bar{e}}_{K}+4K^{\bar{e}}_{\bar{A}}-
2 {g_{10}^2} (0.05 F_{11}(m_{\lambda_G},Q) \nnu &&
+3F_{11}(m_{\lambda_J},Q) +F_{11}(m_{\lambda_F},Q)
+6F_{11}(m_{\lambda_X},Q))\eea \bea K^{\bar{e}}_{T}&=&\sum_ {a =
1}^{\mbox{7}}\biggr(\bar{h}V^T_{1a}-2\bar{f}V^T_{2a}-2\sqrt{2} i
\bar{f}V^T_{4a}+\sqrt{2}\bar{g}V^T_{6a}-\sqrt{2} i
\bar{g}V^T_{7a}\biggr)^*
\biggr(\bar{h}V^T_{1a}-2\bar{f}V^T_{2a}\nnu &&-2\sqrt{2} i
\bar{f}V^T_{4a}-\sqrt{2}\bar{g}V^T_{6a}+\sqrt{2} i
\bar{g}V^T_{7a}\biggr) F_{11}(m^T_{a}, Q)\\
K^{\bar{e}}_{\bar{H}}&=&\sum_ {a =
2}^{\mbox{6}}\biggr(-\bar{h}U^H_{1a}-2\sqrt{3}i\bar{f}U^H_{2a}+
 \bar{g}U^H_{5a}+\sqrt{3} i \bar{g}U^H_{6a}\biggr)^*
\biggr(-\bar{h}U^H_{1a}-2\sqrt{3}i\bar{f}U^H_{2a}\nnu &&-
 \bar{g}U^H_{5a}-\sqrt{3} i \bar{g}U^H_{6a}\biggr)
F_{11}(m^H_{a}, Q)\\ K^{\bar{e}}_{\bar{F}}&=&\sum_ {a = 1,a\neq
3}^{\mbox{4}}\biggr(4i\bar{f}U^F_{1a}-
 2\bar{g}U^F_{4a}\biggr)^*\biggr(4i\bar{f}U^F_{1a}+
 2\bar{g}U^F_{4a}\biggr) F_{11}(m^F_{a}, Q)\\
K^{\bar{e}}_{\bar D}&=&\sum_ {a =
1}^{\mbox{3}}\biggr(4\bar{f}U^D_{2a}-
 2\bar{g}U^D_{3a}\biggr)^*\biggr(4\bar{f}U^D_{2a}+
 2\bar{g}U^D_{3a}\biggr) F_{11}(m^D_{a},Q)\\
K^{\bar{e}}_{K}&=&\sum_ {a =
1}^{\mbox{2}}\biggr(4i\bar{f}V^K_{1a}+
 2i\bar{g}V^K_{2a}\biggr)^*\biggr(4i\bar{f}V^K_{1a}-
 2i\bar{g}V^K_{2a}\biggr) F_{11}(m^K_{a}, Q)\\
K^{\bar{e}}_{\bar{A}}&=& (2\sqrt{2}i\bar{f})^*(2\sqrt{2}i\bar{f})
F_{11}(m^A, Q)\eea

\bea ({ {16\pi^2}}) {\cal K}^{\bar \nu}&=&
3K^{\bar{\nu}}_{T}+2K^{\bar{\nu}}_{ H}+K^{\bar{\nu}}_{\bar
F}+6K^{\bar{\nu}}_{\bar
E}+3K^{\bar{\nu}}_{J}+4K^{\bar{\nu}}_{\bar{G}}- 2 {g_{10}^2} (1.25
F_{11}(m_{\lambda_G},Q) \nnu && +3F_{11}(m_{\lambda_J},Q)
+F_{11}(m_{\lambda_F},Q) +6F_{11}(m_{\lambda_E},Q))\eea \bea
K^{\bar{\nu}}_{T}&=&\sum_ {a =
1}^{\mbox{7}}\biggr(-\bar{h}V^T_{1a}+2\bar{f}V^T_{2a}-2\sqrt{2} i
\bar{f}V^T_{4a}-\sqrt{2}\bar{g}V^T_{6a}-\sqrt{2} i
\bar{g}V^T_{7a}\biggr)^* \biggr(-\bar{h}V^T_{1a}\nnu
&&+2\bar{f}V^T_{2a}-2\sqrt{2} i
\bar{f}V^T_{4a}+\sqrt{2}\bar{g}V^T_{6a}+\sqrt{2} i
\bar{g}V^T_{7a}\biggr) F_{11}(m^T_{a}, Q)\\
K^{\bar{\nu}}_{H}&=&\sum_ {a =
2}^{\mbox{6}}\biggr(\bar{h}V^H_{1a}+2\sqrt{3}i\bar{f}V^H_{2a}-
 \bar{g}V^H_{5a}-\sqrt{3} i \bar{g}V^H_{6a}\biggr)^*
\biggr(\bar{h}V^H_{1a}+2\sqrt{3}i\bar{f}V^H_{2a}+
 \bar{g}V^H_{5a}\nnu &&+\sqrt{3} i \bar{g}V^H_{6a}\biggr)
F_{11}(m^H_{a}, Q)\\ K^{\bar{\nu}}_{\bar{F}}&=&\sum_ {a = 1,a \neq
3}^{\mbox{4}}\biggr(-4i\bar{f}U^F_{1a}+
 2\bar{g}U^F_{4a}\biggr)^*\biggr(-4i\bar{f}U^F_{1a}-
 2\bar{g}U^F_{4a}\biggr) F_{11}(m^F_{a}, Q)\\
K^{\bar{\nu}}_{\bar E}&=&\sum_ {a = 1,a \neq
5}^{\mbox{6}}\biggr(-4\bar{f}U^E_{2a}+
 2\bar{g}U^E_{6a}\biggr)^*\biggr(-4\bar{f}U^E_{2a}-
 2\bar{g}U^E_{6a}\biggr) F_{11}(m^E_{a},Q)\\
K^{\bar{\nu}}_{J}&=&\sum_ {a = 1,a\neq
4}^{\mbox{5}}\biggr(-4i\bar{f}V^J_{1a}-
 2i\bar{g}V^J_{5a}\biggr)^*\biggr(-4i\bar{f}V^J_{1a}+
 2i\bar{g}V^J_{5a}\biggr) F_{11}(m^J_{a}, Q)\\
K^{\bar{\nu}}_{\bar{G}}&=& \sum_ {a = 1}^{\mbox{5}}
(-2\sqrt{2}i\bar{f})^*(-2\sqrt{2}i\bar{f}) |U^G_{5a}|^2
F_{11}(m^G_{a}, Q)\nonumber\eea

\bea ({ {16\pi^2}}) {\cal K}^u
&=&K^{u}_{\bar{T}}+K^{u}_{T}+K^{u}_{\bar
H}+K^{u}_{H}+\frac{8}{3}K^{u}_{C}+\frac{8}{3}K^{u}_{\bar
C}+K^{u}_{\bar{E}}+K^{u}_{\bar D}+3K^{u}_{\bar
P}\nonumber\\&&+12K^{u}_{ P}+48K^{\bar{u}}_{W}+4K^{\bar{u}}_{\bar
L} - 2 {g_{10}^2} (0.05F_{11}(m_{\lambda_G},Q)
+F_{11}(m_{\lambda_J},Q) \nnu
&&+3F_{11}(m_{\lambda_X},Q)+3F_{11}(m_{\lambda_E},Q))=
({{16\pi^2}}) {\cal K}^d \eea  \bea K^{u}_{\bar{T}}&=&\sum_{a =
1}^{\mbox{7}}\biggr(-\bar{h}U^T_{1a}-2\bar{f}U^T_{2a}+\sqrt{2}
 \bar{g}U^T_{6a}\biggr)^*
\biggr(-\bar{h}U^T_{1a}-2\bar{f}U^T_{2a}\nnu &&-\sqrt{2}
 \bar{g}U^T_{6a}\biggr) F_{11}(m^T_{a}, Q)\\
K^{u}_{T}&=&2\sum_ {a =
1}^{\mbox{7}}\biggr(\bar{h}V^T_{1a}+2\bar{f}V^T_{2a}\biggr)^*
\biggr(\bar{h}V^T_{1a}+2\bar{f}V^T_{2a}\biggr) F_{11}(m^T_{a},
Q)\eea \bea  K^{u}_{\bar{H}}&=&\sum_ {a =
2}^{\mbox{6}}\biggr(-\bar{h}U^H_{1a}+\frac{2i}{\sqrt{3}}\bar{f}U^H_{2a}-
 \bar{g}U^H_{5a}+\frac{i}{\sqrt{3}} \bar{g}U^H_{6a}\biggr)^*
\biggr(-\bar{h}U^H_{1a}+\frac{2i}{\sqrt{3}}\bar{f}U^H_{2a}\nnu &&+
 \bar{g}U^H_{5a}-\frac{i}{\sqrt{3}} \bar{g}U^H_{6a}\biggr)
F_{11}(m^H_{a}, Q)\\  K^{u}_{H}&=&\sum_ {a =
2}^{\mbox{6}}\biggr(\bar{h}V^H_{1a}-\frac{2i}{\sqrt{3}}\bar{f}V^H_{2a}+\bar{g}V^H_{5a}-
\frac{i\bar{g}}{\sqrt{3}}V^H_{6a}\biggr)^*\biggr(\bar{h}V^H_{1a}
-\frac{2i}{\sqrt{3}}\bar{f}V^H_{2a}-\bar{g}V^H_{5a}\nnu
&&+\frac{i\bar{g}}{\sqrt{3}}V^H_{6a}\biggr)
F_{11}(m^H_{a}, Q)\\
 K^{u}_{C}&=&\sum_ {a =
1}^{\mbox{3}}\biggr(-4\bar{f}V^C_{2a}-
 2\bar{g}V^C_{3a}\biggr)^*\biggr(-4\bar{f}V^C_{2a}+
 2\bar{g}V^C_{3a}\biggr) F_{11}(m^C_{a}, Q) \\ K^{u}_{\bar{C}}&=&\sum_ {a =
1}^{\mbox{3}}\biggr(4\bar{f}U^C_{1a}+
 2\bar{g}U^C_{3a}\biggr)^*\biggr(4\bar{f}U^C_{1a}-
 2\bar{g}U^C_{3a}\biggr) F_{11}(m^C_{a}, Q)\\
K^{u}_{\bar E}&=&\sum_ {a = 1,a\neq 5}^{\mbox{6}}
\biggr(-4\bar{f}U^E_{2a}-
 2\bar{g}U^E_{6a}\biggr)^*\biggr(-4\bar{f}U^E_{2a}+
 2\bar{g}U^E_{6a}\biggr) F_{11}(m^E_{a},Q)\\
K^{u}_{\bar D}&=&\sum_ {a = 1}^{\mbox{3}}\biggr(4\bar{f}U^D_{2a}+
 2 \bar{g}U^D_{3a}\biggr)^*\biggr(4\bar{f}U^D_{2a}-
 2 \bar{g}U^D_{3a}\biggr) F_{11}(m^D_{a}, Q) \\
 K^{u}_{\bar P}&=&\sum_ {a =
1}^{\mbox{2}}\biggr(2 \sqrt{2}\bar{f}U^P_{1a}-
 \sqrt{2}\bar{g}U^P_{2a}\biggr)^*\biggr(2 \sqrt{2}\bar{f}U^P_{1a}+
 \sqrt{2}\bar{g}U^P_{2a}\biggr) F_{11}(m^P_{a}, Q) \\
K^{u}_{P}&=&2\sum_ {a = 1}^{\mbox{2}}
\biggr(-\frac{\bar{g}}{\sqrt{2}}\biggr)^*\biggr(\frac{\bar{g}}{\sqrt{2}}\biggr)
|V^P_{2a}|^2F_{11}(m^P_{a}, Q)\\ K^{u}_{W} &=&
(\sqrt{2}\bar{f})^*(\sqrt{2}\bar{f})
F_{11}(m^W, Q)\\ K^{u}_{\bar L}&=&\sum_ {a = 1}^{\mbox{2}} (-\sqrt{2}\bar{g})^*(\sqrt{2}\bar{g})
|U^L_{2a}|^2F_{11}(m^L_{a}, Q)\eea

\bea ({ {16\pi^2}}) {\cal K}^{e}&=& 3K^{e}_{\bar T}+K^{e}_{\bar
H}+K^{e}_{ H}+3K^{e}_{ D}+3K^{e}_{E}+ 9 K^{e}_{\bar
P}+K^{e}_{F}+12K^{e}_{\bar O}\nnu &&- 2 {g_{10}^2} (0.45
F_{11}(m_{\lambda_G},Q)  +3F_{11}(m_{\lambda_J},Q)
+3F_{11}(m_{\lambda_X},Q) +3F_{11}(m_{\lambda_E},Q))\nnu &&=({
{16\pi^2}}) {\cal K}^{\nu}\eea  \bea K^{e}_{\bar T}&=&\sum_ {a =
1}^{\mbox{7}}\biggr(-\bar{h}U^T_{1a}-2\bar{f}U^T_{2a}-\sqrt{2}\bar{g}U^T_{6a}\biggr)^*
\biggr(-\bar{h}U^T_{1a}-2\bar{f}U^T_{2a}+\sqrt{2}\bar{g}U^T_{6a}\biggr)\nnu
&& F_{11}(m^T_{a}, Q)\\ K^{e}_{\bar{H}}&=&\sum_ {a =
2}^{\mbox{6}}\biggr(\bar{h}U^H_{1a}+2\sqrt{3}i\bar{f}U^H_{2a}+
 \bar{g}U^H_{5a}+\sqrt{3} i \bar{g}U^H_{6a}\biggr)^*
\biggr(\bar{h}U^H_{1a}+2\sqrt{3}i\bar{f}U^H_{2a}\nnu &&-
 \bar{g}U^H_{5a}-\sqrt{3} i \bar{g}U^H_{6a}\biggr)
F_{11}(m^H_{a}, Q)\\ K^{e}_{H}&=&\sum_ {a =
2}^{\mbox{6}}\biggr(-\bar{h}V^H_{1a}-
 2 \sqrt{3}i\bar{f}V^H_{2a}- \bar{g} V^H_{5a}-i
\sqrt{3} \bar{g}V^H_{6a}\biggr)^*\biggr(-\bar{h}V^H_{1a}-
 2 \sqrt{3}i\bar{f}V^H_{2a}+ \bar{g} V^H_{5a}\nnu &&+i
\sqrt{3} \bar{g}V^H_{6a}\biggr) F_{11}(m^F_{a}, Q)\\
 K^{e}_{ D}&=&\sum_ {a =
1}^{\mbox{3}}\biggr(4\bar{f}V^D_{1a}+
 2\bar{g}V^D_{3a}\biggr)^*\biggr(4\bar{f}V^D_{1a}-
 2\bar{g}V^D_{3a}\biggr) F_{11}(m^D_{a},Q)\\
 K^{e}_{E}&=&\sum_ {a =
1,a \neq 5}^{\mbox{6}}\biggr(-4\bar{f}V^E_{1a}-
 2\bar{g}V^E_{6a}\biggr)^*\biggr(-4\bar{f}V^E_{1a}+
 2\bar{g}V^E_{6a}\biggr) F_{11}(m^E_{a}, Q)\\
 K^{e}_{\bar P}&=&\sum_ {a =
1}^{\mbox{2}}\biggr(2 \sqrt{2}\bar{f}U^P_{1a}+
 \sqrt{2}\bar{g}U^P_{2a}\biggr)^*\biggr(2 \sqrt{2}\bar{f}U^P_{1a}-
 \sqrt{2}\bar{g}U^P_{2a}\biggr) F_{11}(m^P_{a}, Q)\\
 K^{e}_{F}&=& \sum_{a=1,a\neq 3}^{\mbox{4}}(-2\bar{g})^*(2\bar{g})|V^F_{4a}|^2 F_{11}(m^F_a,
Q)\\K^{e}_{\bar O}&=& (2i\bar{f})^*(2i\bar{f}) F_{11}(m^O, Q) \eea
Here $g_{10}$ is the SO(10) gauge coupling and
\[ \bar h= 2 \sqrt{2} h \quad ;  \quad \bar g= 2 \sqrt{2} g   \quad ;  \quad \bar f= 2 \sqrt{2} f  \]
 Due to mixing among the several Higgs (6 pairs) from $\mathbf{10,120}$(2 pairs)$\mathbf{ \oot,126,210}$ SO(10) Higgs
multiplets calculation for the correction factor to the light Higgs doublet
lines  $H,{\overline H}$ is  much more tedious than the matter
lines. The MSSM Higgs is linear combination of SO(10) Higgs multiplets.
 The couplings of the GUT field doublets. Having the Pati -Salam decomposition of trilinear Yukawa terms
 \cite{ag1,ag2,nmsgut} in hand we need to do MSSM decomposition. The we need to recognize those
 terms which forms singlet with the MSSM Higgs field. The NMSGUT spectra is named after 26 letters of alphabet
 \cite{ag1,ag2} and there are precisely 26 different combinations of GUT multiplets which form
 singlets with the MSSM $H [1,2,1]$ and similarly 26 with ${\overline H}[1,2,-1]$.
 After summing them we get \bea (16 \pi^2){\cal K}_{H}
&=&8K_{R\bar C}+
 3K_{J\bar D}+3K_{E\bar J}
 + 9K_{X \bar P} +3K_{X\bar T}
  + 9K_{P\bar E}+
 3K_{T\bar E}
 + 6K_{Y\bar L}\nonumber\\&&
 + K_{VF}
 + 8K_{C\bar Z}
 + 3K_{D\bar I}+
 24K_{Q\bar C}
 + 9K_{E\bar U}
 + 9K_{U\bar D}
 + 6K_{L\bar B}
 + 3K_{K \bar X}\nonumber\\&&
 + 6K_{B\bar M}
 + 18K_{W\bar B}
 + 18K_{Y\bar W}
 + 3K_{V\bar O}
 + 6K_{N\bar Y}
 + K_{\bar V\bar A}
 + 3K_{HO}\nonumber\\&&
 + 3K_{S\bar H}
 + K_{H\bar F}+ K_{G\bar H}\eea
Here the number in front of correction factor K is the
multiplicity factor (color, SU(2) multiplicity etc.). Similarly
for $\overline H$ conjugated pairs run inside the loop (unless it
is a real irrep). Here we present the expression for one of these
and rest of the expressions for correction factors to Higgs line
can be found in \cite{BStabHedge, charanthesis}.

\bea
  K_{R\bar C}
  & = &\sum_ {a = 1}^{\mbox{d(R)}}\sum_ {a' =
  1}^{\mbox{d(C)}} \biggr| \biggr(
  \frac {i\kappa} {\sqrt {2}} V^R_{2a} U^C_{3a'}-\gamma V^R_{1a} U^C_{2a'} + \frac {\gamma } {\sqrt {2}}V^R_{2a} U^C_{2a'} -
  \bar\gamma V^R_{1a}U^C_{1a'} - \frac {\bar\gamma } {\sqrt{2}}V^R_{2a}U^C_{1a'} \biggr) V^H_{11} \nonumber \\ &&
   +\biggr(\frac {2\eta } {\sqrt{3}} V^R_{1a}U^C_{2a'}- \sqrt{\frac{2}{3}}\eta V^R_{2a} U^C_{2a'} + \frac {i \bar\zeta}
    {\sqrt{6}} V^R_{2a} U^C_{3a'}\biggr) V^H_{21} +\biggr(\frac {2 \eta} {\sqrt{3}} V^R_{1a}U^C_{1a'} +\sqrt{\frac{2}{3}}\eta
    V^R_{2a} U^C_{1a'} \nonumber \\ &&+ \frac {i\zeta}{\sqrt{6}} V^R_{2a} U^C_{3a'}\biggr) V^H_{31} +\biggr( \frac {\zeta}
    {\sqrt{2}} V^R_{2a} U^C_{2a'}-\frac {i \rho} {3\sqrt {2}} V^R_{2a} U^C_{3a'} +
   \frac {\bar\zeta}{\sqrt{2}} V^R_{2a} U^C_{1a'}\biggr) V^H_{51} -\biggr(\frac {i\zeta} {\sqrt {6}} V^R_{2a} U^C_{2a'}\nonumber \\ && + \frac {i\bar\zeta} {\sqrt{6}} V^R_{2a} U^C_{1a'} -
  \frac {\rho} {3\sqrt{3}}V^R_{1a} U^C_{3a'}\biggr) V^H_{61} \biggr| ^2 F_{12}(m^R_{a}, m^C_{a'}, Q)\eea

\section{Appendix B: Example fermion fit with threshold effects}
\begin{table}
 $$
 {\small\begin{array}{|c|c|c|c|}
 \hline
 {\rm Parameter }&{\rm Value} &{\rm  Field }& {\rm Masses}\\
 &&{\rm}[SU(3),SU(2),Y]&( {\rm Units ~of 10^{16} GeV} )\\ \hline
       \chi_{X}&  0.133           &A[1,1,4]&      1.36 \\ \chi_{Z}&
    0.056
                &B[6,2,{5/3}]&            0.097\\
           h_{11}/10^{-6}&  3.96         &C[8,2,1]&{      0.93,      5.17,      7.45 }\\
           h_{22}/10^{-4}&  3.61    &D[3,2,{7/ 3}]&{      0.29,      6.07,      9.09 }\\
                   h_{33}&  0.018     &E[3,2,{1/3}]&{      0.11,      0.75,      2.60 }\\
 f_{11}/10^{-6}&
 -0.013+  0.159i
                      &&{     2.60,      4.85,      8.23 }\\
 f_{12}/10^{-6}&
 -1.022-  1.812i
          &F[1,1,2]&      0.19,      0.65
 \\f_{13}/10^{-5}&
  0.072+  0.339i
                  &&      0.65,      4.30  \\
 f_{22}/10^{-5}&
  6.554+  4.376i
              &G[1,1,0]&{     0.025,      0.20,      0.76 }\\
 f_{23}/10^{-4}&
 -0.734+  2.351i
                      &&{     0.77,      0.77,      0.85 }\\
 f_{33}/10^{-3}&
 -1.273+  0.516i
              &h[1,2,1]&{     0.33,      2.67,      5.57 }\\
 g_{12}/10^{-4}&
  0.128+  0.19i
                 &&{      7.65,     17.54 }\\
 g_{13}/10^{-5}&
 -9.543+  2.823i
     &I[3,1,{10/3}]&      0.36\\
 g_{23}/10^{-4}&
 -1.64-  0.628i
          &J[3,1,{4/3}]&{     0.30,      0.39,      1.44 }\\
 \lambda/10^{-2}&
 -4.691-  0.149i
                 &&{      1.44,      5.01 }\\
 \eta&
 -0.249+  0.068i
   &K[3,1, {8/ 3}]&{      1.73,      5.14 }\\
 \rho&
  1.175-  0.297i
    &L[6,1,{2/ 3}]&{      1.79,      2.60 }\\
 k&
 -0.018+  0.058i
     &M[6,1,{8/ 3}]&      1.95\\
 \zeta&
  1.296+  0.951i
     &N[6,1,{4/ 3}]&      1.88\\
 \bar\zeta &
  0.224+  0.589i
          &O[1,3,2]&      3.14\\
       m/10^{16}GeV&  0.010    &P[3,3,{2/ 3}]&{      0.49,      4.65 }\\
     m_\Theta/10^{16}GeV&  -2.55e^{-iArg(\lambda)}     &Q[8,3,0]&     0.31\\
             \gamma&  0.39        &R[8,1, 0]&{      0.10,      0.38 }\\
              \bar\gamma& -2.45     &S[1,3,0]&    0.44\\
 x&
  0.85+  0.51i
         &t[3,1,{2/ 3}]&{      0.38,      1.16,      1.67,      3.05  }\\\Delta_X^{tot},\Delta_X^{GUT}&      0.80,      0.86 &&{      5.18,      5.70,     20.56 }\\
                                \Delta_{G}^{tot},\Delta_{G}^{GUT}&-20.52,-23.43           &U[3,3,{4/3}]&     0.38\\
      \Delta\alpha_{3}^{tot}(M_{Z}),\Delta\alpha_{3}^{GUT}(M_{Z})& -0.012, -0.002               &V[1,2,3]&     0.26\\
    \{M^{\nu^c}/10^{12}GeV\}&{0.0002,    2.33,   81.40    }&W[6,3,{2/ 3}]&              2.50  \\
 \{M^{\nu}_{ II}/10^{-8}eV\}&   0.005, 42.93,       1496.83               &X[3,2,{5/ 3}]&     0.09,     2.83,     2.83\\
                  M_\nu(meV)&{1.171,    7.11,   40.21    }&Y[6,2, {1/3}]&              0.11  \\
  \{\rm{Evals[f]}\}/ 10^{-6}&{0.004,   40.69, 1418.7         }&Z[8,1,2]&              0.38  \\
 \hline\hline
 \mbox{Soft parameters}&{\rm m_{\frac{1}{2}}}=
             0.000
 &{\rm m_{0}}=
         12860.4
 &{\rm A_{0}}=
         -1.98 \times 10^{   5}
 \\
 \mbox{at $M_{X}$}&\mu=
          1.72 \times 10^{   5}
 &{\rm B}=
         -1.5 \times 10^{  10}
  &{\rm tan{\beta}}=           50.0\\
 &{\rm M^2_{\bar H}}=
         -2.9 \times 10^{  10}
 &{\rm M^2_{  H} }=
         -2.9 \times 10^{  10}
 &
 {\rm R_{\frac{b\tau}{s\mu}}}=
  5.64
  \\
 Max(|L_{ABCD}|,|R_{ABCD}|)&
          7.7 \times 10^{ -22}
  {\,\rm{GeV^{-1}}}&& \\
 \hline\hline
 \mbox{Susy contribution to}&&&
 \\
 {\rm \Delta_{X,G,3}}&{\rm \Delta_X^{Susy}}=
            -0.053
 &{\rm \Delta_G^{Susy}}=
             2.91
 &{\rm \Delta\alpha_3^{Susy}}=
            -0.01
 \\
 \hline\end{array}}
 $$
\caption{ \small{Column 1 contains values   of the
NMSGUT-SUGRY-NUHM  parameters at $M_X$
  derived from an  accurate fit to all 18 fermion data and compatible with RG constraints.
 Unification parameters and mass spectrum of superheavy and superlight fields are  also given.
 The values of $\mu(M_X),B(M_X)$ are determined by RG evolution from $M_Z$ to $M_X$
 of the values determined by the EWRSB conditions.}}  \label{table a} \end{table}
 \begin{table}
 $$
 {\small\begin{array}{|c|c|c|c|c|}
 \hline
 &&&&\\
 {\rm  Parameter }&{\rm Target} =\bar O_i & {\rm Uncert.= \delta_i}&{\rm Achieved= O_i}& {\rm Pull =\frac{(O_i-\bar O_i)}{\delta_i}}\\
 \hline
    y_u/10^{-6}&  2.035847&  0.777694&  2.035834& -0.000017\\
    y_c/10^{-3}&  0.992361&  0.163740&  0.994253&  0.011560\\
            y_t&  0.350010&  0.014000&  0.350076&  0.004715\\
    y_d/10^{-5}& 10.674802&  6.223410& 10.374090& -0.048320\\
    y_s/10^{-3}&  2.024872&  0.955740&  2.118158&  0.097606\\
            y_b&  0.340427&  0.176682&  0.349778&  0.052924\\
    y_e/10^{-4}&  1.121867&  0.168280&  1.122417&  0.003267\\
  y_\mu/10^{-2}&  2.369435&  0.355415&  2.364688& -0.013356\\
         y_\tau&  0.474000&  0.090060&  0.471211& -0.030967\\
             \sin\theta^q_{12}&    0.2210&  0.001600&    0.2210&            0.0009\\
     \sin\theta^q_{13}/10^{-4}&   30.0759&  5.000000&   30.0765&            0.0001\\
     \sin\theta^q_{23}/10^{-3}&   35.3864&  1.300000&   35.3924&            0.0046\\
                      \delta^q&   60.0215& 14.000000&   60.0469&            0.0018\\
    (m^2_{12})/10^{-5}(eV)^{2}&    4.9239&  0.521931&    4.9233&           -0.0012\\
    (m^2_{23})/10^{-3}(eV)^{2}&    1.5660&  0.313209&    1.5664&            0.0011\\
           \sin^2\theta^L_{12}&    0.2944&  0.058878&    0.2931&           -0.0217\\
           \sin^2\theta^L_{23}&    0.4652&  0.139567&    0.4622&           -0.0220\\
           \sin^2\theta^L_{13}&    0.0255&  0.019000&    0.0260&            0.0252\\
 \hline
                  (Z_{\bar u})&   0.972582&   0.972763&   0.972764&\\
                  (Z_{\bar d})&   0.967473&   0.967657&   0.967659&\\
                (Z_{\bar \nu})&   0.946651&   0.946835&   0.946838&\\
                  (Z_{\bar e})&   0.961973&   0.962151&   0.962154&\\
                       (Z_{Q})&   0.983138&   0.983334&   0.983336&\\
                       (Z_{L})&   0.967422&   0.967617&   0.967619&\\
              Z_{\bar H},Z_{H}&        0.000480   &        0.001284    &{}&\\
 \hline
 \alpha_1 &
  0.202+  0.000i
 & {\bar \alpha}_1 &
  0.134-  0.000i
 &\\
 \alpha_2&
 -0.481-  0.632i
 & {\bar \alpha}_2 &
 -0.518-  0.285i
 &\\
 \alpha_3 &
  0.011-  0.356i
 & {\bar \alpha}_3 &
 -0.360-  0.286i
 &\\
 \alpha_4 &
  0.362-  0.147i
 & {\bar \alpha}_4 &
  0.497+  0.328i
 &\\
 \alpha_5 &
 -0.0160-  0.045i
 & {\bar \alpha}_5 &
  0.054-  0.229i
 &\\
 \alpha_6 &
 -0.001-  0.217i
 & {\bar \alpha}_6 &
  0.019-  0.105i
 &\\
  \hline
 \end{array}}
 $$
 \caption{\small{Solution 2: Fit   with $\chi_X=\sqrt{ \sum_{i=1}^{17}
 (O_i-\bar O_i)^2/\delta_i^2}=
    0.1326
 $. Target values,  at $M_X$ of the fermion Yukawa
 couplings and mixing parameters, together with the estimated uncertainties, achieved values and pulls.
 The eigenvalues of the wavefunction renormalization increment  matrices $Z_{i}$ for fermion lines and
 the factors for Higgs lines are given.
 The Higgs fractions $\alpha_i,{\bar{\alpha_i}}$ which control the MSSM fermion Yukawa couplings  are also
 given. Right handed neutrino threshold  effects   have been ignored.
  We have truncated numbers for display although all calculations are done at double
 precision.}}
 \label{table b} \end{table}
 \begin{table}
 $$
{\small \begin{array}{|c|c|c|c|}
 \hline &&&\\ {\rm  Parameter }&{\rm SM(M_Z)} &{\rm m^{GUT}(M_Z)} &{\rm m^{MSSM}=(m+\Delta m)^{GUT}(M_Z)} \\
 \hline
    m_d/10^{-3}&   2.90000&   1.05215&   2.80332\\
    m_s/10^{-3}&  55.00000&  21.48237&  57.23281\\
            m_b&   2.90000&   2.77488&   2.94586\\
    m_e/10^{-3}&   0.48657&   0.48189&   0.48468\\
         m_\mu &   0.10272&   0.10148&   0.10207\\
         m_\tau&   1.74624&   1.73337&   1.73251\\
    m_u/10^{-3}&   1.27000&   1.09833&   1.27302\\
            m_c&   0.61900&   0.53640&   0.62171\\
            m_t& 172.50000& 146.22372& 172.58158\\
 \hline
 \end{array}}
 $$

 \caption{\small{ Values of standard model
 fermion masses in GeV at $M_Z$ compared with the masses obtained from
 values of GUT derived  Yukawa couplings  run down from $M_X^0$ to
 $M_Z$  both before and after threshold corrections.
  Fit with $\chi_Z=\sqrt{ \sum_{i=1}^{9} (m_i^{MSSM}- m_i^{SM})^2/ (m_i^{MSSM})^2} =
0.0557$.}}
 \label{table c}\end{table}
 \begin{table}
 $$
 {\small\begin{array}{|c|c|c|c|}
 \hline
 {\rm  Parameter}  &{\rm Value}&  {\rm  Parameter}&{\rm Value} \\
 \hline
                       M_{1}&            246.41&   M_{{\tilde {\bar {u}}_1}}&          12822.53\\
                       M_{2}&            590.18&   M_{{\tilde {\bar {u}}_2}}&          12822.49\\
                       M_{3}&           1200.01&   M_{{\tilde {\bar {u}}_3}}&          48248.96\\
     M_{{\tilde {\bar l}_1}}&          11957.95&               A^{0(l)}_{11}&        -137311.14\\
     M_{{\tilde {\bar l}_2}}&          11961.97&               A^{0(l)}_{22}&        -137158.88\\
     M_{{\tilde {\bar l}_3}}&          38556.41&               A^{0(l)}_{33}&         -93057.53\\
        M_{{\tilde {L}_{1}}}&          15324.76&               A^{0(u)}_{11}&        -147185.28\\
        M_{{\tilde {L}_{2}}}&          15326.33&               A^{0(u)}_{22}&        -147183.69\\
        M_{{\tilde {L}_{3}}}&          30130.38&               A^{0(u)}_{33}&         -81454.63\\
     M_{{\tilde {\bar d}_1}}&          11245.04&               A^{0(d)}_{11}&        -138168.65\\
     M_{{\tilde {\bar d}_2}}&          11246.12&               A^{0(d)}_{22}&        -138165.70\\
     M_{{\tilde {\bar d}_3}}&          49308.99&               A^{0(d)}_{33}&         -76263.17\\
          M_{{\tilde {Q}_1}}&          13440.51&                   \tan\beta&             50.00\\
          M_{{\tilde {Q}_2}}&          13440.94&                    \mu(M_Z)&         155715.41\\
          M_{{\tilde {Q}_3}}&          48976.61&                      B(M_Z)&
          4.1869 \times 10^{   9}
 \\
 M_{\bar {H}}^2&
         -2.5331 \times 10^{  10}
 &M_{H}^2&
         -2.5545 \times 10^{  10}
 \\
 \hline
 \end{array}}
 $$
  \caption{ \small { Values (in GeV) of the soft Susy parameters  at $M_Z$
 (evolved from the soft SUGRY-NUHM parameters at $M_X$).
 The  values of soft Susy parameters  at $M_Z$
 determine the Susy threshold corrections to the fermion Yukawas.
 The matching of run down fermion Yukawas in the MSSM to the SM   parameters
 determines  soft SUGRY parameters at $M_X$. Note the  heavier third
 sgeneration.  The values of $\mu(M_Z)$ and the corresponding soft
 Susy parameter $B(M_Z)=m_A^2 {\sin 2 \beta }/2$ are determined by
 imposing electroweak symmetry breaking conditions. $m_A$ is the
 mass of the CP odd scalar in the Doublet Higgs. The sign of
 $\mu$ is assumed positive.}}
 \label{table d}\end{table}
 \begin{table}
 $$
 {\small\begin{array}{|c|c|}
 \hline {\mbox {Field } }& {\rm Mass(GeV)}\\
 \hline
                M_{\tilde{G}}&           1200.01\\
               M_{\chi^{\pm}}&            590.18,         155715.46\\
       M_{\chi^{0}}&            246.41,            590.18,         155715.44    ,         155715.44\\
              M_{\tilde{\nu}}&         15324.618,         15326.183,         30130.304\\
                M_{\tilde{e}}&          11958.03,          15324.84,          11961.76   ,          15326.63,          30125.09,          38560.60  \\
                M_{\tilde{u}}&          12822.48,          13440.40,          12822.42   ,          13440.85,          48227.49,          48998.14  \\
                M_{\tilde{d}}&          11245.07,          13440.65,          11246.13   ,          13441.10,          48865.41,          49419.24  \\
                        M_{A}&         457636.54\\
                  M_{H^{\pm}}&         457636.55\\
                    M_{H^{0}}&         457636.54\\
                    M_{h^{0}}&            125.00\\
 \hline
 \end{array}}
 $$
 \caption{\small{ Spectra of supersymmetric partners calculated ignoring generation mixing effects.
 Inclusion of such effects   changes the spectra only marginally. Due to the large
 values of $\mu,B,A_0$ the LSP and light chargino are  essentially pure Bino and Wino($\tilde W_\pm $).
   The light  gauginos and  light Higgs  $h^0$, are accompanied by a light smuon and  sometimes  selectron.
 The rest of the sfermions have multi-TeV masses. The mini-split supersymmetry spectrum and
 large $\mu,A_0$ parameters help avoid problems with FCNC and CCB/UFB instability\cite{kuslangseg}.
 The sfermion masses  are ordered by generation not magnituide. This is useful in understanding the spectrum
  calculated including generation mixing effects.
  }}\label{table e}\end{table}

\begin{table}
 $$
{\small \begin{array}{|c|c|}
 \hline {\mbox {Field } }&{\rm Mass(GeV)}\\
 \hline
                M_{\tilde{G}}&           1200.22\\
               M_{\chi^{\pm}}&            590.28,         155704.39\\
       M_{\chi^{0}}&            246.44,            590.28,         155704.37    ,         155704.37\\
              M_{\tilde{\nu}}&          15324.61,          15326.19,         30133.937\\
                M_{\tilde{e}}&          11958.04,          11961.80,          15324.83   ,          15326.64,          30128.73,          38566.31  \\
                M_{\tilde{u}}&          12822.35,          12822.41,          13440.35   ,          13459.00,          48229.64,          48995.86  \\
                M_{\tilde{d}}&          11245.00,          11246.06,          13440.60   ,          13459.25,          48864.06,          49420.94  \\
                        M_{A}&         457783.97\\
                  M_{H^{\pm}}&         457783.98\\
                    M_{H^{0}}&         457783.97\\
                    M_{h^{0}}&            125.02\\
 \hline
 \end{array}}
 $$
 \caption{\small{Spectra of supersymmetric partners calculated including  generation mixing effects.
 Inclusion of such effects   changes the spectra only marginally. Due to the large
 values of $\mu,B,A_0$  the LSP and light chargino are  essentially pure Bino and Wino($\tilde W_\pm $).
  Note that the ordering of the eigenvalues in this table follows their magnitudes, comparison
 with the previous table is necessary to identify the sfermions.}}
 \label{table g}\end{table}
\newpage
\chapter{Inflationary Cosmology}
\section{Introduction}
According to standard Big Bang theory our universe was born 15
billion years ago in a extremely hot and infinite dense state.
With rapid expansion it became cold. It has many surprising and
remarkable predictions e.g. the formation of nuclei which results
in primordial abundances of light elements. Later formation of
neutral atoms which results in the free motion of the cosmic
background photons.When neutral atoms formed the universe became
transparent to photons so they are traveling to us from the time
of recombination till today. This is also known as surface of last
scattering. Big Bang theory became popular with the observation
\cite{wmap,COBE} of cosmic microwave background radiation (CMBR).
However on closer examination it was realized that there are
problems in the Big Bang scenario such as existence of unwanted
relics expected from the particle physics theories and relevant at
big bang times (gravitinos, monopoles, domain walls etc.) and very
robust initial condition
problems (flatness problem, horizon problem etc.).\\
 The idea of inflation \cite{guth,Albrecht} was introduced to solve the problems
 of standard Big Bang. Inflationary cosmology is widely accepted and vastly studied now a days.
 It has become one of the cornerstones of modern cosmology. Not
 only solving the puzzles of Big Bang, inflation can also make
 predictions on the large scale structure \cite{primordstruct1,primordstruct2,primordstruct3}
 of our universe on the basis of quantum
  fluctuations which grew via gravitational instability and are the seeds
  for the large scale structure of present day universe.

 \section{Big Bang Cosmology}
 The Big Bang theory is based upon the cosmological principle
 which states that the our universe is almost homogeneous and isotropic at least on large
 scales ($>$ 1 Megaparsec $\approx$ $10^{24}$ cm). It is also confirmed from observations on CMB temperature
 which is almost same at different parts of the sky.
 However if we look at the sky on smaller scales, they are highly
 inhomogeneous i.e. the material is clumped into stars, galaxies
 and clusters of galaxies. It can be thought to have originated from a distribution
 which was homogeneous in past but after progressive clumping due to gravitational attraction
 over time becomes inhomogeneous. So the dynamics of observable
 universe can be broken into two parts. The large scale behavior
 is homogeneous and isotropic and on this background are imposed
 small scale irregularities which act as small perturbations on
 the evolution of the universe which looks homogeneous on scales greater than Megaparsec.\\
 The homogeneous and isotropic universe on large scale can be described by
 Friedmann-Robertson-Walker (FRW) metric \cite{introcomsmo}:
 \bea ds^2= -dt^2+a^2(t)\biggr[\frac{dr^2}{1-kr^2}+r^2(d\theta^2+sin^2\theta
 d\phi^2)\biggr]\eea
 Here t is time coordinate. r,$\theta$, $\phi$ are polar
 coordinates. $k$ is spatial curvature constant. The values $k$= +1,0,-1
 correspond to closed, flat and open universes respectively. a(t) is scale factor which determines the rate of
 expansion. It distinguishes between the physical distance (it changes due to expansion) and
 comoving distance (remains same during expansion) between two
 points as:
 \bea d_{physical}= a(t)~d_{comoving}\eea
 The expansion of the universe depends upon the material contained
 in it and governed by the Friedmann
 equations:
 \bea H^2&=& \frac{8 \pi \rho}{3 M_{Pl}^2}-\frac{k}{a^2}\label{friedmann}\eea
Here H=$\frac{\dot{a}}{a}$ is Hubble parameter and $\rho$ is the
density of material. $M_{pl}$= 2.43 $\times 10^{18}$ GeV is Planck
scale. However this equation can be best utilized if we know the
evolution of density of material in the universe. This is given by the
fluid equation derived from the first law of thermodynamics relating the rate
of change of the density $\dot{\rho}$ to the density $\rho$ and pressure $p$ and
Hubble parameter H(t). \bea \dot{\rho}+3H(\rho+p)=0
\label{fluid}\eea The first term of bracket denotes decrease in
energy density due to expansion and second term denotes decrease
in energy due to work done by pressure for expansion. The two
equations (\ref{friedmann}), (\ref{fluid}) can be combined to form
a more convenient equation known as the acceleration equation: \bea
\frac{\ddot{a}}{a}=-\frac{4\pi}{3M_{pl}^2}(\rho+3p)\label{acc}\eea
Notice that the acceleration equation doesn't contain $k$.

 For flat universe ($k$=0) the Friedmann equation can be solved to
 yield classical cosmological solutions which corresponds to
 universe dominated by matter, radiation and cosmological
 constant.
 \bea p &= & 0; \quad \quad \rho\propto a^{-3}; \quad a(t) \propto
 t^{\frac{2}{3}} \quad \text{Matter~domination} \nonumber\\&&
p=-\frac{\rho}{3}; \quad \quad \rho\propto a^{-4}; \quad a(t)
\propto
 t^{\frac{1}{2}}\quad \text{Radiation~domination} \nonumber\\&&
 p=-\rho; \quad \quad \rho\propto a^{0}; \quad a(t) \propto
 \exp^{Ht}; \quad \text{Cosmological~constant} \, \lambda \eea

For flat universe and given H, the density of universe equals to
critical density given as:
 \bea \rho_c = \frac{3 M_{pl}^2 H^2}{8 \pi}\eea
 and density of universe expressed as fraction of critical
 density is known as density parameter $\Omega_c$:
 \bea \Omega_c \equiv \frac{\rho}{\rho_c}\eea
 The present value of Hubble parameter is given as
 \bea H_0 = 100 h ~\text{Km} ~\text{s}^{-1} ~\text{Mpc}^{-1}\eea
 Where the current measured value of $h$ lie in range (0.70-0.76).
Hubble parameter is an important scale parameter of Big Bang
theory. $H^{-1}$ gives the Hubble time and $cH^{-1}$=$H^{-1}$ (in
natural units $c$=1) is Hubble length or radius. Hubble time is the
time taken by universe to expand appreciably and distance
traveled by light during that time is Hubble radius. Thus Hubble
radius now determines the size of observable universe. The present
critical density is
 \bea \rho(t_0)=1.88 h^2 \times 10^{-29} \text{gm} ~\text{cm}^{-3} \eea
\section{Shortcomings of Big Bang}
Despite various successes of Big Bang model it is plagued with
some basic initial condition problems which are:
\begin{itemize}
\item {\bf Flatness Problem:} This one is easiest to understand.
From equation (\ref{friedmann}), the total density
parameter is given as: \bea |\Omega_{tot}-1|=\frac{|k|}{a^2
H^2}\eea Present density of universe is close to critical density,
so $\Omega_{tot}$ $\approx$ 1. But
during Big Bang evolution $a^2 H^2$ decreases with time, so
$\Omega_{tot}$ moves away from one. This can be understood in the
following way. During radiation domination $a^2 H^2 \propto
t^{-1}$ and in matter domination $ a^2 H^2 \propto t^{-2/3}$. So
we have \bea  |\Omega_{tot}-1| &\propto& t \quad \quad \text{radiation
~domination}\nonumber\\ |\Omega_{tot}-1| &\propto& t^{2/3} \quad
\quad \text{matter~domination}\eea So in both cases
$|\Omega_{tot}-1|$ is an increasing function of time. For example to obtain the
present day density of universe we require $|\Omega_{tot}-1| < 10^{-16}$ at the time of nucleosynthesis. Such a
fine tuned initial condition seems very implausible and the necessity of assuming it is known as the
flatness problem.

 \item {\bf Horizon Problem:} The
horizon problem refers to the communication between different
regions of universe which are casually disconnected. The cosmic
microwave radiation from two distant places are found to be at
nearly equal temperature ( to a precision of one part in 100,000
\cite{wmap}). It implies universe came into thermal equilibrium
due to interaction between different regions of universe which
nave been causally separated through out the Big Bang. Thus Big
Bang has no explanation for such homogeneity and thermal
equilibrium. The light we see from the opposite sides of the sky
has been traveling towards us since the time very close to the
time of Big Bang itself. Since the light has only just reached us,
so it can not possibly have made it all the way across to the
opposite side of the sky. Therefore, there has not been time for
two regions on opposite sides of the sky to interact in any way,
and so one can not claim that the regions have come to the same
temperature by interaction.\\
Another important issue which adds to this problem is that the
microwave background radiation is not perfectly isotropic, but
instead exhibits small fluctuations, as observed by COBE
\cite{COBE} and WMAP \cite{wmap} satellite. These irregularities
are thought to represent the `seeds' from which the structure in
the Universe grows. In the standard Big Bang theory one does not
have a mechanism allowing the generation of such kinds of seed
perturbation.

  \item {\bf
Relic Particles problem:} Another important issue with Big Bang is
production of unwanted relic (quasi-stable) particles e.g. magnetic monopoles,
supersymmetric particles like gravitinos, moduli fields related to
superstrings, domain walls etc. which particle theories predict
at high energy. The idea is that if these are produced in early
universe then they are expected to decay or annihilate so slowly
that relic populations should be detectable or even dominate the energy density
today. Current Universe is far too cold to produce the reactions
required to make these particles but if they had been produced in
the early Universe we would expect some of them to still be
detectable today.This is contrary to the observations, so we
need a mechanism to get rid of these unwanted particles.
\end{itemize}
\section{The idea of Inflation}
The two major problems of flatness and horizon come from the fact
that the comoving Hubble radius $(aH)^{-1}$ is strictly increasing
during Big Bang expansion. So what will happen if one inverts this
behavior? The concept of inflation is based upon the idea that
the comoving Hubble radius is decreasing during expansion of the
universe. To achieve this one can define inflation as an era when
$\ddot a > 0 $ i.e. accelerating expansion. \bea \ddot a >0 \quad
\Rightarrow \frac{d \dot a}{dt}>0 \quad \Rightarrow
\frac{d(aH)^{-1}}{dt} = \frac{\ddot{a}}{\dot{a}^2} < 0 \eea So according to above equation the
Hubble radius, measured in comoving co-ordinates, which determines
the size of observable universe decreases during inflation. Also
from equation (\ref{acc}), we have \bea p< -\frac{\rho}{3} \label{infcond}\eea Evolution dominated by a
cosmological constant with equation of state
p=-$\rho$ and giving exponential expansion $a(t) ~\propto~
e^{Ht}$.\\
The successes of Big Bang theory depend upon conventional
evolution which is non inflationary so one can not allow inflation
to go on forever. Inflation must come to an end followed by
conventional Big Bang behavior, so it acts as an add-in to the
standard Big Bang theory for its better performance.
\subsection{Standard Big Bang problems revisited}
\begin{itemize}
\item {\bf Flatness Problem:} The solution for this problem lies
in the definition of inflation more or less. As it is clear from
Friedmann equation, \bea |\Omega_{tot}-1|=\frac{|k|}{a^2 H^2}\eea
 the condition for inflation is to achieve an initial  evolution which drives
 $\Omega_{tot}$ very close to 1 rather than away from one.
 The aim is to use inflation not just to force $\Omega_{tot}$ close to one but in fact
to make it so extraordinarily close to 1 that the post
inflationary period can't make it move away from 1. Another way
to understand it is that during inflation the size of the portion of
the Universe we can observe, given roughly by the Hubble length
$cH^{-1}$ (or $H^{-1}$ in natural units c=1) does not change while
it inflates. So, we are unable to notice the curvature of the
surface. This is in contrast to the Big Bang scenario where the
distance we see increases more quickly so we can see more of the
curvature as time passes.
 \item {\bf Horizon Problem:} Inflation increases the size of the Universe so fast that
 it keeps the Hubble scale fixed. This means that small patch of the Universe
 expanded to much larger than the size of our presently
observable Universe. Then microwaves coming from  two opposite
sides of the sky really are at same temperature because the
distance between these two regions of the Universe we see after
(even long after) inflation is much smaller than the distance
before inflation started i.e. they were once causally connected in
past. Equally, these causal processes provide the opportunity to
generate irregularities in the Universe which can lead to
structure formation.

 \item {\bf
Relic Particles problem:} The inflationary solution to this problem is
 that the rapid expansion of the inflationary period
dilutes the unnecessary relic particles, because the energy
density during inflation decreases more slowly than the relic
particle density. But the idea can only work if after inflation,
the energy density of the Universe can be converted into ordinary
 matter (via a process known as reheating) without recreating these relics.
  But reheating should be such that the temperature
 of the universe never gets so high that these unwanted particles are created again.
  The reheating process should generate only those particles that the present universe contains.
   A successful reheating scenario will allow us to get back into
   the standard hot big bang conditions required for the success
   of nucleosynthesis and the microwave background radiation.
\end{itemize}
\section{Ingredients of inflationary expansion}
In the previous section we discussed about how inflation is
successful in curing the diseases of Big Bang theory. But we need
to have a mechanism to generate such kind of expansion. This can
be provided by scalar fields. Scalar fields describe spin zero
particles in particle physics and play an important role in
symmetry breaking e.g. Higgs scalar field is responsible for
electroweak symmetry breaking. But it can be any other scalar
responsible for breaking of other symmetries like GUT or
Supersymmetry etc.\\
 Let us take any general scalar field $\phi$ and call it inflaton
 field. The energy density and pressure for the inflaton can be
 calculated from its energy momentum tensor by considering it a
 perfect fluid.
 \be \rho_{\phi} = \frac{1}{2}\dot{\phi}^2 +
 V(\phi)\label{rhophi}\ee
 \be p_{\phi} = \frac{1}{2}\dot{\phi}^2 - V(\phi)\label{pphi}\ee
In both equations the first term is kinetic energy and second term is
potential energy. Some simple
examples of inflationary potentials are: \bea V(\phi) &=&
\frac{1}{2} m^2\phi^2 \quad \quad \text{massive~scalar}
\\ V(\phi) &=& \lambda\phi^4 \quad \quad \text{self~interacting~scalar~field} \eea
 The inflationary literature contains literally thousands of suggestions for inflaton
 candidates and inflation potential. We discuss a class of models in next section. Here we will
 consider the general form of V($\phi$).\\
  The dynamics of scalar field can be obtained by substituting
  equations (\ref{rhophi}) and (\ref{pphi}) into Friedmann equation
  (\ref{friedmann})
  and fluid equation (\ref{fluid}). It yields,
  \be H^2 = \frac{8 \pi}{3 M_{pl}^2}[V(\phi)+\frac{1}{2}\dot{\phi}^2] \label{dynamics1}\ee
   \be \ddot{\phi} +3H\dot{\phi} = - \frac{dV}{d\phi} \label{dynamics2}\ee
  Here the curvature term is ignored as it becomes insignificant
  once inflation starts.\\
 From the equation (\ref{infcond}), the condition for inflation in terms of K.E. and P.E.is
 given as that for negative pressure \be \frac{\dot{\phi}^2}{2} < V(\phi)\ee So basically the
 potential energy is responsible for inflation. It is a measure
of internal energy associated with the inflaton. A crucial
ingredient which permits inflation is that the inflaton field configuration is far from the minimum
in potential energy configuration during inflation.\\

When potential term dominates, inflation is assured. However a sufficient number of e-folds
is possible only if potential is flat enough so that inflaton rolls slowly thus allowing
sufficient inflation. Also potential must have some minimum where
inflaton can settle down after the end of inflation.\\
 To solve the equations (\ref{dynamics1},\ref{dynamics2}) a standard scheme is
 followed known as Slow Roll approximation (SRA). It assumes that
 the kinetic term in Eqn. (\ref{dynamics1}) and acceleration term in Eqn. (\ref{dynamics2})
  can be neglected to obtain a set of simpler equations,
\bea H^2 &\simeq& \frac{8 \pi}{3
  M_{pl}^2} V(\phi)\\
  3H\dot{\phi} &\simeq& - \frac{dV}{d\phi} \label{simpledynamics}\eea
the most important slow roll parameters are defined as: \bea \epsilon
(\phi) = \frac{M_{pl}^2}{2}(\frac{V'}{V})^2 \quad;\quad
\eta(\phi)= M_{pl}^2(\frac{V''}{V}) \eea $\epsilon$ measures the
slope of potential and $\eta$ measures the curvature of
inflationary potential. The required conditions for slow roll are
\bea \epsilon << 1 \quad; \quad |\eta| << 1 \eea Inflation comes
to an end when $|\eta|$ $\approx$ 1. \\
However whether the inflaton has produced enough inflation or not
is a question of concern. The amount of inflation is measured in
terms of logarithm of amount of expansion termed as ``number of
e-folds" given as: \bea N= ln\frac{a(t_{end})}{a(t_{initial})} =
\int_{t_i}^{t_e} H dt = -\frac{2}{M_{pl}^2}\int_{\phi_i}^{\phi_e}
\frac{V}{V'}d \phi\eea Here $\phi_{i}$, $\phi_{e}$ are the
inflaton field value at the starting of inflation (when $\epsilon
,|\eta| << 1$) and at the end of inflation (when $|\eta| \approx $
1) and can be calculated using equation of motion. However the
number of efolds of importance are the number left when
representative scale of the present day cosmos(pivot scale $k$
$\sim$ .002 $\text{Mpc}^{-1}$ \cite{wmap}) leaves the comoving horizon.
The importance of this scale lies in the fact that the CMB
observations are made with least uncertainty at this scale. The
minimum number of efolds required to have sufficient inflation is
50-70.
It corresponds to an expansion by a factor $10^{16}$. However it
should be noted that the number of `observable efolds'
spanning the conceivably observable scale is limited to about 9.
The rest are inferred from the necessity of flatness and causal connection.\\
The simple picture of inflation driven by single scalar field
makes a definite set of predictions for the form of primordial
cosmological fluctuations. The quantum fluctuations generated by a
single scale scalar field should be gaussian, adiabatic and scale
invariant i.e. $n_s$ (spectral index)=1. Scale invariance comes
from the fact that in the limit of de-Sitter space, Hubble radius
remains constant during inflation. However the slow roll inflation
is quasi de-Sitter, so the perturbation spectra is partly, not
exactly scale invariant. The expression for spectral index and
power spectrum of scalar perturbations is given by \bea
n_{s}&=&1-6\epsilon(\phi_{CMB})+2\eta(\phi_{CMB})\\&& P_R
=\frac{H}{m_{pl}\sqrt{\pi \epsilon}}\biggr|_{\phi=\phi_{CMB}}\eea
Here $\phi_{CMB}$ is the field value at which perturbations
relevant to the CMB spectrum  are observed today. From the WMAP
7-year data \cite{wmap} the values of spectral index and power
spectrum are
 \bea P_R= (2.43 \pm 0.11) \times 10^{-9}\quad;\quad n_s=
0.967 \pm 0.014 \eea Another contribution to CMB spectrum
perturbation comes from gravitational waves which is known as
tensor mode of perturbations. Although their contribution was
usually considered negligible negligible as compared to scalar
perturbation, after announcement of BICEP2 result \cite{BICEP2}, a
large value of r $\approx$ .20$\pm$ 0.05 is now under active
consideration . We will discuss it in Chapter 6. The ratio of
scalar to tensor perturbation is defined as \bea r=
\frac{P_R}{P_T}=16 \epsilon({\phi_{CMB}})\eea So one must also
calculate the value of $r$ while considering constraints from power
spectrum and spectral index values from observational data.
\section{Inflection point Inflation Along MSSM Flat Directions}
Many types of inflationary models exist in literature. But most of
them explain inflation considering a generic scalar field and are
not connected to the SM of particle physics.
   Models where inflation is driven not by a generic  scalar field,  but by an inflaton
which is tied to the Standard Model gauge group and spectrum are
important and carry an obvious appeal. These models can explain
reheating into the Standard Model degrees of freedom as
 required for the success of Big Bang Nucleosynthesis, after the end of inflation.
 In this section we focus on the inflation models based on
inflection points in the inflationary potential. The idea of
inflection point inflation relies on the fact that at an
inflection point the inflationary potential is flat required for
the achievement of slow roll inflation. In this context the
suggestion
\cite{MSSMflat1,MSSMflat2,MSSMflat3,MSSMflat4,MSSMflat5,MSSMflat6}
that inflation can be embedded within the Minimal Supersymmetric
Standard Model (MSSM) is an attractive scenario. These types of
models are based on slow roll inflation associated with ``flat
directions'' in the MSSM field space along which the D-term
potential vanishes.
 A well known theorem \cite{holoflat1,holoflat2,holoflat3,holoflat4}
allows one to use holomorphic gauge invariants formed from
 chiral superfields as coordinates for the D-flat manifold of the scalar field space of SUSY gauge theories. The flat
directions are lifted by supergravity generated soft supersymmetry
breaking terms and by  non
  renormalizable terms in the MSSM effective superpotential.\\
At renormalizable level, supersymmetric theories can have
large number of directions in the space (known as moduli space
collectively) of scalar fields, where the D-term contribution to
the scalar potential is zero identically. These degeneracies are
protected from quantum perturbations due to non-renormalization
theorem  but can be lifted by non-perturbative effects. In MSSM
the flat
directions are condensates of sleptons, squarks and Higgs fields.\\
There are two types of flat directions in MSSM, D-flat and
F-flat directions. The field values where the potential will
vanish can be calculated by solving the equations for D and F term
vanishing conditions: \bea D^a= \Phi_i^* T^a \Phi_i=0 \quad;\quad
F_i= \frac{\partial W}{\partial \Phi_i}=0 \eea Here $\Phi_i$ is
chiral superfield. $T^a$ are generators for the MSSM gauge group.
The field condensate which obeys these equations simultaneously
are called D and F-flat direction. Important thing is
correspondence between the flat direction and gauge invariant
holomorphic monomials in term of chiral superfield $\Phi_i$. The
parameter which moves along flat direction is the scalar component
of the gauge invariant
polynomial of chiral superfields.\\
Now we discuss a D-flat direction which is our interest. Consider
a flat direction from leptonic sector. The potential in this
sector in the absence of Susy breaking terms is given as \bea V=
\frac{g^2}{2}\sum_{i=1}^{3}(\sum_{A}\tilde{L}_A^{\dag}T_i\tilde
{L}_A)^2+\frac{g'^2}{2}(\sum_{A}(\tilde{L}_A^{\dag} \frac{Y}{2}
\tilde{L}_A)^2 +(\tilde{e}_A^{\dag} \frac{Y}{2}
\tilde{e}_A)^2)\eea Where $T_i$ are usual Pauli generators for
SU(2) and $Y$ is hypercharge. One can easily check that the
potential vanishes for a set of flat directions of the form  \bea
\tilde L_1 =
\begin{pmatrix} \phi \cr 0
\end{pmatrix};\quad
 \tilde L_2 = \begin{pmatrix} 0 \cr \phi \end{pmatrix}; \quad \tilde{e}_3 = \phi \eea
Where $\phi$ is the vev along flat direction (it
will act as inflaton in case of inflation along flat direction).
Each such type of flat direction is labelled by gauge invariant
polynomial $\tilde{L}_i \tilde{L}_j\tilde{e}_k$. In the next
section we discuss the inflection point inflation along one
such gauge invariant flat direction of MSSM extended with
$U(1)_{B-L}$ gauge group.
\section{ Dirac Neutrino Inflation
connection and fine tuning issue} 
\begin{figure}
\centering
\includegraphics[scale=0.4]{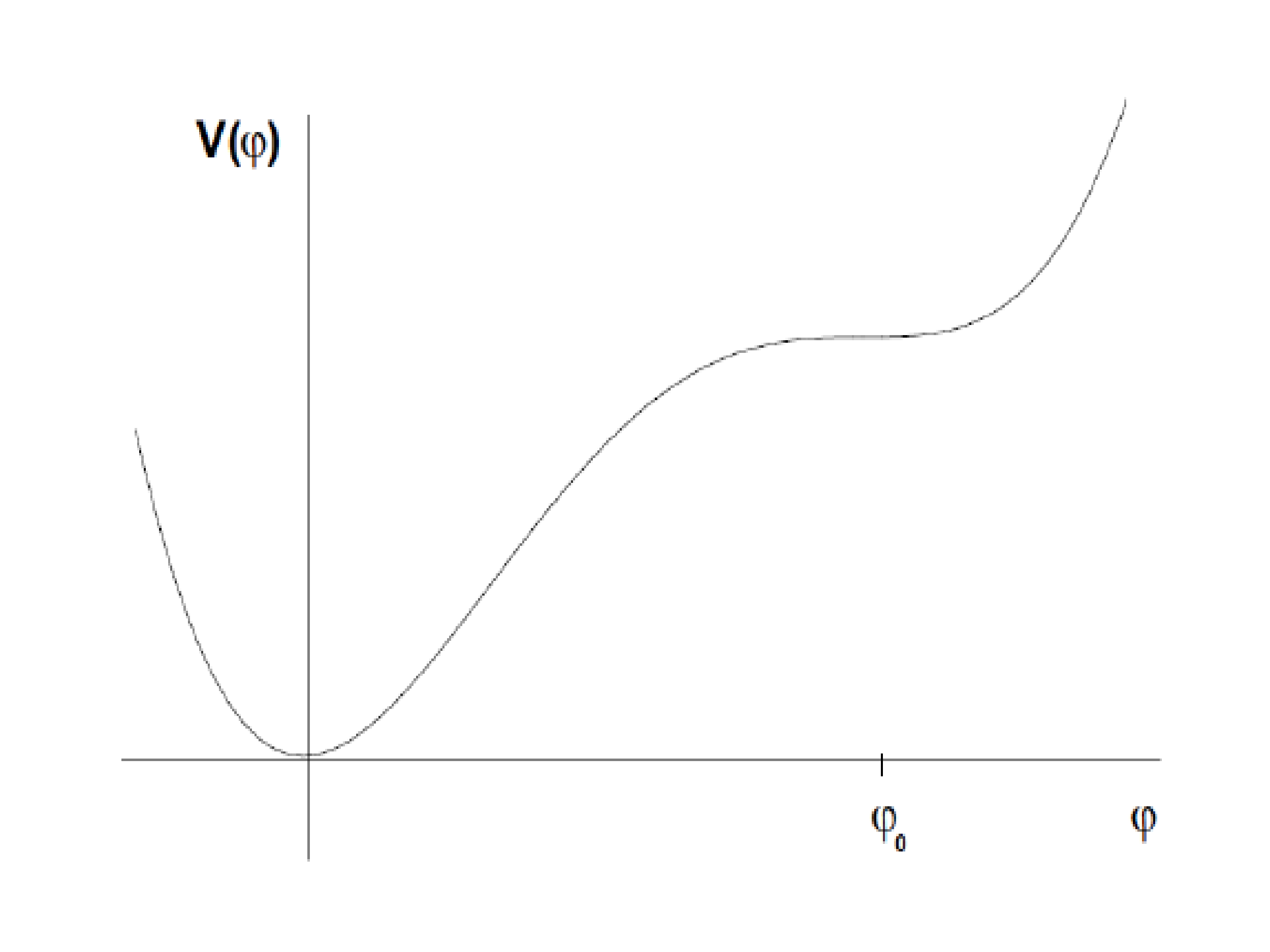}
\caption{The variation of potential with $\phi$. The potential is
flat near inflection point at $\phi_0$ where inflation
occurs.}\label{flatpotential}
\end{figure}

In \cite{akm,hotchmaz} a
connection between smallness of neutrino mass and inflation has
been made. They considered the inflaton to be a gauge invariant
combination of righthanded sneutrino, left handed slepton and
Higgs field, which generates a flat direction along which
inflation will take place on the condition of neutrino Yukawa
being very small $h_{\nu}$ $\sim$ $10^{-12}$. Such types of models
are able
to produce the observed anisotropy and power spectrum of CMB.\\
 They considered MSSM with right handed neutrino and gauge
 symmetry extended to $SU(3)_C \times SU(2)_L \times U(1)_Y \times
 U(1)_{B-L}$. Here B and L are baryon and lepton number. Neutrino
 is considered to be a Dirac particle. Recall that the nature of neutrino i.e.
 whether it is a Dirac or Majorana fermion is
 still under debate. One can introduce small masses of O(0.1 eV)
 corresponding to the atmospheric neutrino oscillations detected by
 Super-Kamiokande experiment \cite{oscillation2} by two way. Either by a seesaw
 mechanism with large right handed neutrino Majorana masses and neutrino Yukawa
 couplings of same magnitude as charged fermions
 or by Dirac masses with small right handed neutrino Majorana masses and small Yukawa couplings.
 Here we review the second case as considered by \cite{akm,hotchmaz}.\\
 In the previous section we noted that MSSM has many flat
 directions made up of squarks, slepton and Higgs fields. In \cite{akm,hotchmaz} it was proposed that
  the D-flat direction along which inflation will take place
 is $NH_{u}L$ (where N is the conjugate neutrino of left chiral superfield) as it is gauge invariant under the gauge group
 considered.
Taking neutrinos in mass diagonal basis the flat direction
$NH_{u}L$ is specified by the field configuration \bea H_u=
\begin{pmatrix}0 \cr \frac{\phi}{\sqrt{3}} \end{pmatrix}; \quad
 \tilde L = \begin{pmatrix}\frac{\phi}{\sqrt{3}} \cr 0 \end{pmatrix}; \quad \tilde{\nu}^C = \frac{\phi}{\sqrt{3}}\eea
 After
including the soft terms (mass terms and trilinear terms), the
renormalizable inflation potential along flat direction has form
 \bea V(\phi)=\frac{m_{\phi}^2}{2} \phi^2+\frac{A h}{6 \sqrt{3}} cos(\theta + \theta_h+\theta_A)\phi^3+
 \frac{h^2}{12}\phi^4 \eea
Here $m_{\phi}^2=(m_{\tilde N}^2+m_{H_u}^2+m_{\tilde{L}}^2)/3$.
Let $\phi$, $\theta$ denotes the radial and angular component of
flat direction: $\phi$= $\phi_{R}+i\phi_{I}$=$\sqrt{2} \phi
~exp(i\theta)$. $\theta_h$ and $\theta_A$ are phases of Yukawa
coupling $h$ and the A-term respectively. The potential can be
minimized when cos($\theta + \theta_h+\theta_A$)=-1. The potential
has global minima at $\phi$=0 and local minima at $\phi_0$ $\sim$
$m_{\phi}/h$. Successful inflation around $\phi_0$  and a graceful
exit from this local minima after the end of inflation is only
possible if \bea  A =4 m_{\phi} \label{finetuning}\eea Then at
$\phi_0$ the first and second derivative of V($\phi$) vanishes
(inflection point) and potential becomes extremely flat (shown in
figure \ref{flatpotential}).
  However at low scale symmetry breaking, the condition given by
  equation (\ref{finetuning}) needs to be achieved to a great
  accuracy and is thus a challenge to overcome. At higher scale of supersymmetry
  breaking severity of fine tuning can be reduced but is still potentially problematic.
  This problem is known as fine tuning problem is inflection point
  inflation models \cite{lythdimo}.\\
  In \cite{hotchmaz} the constraints on inflation potential
  parameters are formulated on the basis of WMAP 7-Year data.
  The fine tuning parameter $\beta$ is defined as
\bea \delta=\frac{A^2}{16 m_{\phi}^2}=1-\frac{\beta^2}{4}\eea Here
$\beta$ can be real or imaginary according to $\delta <$1 or
$\delta>$ 1. \\
 Authors of \cite{hotchmaz} showed the allowed range of parameters
 $m_{\phi}$ and h along with the allowed values of fine tuning parameters,
 imposing constraint that the inflaton VEV should be sub-Planckian i.e $\phi_0$
 $<$ 0.1 $M_{Pl}$ via drawing contour plots between $m_{\phi}$ and $h$.

 The conclusions drawn are that for low inflaton masses,
 the coupling $h$ is required to be very small $\sim 10^{-12}$ for the
model to simultaneously match the spectrum and e-fold constraints.
Such small coupling then gives small masses $\sim$ 100 MeV to the
third neutrino through Dirac term in superpotential. The inflaton
mass is controlled by soft term masses which are in range 1-10
TeV. Then the required value of fine tuning parameter is $\Delta$
$\leq$ $10^{-20}$. This is a severe tuning of Susy breaking
parameters and also unstable since the A terms violates Susy and
are thus not protected from the large radiative corrections.

\section{Discussions}
In this chapter starting from the Big bang theory and introduction
of inflation, we discussed recipe for successful inflation. The
inflection point inflationary models based on gauge invariant flat
directions of MSSM or extended MSSM present an attractive scenario
for inflation linked to known physics of MSSM particles. However
these models are plagued with the fine tuning problems. To reduce
the fine tuning issues several mechanisms have been suggested which
involves invocation of dynamics of other scalar fields
\cite{hotchmaz}. However we found in \cite{SSI} that the
Majorana-neutrino inflation can be a better scenario compared to
Dirac-neutrino-inflation case. The fine tuning conditions are less
severe in Majorana-neutrino inflation and are imposed on the
superpotential parameters rather than soft Susy breaking
parameters. So they are radiatively stable . We present the
details in next Chapter.

\newpage\thispagestyle{empty} \mbox{}\newpage

\chapter{Supersymmetric Seesaw Inflation}
\section{Introduction}
The seesaw mechanism \cite{seesaw} provides an attractive
alternative to a fine tuned Yukawa coupling for explaining small
observed active neutrino masses. Supersymmetric Unified theories
which contain a renormalizable Type I seesaw mechanism for small
neutrino masses compatible with neutrino oscillations, can also
provide slow roll inflection point inflation along a flat
direction associated with a gauge invariant combination of the
Higgs, slepton and right handed sneutrino superfields. Non zero
neutrino masses in the milli-eV range are interpreted as a strong
signal of physics beyond the SM. In the Dirac case light neutrino
masses can be understood by taking highly suppressed
 Yukawa  couplings, ($ {\cal L}= y_\nu {\overline N} H_u L  +.... ; \, y_\nu\sim m_\nu/M_W\sim 10^{-12}$ )
  which are seven or more orders of magnitude smaller than the charged fermion Yukawa
  couplings. Also these masses should be accompanied by highly suppressed right
handed Majorana neutrino masses, $ M_{\nu^c}   \sim 0.1 eV$ or
less. In contrast to it,  the Majorana case, where one can
  generate  small  effective (Type I seesaw) neutrino Majorana masses ($m_\nu\sim (m_{\nu}^D)^2/M_{\nu^c}$)
  for the left handed neutrinos does not require such unnatural assumptions.
  This is possible only if the right handed neutrino  masses $M_{\nu^c}$ take the
  large values allowed by their vanishing SM gauge charges. In this case
  the large Majorana mass of right handed neutrino
   will give required small masses of the neutrinos (due to inverse
  dependence) rather than ultra small Yukawa couplings as required in the Dirac mass
   case.

In the last section of the previous chapter we discussed about a
connection \cite{akm,hotchmaz} made between the smallness of the
(Dirac) neutrino masses and  flatness of the inflaton potential
within the MSSM extended by the addition of $U(1)_{B-L}$ gauge
group and right handed neutrinos. The inflaton field was a gauge
invariant $D$-flat direction, $N H_{u}L$, where $N$ is the right
handed sneutrino, $H_{u}$ is the MSSM Higgs which gives masses to
the up-type quarks, and $L$ is the slepton field. The gauge
invariant superpotential term $y_{\nu} N H_{u}L $ generates the
tiny (Dirac) neutrino masses due to the tiny neutrino Yukawa
coupling ($y_\nu\sim 10^{-12}$). Along with soft trilinear and
bilinear supersymmetry breaking terms of mass scale $\sim 100
~GeV$  to $10~  TeV $,  the associated renormalizable inflaton
potential can then be fine tuned to achieve inflection point
inflation consistent with WMAP 7 year data \cite{wmap}.

Given the attractiveness of the seesaw scenario relative to tuned Yukawas  for Dirac neutrino masses, it is natural to
ask whether it too supports inflectionary inflation.
 At first sight, it seems difficult to achieve conditions of
 the neutrino-inflaton scenario in this prospect i.e  ultra small
superpotential couplings, and TeV scale  trilinear/mass terms.
 Generic Type I seesaw are based upon a mechanism where large right handed neutrino
Majorana masses are generated by breaking of $B-L$ symmetry by
vevs $V_{B-L}>> TeV $. An inflaton involving the right handed
sneutrino  will then have (supersymmetric)  mass as large as the
righthanded neutrino mass since Susy breaking scale is smaller
than $B-L$ symmetry breaking scale . If one considers normal
hierarchy for neutrino masses, then the third generation light
neutrino masses $m_{\nu}\sim y_3^2 v_{EW}^2/M_{\nu^c}\sim 0.1 eV$
can be achieved for $y_3\sim 1, M_{\nu^c} \sim 10^{15} $ GeV. At
first glance it seems that such large couplings and masses would
completely finish the needed flatness of the inflationary
potential. But there is a way out of it. Three generations of
neutrinos and their superpartners are in play, so there is wide
scope for much smaller superpotential couplings. One possibility
that we considered is that the neutrino Yukawa coupling
eigenvalues can have the typical values associated with up type
fermions while off diagonal components matched the tiny Majorana
couplings in smallness. Off diagonal flat directions ($N_A H L_B,
A\neq B$=1,2,3) can serve just as well as diagonal ones. Later we
will discuss others allowed by high scale inflation.

 The famous
Leptogenesis \cite{leptogen} scenario strongly favors right handed
neutrino masses in the range $10^{6} $ to $10^{12}$ GeV. So for
B-L breaking scale $V_{B-L}\sim M_X> 10^{16}$ GeV the
superpotential couplings $f_A, A=1,2,3$ (we work in a basis where
these couplings are diagonal associated with the superpotential
terms), which generate right handed neutrino masses
$M_{\nu^c_A}\sim f_A V_{B-L}$ \cite{MSLRM1,MSLRM2,MSLRM3}, come
out to be very small ($ f_{A}\sim 10^{-9}$ to $10^{-4})$. Thus the
essential ingredients for an inflation in the Type I seesaw
scenario are potentially present justifying detailed analysis.

\section{Generic  renormalizable inflection point inflation }
In this section we present our analytic formulae for the relations
required among the parameters of the generic renormalizable
inflection point inflation (GRIPI) potential for successful
inflation and fulfilling constraints from WMAP 7-year data. In
contrast to the previous analysis \cite{hotchmaz} ( opaque
graphical methods) our calculations are more rigorous and and can
easily be interpreted. The results are viable for any GRIPI model.
A GRIPI model can be formulated in terms of a single complex field
$\varphi$. After extremizing with respect
 to the angular degree of freedom (which has positive curvature
 and cannot support inflation) one is left with the  potential for
a real  degree of freedom $\phi$ in  the complex scalar
 inflaton field $\varphi$
\bea V= {\frac{h^2}{12}} \phi^4  -  {\frac{A h}{6\sqrt{3}}} \phi^3
  + {\frac{M^2}{ 2}}\phi^2 \eea
Here $A,h, \phi$ are real and $A,h$ positive without loss of
generality.

The fine tuning condition between A and M can be expressed in
terms a parameter $\Delta$ given as $A=4 M{\sqrt{ 1-\Delta }}$
($\Delta=\beta^2/4$ in the notation of \cite{hotchmaz}). The
inflection point at  \bea \phi_0 =\frac{\sqrt{3} M}{h}(1-\Delta
+O(\Delta^2)) \qquad :\qquad V''(\phi_0)=0\eea is also a saddle
point ($V'(\phi_0)=0$) when $\Delta =0$. For small $\Delta$  \bea
V(\phi_0)&=&V_0=\frac{M^4}{4 h^2} (1 + 4 \Delta)\qquad;\qquad
V'(\phi_0)=\alpha=\frac{\sqrt{3} M^3 \Delta}{ h} \nonumber\\
V'''(\phi_0)&=&\gamma=\frac{2 M h}{\sqrt{3}}(1-2
\Delta)\label{leadingV}\eea  For a tiny coupling $h $, $V_0\
>> M^4$ and   $\phi_0>>M $. The $\gamma$ is quite
 small due to the smallness of $h$ and $\alpha$ is small
 (but non-zero \cite{lythdimo}) because of the tuning condition.
 So with such a large vacuum energy flatness of potential
 around $\phi_0$ will allow the inflaton $\phi$ to roll slowly
 with small field velocity in the vicinity of $\phi_0$. The  small change
 in the field value during this expansion is given as $\Delta
\phi \sim V_0/\gamma M_p^2$ below $\phi_0$.  Around the inflection
point $\phi_0$, we can   write  the inflection point inflation
potential in the form \bea
V(\phi)=V_0+\alpha(\phi-\phi_0)+\frac{\gamma}{6}(\phi-\phi_0)^3
+{\frac{h^2}{12}}(\phi-\phi_0)^4   \label{eq-Vinf-gen} \eea The
last term is necessarily negligible, since $h^2 $ is very small by
assumption.\\
 The slow roll parameters are defined as($ M_{p }= 2.43 \times 10^{18} \,GeV$)  \bea
 \eta(\phi)&=&\frac{M_p^2 V''}{V} \simeq {\frac{M_p^2}{V_0}}\gamma (\phi-\phi_0)\nonumber
\\ \epsilon(\phi)&=&\frac{M_p^2}{2} (\frac{V'}{V})^2 \simeq(\alpha
+\frac{\gamma}{2}(\phi-\phi_0)^2)^2({\frac{ M_p^2}{2V_0^2}})
\nonumber
 \\  \xi&=&\frac{M_p^4V' V'''}{V^2} \simeq \frac{M_p^4 \alpha \gamma}{V_0^2}
  \label{slrlprm}\eea
The small first and third Taylor coefficients $\alpha,\gamma$
determine \cite{lythdimo,lythstew,liddleleach} the measured
inflation parameters ($P_R,n_s$) at the field value $\phi_{CMB}$
at the time of horizon  entry of the ``pivot'' momentum scale
($k_{pivot}=0.002$ Mpc${}^{-1}$). The termination of the slow roll
 \cite{lythdimo,liddleleach} is fixed on the basis of an overall
cosmogonic scenario and the consistency of the slow roll
approximation ($\eta(\phi_{end})\approx 1$).
 The field value $\phi_{cmb}$ has theoretical importance only. It is the number
($N_{CMB}=N(\phi_{CMB})$) of e-folds of inflation which are left
to occur after field value $\phi_{CMB}$ is reached at the time when
the fluctuation scale of interest ($k_{pivot}$) left the comoving
horizon ( i.e $k=a_k H_k$) during inflation that has physical
significance. This number is determined by the overall history
(inflation, reheating, radiation dominance, matter dominance, BBN
etc.) of the Universe from primordial times\cite{liddleleach}. The
observed perturbations in CMB require
 $ 40<N_{CMB} < 60$ and this restricts the inflation exponents to a great extent.

The field value at the end of slow roll inflation $\phi_{end}$ is
defined as the value where \bea \eta(\phi_{end})\simeq
\frac{\gamma(\phi_{end}-\phi_0) M_p^2}{V(\phi_0)}\simeq 1  \eea
which gives \bea\phi_0-\phi_{end}=\frac{V_0}{\gamma
M_p^2},\label{phiend}\eea Then in the slow roll approx
$\ddot\phi\simeq 0 , \dot\phi=-V'(\phi)/3 H$,
 where $H=\sqrt{V(\phi_0)/(3 M_p^2)}$ is the (constant) inflation
  rate during slow roll inflation, one has
   \bea N(\phi)&=& -3 \int_{\phi}^{\phi_{end}}\frac{H^2}{V'(\phi)}
   d\phi \nonumber \\
    &=& \sqrt{\frac{2}{ \alpha\gamma}}\frac{V_0}{M_p^2}
    \big(\arctan\sqrt{\frac{\gamma}{2\alpha}}(\phi_0-\phi_{end})-
    \arctan\sqrt{\frac{\gamma}{2\alpha}}(\phi_0-\phi )\big)\label{Nphi}\eea
and conversely

\bea  \phi(N)&=&{\frac{\phi_{end} +\phi_0 (\phi_0-\phi_{end})
    \sqrt{\frac{\gamma}{ 2\alpha}}
     \tan\sqrt{\frac{\alpha\gamma}{2}}\frac{N M_p^2}{V_0}+
     \sqrt{\frac{ 2\alpha}{\gamma}}
     \tan\sqrt{\frac{\alpha\gamma}{2}}\frac{N M_p^2}{V_0}}
     { 1+(\phi_0-\phi_{end})
    \sqrt{\frac{\gamma}{ 2\alpha}}
     \tan\sqrt{\frac{\alpha\gamma}{2}}\frac{N M_p^2}{V_0}}}
    \eea
This inversion of the function $N(\phi)$  is derived without
assuming that $\phi_{end}<<\phi(N)$ \cite{lythdimo}.

 The observed CMB data \cite{wmap} put constraints on the power
spectrum and spectral index for modes around the pivot scale.
However the number of e-folds remaining at the time when pivot
scale left the comoving horizon during inflation can be estimated
using the standard Big Bang thermal cosmogony along with estimates
from the reheating behavior. This gives\cite{liddleleach} \eq {N
\subs{pivot}=
 65.5+ \ln\frac{\rho_{rh}^{\frac{1}{12}}V_0^{\frac{1}{6}}}{\mpl }}
where $\rho_{rh}$ is the energy density after reheating and $V_0$
the potential value during inflation.

 The analysis is done
following the steps mentioned in the 4th section of the previous
chapter. For the sake of completeness we present the formula for
power spectrum and spectral index in terms of our
parameterizations. The slow roll inflation formula for the power
spectrum of the mode that is leaving horizon when the inflaton
rolls to $\phi$ is(\cite{lythstew}) \eq{P_R(\phi)=\frac{ V_0}{24
\pi^2 \mpl^4\epsilon(\phi)},\label{eq-inf-PR}} and the
corresponding spectral index and it's variation with momentum is
 \bea n_s(\phi)&\equiv&
 1+2\eta(\phi)-6\epsilon(\phi) \label{eq-inf-ns}\nonumber \\
  {{\cal D}_k}(n_s)&=&\frac{kdn_s(\phi)}{dk} =-16\epsilon\eta + 24
\epsilon^2 + 2 \xi^2\eea The ratio of tensor to scalar
perturbations $r=\frac{P_T}{P_R}=16\epsilon.$\\ In practice,
$\epsilon,\xi$ are so tiny in the small region around $\phi_0$
where slow-roll inflation occurs, that their contribution to $n_s$
is negligible. Thus ${\cal D}_k(n_s)$ is negligible i.e. the
spectral index is scale invariant in the observed range, as is
allowed by observation so far.

To search for sets of potential parameters
 $M,h,\Delta$ compatible with $P_R,n_S,N_{CMB} $  in their
 allowed ranges one may proceed as follows. First one uses the
 chosen (within experimental range)
  values of $P_R,n_s$  and given $M,h$ to
   define
  \bea \epsilon_{CMB}&=&\frac {V_0}{24 \pi^2 M_p^4 P_R}\nonumber \\
  \eta_{CMB} &=& \frac{(n_s-1)}{2}\label{epsetacmb}\eea
From  $\epsilon_{CMB},\eta_{CMB}$ one may deduce
  $\alpha_{CMB},\phi_{CMB}$    using the
 eqns.(\ref{slrlprm}) \bea\phi_{CMB}&=&\phi_0 +
{\frac{V_0 \eta_{CMB}}{ \gamma  M_p^2}} \nonumber\\
\alpha_{CMB}&=& \sqrt{2  \epsilon_{CMB}} {\frac{V_0}{M_p}}
-{\frac{V_0^2\eta_{CMB}^2}{2 \gamma  M_p^4} } \eea
 The required fine-tuning $\Delta$  is then  \be \Delta= {\frac{h\alpha_{CMB}}{\sqrt{3}M^3}}=(\frac{M}{4 h M_p})^4
 (\frac{16 h^2 M_p}{3 \pi M {\sqrt{P_R}}}-(1-n_s)^2)\ee
 $\alpha_{CMB},\Delta$ should emerge real and positive
 and using $\{\alpha_{CMB},\phi_{CMB}\}$ in the formula for
$N_{CMB}$ one should obtain a sensible value  in the range
$N_{CMB}=51\pm 5$. Positivity of $\Delta$ (a local minimum
develops if  $\Delta$ is negative leading to eternal inflation)
requires \be h^2 \geq M \frac{3 \pi(1-n_s)^2\sqrt{P_R}}{16 M_P}\ee
Using eqns.(\ref{leadingV},\ref{phiend}\ref{epsetacmb}) in
eqn(\ref{Nphi}) we have \bea N_{CMB}&=& \frac{1}{z} \arctan\frac{2
z (1+ n_s)}{8 z^2 + 1 - n_s}\label{NCMB}\\z&=& (\frac{h^2 M_p}{3
\pi M\sqrt{P_R}} - \frac{(1-n_s)^2}{16})^{\frac{1}{2}}
\label{z}\eea By solving eqn.(\ref{NCMB}) for $z=z_0(N_{CMB},n_s)$
one obtains the general relation between $h$ and $M$ : \bea
\frac{h^2}{M}&=& \frac{3 \pi \sqrt{P_{R}}}{M_{P}}
(z_0^2(N_{CMB},n_s)+\frac{(1-n_s)^2}{16})\eea and then
\bea\Delta&=& \frac{16 M^2 z_0^2(N_{CMB},n_s) }{9 \pi^2 M_P^2 P_R
((1-n_s)^2+ 16 z_0^2(N_{CMB},n_s))^2  }\eea Where
$z_0(N_{CMB},n_s)$ is the solution of eqn (\ref{NCMB}). An
excellent approximation to the required function in the region of
interest in the $N_{CMB},n_s$ plane is given by the Taylor series
around $n_s^0=0.967,N_{C}^0=50.006$ :\bea &z_0(N_{CMB},n_s)=.0238
-0.0006 (N_{CMB}-N_C^0)+ 0.2407 (n_s- n_s^0) \nonumber\\
&+ 0.000022 \frac{(N_{CMB}-N_C^0)^2}{2}
 -3.70875 \frac{(n_s-n_s^0)^2}{2}+ 0.002353 (N_{CMB}-N_C^0)(n_s-
n_s^0)\nonumber\\& -.0000015 \frac{(N_{CMB}-N_C^0)^3}{6}+
 8.79982 \frac{(n_s-n_s^0)^3}{6}
 -0.000788 \frac{(N_{CMB}-N_C^0)^2 (n_s-n_s^0)}{2}\nonumber\\
& -0.7536 \frac{(N_{CMB}-N_C^0) (n_s-n_s^0)^2}{2}\eea

\begin{figure}
\centering
\includegraphics[scale=0.8]{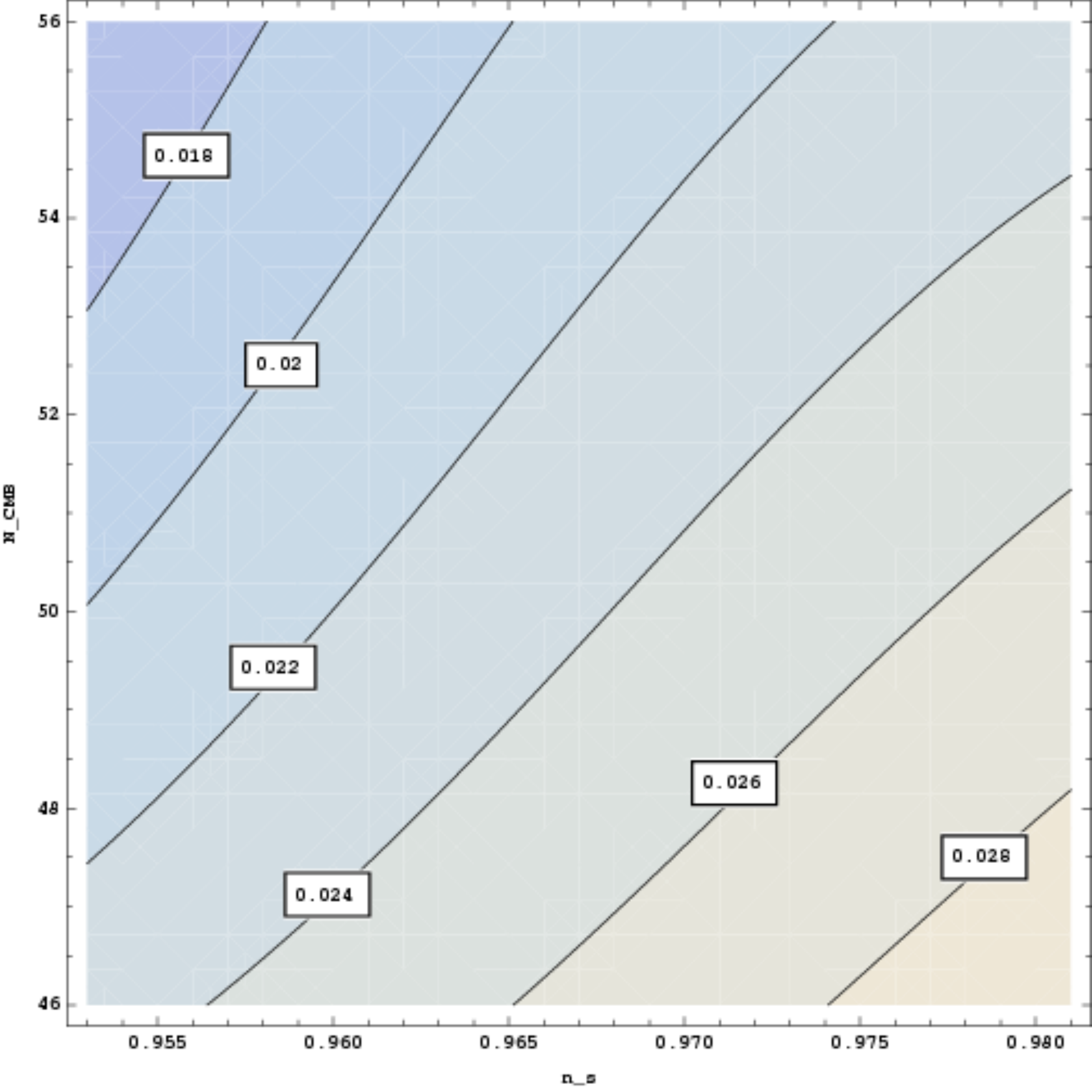}
\caption{$z_0 $ contours in the $N_{CMB},n_s$ plane. The variation
shown contributes to the small range of permitted magnitudes   for
$h^2/M, \Delta/M^2$ etc.} \label{z0plot}\end{figure}
In Fig.
\ref{z0plot} we plot the contours of $z_0(N_{CMB},n_s)$ in the
$N_{CMB},n_s$ plane. It is clear that the variation of $z_0$ is
rather small scaled. So for the plausible range $46<N_{CNB}<56$
one obtains a tight constraint on the exponents in the relation
between h,$\Delta$ and M: \be h^2 \sim 10^{-24.95 \pm 0.17}
(\frac{M}{GeV})\,\,\,\,;\,\,\,\ \Delta \sim 10^{-28.17 \pm
.13}(\frac{M}{GeV})^2\label{hDeltavsM}\ee Notice that the
approximate expressions give $\frac{h^2}{M} \approx
\frac{10^{-21.8}}{N_{CMB}^2}$  which is effectively the same!
\begin{figure}
\includegraphics[scale=0.8]{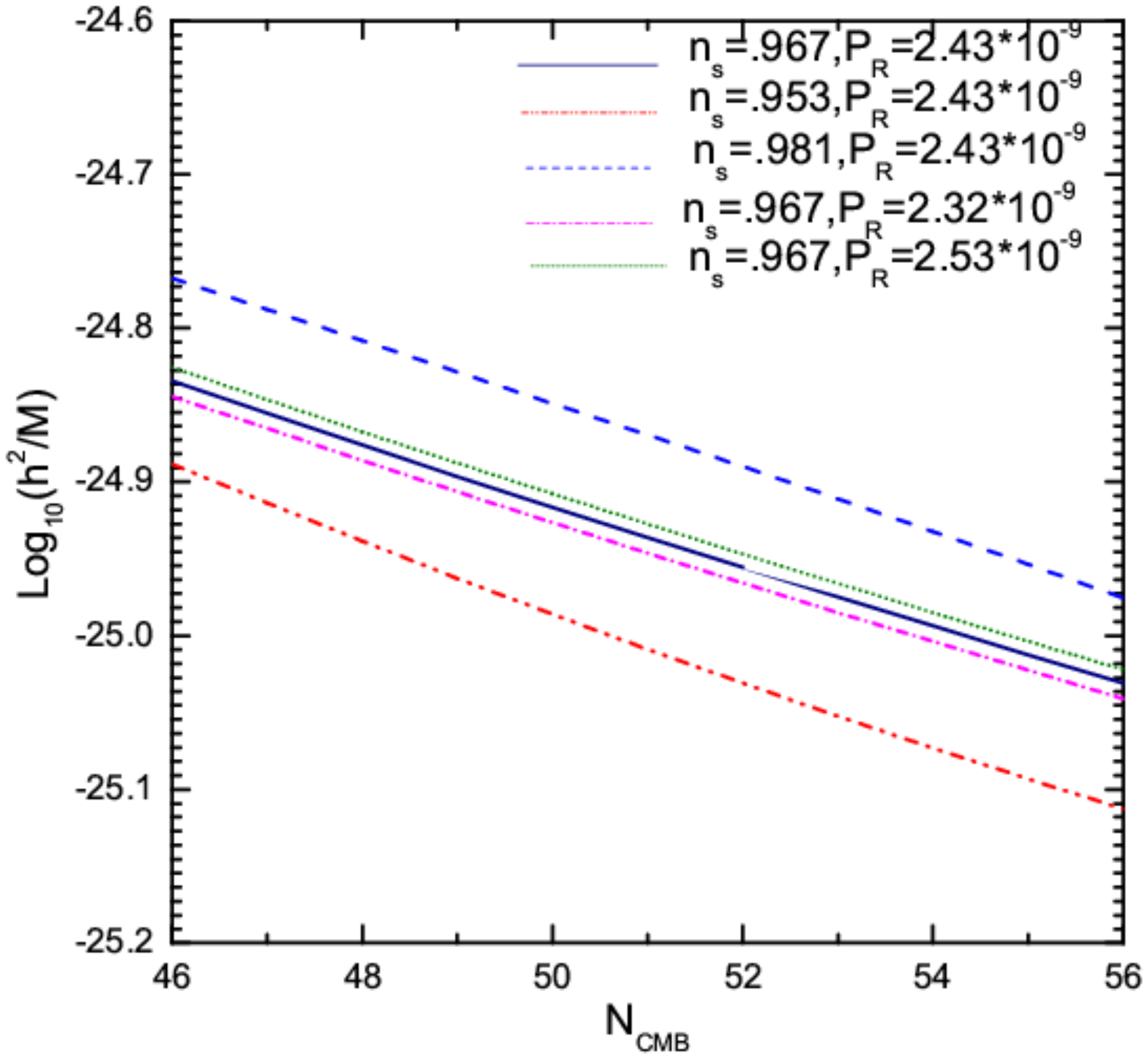}
\caption{Variation of exponent of $\frac{h^2}{M}$ with $N_{CMB}$
for different values of $n_s,P_R$.} \label{hsqbym}\end{figure}
\begin{figure}
\includegraphics[scale=0.8]{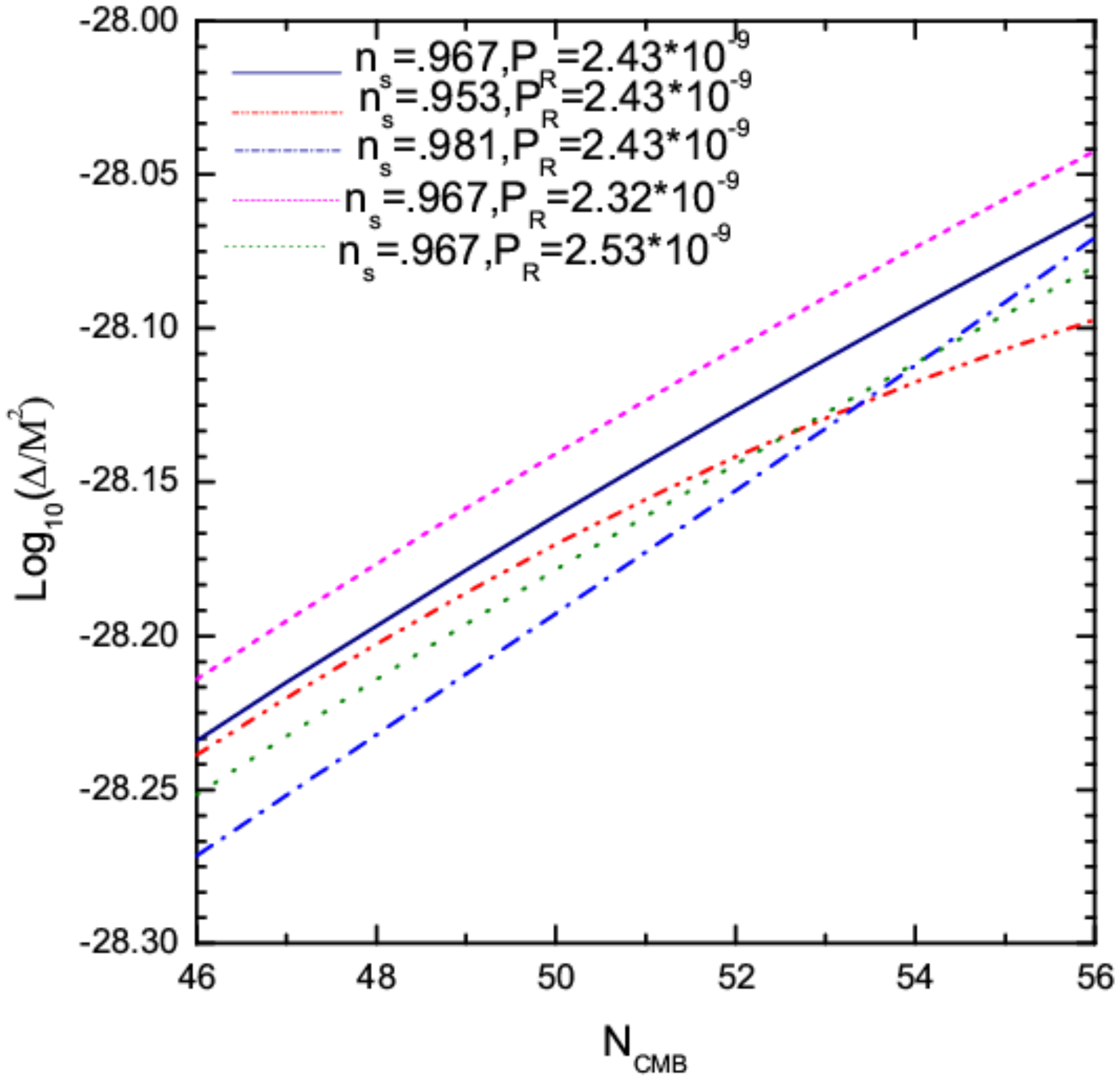}
\caption{Variation of exponent of $\frac{\Delta}{M^2}$ with
$N_{CMB}$ for different values of $n_s,P_R$.}
\label{delbymsq}\end{figure}
We estimate the maximum variations in
the exponents  which corresponds to the quoted errors in the WMAP
7- year data \cite{wmap} from the graphs in  Fig. \ref{hsqbym} and
Fig. \ref{delbymsq}. The formulae we derive are applicable to any
single inflaton theory with a renormalizable potential.

However a clearer qualitative understanding results from noticing
that for $N_{CMB} \sim 50$, $Z_{0}\approx \frac{1.2}{N_{CMB}}$
solves eqn.(\ref{NCMB}) to a good approximation. Then
eqn.(\ref{z}) gives \be \frac{h^2}{M} \approx \frac{3 \pi}{M_{P}}
\frac{\sqrt{P_R}}{N_{CMB}^2} \approx \frac{2.75 \times
{10^{-22}}}{N_{CMB}^2} \approx 10^{-25} GeV^{-1}\ee \be
\frac{\Delta}{M^2} \approx \frac{4.14 \times 10^{-34}}{N_{CMB}^2
P_R} \approx 10^{-28.2} GeV^{-2}\ee These are effectively the same
results that we derive from eq. \ref{hDeltavsM}.

Using these results on parameters of GRIPI, we can find
constraints on the parameters of potential required for Susy
seesaw inflation. The details are presented in next section.

\section{Supersymmetric seesaw inflaton model}

The requirements of the Supersymmetric seesaw inflation scenario
 can be fulfilled by considering a
model with gauge group $ SU(3)\times SU(2)\times U(1)_R\times
U(1)_{B-L} $ and the field content of the MSSM with some
additional superfields, right handed Neutrino chiral multiplet $N
[1,1,-1/2,1]$ and a field ${ S}[1,1,1,-2]$. The vev of  $S$ field
 generates the large Majorana masses $M_\nu$ ($10^6-10^{14}$ GeV) for the conjugate
neutrinos $\nu^c_A \equiv N_A$ through a renormalizable
superpotential coupling $3 \sqrt{2} f_{AB} S \nu^c_A \nu^c_B$.
Additional fields $\Omega_i$  are required which helps to fix
 the vev of $S$ as present in Minimal Supersymmetric
 Left Right Models (MSLRMs)\cite{MSLRM1,MSLRM2,MSLRM3} and in GUTs that embed them
\cite{rpar3,MSLRM1,MSLRM2,MSLRM3,nmsgut,nmsgutIII}. Soft
supersymmetry breaking terms are of the supergravity type (i.e
universal trilinear couplings, scalar masses, gaugino masses
except for Higgs (Non Universal Higgs Masses(NUHM) scenario). The
other essential component of the scenario are neutrino  Dirac mass
generating Yukawa couplings $y_{AB}, A,B=1,2,3$ in the
superpotential.  These couple the right handed neutrinos to the
Left chiral lepton doublets $L_A=\begin{pmatrix}\nu & e
\end{pmatrix}_A^T, A=1,2,3$. $L_A$ transform  as $L[1,2,0,-1]$ and
the up type Higgs doublet type field as $H[1,2,1/2,0]$ so that
$y_{AB}N L_A H_B$ is a gauge invariant term in the Superpotential.
Each doublet of this type present in the underlying theory must
have its complementary doublet transforming as e.g.
$\overline{H}$[1,2,-1/2,0] to cancel anomalies. The relevant flat
direction is assumed to extend out of the minimum of the
supersymmetric potential corresponding to the breaking of the
gauge group down to the MSSM symmetry  \bea SU(3)\times
SU(2)\times U(1)_R\times U(1)_{B-L} \rightarrow SU(3)\times
SU(2)\times U(1)_Y\eea This leads to a Type I seesaw
plus MSSM (SIMSSM) effective theory.\\
The fields $\tilde N$ (the chosen conjugate sneutrino ), $\tilde
\nu $( chosen left sneutrino flavor from a Lepton doublet L with
suitable Yukawa couplings) and the light neutral  Higgs field
(from the doublet $H$ with $Y=+1$) may be parameterized in terms
of the flat-direction associated with the gauge invariant $NLH$ as
\begin{equation}
\tilde{N}= \tilde{\nu}=h_0=\frac{\varphi}{\sqrt{3}}=\phi
e^{i\theta};~~~\phi\geq 0,~~ \theta\in[0,2 \pi)\, \label{inflaton}
\end{equation}
The additional fields $\Omega_i$, (for example a, $\omega$, p in
NMSGUT (discussed in chapter 2)), are assumed to be coupled to $S$
in such a way that extremization of the SUSY potential using
$F_{\Omega_i}=0,~~D_\alpha|_{\phi=0}=0 $ fixes the vev of S:
$<S>=\bar\sigma/\sqrt{2}$ without constraining the inflaton field
$\varphi$. This is also true in the Minimal Susy LR
models\cite{MSLRM1,MSLRM2,MSLRM3}.

The vanishing of the  $D$-term  for the $B-L$ generator requires
$\Omega_i$ to include the conjugate field(s) ${\overline{S}}
[1,1,-1, 2]$ which have a vev of equal magnitude as $S$ in order
to preserve SUSY through the symmetry breaking down to the MSSM
symmetry at high scales. This is also a feature of  MSLRMs and
R-parity preserving GUTs
\cite{MSLRM1,MSLRM2,MSLRM3,rpar3,aulmoh,ckn,abmsv,ag1,bmsv,nmsgut,nmsgutIII}.
 The gauge invariance of $NLH$ ensures that the $D$-terms for
 the flat direction vanish. Thus at scales $\phi\sim \bar\sigma>>M_{S}$
 where SUSY is exact and the relevant superpotential is given by:
\begin{equation}
    W=    3\sqrt{3} y N   \nu h+ 3 f{\sqrt{2}} S NN +...= y\varphi^3 +f{\sqrt{2}}
     S  \varphi^2 +...
\label{WSIMSSM}\end{equation}
where $ h,f,\bar\sigma$ can be taken real without loss of
generality. The right handed neutrino Majorana mass will be
$M_\nu=6f\sigb $.

Since  the equations of motion of the unperturbed vacuum imply
$<F_S>=0,\\ <S>= \bar\sigma/\sqrt{2}$ this superpotential leads to
a flat direction potential
\begin{eqnarray}
    V_{susy}&=&|3 y \varphi^2 + 2 f\bar\sigma\varphi|^2 +   2 |f\bar
    \varphi^2|^2\nonumber\\
    &=& f^2     \left[ (2+9{\tilde{y}}^2){\phi}^4 +12
    {\tilde{y}}{\phi}^3 \sigb\cos\theta + 4 \sigb^2{\phi}^2\right]
\end{eqnarray}
Here $\tilde{y} =y/f$  and it is clear from the potential that
$\sigb$ sets the inflaton mass scale. Minimizing with respect to
$\theta$ gives $\theta=\pi$. As long as we are interested only in
the dynamics of inflationary (once parameters are tuned to ensure
an inflection point in the flat region where $|\varphi| \sim
{\bar\sigma}$), we can concentrate on just the real part of
$\varphi$ and set $\varphi=-\phi$ with $\phi$ real and positive
near the inflection
 point but free to fall into the well around $\phi=0$  and
 oscillate around that value. The reason of taking real value is that
 the imaginary part $\phi'$ of $\varphi$ has a large curvature $V_{\phi'\phi'}\sim
 {\bar\sigma}^2 $ in the flat region. Since
 $V_{\phi'}|_{\phi'=0}=0$ it is viable to consider  the dynamics in the
 real $\varphi $ plane alone as a leading approximation. The
 effect of disturbance in the $\phi'$ direction when the dynamics is
 initiated with $\phi'\neq 0$ can be studied numerically as a
 correction to the dynamics of the inflaton field $\phi$.

 The susy breaking potential from the $\mu $ term for the Higgs doublets together soft quadratic and cubic terms,
  which we assume to be of the type generated by supergravity, but with non universal Higgs
masses, has form:
\begin{eqnarray}
    V_{soft} &=& \big [A_0 (y \varphi^3  + f \sqrt{2} S\varphi^2)+ h.c\big ] +m_{\tilde f}^2
    \sum_{\tilde f} | {\tilde f}|^2 + m_{ H}^2 |  H|^2  + m_{\bar H}^2 |\bar H|^2  \nnu
    &=& f^2    \left[\ytt\Azt\phi^3 \sigb \cos{3\theta} +  \Azt \sigb^2{ \phi}^2
    \cos{2\theta}+\mzt^2 \sigb^2 { \phi}^2 \right]
    \end{eqnarray}
here $\mzt=m_0/{f\sigb}, \Azt= 2 A_0/{f\sigb} $. The soft mass
 $m_0$ receives contributions from the sfermion and Higgs soft
masses as well as the $\mu$ term \bea m_0^2= \frac{2 m_{\tilde
f}^2 + \overline{m}_H^2}{3}\eea $m_{\tilde f,H}$ are the sfermion
and up type Higgs soft effective masses at the unification scale
($\overline{m}_H^2=m_H^2+|\mu|^2$). Since these masses and $A_0$
should be in the range $10^2-10^5$ GeV while the righthanded
neutrino masses lie in the range $10^{6}-10^{12}$ GeV, it is clear
that $\tilde m_0,\tilde A_0 $ are small parameters. Thus these
terms cannot significantly change $\theta=\pi$ assumed earlier.
The total inflaton potential is then
\eq{V_{tot}=f^2\left((2+9\tilde y^2) \phi^4-(\tilde A_0+12)\tilde
y\bar\sigma\phi^3 +  (\tilde A_0+\tilde
m_0^2+4)\bar\sigma^2\phi^2\right).\label{eq-inf-Vtot}} Thus we
have  a generic quartic inflaton potential of the same type as in
Section $\bf{2}$ but  the parameter values  in the case of Type I
seesaw are quite different from the light Dirac neutrino case.  We
  have the  identification
of parameters \bea h&=&f\sqrt{12(2 + 9 \ytt^2)}\nonumber\\
A&=&\frac{3 f (\Azt +12)\ytt \sigb}{\sqrt{(2 + 9
\ytt^2)}}\nonumber\\
M^2&=& 2 f^2 \sigb^2(4+\Azt +\mzt^2)\nonumber \\
\Delta&=& (1-\frac{A^2}{16 M^2})\nonumber\\
&=&\left(1-\frac{9\tilde y^2(\tilde A_0+12)^2}{32(2+9\tilde
y^2)(\tilde A_0+\tilde m_0^2+4)}\right)\label{paramident}\eea For
seesaw models the natural magnitude for the neutrino Dirac mass
is, $m_{\nu}^D
>1 MeV $ i.e $\, |y_\nu^D| > 10^{-5}$ and then the limit $m_{\nu}<<0.01 eV$
for the lightest neutrino (assuming direct hierarchy) implies
$M_{\nu^c} > 10^6$ GeV. The preferred values for the Susy breaking
scale are smaller than 100  TeV (at most) it follows that the
maximum value of $|\Azt|,|\mzt| \sim 0.1 $ and they could be much
smaller for more typical larger values of the conjugate neutrino
masses $M_{\nu^c} \sim 10^8 $ to $ 10^{12}$ GeV.  It is then clear
from the corresponding range $\Delta\sim 10^{-12} $ to $10^{-4}$
that the coupling ratio $\ytt=y/f$ becomes ever closer to exactly
$\ytt =4/3$ as M increases and even for $M\sim 10^6$ GeV differs
from $1.333 $ only at the second decimal place. Thus to a good
approximation $ h=6{\sqrt{6}} f $. Then it follows from
  the  Eqs.(\ref{hDeltavsM})and (\ref{paramident}) that \bea  f
  &\simeq &
10^{-26.83 \pm 0.17}(\frac{\overline{\sigma}}{GeV})\qquad;\qquad
 M \simeq 10^{-25.38 \pm 0.17}(\frac{\overline{\sigma}}{GeV})^2\nonumber\\
 \Delta &\simeq & 10^{-78.93 \pm 0.47} (\frac{\overline{\sigma}}{GeV})^4\eea
 The range $M\sim
10^{6.6}$ to $10^{10.6}$ GeV  corresponds nicely to  $10^{16}
~GeV< \sigb < 10^{18} ~GeV $:  as is natural in single scale Susy
SO(10)
GUTs\cite{rpar3,aulmoh,ckn,abmsv,ag1,bmsv,nmsgut,nmsgutIII}. $f$
increases with $\sigb$  with values above $10^{-11}$ for
$\bar\sigma > 10^{16} GeV$( which is however achievable in the
NMSGUT only with difficulty).  However in MSLRMs, since there are
no GUT constraints on $\sigb$, so one can have somewhat broad
ranges for these parameters.

 In all relevant cases $\Delta < 10^{-8}$ is required. Thus the above equations imply
that ${\tilde y}^2$ must be close to the value \be {{\tilde
y}_0}^2 = \frac{64}{9}  {\frac {4+\Azt +\mzt^2}{16-8\Azt-32
\mzt^2+\Azt^2}}\ee
  Here $\Azt,\mzt \sim O(M_S/M_{\nu^c})
<<1$,  hence $\tilde y_0$ is rather close to $4/3$ and the
equality is very close for larger $M\sim f\sigb$ since then
$\Azt,\mzt$ are very small.
  So this is the type of fine tuning that supports the advancement of
inflation in SIMSSM models. We see that the of severity of fine
tuning $\beta=\sqrt{\Delta} \sim 10^{-4} - 10^{-6}$ is much less
than the case of the MSSM Dirac neutrino inflaton since there
$\beta\sim 10^{-12} $ to $10^{-10}$ due to the low values of the
inflaton mass. Moreover the fine tuning condition is determined in
terms of superpotential parameters, which are radiatively stable
due to non renormalization theorems. Specially for large $\sigb >
10^{16} ~GeV$ the Type I Susy seesaw can provide a rather
attractive inflationary seesaw with a natural explanation for
neutrino masses and demands weaker tuning conditions on the
radiatively unstable Susy breaking parameters. Also unlike the
chaotic sneutrino inflaton scenario\cite{murayana,elliyana}, no
trans-Planckian vevs are invoked.

 We have viable inflation with M in range $10^{6}-10^{12}$ GeV as
 \be V_0
\sim \frac{M^4}{h^2}\sim (M)^3\times 10^{25} \, GeV \sim
10^{43}-10^{61} \, GeV^4\ee \be H_0 \sim \sqrt{\frac{V_0}{M_P^2}}
\sim 10^{3 }-10^{12} \, GeV \,\,\,\,;\,\,\,T_{max} \sim
V_0^{\frac{1}{4}} \sim 10^{11}-10^{15} \,GeV \ee It is important
to note that with maximum value of inflaton mass M=$10^{12}$ GeV
the ratio of tensor to scalar perturbations r=16 $\epsilon$
$\simeq$ 2(M/$10^{14} GeV)^3$ can be less than $10^{-6}$. It is
not compatible with the measurement of the tensor perturbations
via CMB polarization at r $\sim$ $10^{-3}$ level or larger.
However as M is raised above $10^{13}$
 GeV the assumptions of our analysis break down and the question of compatibility with BICEP2 results
 \cite{BICEP2} must be analyzed afresh. This will be discussed in later Chapters.
\section{Conclusions and discussion}
 In this chapter we showed how Supersymmetric Type-I seesaw models with
  the typical superpotential couplings found in MSLRMs and MSGUTs (NMSGUT) allow an attractive
   and natural implementation of renormalizable inflection point inflation.
    Inflation parameters are tied to seesaw parameter values and the required
     fine tuning is less severe and more stable than in the Dirac neutrino case since it is essentially
     independent of the supersymmetry breaking  parameters and is governed by the physics of
     intermediate scales $\sim 10^{6}- 10^{12} $GeV. In the Dirac
     neutrino case \cite{akm} the opposite it is true and the
     inflation occurs at low scales.

The post-inflationary reheating behavior (to be discussed in next
chapter in details) in the our model
   differs  from the Dirac neutrino case. In our case ( unlike \cite{akm}, where $B-L$ is just broken above the susy scale)
$B-L$ is not a gauge symmetry down to low energies. This can have
important consequences for nucleosynthesis and matter domination
since the heavy right handed neutrinos must find a non-gauge
channel to decay through. This channel can be a Yukawa coupling
since the right-handed neutrinos are singlets of the low energy
(SM) gauge group. The reheating can occur via a mechanism of
``instant preheating''\cite{feldkoflinde}. It ensures efficient
transfer of all
   the inflaton energy into thermalized MSSM plasma within few
   Hubble times after the end of inflation through oscillations  around its true minimum
   after slow roll of a flat direction. However a high
   reheat temperature $T_{rh}\sim 10^{11} - 10^{15} $ GeV  requires a gravitino mass larger than
   about $50\,$ TeV. This is necessary to make the inflation consistent with Nucleosynthesis.
    Such large Supersymmetry breaking scales are  also required by the
  NMSGUT to fit all the fermion data\cite{nmsgut}. The high reheat
  temperatures and the presence of the Higgs in the inflaton
  sit comfortably with the requirements of
  thermal\cite{elliyana} and non thermal Leptogenesis\cite{ahnkolb}.
\newpage

\chapter{Reheating in Supersymmertic Seesaw Inflation Scenario}
\section{Introduction}
Inhomogeneities produced during inflation are proposed to have
been the seeds for the growth of large scale inhomogeneities such
as galaxies observed today
\cite{primordstruct1,primordstruct2,primordstruct3}. After the
inflation the universe undergoes a phase of cooling. But we know
the present matter content can be explained very well by hot Big
Bang theory and its modelling of nucleosynthesis (helium formation
from hydrogen). So there must be some way that after the end of
inflation the inflaton dumps its energy into radiation and massive
particles. The process of converting inflaton energy into the
matter content is known as reheating (it involves particle
creation as well as  thermalization). The resulting temperature of
the universe is called ``reheating temperature''. The success of
any inflationary model depends on whether it can lead to Big-Bang
nucleosynthesis and explain the matter content of universe. For
this purpose one needs a reheating mechanism for inflaton to decay
into known particles in the inflationary model considered.

 In our
model of Supersymmetric seesaw inflation \cite{SSI} the inflaton
is a gauge invariant $D$-flat direction, $N H_{u}L$, where $N$ is
the right handed sneutrino, $H_{u}$ is the MSSM Higgs which gives
masses to the up-type quarks, and $L$ is the slepton field. So
their couplings (gauge and Yukawa) to the visible sector are
known. The mechanism of instant preheating \cite{feldkoflinde} can
work very well in such types of inflation scenarios. We followed
the procedure of previous work \cite{Rouzbh}, where authors have
studied the instant preheating in case of inflation along $LLe$
flat direction. Our scenario is very different in the sense that
in our case inflaton mass scale $\sim$ $10^{6}-10^{12}$ GeV (right
handed neutrino mass) while in their case inflaton mass is $\sim$
0.1-10 TeV (Susy breaking scale). Large inflaton mass in our case
results into high reheat temperature $\sim$ $10^{11}-10^{13}$ GeV.

\section{Instant preheating}
The idea of instant preheating \cite{feldkoflinde} is based upon
the non-perturbative decay of inflaton ( into the bosons and
fermions -collectively called  $\chi$ particles- ) it is directly
coupled to when it is close to the minimum of the potential (at
$\phi$ =0). The $\chi$ particles decay further to  other light
MSSM modes   not coupled directly to the inflaton. This happens
before the inflaton reaches its minimum again because of the large
decay width of the $\chi$ particles when $\phi$ is near maximum
magnitude in its oscillation. A well known  mechanism of
non-perturbative particle production called  parametric resonance
was given in \cite{kolflindebinsky,Brandenberger}. In the case of
parametric resonance several oscillations are required to increase
the energy transfer from inflaton to decay products as only a very
tiny fraction of inflaton energy $\sim$ $10^{-2}g^2$ is
transferred in each oscillation. But in case of instant
preheating, if the $\chi$ particles have strong  interaction with
light modes  which are not directly coupled to inflaton (i.e other
MSSM particles collectively called $\psi$ modes)  then they can
  decay before the inflaton returns  to the potential minimum and  the parametric
resonance will not occur. This does  happen because the mass of
$\chi$ modes depends upon the instantaneous inflaton field vev. At
the time of their production, their mass is zero as $\phi$=0, but
as inflaton rolls back to its maximum value
 the $\chi$ modes become heavy. When they have maximum decay width
 they decay rapidly into  $\psi$ modes.
 With this kind of chain reaction one has  an efficient way to transfer the whole
energy of inflaton into   relativistic particles within  few
oscillations. This process called ``instant preheating".
  $\psi$ particle creation is followed by their   thermalization,
  resulting the final ``reheated'' state.
\section{Supersymmetric seesaw inflaton}
\begin{figure}
\centering
\includegraphics[scale=1.2]{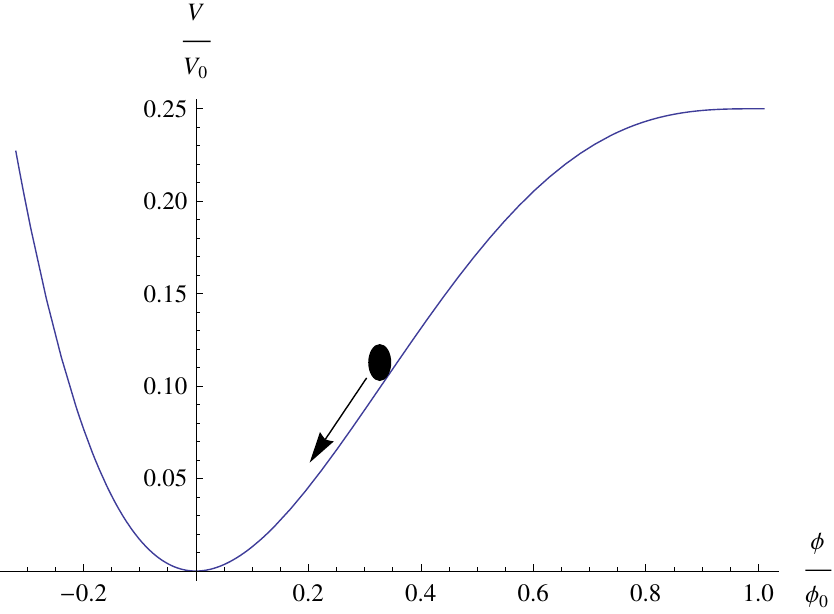}\caption{Rolling of inflaton around minimum of its potential after inflaton.}
\label{poten}\end{figure}

In Supersymmetric Seesaw inflation (SSI) the flat direction is
gauge invariant combination of $\tilde{L}_1$, $H_{u}$,
$\tilde{\nu}^c_3$. The flat direction corresponds to

\bea H_u = \begin{pmatrix} 0 \cr \varphi \end{pmatrix};\quad
\tilde L = \begin{pmatrix} \varphi \cr 0 \end{pmatrix};\quad
\tilde{\nu}^c = \varphi \eea Here $\varphi$ is complex scalar
field and inflaton  $\phi$ corresponds to its real part.
  The inflationary potential has form given in \cite{SSI}:
 \bea V(\phi)=\frac{M^2}{2} \phi^2-\frac{2h}{3 \sqrt{3}} \phi^3+\frac{h^2}{12}\phi^4 \eea
 Here $\phi$ is the inflaton field. At inflection point the value of field is $\phi_0 = \frac{\sqrt{3}M}{h}$ if A$\approx$4M.
 The potential given above can be reparameterized as
 \bea \tilde{V}=\frac{V}{V_0}= \frac{3}{2}\tilde{\phi}^2-2
 \tilde{\phi}^3+\frac{3}{4}\tilde{\phi}^4 \eea
 Where $\tilde \phi= \frac{\phi}{\phi_0}$ and $V_0=\frac{M^4}{h^2}$
 is value of potential energy at inflection point.
A pictorial view of the  inflaton rolling  towards its minimum
after inflation is shown
 in figure \ref{poten}.

 \section{Finding the $\chi$ modes}
 The $\chi $ modes have both gauge and Yukawa couplings but except for the top
 Yukawa (and not even that when $\tan\beta$ is large)
 the gauge couplings dominate the Yukawa couplings. So for simplicity
 we   ignore  the Yukawa coupling contributions  and consider  only gauge terms.
 The first step is to calculate the spectrum in which inflaton
field will decay whenever it crosses the origin i.e. when
$\phi=0$. To calculate this spectrum we need all the terms from
MSSM lagrangian which contains $\tilde{L}$, $H_{u}$,
$\tilde{\nu}^c$ fields. These include interaction with gauge
bosons, fermions and scalars. We calculate them one by one as
follows: \bea H_u =
\begin{pmatrix}H_u^+ \cr H_u^0 \end{pmatrix}=\begin{pmatrix}\phi_1
\cr \phi_2 \end{pmatrix}; \,\,\,\,\,
 \tilde L = \begin{pmatrix}\tilde{\nu} \cr \tilde{e} \end{pmatrix}= \begin{pmatrix}\phi_3 \cr \phi_4 \end{pmatrix};
 \,\,\,\,\, \tilde{\nu}^c = \phi_5\eea
 Here all the $\phi$'s are complex fields given by $\phi=\frac{\phi_R+i\phi_I}{\sqrt{2}}$.
 The inflaton is given by
 \bea
 \varphi= \frac{\phi_{2R}+\phi_{3R}+\phi_{5R}}{\sqrt{3}}\eea
  \subsection{Gauge interactions}
  The gauge interaction terms come from the gauge terms of fields $\tilde{L}$, $H_{u}$,$\tilde{\nu}^c$ which form the inflaton.
  If we recall the field content of SSI then the required fields along with these are S[1,1,1-2] and $\bar{S}$[1,1,-1,2].
  In this case symmetry breaks as:
 \bea SU(3)_C \times SU(2)_L \times U(1)_{R} \times U(1)_{B-L} \rightarrow SU(3)_C \times U(1)_Q \eea
 Here $S$ field gets vev $\sigma$ which breaks the $U(1)_{B-L}$ symmetry.
 The background inflaton field breaks the $SU(2)_L$ symmetry.
The relevant kinetic terms are:
  \bea L_{kinetic}&=& (D_{\mu}H_u)^{\dag}D^{\mu}H_u+(D_{\mu}\tilde L_1)^{\dag}D^{\mu}\tilde L_1
  +(D_{\mu}\tilde{\nu}^c_3)^{\dag}D^{\mu}\tilde{\nu^c}_3+ (D_{\mu}S)^{\dag}D^{\mu}S+
  (D_{\mu}\bar{S})^{\dag}D^{\mu}\bar{S}\nonumber\eea
  Where
  \bea D_{\mu}H_u&=& (\partial_{\mu}- \frac{i}{2}g_{R} C_{\mu}-\frac{i}{2}g_W \sum_{a=1}^3 W_{\mu}^a T^a)H_u\nonumber\\&&
  D_{\mu}\tilde L_1= (\partial_{\mu}+ \frac{i}{2}g_{B-L} E_{\mu}-\frac{i}{2}g_W \sum_{a=1}^3 W_{\mu}^a T^a)\tilde L_1\nonumber\\&&
  D_{\mu}\tilde{\nu}^c_3= (\partial_{\mu}+\frac{i}{2}g_{R} C_{\mu}-\frac{i}{2} g_{B-L} E_{\mu})\tilde{\nu}^c_3\nonumber\\&&
  D_{\mu}S= (\partial_{\mu}-ig_{R} C_{\mu}+ig_{B-L} E_{\mu})S \nonumber\\&&
  D_{\mu}\bar{S}= (\partial_{\mu}+ig_{R} C_{\mu}-ig_{B-L} E_{\mu})\bar{S} \eea
Here $g_w, g_{B-L},g_{R}$ are the gauge couplings and $W_{\mu}^a$, $E_{\mu}$, $C_{\mu}$
are gauge fields corresponding to gauge group $SU(2)_L$, $U(1)_{B-L}$, $U(1)_R$ respectively.
$T^a$ are SU(2) generators:
 \bea T_1=\frac{1}{2}\begin{pmatrix}0 & 1 \cr 1 & 0 \end{pmatrix};\,\,\,\
 T_2=\frac{1}{2}\begin{pmatrix}0 & -i \cr i & 0 \end{pmatrix};\,\,\,\,\
 T_3=\frac{1}{2}\begin{pmatrix}1 & 0\cr 0 & -1 \end{pmatrix}\nonumber\eea
  In terms of inflaton vev and vev for $S$ and $\bar{S}$ fields given,
  \bea <H_u>=\begin{pmatrix} 0 \cr \frac{\varphi}{\sqrt{3}} \end{pmatrix};
  <\tilde L_1>= \begin{pmatrix}\frac{\varphi}{\sqrt{3}}  \cr 0 \end{pmatrix} ;
   <\tilde{\nu}^c_3> =\frac{\varphi}{\sqrt{3}} \\\nonumber
  <S>=\frac{\sigma}{\sqrt{2}};  <\bar S>=\frac{\bar \sigma}{\sqrt{2}} \eea
  one can find the massive gauge modes which will appear due to symmetry breaking.
  According to Higgs mechanism four massive
  gauge bosons will appear. One gets the following mass terms:
\bea L_{kinetic}&=& \frac{1}{12}(g_R C_{\mu}-g_W W_{\mu}^3)^2 \varphi^2 + \frac{1}{3}g_W^2 W_{\mu}^+W^{\mu -} \varphi^2
+\frac{1}{12}(g_{B-L} E_{\mu}-g_W W_{\mu}^3)^2 \varphi^2\nonumber\\&&+
(g_R C_{\mu}-g_{B-L} E_{\mu})^2 (\frac{\varphi^2}{12}+|\sigma|^2)
 \nonumber \label{weak}\eea
 However in this basis of gauge fields, they mix with each other.
 Since the mass matrix is symmetric
  so one needs to find a orthogonal matrix which diagonalize
  the mass matrix of gauge bosons.
  We found that for three different gauge couplings the expressions
   for gauge fields in mass basis
  are quite complicated. For simplification we chose NMSGUT relation
    between gauge couplings:
            $g_{W}=g_{R}=g$ and $g_{B-L}=\sqrt{\frac{3}{2}}g $
 Here g is the $SO(10)$ gauge coupling in the standard unitary normalization.
 After taking these   relations   eq. (\ref{weak})
 can be written as
 \bea L_{kinetic}&=& \frac{1}{3}g^2 W_{\mu}^+W^{\mu -} \varphi^2+ \frac{1}{2}M_{X'}^2 {X'}^2+\frac{1}{2}M_{Z'}^2 Z^{'2}\eea
 Where \bea M_{X'}^2=(14+5a+p)g^2 \frac{\varphi^2}{24}  ;\,\,\,\,\,\,\ M_{Z'}^2=(14+
 5a-p)g^2 \frac{\varphi^2}{24}\eea
 Where $a=\frac{12|\sigma|^2}{\varphi^2}$ and $p=\sqrt{14+5a+25a^2}$.
 $X'_{\mu}$ and $Z'_{\mu}$ are liner combination
 of $X_{\mu}$, $Z_{\mu}$ with coefficients which are functions of $a$ and $p$.
 \bea X_{\mu}'&=& cos\theta X_{\mu}+
 sin\theta Z_{\mu}\nonumber\\
 Z_{\mu}'&=& -sin\theta X_{\mu}+
  cos\theta Z_{\mu}\eea
  Where tan$\theta$= $(6+25a-5p)/8$. $X_{\mu}$ =$\sqrt{2/5}
  C_{\mu}-\sqrt{3/5} E_{\mu}$ is the gauge field which
 gets a large mass when S and $\bar{S}$ field get vevs.
 The orthogonal combination of $X_{\mu}$, say $B_{\mu}$
  (represents gauge field for hypercharge Y/2) mixes with $W_{\mu}^3$ to give the
  $Z_{\mu}$=$\sqrt{\frac{3}{8}} B_{\mu}-\sqrt{\frac{5}{8}} W_{\mu}^3$.
 The orthogonal combination to $Z_{\mu}$ is the photon field $A_{\mu}$
  which remains massless.
 \be  A_{\mu}=\sqrt{\frac{5}{8}} B_{\mu}+\sqrt{\frac{3}{8}} W_{\mu}^3\ee
 Now the masses of $X'$ and $Z'$ contain terms with $|\sigma|^2$ .
 In the SSI scenario  $|\sigma|$  doesn't vary with time,
  so these fields will remain heavy during the oscillation of inflaton.
  They don't participate in the preheating, so only the charged gauge
  bosons are of interest to us.

The massive charged $W_{\pm}$ gauge fields have interaction terms
with scalars through their kinetic terms. The gauge fields decay
into squarks, sleptons and Higgs fields. The relevant terms are:
 \bea L &\supset& \frac{i}{\sqrt{2}} g (\partial \tilde{u}_i W_{\mu}^+ \tilde{d}_i^* -\partial \tilde{d}_i^* W_{\mu}^+
 \tilde{u}_i)+h.c.\nonumber\\&&
 +\frac{i}{\sqrt{2}} g (\partial \tilde{\nu}_{i} W_{\mu}^+ \tilde{e}_{i}^* -\partial \tilde{e}_{i}^* W_{\mu}^+
 \tilde{\nu}_{i})+h.c.\nonumber\\&&
 +\frac{i}{\sqrt{2}} g (\partial H^0_d W_{\mu}^+ H^{-*}_{d} -\partial H^{-*}_{d} W_{\mu}^+
 H^0_d)+h.c.\nonumber\\&&
 +\frac{i}{\sqrt{2}} g (\partial H^+_u W_{\mu}^+ H^{0*}_{u} -\partial H^{0*}_{u} W_{\mu}^+
 H^+_u)+h.c.\nonumber \eea
 Here i runs over the family index. The decay rate for decay
  of a gauge boson with mass
 $M$ to a pair of massless scalars coupled as above is $g^2 M/96\pi$.\\
Similarly the interaction terms of massive gauge fields with
fermions are given by: \bea L &\supset& \frac{i}{\sqrt{2}} g
(u_i^{\dag}\bar{\sigma}^{\mu} W_{\mu}^+ d_i
+\nu_i^{\dag}\bar{\sigma}^{\mu} W_{\mu}^+ e_i)+h.c.\nonumber\\&&
 +\frac{i}{\sqrt{2}} g (\tilde H^0_d \bar{\sigma}^{\mu}W_{\mu}^+ \tilde H^{-}_{d} +\tilde H^{+}_{u} \bar{\sigma}^{\mu} W_{\mu}^+
 \tilde H^0_u)+h.c.\nonumber \eea
 The decay rate for decay of a gauge boson with mass
  $M$ to a pair of massless fermions is $g^2 M/48\pi$.
 The decay rates for $W_{\mu}^{\pm}$ after summing over all channels is given
 by
 \bea \Gamma_{W_{\mu}^{\pm}}=\frac{7 g^3 \phi(t)}{16 \sqrt{3}\pi} \eea

 \subsection{Scalar  interactions}
 Due to breaking of symmetry 2 degrees of freedom of neutral and 2 charged scalars are
 eaten by gauge modes so we are left with four massive scalar fields whose
  masses can be obtained from D-terms of the MSSM potential:
 \bea
 V_D(\phi,\phi^*)&=&\frac{1}{2}\sum_a D^a D^a = \frac{1}{2}(\sum_{a=1}^3 D^a D^a + D^R D^R+D^{B-L}D^{B-L})\\\nonumber&&
 = \frac{1}{2}\sum_{a=1}^3 g_a^2 (\phi^* T^a \phi)^2+g_R^2 (\phi^* T^R \phi)^2+g_{B-L}^2 (\phi^* T^{B-L} \phi)^2\eea

  Now from the D-terms we concentrate on terms linear in the
  inflaton  so that square of this term
   gives us the mass of scalar field and cross terms with bilinears
   give the couplings responsible for the decay of this
    massive scalar to massless scalars.
     We consider only trilinear terms as they are dominant
     over the quartic ones. For example $D_1$ is given as
  \bea D_1 &=& H_u^* T^1 H_u + \tilde{L}^*_1 T^1 \tilde L_1+ \tilde{Q}^*_i T^1 \tilde{Q}_i
  +\tilde{L}^*_{2,3} T^1 \tilde{L}_{2,3}+
\tilde{H}^*_d T^1 \tilde{H}_d \\\nonumber&&
  = \frac{1}{2}(\phi_2^* \phi_1+\phi_1^* \phi_2+\phi_4^* \phi_3+\phi_3^* \phi_4)+
   \tilde{Q}^*_i T^1 \tilde{Q}_i+\tilde{L}^*_{2,3} T^1 \tilde{L}_{2,3}+
H^*_d T^1 H_d \eea After substituting $\phi_{2}=\hat{\phi}_{2}
+\phi$ and $\phi_{3}=\hat{\phi}_{3} +\phi$, where fields with
``$hat$'' represent the fluctuation part,  we get: \bea D_1&=&
\frac{1}{2}(\hat{\phi}_{1}^- +\hat{\phi}_{1}^+ +\hat{\phi}_{4}^-
+\hat{\phi}_{4}^+)\frac{\varphi}{\sqrt{3}}+\tilde{Q}^*_i T^1
\tilde{Q}_i+\tilde{L}^*_{i} T^1 \tilde{L}_{i}+ H^*_d T^1 H_d+
H^*_u T^1 H_u \nonumber\eea Similarly from other D-terms \bea
D_2&=& \frac{-i}{2}(\hat{\phi}_{1}^- -\hat{\phi}_{1}^+
+\hat{\phi}_{4}^- -\hat{\phi}_{4}^+) \frac{\varphi}{\sqrt{3}}+
\tilde{Q}^*_i T^2 \tilde{Q}_i+\tilde{L}^*_{i} T^2 \tilde{L}_{i}+
H^*_d T^2 H_d+ H^*_u T^2 H_u \nonumber\\
D_3&=&\frac{(\hat{\phi}_{3R}-\hat{\phi}_{2R})}{\sqrt{2}}\frac{\varphi}{\sqrt{3}}+
\tilde{Q}^*_i T^3 \tilde{Q}_i+\tilde{L}^*_{i} T^3 \tilde{L}_{i}+
H^*_d T^3 H_d+ H^*_u T^3 H_u \nonumber\eea \bea D_R&=&
\frac{(\hat{\phi}_{2R}-\hat{\phi}_{5R})}{\sqrt{2}}\frac{\varphi}{\sqrt{3}}
+\frac{1}{2}(|H_u|^2-|H_d|^2-|\tilde{\nu}^c_{i}|^2
-|\tilde{u}^c_{i}|^2+|\tilde{d}^c_{i}|^2+|\tilde{e}^c_{i}|^2)\nonumber\\&&
 D_{B-L}=
\frac{(\hat{\phi}_{5R}-\hat{\phi}_{3R})}{\sqrt{2}}\frac{\varphi}{\sqrt{3}}
-\frac{1}{2}(|L_{i}|^2-|\tilde{\nu}^c_{i}|^2-\frac{1}{3}|\tilde{Q}_i|^2+\frac{1}{3}
|\tilde{u}^c_{i}|^2+\frac{1}{3}|\tilde{d}^c_{i}|^2-|\tilde{e}^c_{i}|^2)
\eea The MSSM potential corresponding to these D-terms is given by
 \bea V&=&g^2\biggr(\varphi^2 \frac{\chi^+\chi^-}{3}+\varphi^2 \frac{\chi_1^2}{4}
 + \varphi^2\frac{\chi_2^2}{3}\nonumber\\&&
+ \frac{ \varphi \chi^+}{\sqrt{6}}(\tilde{Q}^*_i (T^1+iT^2)
\tilde{Q}_i+\tilde{L}^*_{i} (T^1+iT^2) \tilde{L}_{i}+ H^*_d
(T^1+iT^2)H_d+H^*_u (T^1+iT^2)H_u)\nonumber\\&&+ \frac{ \varphi
\chi^-}{\sqrt{6}}(\tilde{Q}^*_i (T^1-iT^2)
\tilde{Q}_i+\tilde{L}^*_{i} (T^1-iT^2) \tilde{L}_{i}+ H^*_d
(T^1-iT^2) H_d+H^*_u (T^1-iT^2)H_u)\nonumber\\&&- \frac{ \varphi
\chi_1}{\sqrt{3}}(\tilde{Q}^*_i T^3 \tilde{Q}_i+\tilde{L}^*_{i}
T^3 \tilde{L}_{i}+ H^*_d T^3 H_d+H^*_u T^3 H_u)\nonumber\\&&-
\frac{\varphi \chi_1}{4
\sqrt{3}}(|H_d|^2-|H_u|^2+|\tilde{L}_{i}|^2+\frac{4}{3}|\tilde{u}^c_{i}|^2
-\frac{2}{3}|\tilde{d}^c_{i}|^2-\frac{1}{3}|\tilde{Q}_i|^2-2|\tilde{e}^c_{i}|^2)\nonumber\\&&-
\frac{\varphi \chi_2}{2
\sqrt{3}}(|H_d|^2-|H_u|^2-|\tilde{L}_{i}|^2+\frac{2}{3}|\tilde{u}^c_{i}|^2
-\frac{4}{3}|\tilde{d}^c_{i}|^2+\frac{1}{3}|\tilde{Q}_i|^2+2|\tilde{\nu}^c_{i}|^2)\nonumber\\&&
+\frac{1}{4}\chi_1^2 (-|H_d|^2-2|\tilde{\nu}^c_{i}|^2+|\tilde{L}_{i}|^2-\frac{2}{3}|\tilde{u}^c_{i}|^2
+\frac{4}{3}|\tilde{d}^c_{i}|^2-\frac{1}{3}|\tilde{Q}_i|^2)\nonumber\\&&
+\frac{5}{16}\chi_2^2(|H_d|^2+2|\tilde{\nu}^c_{i}|^2-|\tilde{L}_{i}|^2+\frac{2}{3}|\tilde{u}^c_{i}|^2
-\frac{4}{3}|\tilde{d}^c_{i}|^2+\frac{1}{3}|\tilde{Q}_i|^2)\biggr)
\eea where \bea \chi^-= \frac{\hat{\phi}_{1}^-
+\hat{\phi}_{4}^-}{\sqrt{2}},\,\,\,\,\, \chi^+=
\frac{\hat{\phi}_{1}^+
 +\hat{\phi}_{4}^+}{\sqrt{2}},\,\,\,\,\, \chi_1 =
 \frac{(\hat{\phi}_{2R}-\hat{\phi}_{3R})}{\sqrt{2}}\,\,\,\,\,\,
\chi_2 =
\frac{\hat{\phi}_{2R}+\hat{\phi}_{3R}-\sqrt{6} \hat{\phi}_{5R}}{\sqrt{8}}\nonumber\eea
The decay rate for decay of a massive scalar with mass M into a pair of massless scalars
is $\sigma^2/16 \pi M$. Here $\sigma$ is the coupling of massive scalar with massless scalar.
The decay rates for $\chi^{\pm},\chi_1,\chi_2$ after summing
over all channels is given by
 \bea \Gamma_{\chi^{\pm}}=\frac{7 g^3 \phi(t)}{16 \sqrt{3}\pi};\,\,\,\,\,\
 \Gamma_{\chi_1}= \frac{67 g^3 \phi(t)}{96 \sqrt{2}\pi}\,\,\,\,\,\
 \Gamma_{\chi_2}=\frac{11 g^3 \phi(t)}{16 \sqrt{6}\pi}\eea

\subsection{Fermion interactions}
The fermion interactions of inflaton can be found from the
following part of MSSM lagrangian
\bea L &\supset& -\sqrt{2}g_W \sum_{a=1}^3(\tilde L_1^* \tilde W_a T^a L_1 + H_u^* \tilde W_a T^a \tilde H_u)
- \sqrt{2}g_R(S^* \tilde C \tilde S -\bar{S}^* \tilde C \tilde{\bar{S}}-
 \frac{1}{2}\tilde{\nu}^{c*}_3 \tilde C \nu_3^c \nonumber\\&&+ \frac{1}{2} H_u^* \tilde C \tilde H_u)-
 \sqrt{2}g_{B-L}(-S^* \tilde E \tilde S + \bar{S}^* \tilde E \tilde{\bar{S}}+
  \frac{1}{2}\tilde{\nu}^{c*}_3 \tilde E \nu_3^c- \frac{1}{2} \tilde{L}_1^* \tilde E L_1)\eea
  Putting vevs for inflaton fields, S and $\bar{S}$, we get the mass terms for
  fermions.
\bea L &\supset& - \frac{g_W \varphi}{\sqrt{6}}(\tilde W^3 \nu_1 + \sqrt{2} \tilde W^+ e_1
 +\sqrt{2} \tilde W^- \tilde H_u^+ -\tilde W^3 \tilde H_u^0 )
 -\frac{g_R \varphi}{\sqrt{6}}(\tilde H_u^0-\nu_3^c )\tilde{C} \nonumber \\&&
 -g_{B-L}(\nu_3^c-\nu_1)\tilde{E}\frac{ \varphi}{\sqrt{6}} -g_{R}(\tilde{S}-\tilde{\bar{S}})\tilde{C} |\bar{\sigma}|
  -g_{B-L}(-\tilde{S}+\tilde{\bar{S}})\tilde{E} |\bar{\sigma}|\nonumber\\&&
 =  -\frac{g_W \varphi}{\sqrt{3}}( \tilde W^+e_1 + \tilde W^- \tilde H_u^+ )-
   (g_W \tilde W^3- g_{B-L} \tilde{E}) \nu_1\frac{ \varphi}{\sqrt{6}}
   -(g_{R} \tilde{C}-g_W \tilde W^3)\tilde{H}_u^0 \frac{ \varphi}{\sqrt{6}}
   \nonumber\\&&-(g_{B-L} \tilde{E}-g_R \tilde C )\nu_3^c \frac{
   \varphi}{\sqrt{6}}-(g_R \tilde C- g_{B-L}
   \tilde{E})|\bar{\sigma}|(\tilde{S}-\tilde{\bar{S}})
   +H.C.\nonumber\\&&
   = -\frac{g \varphi}{\sqrt{3}}( \tilde W^+e_1+\tilde W^-\tilde H_u^+
   )+ g\frac{ \varphi}{\sqrt{6}} \frac{\alpha^2+2\sqrt{a}}{4 \sqrt{3a}} \tilde X' \psi_3 +
    g\frac{ \varphi}{\sqrt{6}} \frac{\alpha^2-2\sqrt{a}}{4 \sqrt{3a}}\tilde Z'\psi_4
\nonumber\eea Where \bea \psi_3 &=& \nu_1-\frac{2
\sqrt{3}}{\alpha+ 2\sqrt{a}}\tilde{H}_u^0+\frac{4
\sqrt{3a}}{\alpha^2+2 \sqrt{a} \alpha}\nu_3^c+
\frac{2 \sqrt{6}a}{\alpha^2+2\sqrt{a} \alpha}(\tilde S-\tilde{\bar{S}})\nonumber\\
\psi_4 &=& \nu_1-\frac{2 \sqrt{3}}{\alpha-
2\sqrt{a}}\tilde{H}_u^0+\frac{4 \sqrt{3a}}{\alpha^2-2 \sqrt{a}
\alpha}\nu_3^c+ \frac{2 \sqrt{6}a}{\alpha^2-2\sqrt{a}
\alpha}(\tilde S-\tilde{\bar{S}})\nonumber\eea With $\alpha= 2
\sqrt{\sqrt{3a}+a}$. In form of Dirac spinor the $\chi$ modes are
\bea \Psi_1=  \begin{pmatrix}e_1 \cr \tilde{W}_{\mu}^{+*}
\end{pmatrix} ;\quad \Psi_2=  \begin{pmatrix} \tilde H_u^+ \cr
\tilde{W}_{\mu}^{-*} \end{pmatrix}; \Psi_3=
\begin{pmatrix}\psi_3 \cr \tilde{Z} \end{pmatrix};\quad \Psi_4=
\begin{pmatrix}\psi_4\cr \tilde{Z}' \end{pmatrix}\eea
The Dirac fermions $\Psi_3,\Psi_4$ are heavy as their masses
contain $|\sigma|$ which doesn't vary during oscillations, so they
don't participate in reheating. So we are left with
$\Psi_1,\Psi_2$ only and the interaction terms for these fermion
$\chi$ particles are given as: \bea L &\supset& -\sqrt{2}g
\biggr(\tilde U_{i}^{*}\Psi_1 P_L D_i +\tilde L_{i}^{*}\Psi_1 P_L
N_i
 + H^{0*}_d \Psi_1 \tilde H^{-}_{d} + H^{+*}_{u} \Psi_1 \tilde H^0_u+h.c.\nonumber\\&&
 +\tilde D_{i}^{*}\Psi_2 P_L U_i +\tilde N_{i}^{*}\Psi_2 P_L L_i
 + H^{-*}_d \Psi_2 \tilde H^{0}_{d} + H^{0*}_{u} \Psi_2 \tilde H^+_u+h.c.
 \biggr)\nonumber \eea
The decay rate for decay of a massive fermion with mass M into a
massless scalar and
 its fermionic partner is $g^2 M/32 \pi$.
The decay rates for $\Psi_{1,2}$ after summing over all channels
are
 \bea \Gamma_{\Psi_{1,2}}=\frac{7 g^3 \phi(t)}{16 \sqrt{3}\pi} \eea
 Once we have  all the $\chi$ modes and their decay
  rates (given in table \ref{modes}) we can estimate instant
 preheating in our model.
 \begin{table}[h]
 $$
 \begin{array}{|c|c|c|c|}
 \hline {\mbox {$\chi$ mode } }&{\rm Mass}&{\rm Decay ~Rate}&{ \rm Degrees ~of ~freedom}\\
 \hline
 W_{\mu}^{\pm}& 1/\sqrt{3}g \phi(t) & 7 g^3\phi(t)/16 \sqrt{3}\pi & 6\\
 \chi^{\pm}&  1/\sqrt{3}g \phi(t) & 7 g^3\phi(t)/16 \sqrt{3}\pi & 2\\
 \chi_1 & 1/\sqrt{2}g \phi(t) & 67 g^3\phi(t)/96 \sqrt{2}\pi & 1\\
 \chi_{2}& \sqrt{2/3}g \phi(t) & 11 g^3\phi(t)/16 \sqrt{6}\pi & 1\\
 \Psi^{\pm}& 1/\sqrt{3}g \phi(t) & 7 g^3\phi(t)/16 \sqrt{3}\pi & 8\\
 \hline
 \end{array}$$
 \caption{The $\chi$ modes produced due to non-perturbative decay of
 inflaton.}\label{modes}\end{table}

 \section{Instant Preheating in context of SSI}
 After the end of inflation the inflaton starts
 oscillating around the minimum of potential and
 decays to the $\chi$ modes we calculated in the previous section.
  The mass of these $\chi$ modes is proportional to
 time dependent value of the  inflaton. Let us first consider the charged
 $\chi$ type modes which include charged scalars,
 fermions and gauge bosons (we collectively called them $\chi_1$ modes).
  The energy of such fields is given by
\bea \omega_k &=& \sqrt{k^2+\frac{g^2 \phi(t)^2}{3}}\eea
 It is
convenient to write this equation in terms a of parameter $Q_1$
(resonance parameter) \bea \omega_k &=&\sqrt{k^2+4 M^2 \tau^2
Q_1}; \quad\quad Q_1 = \frac{g^2\dot{\phi}_0^2}{12 M^4} \eea and
$\tau=MT$
 measures the time period of each oscillation in the units of $M^{-1}$.
  Here $\dot{\phi}_0$ is the velocity of
inflaton at the time of zero crossing. The $\chi_1$ modes are
produced as inflaton moves towards origin and adiabaticity
condition is violated i.e. $\dot{\omega_k}
>> \omega_{k}^2$. This condition is true for those modes having
momentum less than, \bea k^2_{max} \simeq g
\frac{\dot{\phi}_0}{\sqrt{3}}\eea The number density of $\chi_1$
modes with momentum $k$ is \cite{kolflindebinsky} \bea n_{\chi_1
,k}&=& exp[\frac{-\pi \sqrt{3}k^2}{g \dot{\phi(t)}}]\nonumber\\&&
            =exp[\frac{-\pi k^2}{2M^2 \sqrt{Q_1}}]\eea
 The total number density is given as
 \bea n_{\chi_1}&=& \int_0^{\infty} \frac{d^3k}{(2\pi)^3}exp[\frac{-\pi \sqrt{3}k^2}{g \dot{\phi(t)}}]
 \nonumber\\&& = \frac{M^3}{2\sqrt{2}\pi^3} Q_1^{\frac{3}{4}} \eea
As soon as the adiabaticity condition is restored the $\chi_1$
modes can decay
 further to the light particles which are not coupled to inflaton directly.
  The decay time of $\chi_1$ modes is given
   as $ t_{dec}^{\chi_1}$ = $\Gamma_{\chi_1}^{-1}$.
    In terms of variable $\tau$ it is be given as
 \bea \tau_{dec}^{\chi_1} = (\frac{8\pi}{7 g^2})^{\frac{1}{2}} Q^{-\frac{1}{4}}_{1} \label{dectime}\eea
 The energy density of $\chi$ modes after zero crossing is given
 by
 \bea \rho_{\chi_1}(\tau^{\chi_1}) &=& \int_0^{\infty}
 \frac{k^2dk}{(2\pi^2)}n_{\chi_1 , k} \omega_{k}(\tau)\nonumber\\&&
 =\frac{Q_1 M^4}{\pi^4} A_1 \exp{A_1}K_1(A_1)\eea
 Where
 \bea A_1 &=& \pi \sqrt{Q_1} \tau^2 \eea
 and $K_1$ is the modified Bessel function of second kind.
 As the $\chi_1$ modes decay when the time reaches to their time of decay,
 so their energy density also decreases.
  \bea \rho_{\chi_1}(\tau) &=& \rho_{\chi_1} exp[-\int_{0}^{\tau} \Gamma_{\chi_1}dt']=
   \frac{Q_1 M^4}{\pi^4} A_1 \exp{A_1}K_1(A_1) exp[-\frac{7 g^2 A_1}{16\pi^2}]\eea
  The amount of energy transferred to light MSSM fields every time the
  inflaton crosses the origin is (calculated at
   $\tau$=$\tau_{dec}^{\chi_1}$)
\bea \bar{\rho}_{\chi_1}&=& \rho_{\chi_1}(\tau_{\chi_1})=3.606
\frac{QM^4}{\pi^4} \eea
   If $\rho_{\phi}$= $\frac{\dot{\phi_0}^2}{2}$ is the inflaton energy
   when it crosses zero,
   then the fraction of inflaton energy transferred by $\chi_1 $ modes
   to relativistic degrees of freedom is
   \bea \frac{ \bar{\rho}_{\chi_1}}{\rho_{\phi}}=.0062 g^2 \label{chi}\eea
 Similarly for $\chi_2, \chi_3$ modes we have,
   \bea Q_2 &=& \frac{g^2\dot{\phi}_0^2}{8 M^4};\,\,\,\,\,\,\,\,\, Q_3 = \frac{g^2\dot{\phi}_0^2}{6 M^4}\eea
  \bea  \tau_{dec,2} = (\frac{48\pi}{67 g^2})^{\frac{1}{2}} Q_{2}^{-\frac{1}{4}};\,\,\,\,\,\, \tau_{dec,3} =
  (\frac{16\pi}{11 g^2})^{\frac{1}{2}} Q_{3}^{-\frac{1}{4}}\eea
  The expressions for number density and energy transferred are given as
  \bea n_{\chi_2}&=& \frac{M^3}{2\sqrt{2}\pi^3}Q_2^{\frac{3}{4}} ;
  \quad  n_{\chi_3}=\frac{M^3}{2\sqrt{2}\pi^3} Q_3^{\frac{3}{4}}\eea

  \bea \bar{\rho}_{\chi_2}&=& 2.883 \frac{Q_2 M^4}{\pi^4} ; \quad
   \bar{\rho}_{\chi_3}= 4.054 \frac{Q_3 M^4}{\pi^4} \eea

  \bea \frac{ \bar{\rho}_{\chi_2}}{\rho_{\phi}}=.0074 g^2 ;
  \quad \frac{ \bar{\rho}_{\chi_3}}{\rho_{\phi}}=.0138 g^2  \label{chi1chi2}\eea
To find out total energy transferred to MSSM degrees of freedom we
need to add contributions from all decay channels into light final
states. It includes 6 degrees of freedom from gauge bosons
($W_{\mu}^{\pm}$), four degrees of freedom from scalar
($\chi^{\pm},\chi_1,\chi_2$) and 8 from fermions ($\Psi_{1,2}$).
The charged degrees of freedom $\chi_1$ which sum to a total of 16
d.o.f lose  energy according to equation (\ref{chi}) and
$\chi_2,\chi_3$ according to eq. \ref{chi1chi2} respectively. So
the total amount of energy transferred to MSSM degrees of freedom
is
   \bea \rho_{rel} = 16 \rho_{\chi_1}+\rho_{\chi_2}+\rho_{\chi_3}\eea
   For g = .72, the amount of energy loss per zero crossing is
   \bea \frac{\rho_{rel}}{\rho_{\phi}}\sim 6.22 \% \eea
Thus every time the inflaton crosses zero (i.e. twice in each
cycle) it losses about $6\%$ of its energy to relativistic
particles. After N oscillations the $\rho_{\phi}$ and $\rho_{rel}$
are given by \bea \rho{\phi}=0.88^{N}\rho_{0}; \quad \rho_{rel}=
(1-0.88^{N})\rho_{0}\eea Here $\rho_0$ is the initial energy
density of inflaton before it starts oscillating.
\section{Coupled evolution of energy density of $\chi$ modes}
The transfer of energy from inflaton to $\chi$ modes and then to
the MSSM degrees of freedom can be depicted with the study of
coupled differential equations for  energy densities of $\phi$
field, $\chi$ modes and relativistic particles.

To illustrate the kind of picture that emerges we first take $M
\sim 10^{6}$ GeV (i.e mass of inflaton is determined by Majorana
mass of first generation right handed neutrino ). Then the Hubble
rate comes out to be is very small:

\bea H=\sqrt{\frac{M^3}{10^{12}}}GeV^{-1} \approx 10^{-3}M \eea In
other words the Hubble time is much greater than the inflaton
oscillation time. So inflaton undergoes many oscillations in one
Hubble time and the Hubble expansion can be ignored while studying
the energy density evolution. \bea
    \dot{\rho}_{\phi} &=& -(\Gamma_{\phi}^1+\Gamma_{\phi}^2+\Gamma_{\phi}^3) \rho_{\phi} \nonumber\\
    \dot{\rho}_{\chi_{1}} &=& \Gamma_{\phi}^1 \rho_{\phi}- \Gamma_{\chi_1} \rho_{\chi_1}  \rho_{\phi}^{1/4} \nonumber\\
    \dot{\rho}_{\chi_2} &=& \Gamma_{\phi}^2 \rho_{\phi}- \Gamma_{\chi_2} \rho_{\chi_2}  \rho_{\phi}^{1/4}\nonumber\\
    \dot{\rho}_{\chi_3} &=& \Gamma_{\phi}^3 \rho_{\phi}- \Gamma_{\chi_3} \rho_{\chi_3}  \rho_{\phi}^{1/4} \nonumber\\
    \dot{\rho}_{rel}&=& (\Gamma_{\chi_{1}} \rho_{\chi_{1}}+ \Gamma_{\chi_2} \rho_{\chi_2}+ \Gamma_{\chi_3} \rho_{\chi_3})
    \rho_{\phi}^{1/4}\eea
Here $\Gamma_{\phi}^1$=0.105 M represents the effective decay rate
of inflaton due to loss of energy to $\chi_{1}$ and estimated by
\be
\frac{\Gamma}{M}=-\frac{Log(\rho_{\phi}(N+1))-Log(\rho_{\phi}(N))}{M(t_{N+1}-t_{N})}\ee
For half of the oscillation $M(t_{N+1}-t_{N})$= 0.5. Similarly
$\Gamma_{\phi}^2$=.008 M and $\Gamma_{\phi}^3$=.014 M are the
effective rates due to decay into $\chi_2$ and $\chi_3$
respectively. $\Gamma_{\chi_1}$= 1.16 $\times 10^4
\rho_{\phi}^{1/4} M$ is the effective decay rate of $\chi_1$
particles into MSSM particles (estimated from equation
(\ref{dectime}) and taking
$(\dot{\phi}_0)^{1/2}$=$(2\rho_{\phi})^{1/4}$). Similarly
$\Gamma_{\chi_2}$= 1.62$ \times 10^4 \rho_{\phi}^{1/4} M$ and
$\Gamma_{\chi_3}$= 1.22$ \times 10^4 \rho_{\phi}^{1/4} M$ are the
effective decay rates for the $\chi_2$ and $\chi_3$ particles
respectively. The differential equations are solved with the
initial conditions $\rho_{\phi}/\rho_{0}$=1,
$\rho_{\chi_1}/\rho_{0}$=$\rho_{\chi_2}/\rho_{0}$=$\rho_{\chi_3}/\rho_{0}$=0
and  $\rho_{rel}/\rho_{0}$=0.

\begin{figure}[h]
\centering
\includegraphics[scale=0.8]{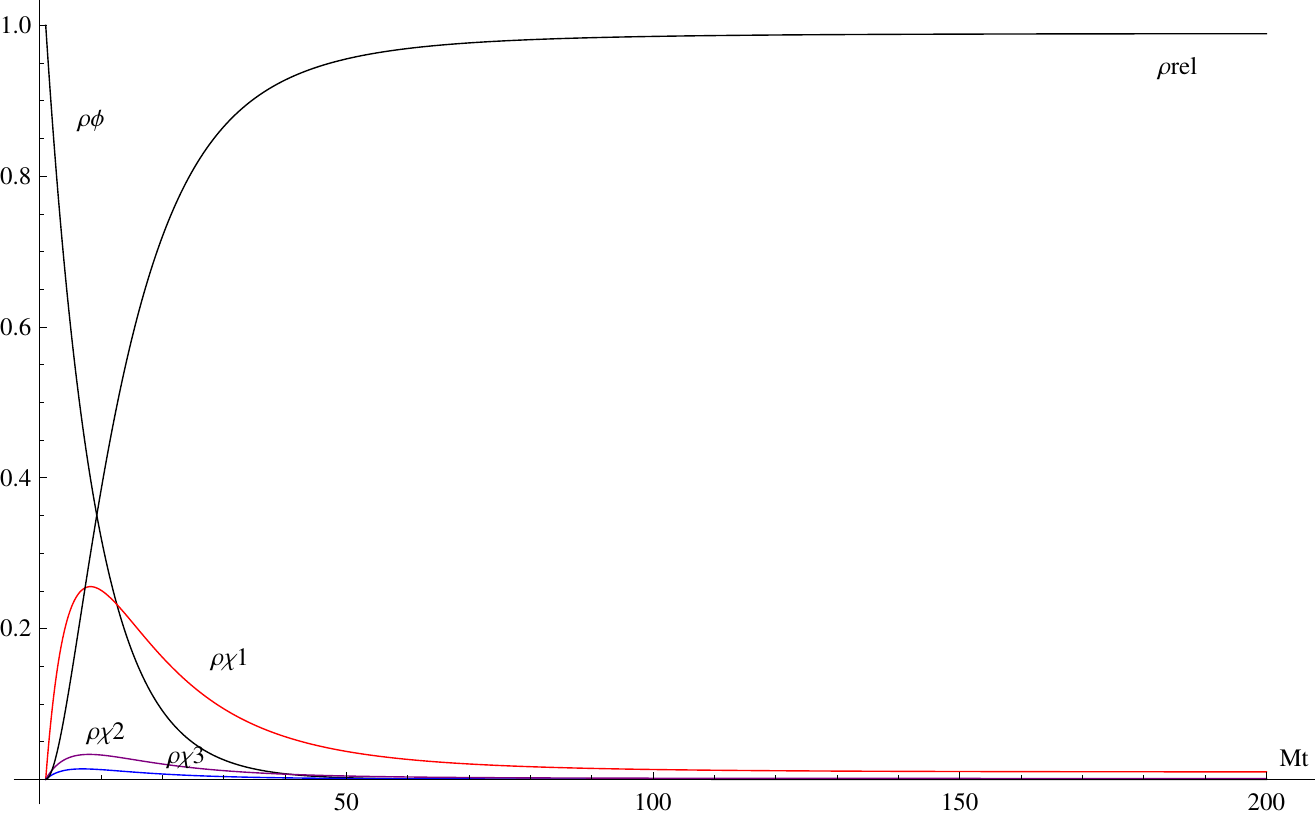}
\caption{Evolution of energy densities of inflaton, $\chi$ modes
and relativistic particles
 normalized to initial energy density of inflaton ($\rho_0$) with Mt.}\label{coupled}\end{figure}

\begin{figure}[h]
\centering
\includegraphics[scale=0.7]{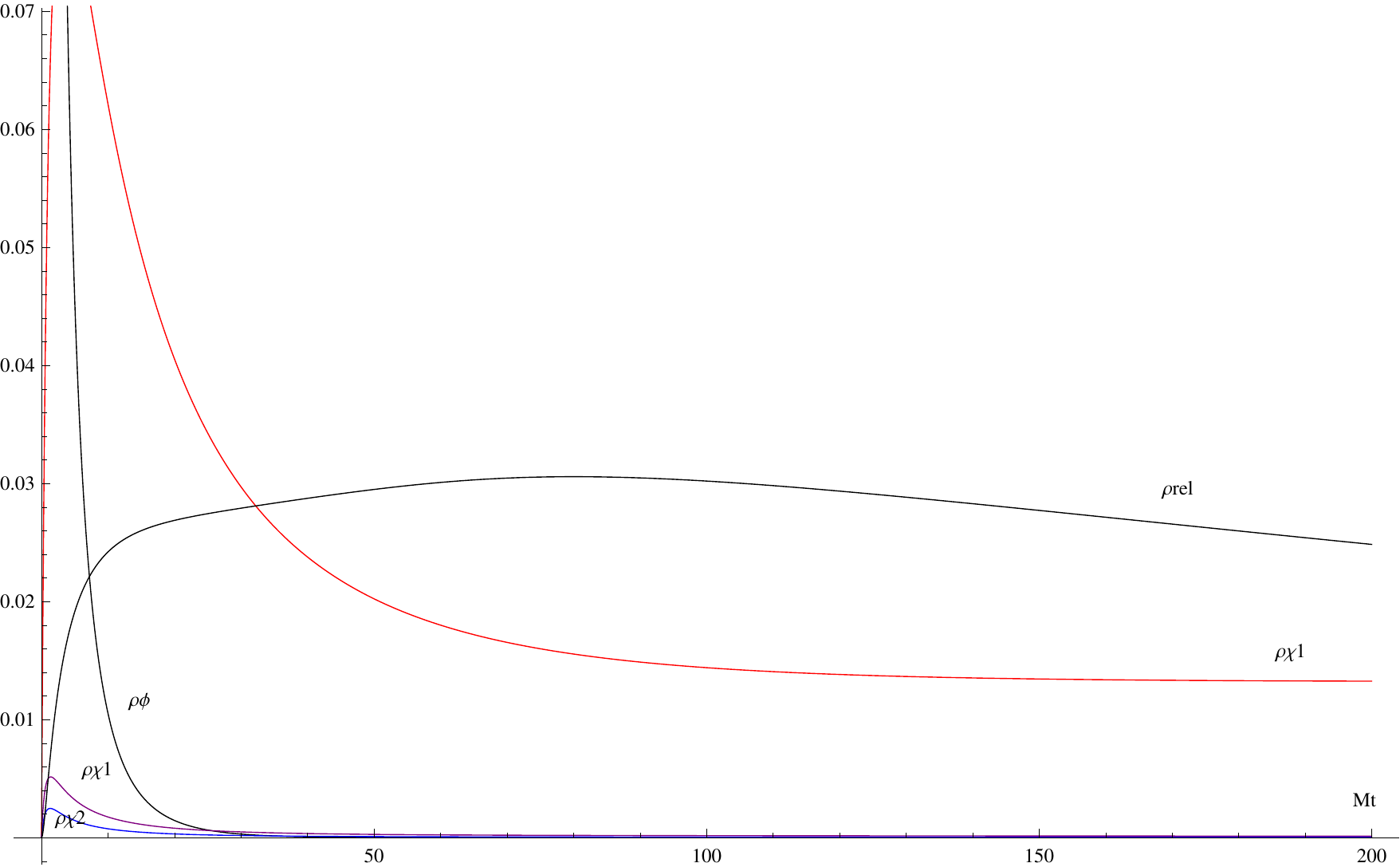}
\caption{Evolution of energy densities of inflaton, $\chi$ modes and
relativistic particles
 normalized to initial energy density of inflaton ($\rho_0$) with $M t$when
  Hubble expansion rate is
 large and comparable to M.}\label{coupled1}\end{figure}

 However If we take inflaton mass $\geq $ $10^{10}$ GeV, the Hubble
 rate comes out to be much
 larger: H $\geq$ $10^{-1}$M i.e. $H^{-1}$ $\leq$ 10 $T_{osc}$.
 In this case the oscillations are strongly damped due to Hubble expansion only.
 The whole inflaton energy is
 not converted into $\chi$ modes and then to relativistic particles
  (see Figure \ref{coupled1}).
 However the initial energy density is so large ($\sim$ $10^{55} GeV^4$)
 even the small fraction of it
 transferred to MSSM degrees of freedom can still give us large reheat
  temperature ($T_{rh}$ $\propto$ $\rho^{1/4}$).

For M=$10^6$ GeV we see from Figure \ref{coupled} that within 100
oscillations the inflaton loses all of its energy into the
relativistic MSSM degrees of freedom. The relativistic particles
produced by this chain process are not in thermal equilibrium at
the time of their production. We study here the preheating part of
the scenario, the complete reheating will take place when the
relativistic particles come into equilibrium by undergoing
subsequent scattering and collision. After their thermalization
    one can obtain   the reheat  temperature of the universe.
    The ultimate symmetry left in this case is
    $SU(3)_C \times U(1)_{Q}$. So we have gluons and photons
    as massless particles which are exchanged during
     the scattering of MSSM particles to bring them in thermal equilibrium.
     The dominant processes
      are 2 $\longrightarrow$ 3 scattering process via gluon and photon
      exchange \cite{davidson}.
       The rate of scattering via exchange
      of massive gauge bosons is suppressed because of large inflaton vev
      induced mass. Their exchange becomes
      effective when the amplitude of inflaton oscillation is quite small
      i.e. just near to the origin.
      When all the MSSM degrees of freedom get thermalized,
      it results in a reheating temperature of the universe, given by
       \be T_{rh}=\frac{30}{\pi^2g_{*}} \rho_{rel}^{1/4} \simeq 10^{11}-10^{13} GeV\ee
       Here $g_{*}$=228.75 are total effective  MSSM degrees of freedom in thermal bath.
       $\rho_{rel}$ $\approx$ $\rho_{0}$=$\frac{M^4}{h^2}$ is the initial energy density of inflaton.
       Such a high reheat temperature can lead to overproduction of gravitinos.
       For successful nucleosynthesis, they should decay
        before the time of  nucleosynthesis  $\tau_N$ $\sim$ 1
        otherwise their late decay will destroy the created nucleons.
         The solution to this problem is
        if gravitino is massive enough to decay before nucleosynthesis \cite{moroi}.
       \bea \tau_{grav} \sim 10^5 \, sec
({\frac{1 \, TeV }{m_{3/2}}})^3 << \tau_N \sim 1 sec \eea We see
that the reheating in SSI scenario favors the scale of
supersymmetry breaking -as indicated by the gravitino mass- should
be  above $ 50 $ TeV. It is interesting that
 such large supersymmetry breaking scales
are preferred by both the NMSGUT \cite{nmsgut,nmsgutIII,BStabHedge}
and the latest data
indicating \cite{Atlas,CMS} light Higgs mass $M_h\sim 125\, GeV$.

\section{Conclusion}
 The idea of instant preheating works well in SSI scenario.
 For small inflaton mass the Hubble expansion rate is $\sim$ $10^{-3}$M
 so inflaton undergoes many osillations in one Hubble time and
 within 100 oscillations it loses all of its energy into the relativistic
 MSSM degrees of freedom which  then   undergo collisions and
 thermalize.
 In this way inflaton can efficiently transfer its whole energy to
 relativistic particles which
  after thermalization results into the reheat temperature $\sim 10^{11}$ GeV.
  For large inflaton mass
  the Hubble rate is comparable to the inflaton mass then only a very small
  fraction of inflaton energy
  is converted into the relativistic degrees of freedom. But the resulting
   reheat temperature is still
  very large  $\sim 10^{13}$ GeV. Such large reheat
  temperature points to a high scale of Susy breaking which is favored in NMSGUT.
  So it points towards
  the idea of embedding of SSI in NMSGUT. We will discuss this idea in detail
  in the next Chapter.
\newpage
\chapter{ Susy Seesaw Inflation and NMSGUT}
\section{ Introduction}
In Chapter 5 we discussed the properties of generic inflection
point inflation potential of the Susy Seesaw inflection (SSI)
model. Since SO(10) NMSGUT is a natural home for Seesaw so we can
embed our SSI model in it. In \cite{SSI} we presented the
conditions on NMSGUT superpotential parameters consistent with the
slow roll inflation conditions and results of inflation parameters
along with fermion fit. Although we failed to achieve the required
number of e-folds in NMSGUT inflation and also the required fine
tuning together with a viable fermion fit, but results in
\cite{SSI} did not include high scale threshold corrections
\cite{BStabHedge} which are crucial determining the SO(10) Yukawas
and hence inflation parameters. During our study of inflation
conditions in NMSGUT after implementing GUT threshold conditions,
the startling announcement from BICEP2 experiment arrived. The
BICEP2 collaboration \cite{BICEP2} claimed the observation of
B-mode polarization in CMB.  If confirmed then it is a direct
evidence of primordial inflation and quantum gravitational effects
\cite{Bhupal,Krauss} because the large B-mode polarization they claim to
have detected could only be due to primordial gravitational waves
which were produced during inflation via quantum effects. They
 claim a large value of tensor to scalar fluctuations
parameter $r=.2^{+.07}_{-.05}$ which indicates a very high scale
of inflation for single field inflation models. Later doubts were
raised regarding their treatment of contamination of polarization
data by foreground dust \cite{PLANCK5,Flauger,Mortonson}. More
polarization data is required to resolve the issue. A final word
is awaited from the PLANCK collaboration. Now for a non zero value
of $r$, from \cite{Lyth} we have\be
V_0^{\frac{1}{4}}=\biggr(\frac{r}{0.1}\biggr)^{1/4}\times 2 \times
10^{16} GeV \label{lyth}\ee Thus the BICEP2 claim favors   single
field inflation models at precisely the MSSM gauge coupling
unification scale ! Also the inflation takes place over a field
interval $\delta \phi~\sim 5 M_{Pl}$ \cite{Lyth,antuschnolde}.
Previously almost all inflation models were designed to avoid such
a large value of $r$ since there appeared little chance of finding
$r \sim 0.1$ as reported by the PLANCK collaboration
\cite{planck}. A large value of $r$ will change the whole scenario
of inflection point inflation models. For such a large field
digression one needs a steep potential rather than a flat
potential. The condition found on the trilinear-Mass term tuning
parameter ($\Delta$ $\approx$ $10^{-28}M^2 \text{GeV}^{-2}$) in
\cite{SSI} stops being a fine tuning condition for $ M > 10^{13}$
GeV and the assumptions of our previous analysis break down.

In this chapter after discussing our first attempt to embed SSI in
NMSGUT (with $M \leq 10^{12}$ GeV when fine tuning is required and
the generic analysis of Chapter 5 holds) we revisit the issue of
supersymmetric seesaw inflation in
 NMSGUT without   any fine tuning of inflation
potential and without making any assumption on the family index of
matter fields in the inflaton LHN condensate. We derive the new
conditions for inflation parameters for slow roll inflation
conditions for a generic renormalizable potential for such large
$r$.

\section{Seesaw inflection and NMSGUT}
 To study the NLH flat direction in NMSGUT we need to consider this flat direction
 in full GUT potential in terms of light (SIMSSM) field vevs.
 Then this flat direction   rolls
  out of the MSGUT vacuum with SIMSSM as effective theory.
 The required fields are the GUT scale vev fields
  $\Omega (p,a,\omega, \sigma ,\bar\sigma)$  and
 the (6) possible components $ h_i,{\bar{h}_i};i=1...6$ of the light MSSM Higgs doublet
 pair $H,{\overline H} $ along with  the  chiral lepton fields $L_A,\nu_A^c,
 A=1,2,3$.  The relevant superpotential is then \cite{aulmoh,ckn,abmsv,ag1,bmsv}
\bea W&=& 2\sqrt{2}(h_{AB}h_1-2\sqrt{3}f_{AB}h_2-g_{AB}(h_5+i
\sqrt{3}h_6))+\bar{h}^T
{\cal{H}}(<\Omega>)h\nonumber\\&&+4\sqrt{2} f_{AB}\bar\sigma
 \bar \nu_{A} \bar\nu_B + W_\Omega(\Omega) \eea
where
  \bea
 W_\Omega(\Omega)&=& m(p^2+3a^2+6\omega^2)+2 \lambda(a^3+3p\omega^2)
 \nonumber\\&&+(M+\eta(p+3a-6 \omega))\sigma \bar\sigma \eea
 and \be \frac{\partial W_{\Omega}}{\partial \Omega }|_{h,\bar \nu,L = 0} =0 \,\,\,\,\,\,\,\,\,\,\,\,\,\,
  D_{\alpha}(\Omega)|_{h,\bar \nu,L = 0}=0\label{omegavac}\ee
 Here  $h_{AB},f_{AB},g_{AB}$ are the usual Yukawa coupling matrices of
the three matter 16-plets to the $\mathbf{10,120,\oot}$ Higgs
multiplets in NMSGUT. Equation (\ref{omegavac}) represents the
MSGUT vacuum \cite{aulmoh,ckn,abmsv,ag1,bmsv}. Out of the 5
diagonal D-terms of SO(10) the charge and color neutral vevs for
$\Omega$ and $\nu,\nu^c,h_0$ correspond to breaking of the
generators $T_{3L},T_{3R},B-L $. The vevs $\Omega$ do not
contribute to the D terms for their generators, so their values
are
 \be
 D_{3L}=\frac{g_{u}}{2}(-\sum_{i=1}^6|h_{i0}|^2+\sum_{A}|\tilde \nu_A|^2)\ee
 \be
D_{3R}=\frac{g_{u}}{2}(\sum_{i=1}^6|h_{i0}|^2-2|h_{40}|^2-\sum_{A}|\tilde
{\bar\nu}_{A}|^2)\ee \be
D_{B-L}=\sqrt{\frac{3}{8}}g_{u}(\sum_{A}(|\tilde
{\bar\nu}_{A}|-|\tilde \nu_A|^2)+2|h_{40}|^2)\ee
 Where from  \cite{aulmoh,ckn,abmsv,ag1,bmsv}
only $h_{4\alpha}=\Phi^{44}_{{\dot{2}}\alpha}$ has
$B-L=+2,T_{3R}=-1/2 $ and thus $T_{3L}=1/2$ while all others have
$T_{3R}=1/2$ and $B-L=0$. $g_u$ is the SO(10) gauge coupling in
the standard unitary normalization. Thus the D-flatness conditions
give
  \be \sum_A|\tilde
\nu_A|^2=\sum_{i}|h_{i0}^2|=\sum_A |\tilde{\bar
\nu}_{A}|^2+2|h_{40}|^2 \ee
We know from the previous
 discussion that in MSGUTs the MSSM
Higgs doublet pair is defined by fine tuning $Det({\cal{H}})\simeq
0$ so that its lightest eigenvalue $\mu \sim 1-100 $ TeV specifies
the $\mu$ term in the superpotential of the SIMSSM : $W=\mu
{\overline{H}}H+...$. Here the doublet pair $H,{\overline{H}}$ is
a linear combination \cite{aulmoh,ckn,abmsv,ag1,bmsv} of the 6
doublet pairs of the NMSGUT:
 \bea  h_i=U_{ij} H_j   \qquad \qquad  \bar{h}_i={\overline{U}}_{ij}
 \overline{H}_j\eea
where $U,{\overline{U}}$ are the unitary matrices that diagonalize
the doublet mass matrix ${\cal{H}}$ such that
${\overline{U}}{\cal{H}} U= Diag\{\mu,M^H_2,....,M^H_6\}$. To
obtain the tree level Yukawa couplings replace $ h_i,\bar{h}_i
\rightarrow \alpha_i H,\bar{\alpha}_i {\bar{H}}$ in the GUT Higgs
doublets ($h_i,\bar{h}_i$) couplings to the matter fermions of the
SIMSSM. Thus in particular the neutrino Dirac coupling is \be
y^{\nu}_{AB}=\tilde h_{AB} \alpha_1-2\sqrt{3}\tilde
f_{AB}\alpha_2-\tilde g_{AB}(\alpha_5+i \sqrt{3}\alpha_6)\ee Here
($(\tilde h_{AB},\tilde g_{AB},\tilde
f_{AB})$=$2\sqrt{2}(h_{AB},g_{AB},f_{AB})$ ) We assumed that only
light Higgs doublets will contribute to inflaton flat direction
since from the $|F_{\bar h}|^2$ terms of the potential it is
evident that the involvement of any other Higgs doublet would lead
to GUT scale rather than conjugate neutrino scale mass for the
inflaton. Also only one generation of sneutrinos $\nu_A$  and
conjugate sneutrinos $\bar{\nu}_B$  contribute to the inflaton
flat direction with A $\neq$ B. In view of the tight upper bounds
on the fermion Yukawas (see eqn.(\ref{hDeltavsM})) we need to
involve the lightest generation of sfermion. Thus we take
$\nu_{A}=\nu_1$. On the other hand we take $\nu^c_A=\nu^c_3 $ to
satisfy the fine tuning condition. So the ansatz for the inflaton
flat direction is \be \tilde{\nu}_1=
\frac{\phi}{\sqrt{3-2|\alpha_4|^2}} \qquad\qquad
 H_{1}^0=\frac{\phi}{\sqrt{3-2|\alpha_4|^2}}  \qquad\qquad \tilde{\bar
\nu}_{3}=\frac{\phi
\sqrt{1-2|\alpha_4|^2}}{\sqrt{3-2|\alpha_4|^2}} \ee Now the F-term
potential corresponding to fields under consideration is \bea
V_{hard}&=& \big[({y^\nu}^\dagger y^\nu)_{11} +
\Gamma(|\tilde{h}_{31}|^2
 + 4|\tilde g_{31}|^2+(y^\nu{y^\nu}^\dag)_{33}) +4|\tilde  f_{33}|^2\Gamma^2 \big ] {\frac{|\phi|^4}{9}} \nonumber
 \\&&+\frac{8}{3\sqrt{3}} \tilde
 f_{33}|y^{\nu}_{31}| |\bar \sigma| {\sqrt{\Gamma}} \text
 Cos(\theta_\phi+\theta_{y^{\nu}_{31}}-\theta_{\bar \sigma})) |\phi|^3 +\nonumber \\&&\big( {\frac {|\mu|^2}{3}} +
 \frac{16}{3}|\tilde f_{33}|^2|\bar \sigma|^2\Gamma\big )|\phi|^2   \eea
 Here $\Gamma=1-2|\alpha_4|^2$. The corresponding Supergravity(SUGRY)-NUHM
 type soft terms in terms of a common trilinear parameter
$A_0$  but different soft mass parameters $\tilde{m}_{\tilde
f}^2,\tilde{m}_{h_i}^2$ for the  16 plets  and the different Higgs
is given as
\bea V_{soft}&=&A_0 W +c.c.+{\tilde{m}}_{ 16}^2|\tilde{\Psi}|^2+\sum_{i}\tilde m^2_{h_i}|h_i|^2\nonumber\\
   &=&2A_0\sqrt{\Gamma}|y^{\nu}_{31}|\frac{|\phi|^3}{3\sqrt{3}}\text Cos(3\theta_\phi+\theta_{y^{\nu}_{31}})
+\frac{4}{3}A_0 \tilde f_{33}|\bar \sigma| \Gamma |\phi|^2\text
Cos(2\theta_\phi+\theta_{\bar\sigma})\nonumber\\&&+
({\widehat{m}}_0^2 -{\frac{|\mu|^2}{3}} )|\phi|^2 \eea where \bea
{\widehat{m}}_0^2=\frac{ \tilde{m}_{16}^2}{3}(1+\Gamma)
+\sum_{i}\frac{ \tilde{m}_{h_i}^2 |\alpha_i|^2}{3}
+{\frac{|\mu|^2}{3}}
 \eea Since $\tilde{m}_{16},
\tilde{m}_{h_i},A_0$ are all   $\sim O( M_S)$, so
 $f_{33}|\bar\sigma|>> M_S$. It implies that the phase
$\theta_\phi$ can be fixed by minimizing the term in $V_{hard}$:
\be \theta_\phi= \pi + \theta_{ \bar\sigma }
-\theta_{y^\nu_{31}}\ee We assume that $\theta_\phi$ is fixed at
this value.  Since the inflationary dynamics is at large values of
$|\phi|$ and $\theta_\phi$ is fixed so we can work with a real
field $\phi$.  Now adding the hard and soft potentials and then
comparing with the generic renormalizable inflaton potential, we
can identify the parameters ($M,h,A$). However for simplicity we
can drop the soft term contribution (in trilinear and quadratic
terms) since they play a negligible role in fine tuning and
inflaton mass being completely determined by $m_{\nu^c} \geq 10^6$
GeV. Inflation parameters can be deduced from superpotential: \be
\label{quartic} h=\frac{2 \sqrt{3}}{(3-2|\alpha_4|^2)}
\big[(y^{\nu \dag} y^{\nu})_{11}+ (|\tilde{h}_{31}|^2+4|\tilde
g_{31}|^2+(y^{\nu}y^{\nu \dag})_{33})\Gamma+\\
4|\tilde f_{33}|^2 \Gamma^2)\big ]^{\frac{1}{2}} \ee \bea
A&=&\frac{48 \sqrt{3}|\tilde f_{33}|
|y^{\nu}_{31}||\bar\sigma|{\sqrt{\Gamma}}}{h (3-2|\alpha_4|^2)^{\frac{3}{2}}}\\
M^2 &=& 32|\tilde f_{33}|^2|\bar
\sigma|^2\frac{\Gamma}{(3-2|\alpha_4|^2)}+2{\mu}^2 \label{Minf}
\eea The fine tuning condition is now determined to be:
\bea|y^{\nu}_{31}|^2&=& \frac{8
}{1-8\Gamma}(\Gamma(|\tilde{h}_{31}|^2+4|\tilde
g_{31}|^2+|y^{\nu}_{32}|^2+|y^{\nu}_{33}|^2)\nonumber\\&&+
|y^{\nu}_{11}|^2+|y^{\nu}_{21}|^2+4|\tilde f_{33}|^2\Gamma^2)\eea

If we tune \be \Gamma \approx 0 \,\,\,\ {\text{i.e}}\qquad\quad
 |\alpha_4|={\frac{1}{\sqrt{2}}} \ee  then it may be possible
 to satisfy it to a good accuracy. This means that the
MSSM doublet H is almost exactly 50\% derived from the doublet in
the 210 plet ! If this condition can be achieved the remaining
tuning condition is only \bea|y^{\nu}_{31}|^2&=& 8
(|y^{\nu}_{11}|^2+|y^{\nu}_{21}|^2 )\eea
We found that this fine
tuning condition is achievable in NMSGUT along fermion fitting.
But to achieve the required number of efolds, the stringent
condition \cite{SSI} on quartic coupling and inflaton Mass
$\frac{h^2}{M}$ $\sim$  $({y^\nu}^\dag y^\nu)_{11}/M_3$ $\approx
10^{-25}$ is quite hard to satisfy. In the present case the third
generation right handed sneutrino mass $\sim$ $|\tilde{f}_{33}|
|\bar\sigma|$
 determines the inflaton mass.
\begin{table}[h]
 $$
 \begin{array}{|c|c|}
 \hline
 {\rm  Parameter }& {\rm  Value }\\
 \hline
 h & 2.44 \times 10^{-4}\\
 M & 3.043 \times 10^{11} GeV\\
 \Delta& 10^{-2}\\
 \Gamma&  4.343 \times 10^{-5}\\
 M^{\nu^c}_3 & 4.86   \times 10^{13} GeV\\
 Log_{10}(h^2/M) &-18.706 GeV^{-1}\\
 N_{CMB}  & 4.78 \times 10^{-4}\\
\hline
 \end{array}
 $$
 \caption{Set of inflation parameters from \cite{SSI}.}
\label{TAB1} \end{table}
In Table \ref{TAB1} we have given the
important set of inflation parameters from the table given in
\cite{SSI}. The third generation right handed neutrino mass comes
out to be $ \sim 10^{13} GeV$ and inflaton mass is $ \sim 10^{11}$
GeV. The value of quartic coupling in eq(\ref{quartic}) is
controlled by the second and third terms
  since they are the
largest ($O(10^{-3})$) \cite{nmsgut}. However we can lower them
with the help of factor $\Gamma \approx 0$. Then the controlling
term will the first one ($(y^{\nu \dag} y^{\nu})_{11}$). But the
same factor appears in the formula for inflaton mass
 Eqn.(\ref{Minf}) (determined by Majorana mass term ($\sim ~|f_{33}||\bar \sigma|$)
for 3rd generation right handed neutrino), so it will reduce the
value of M (from O($10^{13}$ GeV) to O($10^{11}$ GeV)).
Corresponding to this value of M we need quartic coupling
O($10^{-7}$) and $\Delta \approx 10^{-6}$ (approximated using eqn.
(\ref{hDeltavsM})).
 But the actual value of quartic coupling as well as fine-tuning parameter
 is quite large as compared to the required values.
Hence it results in a very small value of $N_{CMB}$ $\sim$
$10^{-4}$. So this simple embedding of SSI in NMSGUT doesn't work.

\section{BICEP2 experiment}
BICEP2 is a telescope mounted at the south pole for background
imaging of cosmic extragalactic polarization. This experiment is
specially designed to detect the signal of primordial
gravitational waves. The gravitational waves active during the
inflationary epoch produce
 polarization in cosmic background radiation. They leave their imprint on the CMB
in terms of a curl or rotation which is known as primordial B-mode
polarization or tensor component. However this type of
polarization can also be produced via gravitational lensing  and
foreground  dust (non- primordial B-mode polarization detected by
South pole telescope (SPT) \cite{SPT}). Scalar density
fluctuations also produce polarization in CMB known as E-mode
polarization or scalar component. The difference in two modes
cannot be detected simply by looking at the variation of CMB
temperature. Moreover the B-mode is weaker than E-mode and thus
difficult to detect. However certain angles in polarization can be
measured and allow one to distinguish between the two modes. The
BICEP2 experiment claimed to measure this difference and has given
their results in terms of tensor to scalar component ratio
$r=.2^{+.07}_{-.05}$ at 3$\sigma$ \cite{BICEP2} level. They also
discarded the possibility of $r$=0 at 7$\sigma$ level. The central
value (0.2) of $r$ is quite large as compared to those
 found by WMAP \cite{wmap} and PLANCK \cite{planck}.
 However the contamination of the polarization data
by foreground dust may not have been correctly accounted for which
can reduce this value.
  So to confirm the results great scrutiny is required.
 If BICEP2 results are confirmed, it will be a big blow to
 a huge class of inflationary models
 and start a new era of modern cosmology. The PLANCK collaboration will report on this
 issue soon using their original data on the foreground dust.

\section{Generic renormalizable inflation potential without fine tuning }
The GUT scale of inflation doesn't require fine tuned flat
 potential. Also for the inflaton mass $\geq$ $10^{13}$ GeV the analysis
  done in chapter {\bf 5} does not hold.
 So first we will find out the condition on inflation parameters of generic
 renormalizable
  potential for slow roll inflation without any fine tuning.
 The general renormalizable inflationary potential is  :
 \bea  V (\phi)=  M^2\frac{\phi^2}{2}- \frac{ Ah}{6 \sqrt{3}}\phi^3+h^4\frac{\phi^4}{12} \eea
 Here M,A,h are real without loss of generality. It's dimensionless form
  \bea \tilde V= \frac{V}{V_0}= \frac{x^2}{2}- \frac{\tilde A x^3}{6 \sqrt{3}}+\frac{x^4}{12} \eea
 where $ x= \frac{\phi}{\phi_0}$, $V_0=\frac{M^4}{h^2}$, $\tilde A=\frac{A}{M} $ and $\phi_0 =\frac{M}{h}$
 is inflaton vev is more convenient for numerical work.

\begin{figure}
        \centering
                \includegraphics[scale =1]{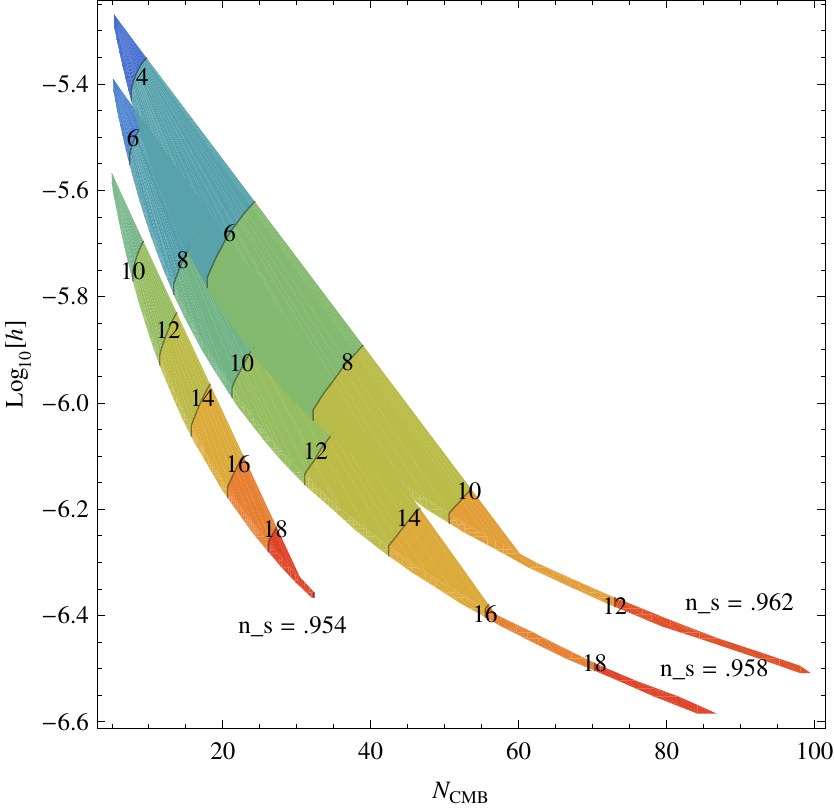}
                \caption{Plot of quartic coupling h and Number of e-folds for
different values of $\omega$ at trilinear dimensionless parameter
$\tilde{A}$=.1. The contour line bears the constant value of
$\omega$ and different colors of contour represent a range for
$\omega$. As we move towards higher values of $\omega$ number of
e-folds increases and quartic coupling
decreases.}\label{fig0.1}\end{figure}

\begin{figure}
        \centering
                \includegraphics[scale =1]{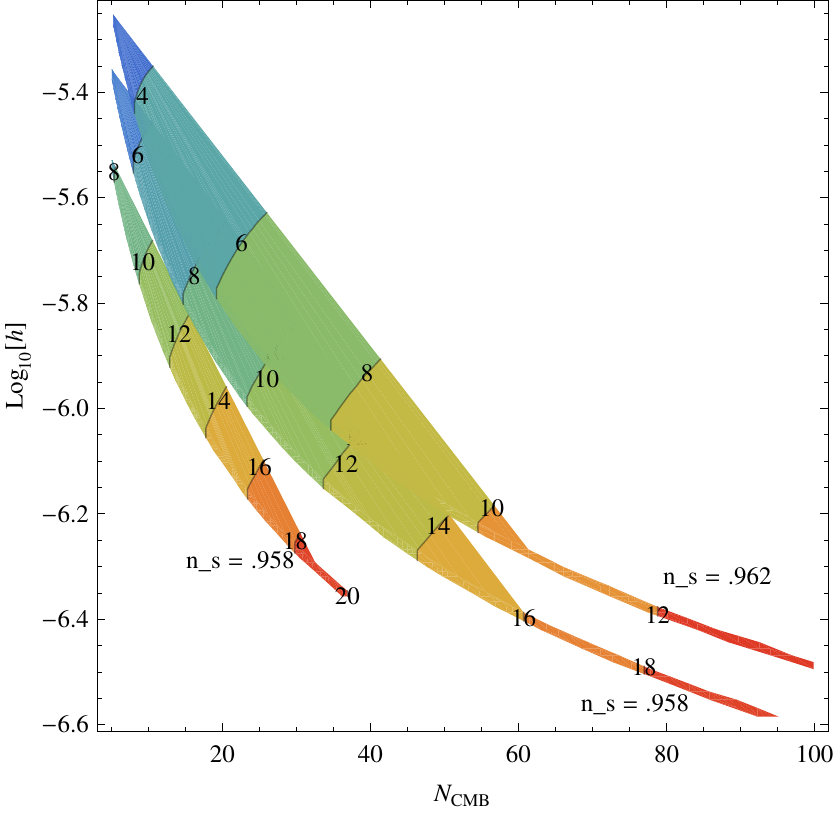}
                \caption{ Same as figure \ref{fig0.1} with $\tilde{A}$=.01}\label{fig0.01}\end{figure}

\begin{figure}
        \centering
               \includegraphics[scale =1]{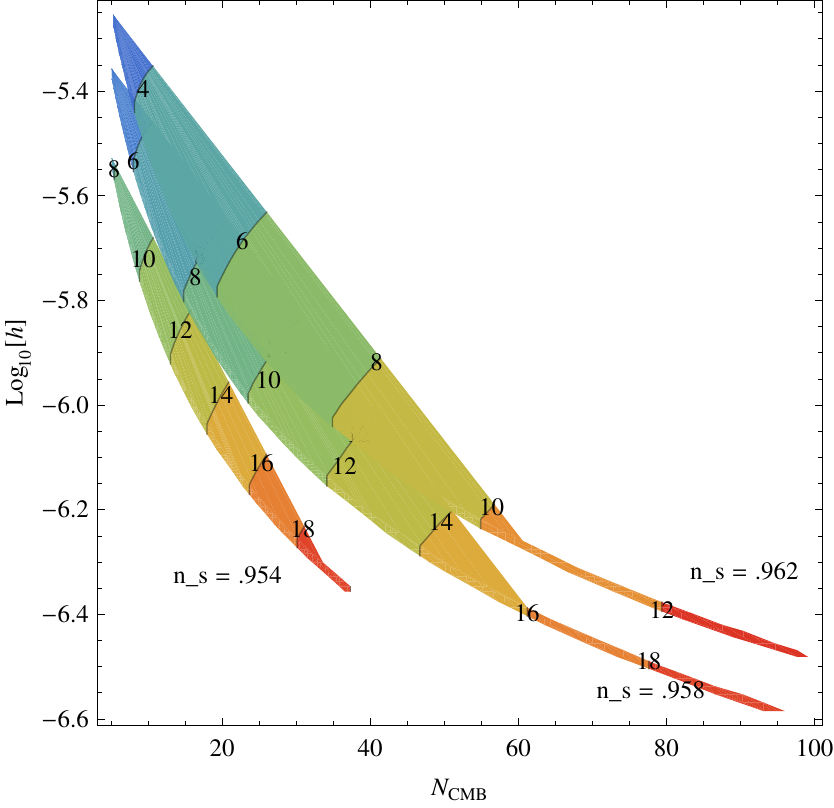}
               \caption{ Same as figure \ref{fig0.1} with $\tilde{A}$=.001}\label{fig0.001}\end{figure}

\begin{figure}
        \centering
               \includegraphics[scale =1]{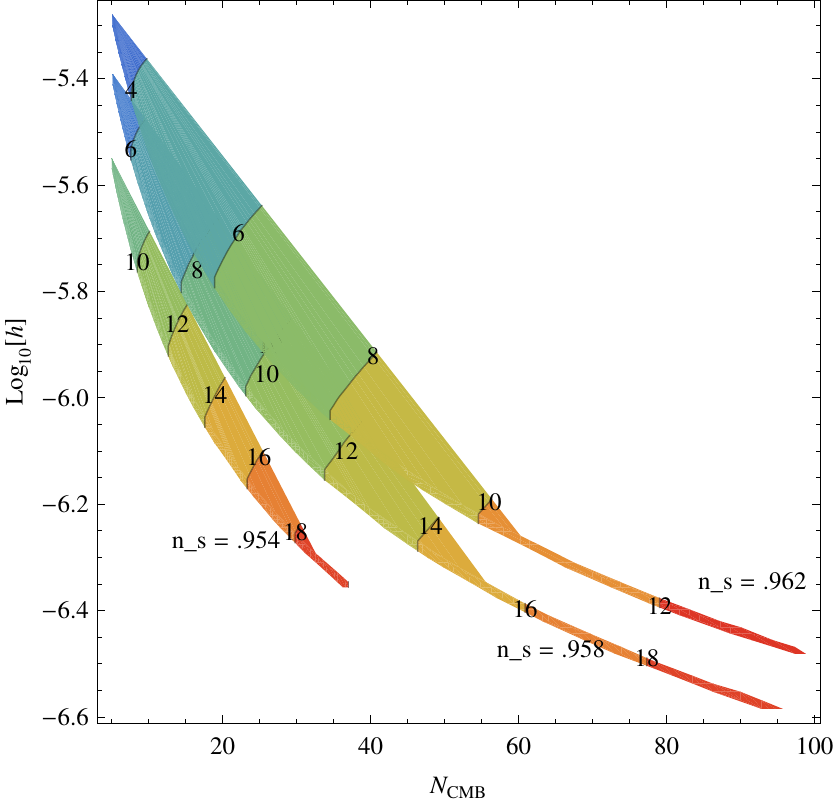}
               \caption{Same as figure \ref{fig0.1} with $\tilde{A}$= 0}\label{fig0}\end{figure}

If we define a new parameter $\omega =\frac{M}{h M_{pl}} $, then
for a given potential the slow roll parameters $\epsilon$ and
$\eta$ are given by: \bea \epsilon = \frac{\tilde{V}_{x}^2}{2
\omega^2 \tilde V}; \,\,\,\,\,\, \eta =
\frac{\tilde{V}_{xx}}{\omega^2 \tilde V} \eea
 Where $M_{Pl}=2.43 \times 10^{18}$ GeV. The slow roll inflation
  formula for Power spectrum, spectral
index  and tensor to scalar ratio are given by: \be P_R=
\frac{\omega^4 h^2 \tilde V}{24 \pi^2 \epsilon};\quad n_s= 1+2
\eta - 6 \epsilon; \quad r=16 \epsilon \label{PRnsandr}\ee The
values for Power spectrum, spectral index given by PLANCK
\cite{planck} are \bea P_R = (2.1977 \pm .103) \times
10^{-9};\,\,\,\,\,\,\ n_s= .958 \pm .008 \label{infdata}\eea and
BICEP2 value \cite{BICEP2} for tensor to scalar ratio,$r$ \bea
r=0.2^{+.07}_{-.05} \label{infr}\eea The number of e-folds of
inflation remaining when pivot scale crosses the horizon can be
found from equation of motion of inflaton field \bea N=
\omega^2\int_{x_{cmb}}^{x_{end}} \frac{\tilde V}{\tilde{V}_x} dx
\eea Integrating this equation gives \bea N&=&\frac{\omega^2}{128}
(-3(\tilde{A}^2-16) \log (-3
   \tilde{A} x+4 x^2+12)-\frac{2 \tilde{A} (3
   \tilde{A}^2-80) \tan ^{-1}(\frac{3 \tilde{A}-8
   x}{\sqrt{192-9\tilde{A}^2}})}{\sqrt{\frac{64}{3}-\tilde{A}^2}}\nonumber\\&&+8 x (2
   x-\tilde{A}))\biggr|_{x_{cmb}}^{x_{end}}\eea
Now we need to find the viable parameter values of M,
  h and $\tilde{A}$ to achieve values given in equation
 (\ref{infdata}) and $N_{efolds}$ $\approx$ 50.
 The procedure we follow is:
 \begin{enumerate}
 \item  Take some value of $r$ and $n_s$ in the allowed range given by Eqn. (\ref{infdata}).
 \item Calculate the value of $\epsilon$ and $\eta$ by Eqns. (\ref{PRnsandr}).
 \item  Then define a ratio $\theta$=$\eta/\epsilon$ and solve for x.
 It gives one real root for x which we take as $x_{cmb}$ (Note that it does not give
 real root for whole range of values given in (\ref{infdata}))
 \item Then the field value ($x_{end}$) is calculated from $\eta$ $\approx$ 0.8.
\item Throwing the values of M,h,A we calculate the value of
$P_R$, $V_0^{1/4}$ and $N_{e-folds}$.
 \end{enumerate}
 In figures \ref{fig0.1}-\ref{fig0} we have shown
 contour plots between quartic coupling h and number of
efolds with contour lines bearing constant value of $\omega$. We
have kept the value of $r$=.2 fixed and three set of values of
spectral index $n_s$=0.954, 0.958, 0.962. So each figure contains
three contour plots. For each figure we vary the trilinear term
parameter $\tilde{A}$ over the values(.1,.01,.001, 0).
 We accepted only those values of M and h for which the
  value of $P_R$ falls in range (1.0947-1.3007)
 $\times 10^{-9}$ and the value of $V_0^{1/4}$ as given by Eqn. (\ref{lyth}).
 By comparing Figures \ref{fig0.1},\ref{fig0.01}, parameter $\tilde{A}$ has only a
 small effect: when we change $\tilde{A}$ from .1 to .01, the plots
shift little bit towards higher values of $N_{CMB}$. However
reducing $\tilde{A}$ further doesn't make any difference. For
$n_s$=.962, we need h $\sim $ $10^{-6.1}$ and $\omega \approx $ 9
to have $N_{efolds}$ $\approx$ 50. This corresponds to inflaton
mass of order $10^{13.1}$ GeV. For $n_s$=.958, to achieve
$N_{efolds}$ $\approx$ 50 we need smaller h $ \sim $ $10^{-6.35}$
and $\omega \approx$ 15. For $n_s$=.954, the required quartic
coupling is less than $10^{-6.4}$ and $\omega$ more than 20. These
estimates thus provide a revised rule of thumb replacing the rule
$h^2 ~\sim ~10^{-25}
 (M/GeV)$ derived earlier in the fine tuned case.
  Note that $\omega$ is controlled by $M/h$ so that for
 $\omega$ $\approx$ 20, $h~\approx ~10^{-6.35}$ corresponds to
 $M~ \approx ~10^{13.1}$ GeV and $h^2/M ~\sim ~10^{-26}
 ~GeV^{-1}$. Which is about an order of magnitude smaller than for
 $M << 10^{13}$ GeV.

\section{NMSGUT inflation after BICEP2}
In Section  7.3 we assumed that only the light Higgs will
contribute to inflaton flat direction so as to keep $M < 10^{13}$
GeV. With M allowed to be large the 5 heavy Higgs pair  with GUT
scale masses   in the  NMSGUT can also be allowed to  contribute
i.e.  the strong condition that only the light Higgs corresponding
to MSSM   contributes to the inflaton condensate can be relaxed.
So in this case the heavy Higgs having masses O($10^{16}$ GEV)
will control the mass of the inflaton. Also no assumption on
family index for s-fermion components of inflaton is made. Our new
ansatz for inflaton is

\bea \tilde{\overline{\nu}}_{A}=b_A \frac{\phi}{\sqrt{3}}
\,\,\,\,\,;\tilde \nu_{A}=c_A \frac{\phi}{\sqrt{3}}\,\,\,\,\,;
h_l= U_{lm}H_m =U_{lm} a_m \frac{\phi}{\sqrt{3}} \eea Here A=1,2,3
(family index) and $l,m$ =1-6 (types of different Higgs). The
parameters $b_A$, $c_A$ and $a_m$ are complex numbers which decide
the fraction
 each field contributes to the inflaton.
 From D-flatness
condition we have \bea \sum_A |c_A|^2=\sum_l |a_l|^2=\sum_A
|b_A|^2 +\sum_l 2 |U_{4l}a_l|^2  \eea To achieve canonically
normalized kinetic term for inflaton field we need: \bea \sum_A
|b_A|^2+\sum_A |c_A|^2+\sum_l |a_l|^2=1 \eea And this can be
achieved in this scenario in the following way:

\bea \sum_A |b_A|^2= \sum_A |c_A|^2=\sum_l |a_m|^2=\frac{1}{3}
 \eea With $ \sum_l U_{4l}a_l=0 $.
Now $y^{\nu}$ is replaced
by this new matrix given by
 \bea
Z_{AB}&=& \tilde{h}_{AB} V_{1l}a_l -2 \sqrt{3} \tilde{f}_{AB}
V_{2l}a_l- \tilde{g}_{AB}(V_{5l}a_l+i \sqrt{3} V_{6l}a_l)\eea
The
inflation parameters are given by
 \bea h^2&=&\frac{4}{3}(|b_A
\tilde{h}_{AB} c_B|^2+12 |b_A \tilde{f}_{AB} c_B|^2+4 |b_A
\tilde{g}_{AB} c_B|^2 +  |b_A Z_{AB}|^2\\&&\nonumber +4|b_A
\tilde{f}_{AB} b_B|^2 + |Z_{AB}c_B|^2)\\ M^2&=& \frac{32}{3}
|\tilde{f}_{AB}
b_B|^2|\bar \sigma|^2+\frac{2}{3} |a_l|^2 m_{H_l}^2\\
A&=&\frac{16}{h}|c^*_A  Z^{\dag}_{AB} \tilde{f}_{BC} b_C \bar
\sigma| \eea
Now we throw the 24=6($b_{A}$)+6($c_A$)+12($a_m$)
real numbers randomly along with 38 parameters of NMSGUT
superpotential
  at $M_X$ while fitting. The dimensionality of space in which our search
  engine will look for the solution is thus 62.  The basic idea is that there can be an
   interplay between $b_A, c_A, a_m$ and SO(10) Yukawas $h_{AB},f_{AB},g_{AB}$ such
   that they can give us small quartic
   coupling h $\sim$ $10^{-6.1}$ or less required for
   50 e-folds along with fermion fitting. Also the mass of inflaton of O($10^{13}$ GeV)
    can be achieved with heavy higgs fields of GUT scale masses.
     An important point worth
    mentioning is that GUT scale threshold corrections  which lower the required SO(10)
    Yukawas so much that $\Gamma_{d=5}^{\Delta B\neq 0}$
   is suppressed  to less than $ 10^{-34}$ yrs  will also control the quartic
   coupling $h\sim10^{-6.3}$. In this way NMSGUT may connect the Baryon
   stability with the primordial inflation.

  \section {Details of Example Fit With Inflation Parameters}
  With these new formulas we tried to find a good fermion fit along
   with sufficient inflation.
  In Appendix A we give the complete set of tables.
  In Table (\ref{taba}) we quote the inflation parameters.
  We are able to achieve quartic
  coupling value h=$1.78 \times 10^{-5}$  and $N_{efolds} \approx $ 1. Note that this
  is an improvement over our earlier attempt by a factor of $10^4$.
  The quartic coupling as well as the mass of inflaton is somewhat larger
  than required for
  successful inflation. The value of $\omega$ ($\approx$ 2) achieved is
  still smaller than required.
 The rest of the parameters are in an acceptable range.

  However the search for inflation conditions along
   with fermion fitting led us to completely new kind of NMSGUT fitting solution!
   We provide an example fermion fit in Tables (\ref{tabb}-\ref{tabg}).
 If we look at Table
   (\ref{tabb}) the value of tan$\beta$=2.58 (which we didn't keep
   fixed but threw randomly
   along with soft parameters) given along with the soft Susy breaking
   parameters at scale  $M_{X}$   is quite small.
    NMSGUT was never hitherto able to achieve fitting with small tan$\beta$!
    This accidental discovery motivates new searches for low tan$\beta$
    solutions in NMSGUT. If we switch off the
     inflation penalties then it should be easier to find solutions.
     The play of GUT scale thresholds
     \cite{BStabHedge} might be responsible for such type of behavior
      as they were not applied before \cite{nmsgut}.
     Along with it one of the Higgs mass square parameter
     $M^2_{H}$= $7.0208 \times 10^{10} GeV^2$ comes out positive,
      which is also a new feature. It results into the more commonly
      encountered  inverted
      s-fermion hierarchy given in Table (\ref{tabd}).
      It  is notable  that all the s-particles are still
      very heavy of O(100 TeV). The rest of the features
      are already explained in Chapter 3.

 \section{Conclusions and discussion}

In the present scenario of inflation in NMSGUT we can achieve $
M=10^{13.0-13.2}$ GeV. The quartic coupling $h^2
=10^{-12.2}-10^{-12.7}$ seems harder to achieve but not
impossible. Because to search in such a high dimensional parameter
space one needs patience for  running of code for months. For such
a large value of inflaton mass the trilinear term comes out to be
very small and plays a minor role. Thus one can even drop that
term from inflation potential. Super Planckian vev defined in
terms of parameter $ \omega$= $\frac{\phi_0}{M_{pl}}$ $\sim$ 15 is
required to achieve  sufficient efolds. However in NMSGUT the
number of e-folds increased from $10^{-4}$ to 1 but is still not
realistic. The BICEP2 results favor inflation at GUT  scale and if
BICEP2 results stand scrutiny then NMSGUT
 can be a desirable candidate to explain inflation.
 Apart from SSI inflation, NMSGUT superpotential
 can have many other flat directions along which inflation
 can take place to evaluate which a detailed study is required.
 One needs to explore the NMSGUT potential to study  evolution
 with different field configurations. An important conclusion
that comes out from this study is that fitting of fermion data is
possible in NMSGUT even with small tan$\beta$ $\approx$ 2.5. It
implies that NMSGUT is viable on the entire range of tan$\beta$,
 thus overturning some assumptions of our previous
work \cite{nmsgut,nmsgutIII,BStabHedge}.

\section{Appendix A}
\begin{table}[h]
 $$
 {\small\begin{array}{|c|c|}
 \hline {\mbox {Inflation Parameter } }&{\mbox {Value}}\\
 \hline
               M &           7.8826 \times 10^{13}GeV \\
               h&            1.7736 \times 10^{-5}\\
       N_{efolds}&           .73\\
             \epsilon &       1.25 \times 10^{-2}\\
                \eta&        1.85 \times 10^{-2}\\
                n_s&         .962  \\
                P_R&          2.1977 \times 10^{-9} \\
                r&             .20\\
                A_0&         1.277 \times 10^{-4}\\
                  V_0^{1/4}&          2.1825 \times 10^{16} GeV \\
                    \omega &          1.82\\
                    |a_m| &             0.999,    4.295 \times 10^{-2}, 6.287 \times 10^{-4}, 3.049 \times
                    10^{-4},\\& 2.752 \times 10^{-4},  5.439 \times
                    10^{-5}\\
                    |b_A| &  0.999 , 3.601 \times 10^{-2}, 4.119 \times
                    10^{-3}\\
                    |c_A| & 0.999,           4.325 \times 10^{-2},    3.209 \times
                    10^{-3}\\
\hline
 \end{array}}
 $$\caption{Inflation parameters along with parameters $b_A, c_A, a_m$.
 Note that $|b_1|,|c_1|,|a_1|$ $\approx$ 1, and rest are very small.}
 \label{taba}\end{table}

\begin{table}
 $$
 {\small\begin{array}{|c|c|c|c|}
 \hline
 {\rm Parameter }&{\rm Value} &{\rm  Field }& {\rm Masses}\\
 &&{\rm [SU(3),SU(2),Y]}& {\rm ( Units\,\,of 10^{16} Gev)}\\ \hline
       \chi_{X}&  0.98           &A[1,1,4]&      0.69 \\ \chi_{Z}&
    0.40
                &B[6,2,{5/3}]&            0.05\\
           h_{11}/10^{-6}& -5.49         &C[8,2,1]&{      0.78,      3.55,      4.33 }\\
           h_{22}/10^{-4}&  4.45    &D[3,2,{7/ 3}]&{      0.35,      4.02,      5.10 }\\
                   h_{33}& -0.02     &E[3,2,{1/3}]&{      0.05,      0.51,      1.64 }\\
 f_{11}/10^{-6}&
  0.05-  0.03i
                      &&{     1.6,      3.41,      4.61 }\\
 f_{12}/10^{-6}&
 -2.51-  0.86i
          &F[1,1,2]&      0.03,      0.33
 \\f_{13}/10^{-5}&
  0.15-  0.095i
                  &&      0.33,      1.98  \\
 f_{22}/10^{-5}&
  5.17+  5.76i
              &G[1,1,0]&{     0.02,      0.12,      0.43 }\\
 f_{23}/10^{-4}&
 -1.23+  1.05i
                      &&{     0.44,      0.44,      0.51 }\\
 f_{33}/10^{-3}&
 -0.42-  0.24i
              &h[1,2,1]&{     0.22,      1.21,      3.68 }\\
 g_{12}/10^{-4}&
  0.11-  0.06i
                 &&{      4.38,     18.50 }\\
 g_{13}/10^{-5}&
  1.79+  4.67i
     &I[3,1,{10/3}]&      0.22\\
 g_{23}/10^{-4}&
 -0.93-  0.02i
          &J[3,1,{4/3}]&{     0.19,      0.28,      0.80 }\\
 \lambda/10^{-2}&
 -4.38-  1.01i
                 &&{      0.80,      2.76 }\\
 \eta&
 -0.26-  0.04i
   &K[3,1, {8/ 3}]&{      0.95,      2.69 }\\
 \rho&
  1.08-  0.38i
    &L[6,1,{2/ 3}]&{      0.97,      1.28 }\\
 k&
  0.25-  0.03i
     &M[6,1,{8/ 3}]&      1.16\\
 \zeta&
  1.08+  1.11i
     &N[6,1,{4/ 3}]&      1.11\\
 \bar\zeta &
 -0.03+  0.65i
          &O[1,3,2]&      2.07\\
       m/10^{16}GeV&  0.005    &P[3,3,{2/ 3}]&{      0.16,      2.65 }\\
     m_\Theta/10^{16}GeV&  -1.11e^{-iArg(\lambda)}     &Q[8,3,0]&     0.20\\
             \gamma&  0.20        &R[8,1, 0]&{      0.05,      0.22 }\\
              \bar\gamma& -4.05     &S[1,3,0]&    0.25\\
 x&
  0.85+  0.41i
         &t[3,1,{2/ 3}]&{      0.20,      0.75,      0.99,      1.91  }\\\Delta_X^{tot},\Delta_X^{GUT}&      0.86,      0.93 &&{      3.13,      4.28,     21.76 }\\
                                \Delta_{G}^{tot},\Delta_{G}^{GUT}&-19.96,-24.02           &U[3,3,{4/3}]&     0.23\\
      \Delta\alpha_{3}^{tot}(M_{Z}),\Delta\alpha_{3}^{GUT}(M_{Z})& -0.012,  0.002               &V[1,2,3]&     0.14\\
    \{M^{\nu^c}/10^{12}GeV\}&{0.004271,    1.02,   18.34    }&W[6,3,{2/ 3}]&              1.93  \\
 \{M^{\nu}_{ II}/10^{-8}eV\}&  .025,  6.03,        108.54               &X[3,2,{5/ 3}]&     0.05,     1.77,     1.77\\
                  M_\nu(meV)&{2.29,    7.21,   38.59    }&Y[6,2, {1/3}]&              0.06  \\
  \{\rm{Evals[f]}\}/ 10^{-6}&{0.13,   29.93,  538.95         }&Z[8,1,2]&              0.22  \\
 \hline\hline
 \mbox{Soft parameters}&{\rm m_{\frac{1}{2}}}=
             0.000
 &{\rm m_{0}}=
        278926.8
 &{\rm A_{0}}=
         -6.74 \times 10^{   5}
 \\
 \mbox{at $M_{X}$}&\mu=
          1.53 \times 10^{   5}
 &{\rm B}=
         -2.68 \times 10^{  10}
  &{\rm tan{\beta}}=            2.5842\\
 &{\rm M^2_{\bar H}}=
         -5.47 \times 10^{  10}
 &{\rm M^2_{  H} }=
          7.02 \times 10^{  10}
 &
 {\rm R_{\frac{b\tau}{s\mu}}}=
  7.9039
  \\
 Max(|L_{ABCD}|,|R_{ABCD}|)&
          5.95 \times 10^{-22}
  {\,\rm{GeV^{-1}}}&& \\
 \hline\hline
 \mbox{Susy contribution to}&&&
 \\
 {\rm \Delta_{X,G,3}}&{\rm \Delta_X^{Susy}}=
            -0.07
 &{\rm \Delta_G^{Susy}}=
             4.05
 &{\rm \Delta\alpha_3^{Susy}}=
            -0.014
 \\
 \hline\end{array}}
 $$
 \caption{\small{Fit : Column 1 contains values   of the NMSGUT-SUGRY-NUHM  parameters at $M_X$
  derived from an  accurate fit to all 18 fermion data and compatible with RG constraints.
 Unification parameters and mass spectrum of superheavy and superlight fields are  also given.
 The values of $\mu(M_X),B(M_X)$ are determined by RG evolution from $M_Z$ to $M_X$
 of the values determined by the EWRSB conditions.}}\label{tabb} \end{table}
 \begin{table}
$$
 {\small\begin{array}{|c|c|c|c|c|}
 \hline
 &&&&\\
 {\rm  Parameter }&{\rm Target =\bar O_i}&{\rm Uncert.= \delta_i}
 &{\rm Achieved= O_i}&{\rm Pull=\frac{(O_i-\bar O_i)}{\delta_i}}\\
 \hline
    y_u/10^{-6}&  2.054931&  0.784984&  2.049459& -0.006971\\
    y_c/10^{-3}&  1.001645&  0.165271&  0.971159& -0.184458\\
            y_t&  0.348829&  0.013953&  0.350779&  0.139806\\
    y_d/10^{-5}&  1.120978&  0.653530&  0.788682& -0.508463\\
    y_s/10^{-3}&  0.212620&  0.100356&  0.186201& -0.263246\\
            y_b&  0.012296&  0.006381&  0.013778&  0.232372\\
    y_e/10^{-4}&  0.053871&  0.008081&  0.053091& -0.096439\\
  y_\mu/10^{-2}&  0.113725&  0.017059&  0.109655& -0.238555\\
         y_\tau&  0.019341&  0.003675&  0.020974&  0.444360\\
             \sin\theta^q_{12}&    0.2210&  0.001600&    0.2210&            0.0045\\
     \sin\theta^q_{13}/10^{-4}&   31.7661&  5.000000&   31.9197&            0.0307\\
     \sin\theta^q_{23}/10^{-3}&   37.3725&  1.300000&   37.3109&           -0.0474\\
                      \delta^q&   60.0259& 14.000000&   59.9487&           -0.0055\\
    (m^2_{12})/10^{-5}(eV)^{2}&    4.6513&  0.493035&    4.6699&            0.0378\\
    (m^2_{23})/10^{-3}(eV)^{2}&    1.4536&  0.290712&    1.4369&           -0.0573\\
           \sin^2\theta^L_{12}&    0.2999&  0.059990&    0.3001&            0.0017\\
           \sin^2\theta^L_{23}&    0.5002&  0.150065&    0.4875&           -0.0844\\
           \sin^2\theta^L_{13}&    0.0286&  0.019000&    0.0190&           -0.5077\\
 \hline
       Eigenvalues(Z_{\bar u})&   0.945693&   0.946064&   0.946065&\\
       Eigenvalues(Z_{\bar d})&   0.939238&   0.939606&   0.939607&\\
     Eigenvalues(Z_{\bar \nu})&   0.907891&   0.908263&   0.908264&\\
       Eigenvalues(Z_{\bar e})&   0.927277&   0.927638&   0.927640&\\
            Eigenvalues(Z_{Q})&   0.960652&   0.961036&   0.961036&\\
            Eigenvalues(Z_{L})&   0.935785&   0.936151&   0.936152&\\
              Z_{\bar H},Z_{H}&        0.112304   &        0.000850    &{}&\\
 \hline
 \alpha_1 &
  0.176+  0.000i
 & {\bar \alpha}_1 &
  0.078+  0.0000i
 &\\
 \alpha_2&
 -0.676-  0.535i
 & {\bar \alpha}_2 &
 -0.409-  0.125i
 &\\
 \alpha_3 &
 -0.048-  0.376i
 & {\bar \alpha}_3 &
 -0.290-  0.269i
 &\\
 \alpha_4 &
  0.006+  0.042i
 & {\bar \alpha}_4 &
  0.517+  0.527i
 &\\
 \alpha_5 &
  0.088+  0.068i
 & {\bar \alpha}_5 &
  0.166-  0.283i
 &\\
 \alpha_6 &
 -0.044-  0.259i
 & {\bar \alpha}_6 &
 -0.003-  0.071i
 &\\
  \hline
 \end{array}}
 $$
 \caption{\small{Fit   with $\chi_X=\sqrt{ \sum_{i=1}^{17}
 (O_i-\bar O_i)^2/\delta_i^2}=
    0.9866
 $. Target values,  at $M_X$ of the fermion Yukawa
 couplings and mixing parameters, together with the estimated uncertainties, achieved values and pulls.
 The eigenvalues of the wavefunction renormalization increment  matrices $Z_i$ for fermion lines and
 the factors for Higgs lines are given.
 The Higgs fractions $\alpha_i,{\bar{\alpha_i}}$ which control the MSSM fermion Yukawa couplings  are also
 given.
  Right handed neutrino threshold  effects   have been ignored.
  We have truncated numbers for display although all calculations are done at double
 precision.}}
\label{tabc} \end{table}
 \begin{table}
 $$
 {\small\begin{array}{|c|c|c|c|}
 \hline &&&\\ {\rm  Parameter }& {\rm SM(M_Z)} & {\rm m^{GUT}(M_Z)} & {\rm  m^{MSSM}=(m+\Delta m)^{GUT}(M_Z)} \\
 \hline
    m_d/10^{-3}&   2.90000&   1.84864&   2.23856\\
    m_s/10^{-3}&  55.00000&  43.64483&  52.85039\\
            m_b&   2.90000&   3.01506&   3.57003\\
    m_e/10^{-3}&   0.48657&   0.52627&   0.52626\\
         m_\mu &   0.10272&   0.10870&   0.10869\\
         m_\tau&   1.74624&   2.07812&   2.07807\\
    m_u/10^{-3}&   1.27000&   1.02974&   1.24699\\
            m_c&   0.61900&   0.48791&   0.59087\\
            m_t& 172.50000& 143.37063& 170.87710\\
 \hline
 \end{array}}
 $$
\caption{\small{Values of standard model
 fermion masses in GeV at $M_Z$ compared with the masses obtained from
 values of GUT derived  Yukawa couplings  run down from $M_X^0$ to
 $M_Z$  both before and after threshold corrections.
  Fit with $\chi_Z=\sqrt{ \sum_{i=1}^{9} (m_i^{MSSM}- m_i^{SM})^2/ (m_i^{MSSM})^2} =
0.4014$.}}
 \label{tabd}%\end{table}
 %\begin{table}
 $$
 {\small\begin{array}{|c|c|c|c|}
 \hline
 {\rm  Parameter}  & {\rm Value}&  {\rm  Parameter}& {\rm Value} \\
 \hline
                       M_{1}&            437.38&   M_{{\tilde {\bar {u}}_1}}&         278803.51\\
                       M_{2}&           1000.00&   M_{{\tilde {\bar {u}}_2}}&         278801.41\\
                       M_{3}&           2350.15&   M_{{\tilde {\bar {u}}_3}}&          25640.30\\
     M_{{\tilde {\bar l}_1}}&         265890.12&               A^{0(l)}_{11}&        -673734.44\\
     M_{{\tilde {\bar l}_2}}&         265888.42&               A^{0(l)}_{22}&        -673732.79\\
     M_{{\tilde {\bar l}_3}}&         265267.68&               A^{0(l)}_{33}&        -673131.35\\
        M_{{\tilde {L}_{1}}}&         283496.59&               A^{0(u)}_{11}&        -482870.96\\
        M_{{\tilde {L}_{2}}}&         283495.79&               A^{0(u)}_{22}&        -482868.74\\
        M_{{\tilde {L}_{3}}}&         283206.48&               A^{0(u)}_{33}&        -295397.84\\
     M_{{\tilde {\bar d}_1}}&         266537.97&               A^{0(d)}_{11}&        -675378.37\\
     M_{{\tilde {\bar d}_2}}&         266537.84&               A^{0(d)}_{22}&        -675377.52\\
     M_{{\tilde {\bar d}_3}}&         266010.83&               A^{0(d)}_{33}&        -612876.05\\
          M_{{\tilde {Q}_1}}&         267304.87&                   \tan\beta&              2.58\\
          M_{{\tilde {Q}_2}}&         267303.72&                    \mu(M_Z)&         173486.65\\
          M_{{\tilde {Q}_3}}&         182583.24&                      B(M_Z)&
          3.3063 \times 10^{   9}
 \\
 M_{\bar {H}}^2&
         -5.2721 \times 10^{  10}
 &M_{H}^2&
         -5.2812 \times 10^{  10}
 \\
 \hline
 \end{array}}
 $$
 \caption{ \small {Values (GeV) in  of the soft Susy parameters  at $M_Z$
 (evolved from the soft SUGRY-NUHM parameters at $M_X$).
 The  values of soft Susy parameters  at $M_Z$
 determine the Susy threshold corrections to the fermion Yukawas.
 The matching of run down fermion Yukawas in the MSSM to the SM  parameters
 determines  soft SUGRY parameters at $M_X$. Note the  heavier third
 sgeneration.  The values of $\mu(M_Z)$ and the corresponding soft
 Susy parameter $B(M_Z)=m_A^2 {\sin 2 \beta }/2$ are determined by
 imposing electroweak symmetry breaking conditions. $m_A$ is the
 mass of the CP odd scalar in the in the Doublet Higgs. The sign of
 $\mu$ is assumed positive. }}
 \label{tabe}\end{table}
 \begin{table}
 \centering
$$
{\small \begin{array}{|c|c|}
 \hline {\mbox {Field } }&Mass(GeV)\\
 \hline
                M_{\tilde{G}}&           2350.15\\
               M_{\chi^{\pm}}&            999.97,         173486.70\\
       M_{\chi^{0}}&            437.37,            999.97,         173486.66    ,         173486.70\\
              M_{\tilde{\nu}}&        283496.581,        283495.788,        283206.471\\
                M_{\tilde{e}}&         265890.12,         283496.59,         265888.42   ,         283495.80,         265267.68,         283206.48  \\
                M_{\tilde{u}}&         267304.86,         278803.51,         267303.71   ,         278801.40,          25639.07,         182583.52  \\
                M_{\tilde{d}}&         266537.97,         267304.87,         266537.84   ,         267303.72,         182583.25,         266010.83  \\
                        M_{A}&          99112.92\\
                  M_{H^{\pm}}&          99112.96\\
                    M_{H^{0}}&          99112.94\\
                    M_{h^{0}}&            125.00\\
 \hline
 \end{array}}
 $$
 \caption{\small{Spectra of supersymmetric partners calculated ignoring generation mixing effects.
 Inclusion of such effects   changes the spectra only marginally. Due to the large
 values of $\mu,B,A_0$ the LSP and light chargino are  essentially pure Bino and Wino($\tilde W_\pm $).
   The light  gauginos and  light Higgs  $h^0$, are accompanied by a light smuon and  sometimes  selectron.
 The rest of the sfermions have multi-TeV masses. The mini-split supersymmetry spectrum and
 large $\mu,A_0$ parameters help avoid problems with FCNC and CCB/UFB instability\cite{kuslangseg}.
 The sfermion masses  are ordered by generation not magnitude. This is useful in understanding the spectrum
  calculated including generation mixing effects.}}\label{tabf}\end{table}

 \begin{table}
 $$
 {\small\begin{array}{|c|c|}
 \hline {\mbox {Field } }&{\rm Mass(GeV)}\\
 \hline
                M_{\tilde{G}}&           2352.37\\
               M_{\chi^{\pm}}&           1000.91,         173433.06\\
       M_{\chi^{0}}&            437.78,           1000.91,         173433.02    ,         173433.06\\
              M_{\tilde{\nu}}&         283206.72,         283496.04,        283496.832\\
                M_{\tilde{e}}&         265267.64,         265888.38,         265890.08   ,         283206.73,         283496.05,         283496.84  \\
                M_{\tilde{u}}&          25103.10,         182549.16,         267304.28   ,         267305.88,         278802.29,         278804.40  \\
                M_{\tilde{d}}&         182548.89,         266011.59,         266538.66   ,         266538.78,         267304.29,         267305.89  \\
                        M_{A}&          99562.44\\
                  M_{H^{\pm}}&          99562.47\\
                    M_{H^{0}}&          99562.46\\
                    M_{h^{0}}&            124.66\\
 \hline
 \end{array}}
 $$
 \caption{\small{Spectra of supersymmetric partners calculated including  generation mixing effects.
 Inclusion of such effects changes the spectra only marginally. Due to the large
 values of $\mu,B,A_0$ the LSP and light chargino are  essentially pure Bino and Wino($\tilde W_\pm $).
  Note that the ordering of the eigenvalues in this table follows their magnitudes, comparison
 comparison with the previous table is necessary to identify the sfermions}}\label{tabg}\end{table}
\newpage

\chapter{ Relic Density calculation for Neutralino Dark Matter}
\section{Introduction} The astrophysical evidences of dark matter
are well known now a days.  From the observations of COBE
\cite{COBE}, WMAP \cite{wmap}, PLANCK \cite{planck} it is evident
that our universe is filled with non bayronic matter other than
proton and neutrons. The total matter density of universe is
$\Omega_m h^2 \simeq $ 0.133 and baryonic density is $\Omega_b h^2
\simeq$ 0.022, which implies the $\Omega_{dm} h^2 \simeq$ 0.11.
The amount of this invisible matter is 4-5 times the visible
matter. Although evidence for existence of dark matter is strong,
 its nature is still unknown. Particle physics should provide some
suitable candidate for dark matter and its interactions with
Standard Model particles. Dark matter can't be baryonic. The CMB
structure that we see today would look different in case of
baryonic DM. Also the abundance of light elements during BBN (big
bang nucleosynthesis) era gives similar constraints on baryonic
content. Another constraint is that DM should be neutral, weakly
interacting and massive. DM candidates with masses $\sim$ 0.1-1
TeV are popularly known as
WIMPs (weakly interacting massive particle) in astro-particle
physics terminology. The Standard Model (SM) contains left handed
neutrino which is stable and neutral. But WMAP has constrained the
neutrino masses to be $\leq$ .23 eV, which gives the $\Omega_{\nu}
h^2 \simeq$ 0.0072. So it  cannot account for the whole
missing matter content. The Minimal Supersymmetric Standard Model
(MSSM), an appealing candidate for new physics, provides a suitable WIMP
candidate for dark matter namely the lightest supersymmetric particle (LSP).
 In MSSM the assumed symmetry R-parity
(introduced to eliminate the terms in Lagrangian which lead to
fast proton decay) makes the LSP stable.  We studied the
possibility of neutralino as dark matter candidate in context of
SO(10) NMSGUT. Since NMSGUT preserves R parity down to weak scale
it leads to LSP as suitable candidate for DM. In NMSGUT the LSP is
pure Bino. In this chapter we present tests of NMSGUT generated
s-spectra by evaluating the associated relic density for LSP. The
relic density for LSP is calculated using publicly available
package DarkSusy \cite{darksusy}.

\section{Evidences for dark matter}
There are many strong and convincing evidences for dark matter from the study of
galactic clusters, gravitational lensing and CMB. We briefly review each of these.
\subsection{Galactic clusters}
Jan Oort \cite{oort}, while observing stellar motions in the
galactic neighborhood, calculated  the velocities of stars by
Doppler Shift. He observed that the galaxies should be three times
more massive than the mass of visible matter contained to hold the
stars from escaping. So there is some missing matter.

 In
astronomy, the mass and size of galaxy can be determined by the
Virial theorem. While studying galaxy dynamics to find mass of
galaxy, the rotational velocity of gas and stars is used. Then the
K.E. and P.E of a galactic cluster of radius R and mass $M_c$ can
be related as  \be T=-\frac{1}{2}V \Rightarrow \frac{3
\sigma^{2}}{2} =\frac{GM_{c}}{R}\Rightarrow
M_{c}=\frac{3\sigma^{2} R}{2 G} \ee Here $\sigma$ represent
dispersion of galactic velocities using Doppler shift of light
from its galaxies, G=gravitational constant. Fritz Zwicky
\cite{zwicky} in his study of Coma cluster independently estimated
the mass of each galaxy in the cluster based on their luminosity
and added up all of galaxy masses to get a total luminous cluster
mass i.e. $M_{L}$ and obtained $ M_{c} = 400 M_{L}$. Thus majority
of the mass on the coma cluster is non luminous.
\subsection{Galactic rotation curve}
The presence of dark matter in spiral galaxies can be studied from
galactic rotation curves \cite{rubin}.
\begin{figure}
\centering
\includegraphics[scale=.3]{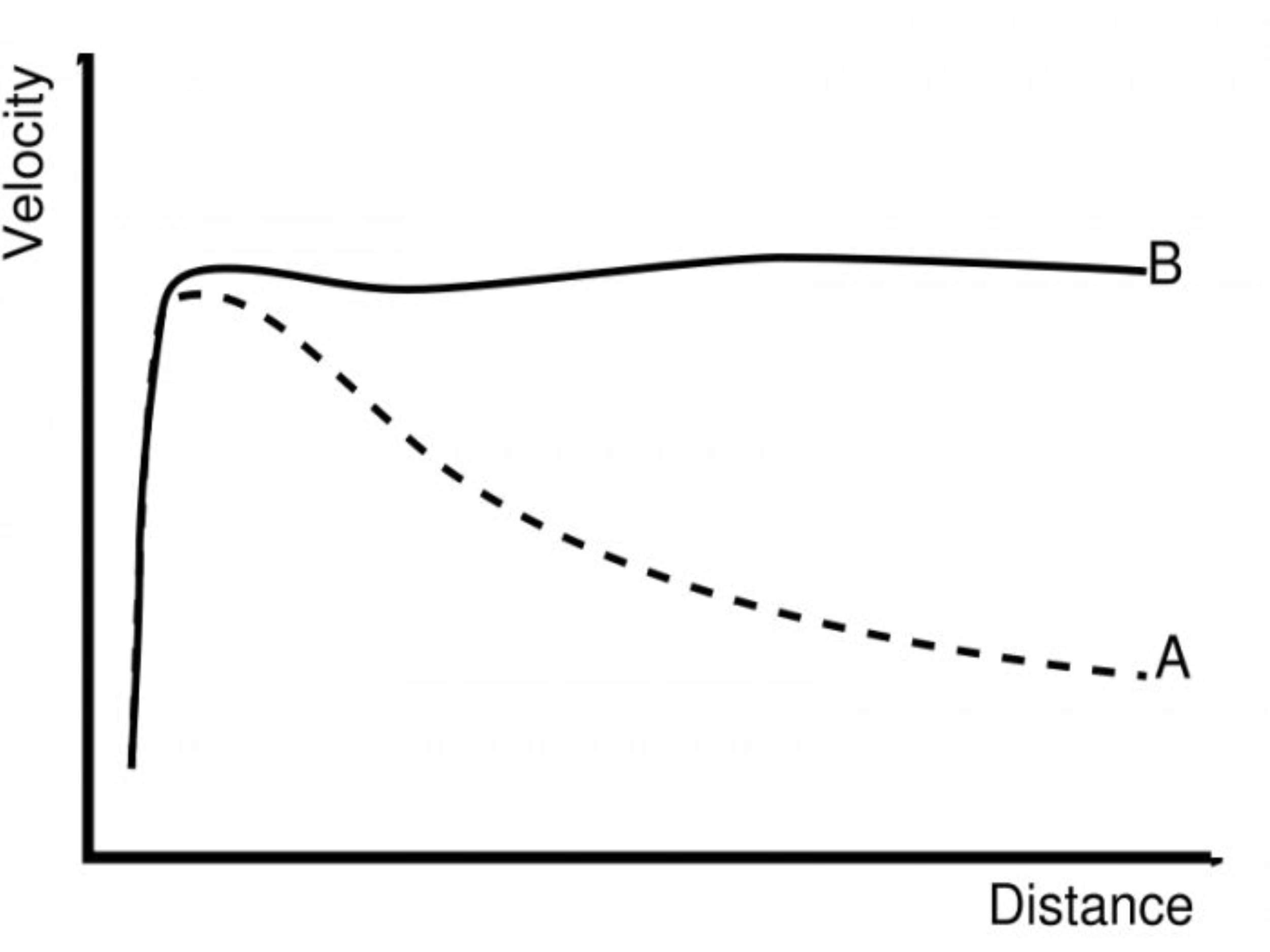}
\caption{Rotation curve of a typical spiral galaxy. Predicted A
and observed B.} \label{rotcur}\end{figure}
 According to Newton's
law of gravity, the stellar velocity should decrease with
increasing distance from center if the mass in the galaxy is
confined to its central region.\be v=\sqrt{\frac{GM}{R}} \ee But
from galactic rotation curves (fig. \ref{rotcur}) it is found that
stellar velocity remains constant with increasing distance. Thus
from flat galactic rotation curves one can say that there must be
some non luminous mass associated with each galaxy extending out
to a great distance beyond the central luminous region of the
galaxy
 which is still undiscovered.
\subsection{Gravitational lensing}
 The most recent evidence of dark matter came from gravitational
lensing by Bullet cluster \cite{bullet}. The Bullet
Cluster(IE0657-56) consists of two colliding clusters of galaxies.
The speed and shape of the 'Bullet' and other information from
various telescopes suggests that the smaller cluster passed
through the core of the larger one about 150 million years ago.
These two enormous objects collided at speeds of several million
miles an hour. The force of this event was so great that it
separated the normal matter in the form of hot gas away from dark matter.\\
Gravitational lensing by Bullet cluster can be the best evidence
of dark matter to present date.
 This technique can be used to determine the location of mass in a galaxy
cluster. Gravity from mass in the galaxy cluster distorts light
from background galaxies. Due to lensing, two distorted images of
one background galaxy are seen above and below the real location
of the galaxy. By looking at the shapes of many different
background galaxies, it is possible to make a map showing where
the gravity and therefore the mass in the cluster is located.
Largest lensing is in the region where dark matter is present.
Bullet cluster contain galaxies($\sim2\%$ of the mass),
intergalactic plasma($\sim10\%$ of the mass), and dark
matter($\sim88\%$ of the mass).
\subsection{Cosmic microwave background}
The above observations give strong evidence of localized dark matter but not of its total amount in the universe.
 However this information can be extracted from the study of cosmic microwave background radiation.
 CMB radiation was emitted 13.7 billion
years ago, only a few thousand years after the big bang. As the
Universe cooled, neutral hydrogen
 atoms formed and photons decoupled from matter. CMB photons interact weakly with the hydrogen
 allowing them to travel long distances($\sim Mpc $) in straight lines i.e. without scattering (surface of last scattering).
 The WMAP data \cite{wmap} of CMB put constraints on the baryonic and matter content of universe:
 \be \Omega_{m} h^2=0.14 \pm 0.02; \,\,\,\,\, \Omega_{b} h^2=0.024 \pm 0.001 \ee
During the Big-bang nucleosynthesis (BBN) period (from a few
seconds to a few minutes after the Big
 Bang in the early hot universe) neutrons and protons
 fused together to form deuterium, Helium and trace amounts of
 lithium and other light elements. Other heavy elements were
 produced later in stars. BBN limits the average baryonic content
 of the universe. BBN is the largest source of deuterium
  in the universe as any deuterium found or produced in
 stars immediately fused to helium.
 By considering the  deuterium to hydrogen ratio of regions with low levels of elements heavier than lithium, we are
 able to determine the D/H abundance directly after BBN.
 D/H ratio is heavily dependent on the overall density of baryons in the
  universe so measuring the D/H abundance gives the overall baryon abundance
  which is consistent with the CMB observations. The global dark matter density
  can be found by subtracting the baryon density from the total matter density.
  \be \Omega_{dm} h^2=0.11 \pm 0.021\ee

\section{Detection mechanism}

\subsection{Direct detection experiments}
The idea of direct detection is simple. If our galaxy is filled
with some non luminous matter then motion of our planet
through this matter will cause a recoil of nuclei of detector due
to elastic scattering of dark matter with the detector material.
The basic goal of direct detection experiments is to measure the
energy deposited after the recoil of WIMPs. The detectors, located
far underground to reduce the effect of background cosmic rays,
are sensitive to WIMP's that stream through the earth and interact
with nuclei in the detector. The recoiling nucleus deposits energy
in the form of ionization, heat and light, that is detected. After
an elastic collision with WIMPs of mass m, a nucleus of mass M
recoils with energy given as \be E=\left(\frac{\mu^{2}
v^{2}}{M}\right)(1-\cos\theta) \ee where $\mu=\frac{mM}{m+M}$ is
the reduced mass, v is the relative velocity of WIMPs w.r.t
nucleus and $\theta$ is the scattering angle in the center of mass
frame. Some experiments \cite{damalibra} find annual modulation in the dark matter
signal because of the relative motion of the earth through the
dark matter halo. For the simplified assumptions about the
distribution of dark matter in the halo, flux is maximum in June
and minimum in December. Annual modulation is a powerful signature
for dark matter because most background signals e.g radioactivity
don't exhibit such time dependence. Another method is to detect directional
dependence of nucleus recoil (away from the direction of Earth motion)
which is more powerful than annual modulation.

\subsection{Indirect detection experiments}
Indirect detection experiments are based on detection of the
particles produced due to annihilation of dark matter. These
include neutrinos, gamma rays and antimatter (positrons).

 Gamma
rays from WIMPs annihilation are more readily produced in galactic
center. These can be produced by two processes.
 First is WIMPs annihilates into quark and anti quark pair, which further produced a
particle jet from which stream of gamma rays is released. Second
is the direct annihilation of WIMPs into gamma rays. Typical WIMPs
masses of O(100) GeV produce very high energy gamma rays. Such a
gamma ray line would be a direct indication of the dark matter but
the detection is not that easy since the production of gamma rays
from other sources is not well understood. The Fermi Gamma Ray
Space Telescope (NASA's) \cite{fermilat}, designed to capture
gamma rays from the center of our own galaxy can provide the best
evidence for the existence of Dark Matter.

 If the galactic halo is
filled with the dark matter then it will scatter off nuclei in the
sun and the earth. The probability of scattering is small and the
recoil velocity is smaller than the escape velocity so they get
gravitationally bound to sun or earth. With time they will undergo
multiple scattering and get settled inside the core and can
annihilate with another WIMPs to quarks, leptons and Higgs bosons.
 But these particles get absorbed almost immediately. However if annihilation
  of WIMPs produces energetic muon neutrinos then they can travel long
  distances and can be detected by neutrino detectors. Inside the detector
   muon neutrino suffers a charged current event and produces a muon.
   These neutrinos can easily be distinguished from solar, atmospheric
   neutrinos as they are more massive (O(GeV) because they are produced from WIMPs)
   than the previous ones (O(MeV)). Moreover the background signal from solar,
   atmospheric neutrinos over from the WIMP annihilation can easily be removed.

    Antimatter can be an excellent signal of DM because
 production of antimatter by WIMP annihilation.
  For example: WIMP annihilation can produce positrons through secondary processes
  such as $W^{+},W^{-}$ and ZZ where $W^{+}\rightarrow$ $e^{+}\nu_{e}$. One needs to study the flux of
 antimatter particles over the entire galactic halo rather than on
 particular areas because the products are charged and affected by
 magnetic fields within space and lose energy very soon so we can't say
 where the annihilation took place unlike gamma rays and
 neutrinos.

\section{Neutralino as a cold DM}
Beyond Standard Model theory, Supersymmetry gives us a leading
 cold dark matter candidate. In Minimal Supersymmetric Standard Model(MSSM) to eliminate
 new B,L violating Yukawa couplings in the renormalizable superpotential we add a new
discrete($Z_{2}$)symmetry. This new symmetry ia called R parity
which gives us the best possible cold dark matter candidate i.e.
Light Supersymmeteric Particle(LSP)\cite{martin}. R parity is
defined as:
 \be           R = (-1)^{3(B-L)+2s} \ee
 where $s$ is spin of particle.
 All SM particles have R=1 while all supersymmeteric
 particles have R=-1. Because of this R parity, single SUSY particles cannot decay into
just standard model particles. If R parity is conserved, all
the supersymmeteric particle will eventually decay into lightest
susy particle (LSP). The LSP can't decay further but it can
annihilate. This makes supersymmetery an interesting theory from
 astrophysical point of view. In MSSM we have three color and charge neutral particles sneutrino, gravitino and
 neutralino. Here we will discuss only about the possibility of
 neutralino only. In MSSM the neutral higgsinos $
(\tilde{H_{u}^{0}}, \tilde{H_{d}^{0}
 })$ and the neutral gauginos $ (\tilde{B},\tilde{W}^{0}) $ combine to form
 four mass eigenstates called neutralinos. In general neutralinos can be expressed as:
\be \chi= a \tilde{B}+b \tilde{W}^0+c\tilde{H_{u}^{0}}+d
\tilde{H_{d}^{0}}\ee The coefficients a,b,c,d can be found by diagonalizing
the mass matrix for neutralinos:
  \be M_{\tilde N}= \begin{pmatrix} M_1 & 0 &
  \frac{-g'v_d}{\sqrt{2}}& \frac{g'v_u}{\sqrt{2}} \cr
0 & M_2 &
  \frac{g v_d}{\sqrt{2}}& \frac{-g v_u}{\sqrt{2}} \cr

  \frac{-g'v_d}{\sqrt{2}}& \frac{g v_d}{\sqrt{2}} & 0 & -\mu \cr
  \frac{g'v_u}{\sqrt{2}}& \frac{-g'v_u}{\sqrt{2}} & -\mu & 0
  \end{pmatrix}\ee
Here $M_1$ and $M_2$ (deduced from $m_{1/2}$ by RG
evolution from the Planck/GUT scale to to Susy scale $\sim$ TeV) are the soft susy breaking terms giving
mass to the U(1) and SU(2) gauginos respectively. $g^{'}$ and g are
gauge coupling corresponding to U(1) and SU(2) group. $\mu$ is
electroweak symmetry breaking parameter. $v_u, v_d$ are Higgs vevs
determined by tan $\beta$. So the mixing and masses of the
neutralinos depend upon $m_{\frac{1}{2}}, \mu $ and tan$\beta$.
After diagonalizing, by
 convention the mass eigen-values can be labelled in ascending order so that$\ m_{\chi_{1}^{0}} <
 m_{\chi_{2}^{0}}<m_{\chi_{3}^{0}}<m_{\chi_{4}^{0}}$. The
 lightest neutralino $\ \chi_{1}^{0} $ is assumed to be the
 desired candidate for dark matter. If it is produced in the early universe,
 then it may exist in the present day universe with sufficient
 density (relic density) to act as dark matter.

\section{Relic density}
The study of thermodynamics of early universe is necessary to
understand the particle creation. Then solution of Boltzmann
equations which governs the evolution of the different particle densities,
tell us the relic density for any particular species
 today. The most important quantity to calculate is the
interaction rate. If we consider WIMP (LSP) to be dark matter then due to
assumed conservation of R-parity it cannot decay but it can annihilate or co-annihilate
(i.e. produce normal matter by collision with other R-parity odd superparticle).
So we are interested in the calculation of
relic density of LSP dark matter whose number density varies due to
annihilation (coannihilation). The annihilation rate is given as
\bea \Gamma = n \sigma v. \eea Here
n is the number density, $\sigma $ is annihilation cross section and $v$ is
relative velocity of particles. As long as $\Gamma > H $(Hubble
constant), the particles remain in thermal equilibrium with the
rest of the universe. But when $\Gamma < H$, the particles are
said to be decoupled i.e. they move freely.
The evolution of number density of a particle in early universe is governed by its Boltzmann equation:
 \be \frac{dn}{dt}+3Hn= -<\sigma
v>[n^2-n_{eq}^2] \ee One can solve this equation by standard approximation methods
 however \cite{griestseckel} pointed out 3 major improvements required while
 calculating the relic density from this equation:
\begin{enumerate}
\item  \textbf{Relic density from coannihilation} If the supersymmetric theory
contains a superparticle slightly more massive than the LSP then the relic density of
LSP is not determined by only annihilation rates but one must also consider the coannihilation
channels wherein the LSP and the nearly degenerate superparticle ``co-annihilate'' to ordinary matter.
\item \textbf{Annihilation into forbidden channels} This case deals with
the annihilation of LSP into particles which are more massive than the
LSP. Since at the time of freeze out LSPs are thermally distributed and
 annihilation into heavier particle can occur. \cite{griestseckel} showed that if the heavier
supersymmetric particles are only 5-15$\%$ more massive, then these
channels can dominate the annihilation cross section and determine
the relic density. for example, annihilation into higgs, $
b\overline{b}$, $t\overline{t}$, $W^{+}W^{-}$ etc.
\item\textbf{Annihilation near poles} In this case the relic density is greatly affected
if the annihilation occurs near a pole. This can happen if the mass of neutralino is just
 near to half of the mass of exchanged particle during annihilation.
 These exceptions basically alter the way one calculates the thermal averaging of cross-section and hence the relic density.
\end{enumerate}
We review here the calculation of relic density as given in \cite{gondgel,edsjo}
  considering all the factors mentioned above.\\
Let us consider that N supersymmetric particles
$\chi_{i}$(i=1,...,N) with masses $m_{i}$ annihilate. Assume that
$\chi_{1}$ with mass $m_{1}$ is lightest, $\chi_{2}$ is
the second lightest and so on. Evolution of the $\chi_{i}$ can be
determined by set of N Boltzmann equations:\\
\be \begin{split} \label{eu2}
\frac{dn_{i}}{dt}&=-3Hn_{i}- \sum_{j=1}^N[<\sigma_{ij}v_{ij}>(n_{i}n_{j}-n_{i}^{eq}n_{j}^{eq})\\
&-\sum_{j\neq i}(<\sigma_{Xij}^{'}v_{ij}>(n_{i}n_{X}-n_{i}^{eq}n_{X}^{eq})-
<\sigma_{ij}^{'}v_{ij}>(n_{j}n_{X}^{'}-n_{j}^{eq}n_{X}^{eq})\\
&-\sum_{j\neq i}[\Gamma_{ij}(n_{i}-n_{i}^{eq})-\Gamma_{ji}(n_j-n_j^{eq})]\end{split} \ee
The term with Hubble rate H on the R.H.S  represents decrease in density
due to the expansion of the universe. The second term
describes $\chi_{i}\chi_{j}$ annihilations into standard model
particles $\chi_{i}\chi_{j}$ $\rightarrow$ X$X^{'}$ with cross section \be
\sigma_{ij}=\sigma(\chi_{i}\chi_{j}\rightarrow XX^{'}) \ee The
third term represents conversions with the cosmic backgrounds
($\chi_{i}\rightarrow$$\chi_{j}$) which changes $n_{i}$ if i$\neq$j and have cross section
\be \sigma^{'}_{ij}=\sigma(\chi_{i}X\rightarrow \chi_{j}X^{'}) \ee
The last term represents decay of $\chi_{i}$ particles with decay rate
\be
\Gamma_{ij}\rightarrow\Gamma(\chi_{i}\rightarrow\chi_{j}XX^{'})
\ee  $v_{ij}$ is the relative velocity defined as:
\be v_{ij}=\frac{\sqrt{(P_i.P_j)^2-m_i^2m_j^2}}{E_iEj}\ee
$n_i^{eq}$ is the equilibrium number density for $i^{th}$ particle given as:
 \be n^{eq}_i= \frac{g_i}{2\pi}^3\int d^3p_if_i \ee
 Here $p_i$ is the three-momentum of particle and $f_i$ is the phase space distribution
  function defined in Maxwell-Boltzmann approximation as
 \be f_i= e^{\frac{-E_i}{T}} \ee
And finally the $<\sigma_{ij}v_{ij}>$ in this thermal distribution is defined as
\be <\sigma_{ij}v_{ij}> = \frac{\int d^3p_i d^3 p_j f_i f_j \sigma_{ij}v_{ij}}{\int d^3p_i d^3 p_j f_i f_j}\ee
Since all the $\chi_{i}$, which are heavier than LSP will
eventually decay into $\chi_{1}$, the total density of $\chi_{i}$
particles is, n=$\sum_{i=1}^{N}$$n_{i}$. Therefore the sum of equation
(\ref{eu2}) over i gives:
\be \label{eu3}
\frac{dn}{dt}=-3Hn-\sum_{ij=1}^{N}<\sigma_{ij}v_{ij}>(n_{i}n_{j}-n_{i}^{eq}n_{j}^{eq})
\ee
Since the  scattering rate of supersymmetric particles in thermal
background is much faster than their annihilation rate because $\sigma_{ij}^{'}$
$\sim$ $\sigma_{ij}$ and when $n_{X}$ $>>$ $n_{i}$, then due to relativistic nature of X particles
 and non relativistic nature of $\chi$ particles
lead them to suppressed by Boltzmann factor. So $\chi_{i}$ distribution remain in
thermal equilibrium and we have
\be \frac{n_i}{n}\simeq \frac{n_i^{eq}}{n^{eq}}\ee
After putting it in equation (\ref{eu3}), we get
\be
\frac{dn}{dt}=-3Hn-<\sigma_{eff} v>(n^2-n^2_{eq})
\label{eu4}\ee
Where
\be <\sigma_{eff} v>= \sum_{ij} <\sigma_{ij}v_{ij}>\frac{n_i^{eq} n_j^{eq}}{n^{eq}n^{eq}}
\ee
The numerator (let us denote it by A) in the above equation is the annihilation rate per unit
 volume and it can be written in a more convenient way as
\be  A= \sum_{ij} <\sigma_{ij}v_{ij}> n_i^{eq} n_j^{eq}= \sum_{ij}\int W_{ij}
\frac{g_i f_i d^3 p_i}{(2\pi)^3}\frac{g_j f_j d^3 p_j}{(2\pi)^3}\ee
  Here $g_i$ are internal degrees of freedom of particle. $W_{ij}$ is Lorentz invariant and related to cross section as
  \be  W_{ij}= 4p_{ij}\sqrt{s}\sigma_{ij}= 4E_iE_j \sigma_{ij}v_{ij}\ee
  Here $p_{ij}$ is the momentum of $i^{th}$ particle w.r.t $j^{th}$ or vice verse in the center
  of mass frame of both particles and given as

  \be p_{ij}= \frac{[s-(m_i+m_j)^2]^{1/2}[s-(m_i-m_j)^2]}{2\sqrt{s}} \ee
 After following steps given in \cite{gondgel} one can find effective form of  $<\sigma_{eff} v>$ given as
\be <\sigma_{eff} v>= \frac{\int_{0}^{\infty}dp_{eff}~p_{eff}^2 W_{eff} K_1(\frac{\sqrt{s}}{T})}{m_1^4T[\sum_i \frac{g_i}{g1}\frac{m_i^2}{m_1^2} K_2(\frac{\sqrt{m_i}}{T})]^2} \label{sigmaeff}\ee
Here $p_{eff}= \sqrt{s-4m_1^2}/2$ and T is the temperature. $K_1$ is the modified Bessel function of second kind of order 1 and $k_{2}$ is of order 2. $W_{eff}$ has form
\be W_{eff}=\sum_{ij}\frac{p_{ij}}{p_{eff}}\frac{g_{i}g_{j}}{g_{1}^2} W_{ij}\ee
The eq. (\ref{eu4}) can be written in a more convenient way if we put $Y=\frac{n}{s}$ in this equation. Here s is entropy density. We have
\be \label{eu11} \frac{dY}{dt}= -<\sigma
v>s[Y^{2}-Y_{eq}^{2}]\ee
Since the right hand side depends upon temperature T so it is convenient to make the T as a independent variable instead of time. Define a new variable $x_f$=$m_1/T$ then this equation can be written as
 \be \label{eu11} \frac{dY}{dx}= -\sqrt{\frac{\pi}{45G}}\frac{g_{*}^{1/2}m_1}{x^2}<\sigma_{eff}v>[Y^{2}-Y_{eq}^{2}]\ee
 Where \be Y_{eq}=\frac{45x^2}{4\pi^4 h_{eff}(T)}\sum_{i} g_i(\frac{m_i}{m_1})^2 K_2(x\frac{m_i}{m_1})\ee
 and $g_{*}$ is the total number of relativistic degrees of freedom given as
\be g_{*}^{1/2} = \frac{h_{eff}}{\sqrt{g_{eff}}} (1+\frac{T}{3 h_{eff}}\frac{dh_{eff}}{dT}) \ee
$g_{eff}$ and $h_{eff}$ are the effective energy and entropy degrees of freedom. $g_{eff}$,
 $h_{eff}$ and $g_{*}^{1/2}$ are function of temperature and can be calculated at QCD phase
 transition temperature $T_{QCD}$=150 MeV as given in \cite{gondgel}.

\begin{figure}[h]
\centering
\includegraphics[scale=.5]{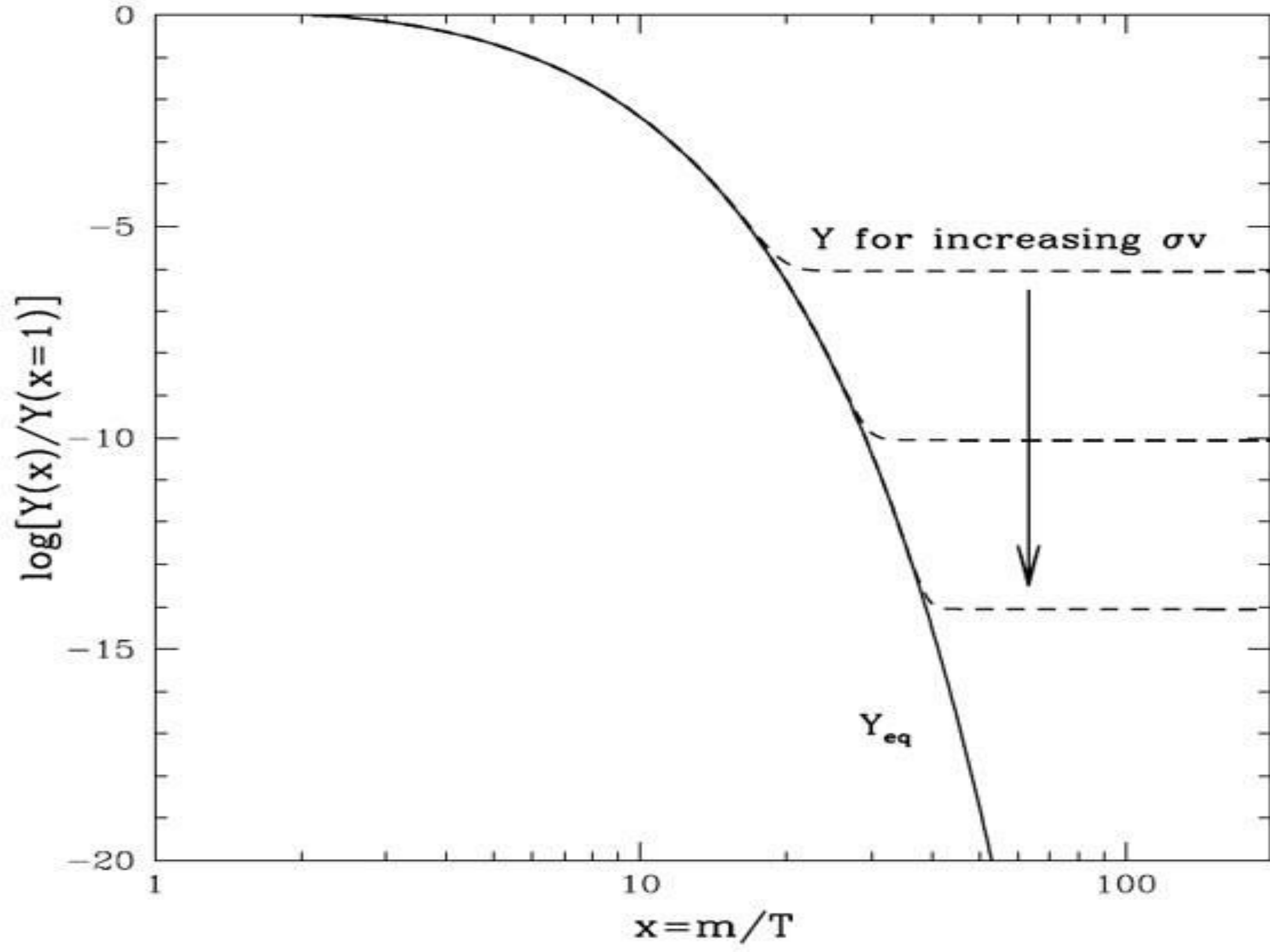}
\caption{The dashed line is the actual abundance, and the solid
line is in equilibrium abundance which is boltzmann factor
suppressed.} \label{boltzman}\end{figure}

Fig.\ref{boltzman} plots an
analytic approximation to the Boltzmann equation and illustrates
several key points. At freeze out, the actual abundance leaves the
equilibrium value, and remains essentially constant, the
equilibrium value continues to decrease.
 The larger the annihilation cross section, lower the value of
 relic abundance Y.
 Using freeze-out
approximation and integrating this differential equation between
x=0 to $x=m_1/T_0$, where $T_0$ is the present day CMB
temperature, leads to the formula for the relic density
\cite{gondgel}: \be \Omega h^{2}
=2.755\times10^{8}\frac{m_{1}}{GeV} Y_0 \label{eu20} \ee

\section{DarkSusy}
DarkSusy \cite{darksusy} is a publicly available package for the
neutralino dark matter calculations. It a code written in FORTRAN
77. With the help of DarkSUSY one can calculate the present relic
density for neutralino. While calculating the relic density
DarkSUSY takes care of resonances, annihilation thresholds and
coannihilations which can have large effect on the value of relic
density. Also it provides the rates from direct and indirect
detection methods. Our interest in this study is calculation of
relic density for NMSGUT generated spectra. We use  version 5.1.1
of DarkSusy for this purpose. It basically provides a library
which can be interfaced with one's own code. However for testing
it can also be used by giving it sample set of model parameters.
Let us discuss the various steps how it calculates the relic
density for a particular model.

 The main function is DSMAIN inside
the test directory. One can enter model parameters during
interactive run by hand (for MSSM, mSUGRA model) or can provide a
SLHA file \cite{slha}. Then it calls ISASUGRA (from ISAJET) to
calculate s-particle spectrum and checks whether the model
parameters are consistent or not. However one can give the low
energy spectrum. In our case, we gave it the tree level spectrum
(with loop corrected Higgs mass) which NMSGUT calculates and
directly perform the task of calculating relic density. The
subroutine which calculates relic density is the function
DSRDOMEGA. All the main routine files which form the DarkSusy
library are given inside the SRC directory. The steps how it calls
main subroutines and uses functions while calculation of relic
density in DarkSusy are given below.

\begin{enumerate}
\item {\bf DSRDOMEGA} It is function which returns relic density $\Omega h^2$ for MSSM
neutralino using subroutine  DSRDENS. It uses a switch which controls whether
one wants to include co-annihilation channels or not. It also
checks for resonances and thresholds looking at the mass of each
 particle which need to be included .

\item {\bf DSRDENS} It is a subroutine
 which calculates the relic density in the units of
critical density time $h^2$. To calculate relic density it
requires invariant annihilation rates for which it requires
function DSANWX.

\item{\bf DSANWX} It is a function which
 calculates the neutarlino annihilation invariant rates. It
requires the value of momentum ($p_{eff}$) of two annihilating particles in their center of momentum frame.

\end{enumerate}
DarkSusy takes care of thresholds and resonances through
variation of $W_{eff}$ and $p_{eff}$. Since the equation (\ref{sigmaeff})
is independent of the temperature so $W_{eff}$ and $p_{eff}$ can be tabulated for each model.
The value of $p_{eff}$ should be such that it includes all resonances, thresholds and coannihilation.
The thermal average of $W_{eff}$ is adjusted by factor $K_{1}p_{eff}^2$ and at resonances
and thresholds there is exponential decay of $K_{1}$ for large $p_{eff}$. DarkSusy calculates
 the thermal average via integrating over $p_{eff}$ using adaptive Gaussian integration, using
 spline interpolation in ($p_{eff},W_{eff}$) table and splits the integration interval at sharp thresholds.
\section{NMSGUT and relic density calculation with DarkSusy}
 In this section, we
present results for relic density of lightest neutralino (Bino in this case) calculated with
DarkSusy for two example sets of mass spectra taken
from \cite{nmsgut} compatible with SM fermion data. The sparticles
are mostly heavy O(10-50 TeV) except for gauginos and in one of these
spectra Smuon is also light O(100GeV). The input i.e. low energy MSSM data is given in SLHA (Susy Les
Houches Accord)\cite{slha} format. The input spectrum is shown in
Table \ref{table:massmat}
and outputs are shown in Table \ref{table:relic}.\\
Results for the relic density with and without including
coannihilation channels are shown. From the output it is clear
that we got large value of relic density for Soln.1 and reasonable
value for Soln.2. Also in Soln.1 all the particles have masses
much heavier than Neutralino 1($\chi_{1}^{0}$), so no
coannihilation channels are possible, so relic density with and
without co-annihilation is almost same. In Soln.2 there is one
particle right handed Smuon($\widetilde{\mu}_R$) whose mass is
comparable to that of $\chi_{1}^{0}$, So coannihilation channels
are possible which result into further decrease in relic density.

Approximate value of relic density given in \cite{jungman} by
taking $g_{*}^{1/2}=10$, $x_{f}=25$, $M_{pl}=10^{19}$ gives us:
\be \Omega h^{2}= \frac{3 \times 10^{-27} cm^{3}s^{-1}}{<\sigma
v>} \label{approxsig}\ee The values of $<\sigma v>$ for individual
annihilation channels are given in Table \ref{table:sol1} and for
coannhilation channels are given in Table \ref{table:sol2}.

\begin{table}[!h]
\centering
\begin{tabular}{ |c| c| c| c|} % centered columns (4 columns)
\hline\hline %inserts double horizontal lines
  PDG code\ & Mass(Soln.1)\ & Mass(Soln.2) \ & Particles   \ \\ [0.5ex] % inserts table
%heading
\hline % inserts single horizontal line
         24 &   80.32 &  80.326      &                   $  W^{+}$  \\
        25  &  125.000 &   123.99   &                 $h_{0} $ \\
        35  &  457636.54 &  377025.28   &             $H_{0}$ \\
        36  &  457636.54  & 377025.29          &           $A_{0}$ \\
        37  &  457636.55  & 377025.30     &          $H^{+}  $ \\
    1000021 &   1200.000      &       1000.13    &        $ \widetilde{g}$  \\
    1000022  &  {\bf 246.4066}           &   {\bf210.0974}  & $\chi_{1}^{0} $  \\
    1000023  &  590.178           & 569.808     &   $\chi_{2}^{0}$   \\
    1000024   & 590.178            &  569.8083      &  $\chi_{1}^{+}$ \\
    1000025   & 155715.44     &    125591.2   &$\chi_{3}^{0} $   \\
    1000035   & 155715.44     &    125591.2   & $ \chi_{4}^{0}$   \\
    1000037   & 155715.46       & 125591.22    &  $\chi_{2}^{+}$    \\
    1000001   & 11245.066       &  8402.9938   & $\widetilde{d}_{L}$        \\
      1000002 &  12822.48       &  11271.798        &$ \widetilde{u}_{L}$          \\
    1000003   & 11246.13        &  8401.4765         &$ \widetilde{s}_{L} $           \\
    1000004   & 12822.424       &  11270.628       &$ \widetilde{c}_{L}$      \\
    1000005   & 48865.409       &  40269.187           &$\widetilde{b}_{L} $  \\
    1000006   & 48227.493       &  24607.508       &$ \widetilde{t}_{L}$   \\
    1000011   & 11958.032       &  1761.88     & $ \widetilde{e}_{L} $            \\
    1000012  &  15324.618       &   11270.946 & $ \widetilde{\nu}_{eL} $      \\
    1000013  &  11961.759       &    15258.60 & $\widetilde{\mu}_{L}$  \\
    1000014  &  15326.183       &   15258.32 & $\widetilde{\nu}_{\mu L} $ \\
    1000015  &  30125.093       &    20674.71 & $ \widetilde{\tau}_{L} $  \\
    1000016  &  30130.304       &   21320.059 & $  \widetilde{\nu}_{\tau L}$ \\
    2000001  &  13440.651       &  11272.10    & $ \widetilde{d}_{R} $  \\
    2000002  &  13440.398       &  14446.76    &  $ \widetilde{u}_{R}$  \\
    2000003  &  13441.103       &  11270.09    & $\widetilde{s}_{R} $  \\
    2000004  &  13440.845       &   14445.80  &  $ \widetilde{c}_{R} $  \\
    2000005  &  49419.242       &  51845.90   &   $  \widetilde{b}_{R} $  \\
    2000006  &  48998.142      &   40275.875    & $ \widetilde{t}_{R} $ \\
    2000011  &  15324.839      &  15308.291     &  $ \widetilde{e}_{R}$ \\
    2000013  &  15326.634       & {\bf211.56859}    &   $  \widetilde{\mu}_{R}$\\
    2000015  &  38560.601       & 21419.559     &    $\widetilde{\tau}_{R}$  \\ [1ex]
    \hline
 \end{tabular}
 \caption{The input sparticle spectrums in SLHA format to DarkSusy.}\label{table:massmat}
 \end{table}

\begin{table}[!h]
 \centering
\begin{tabular}{|c| c|c|} % centered columns (4 columns)
\hline\hline %inserts double horizontal lines
 $\Omega h^{2}$  \ & Soln.1 \ & Soln.2  \\ [0.5ex] % inserts table
%heading
\hline % inserts single horizontal line
With coannihilation & 97182.8934 & 0.051  \\

Without coannihilation                & 97182.8934  & 1.285   \\
\hline
\end{tabular}
\caption{Relic Density of neutralino.} \label{table:relic}
 \end{table}

\begin{figure}[!h]
\begin{center}
\includegraphics[scale=.5]{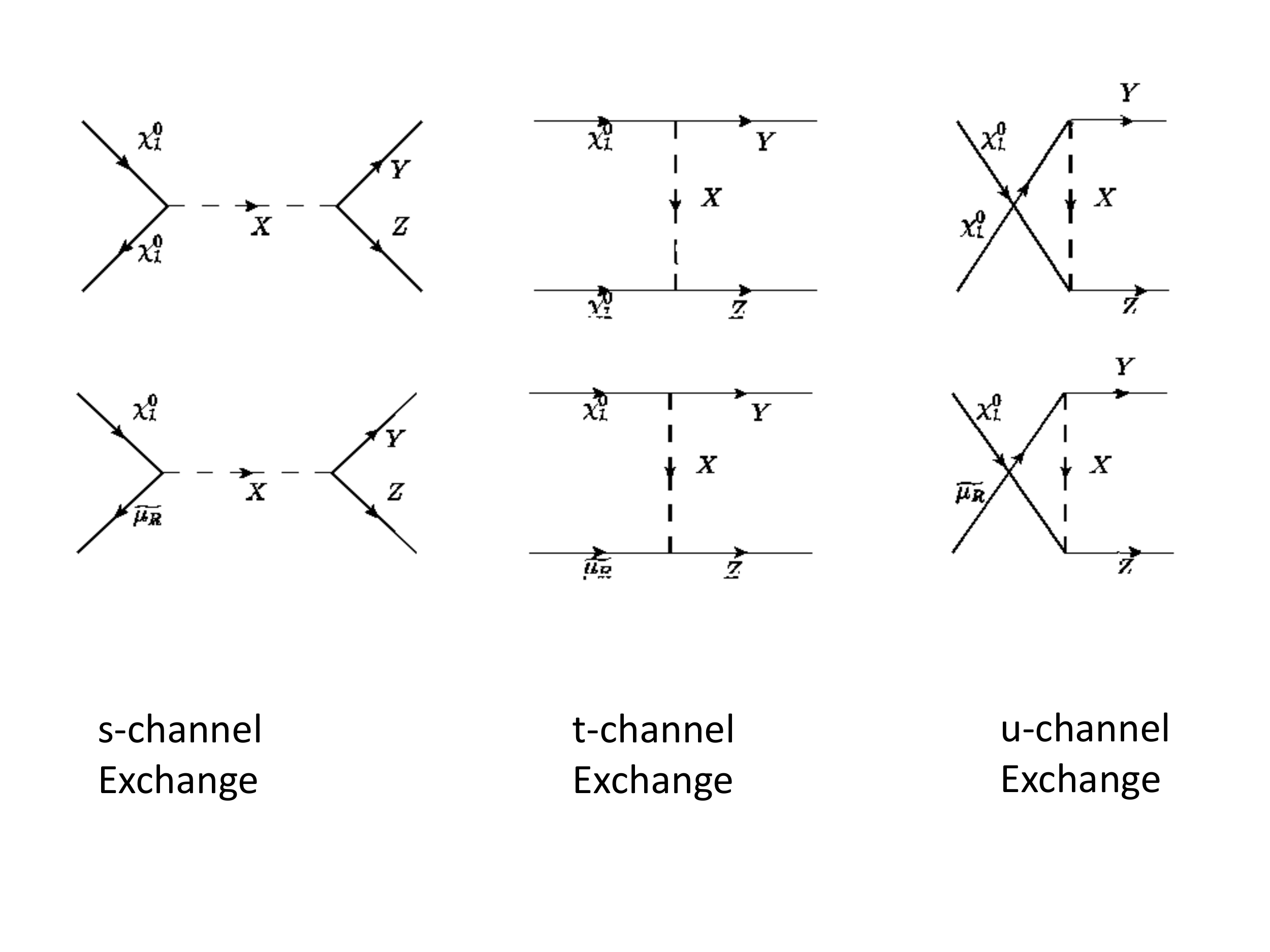} \caption{Neutralino pair
annihilation and Neutralino Smuon coannihilation. See tables
\ref{table:sol1} and \ref{table:sol2} for definitions of
YZ,X.}\label{annchannel}
\end{center}
\end{figure}

 The reason for large value of relic density in the first solution can be explained
 from the diagrams of neutralino annihilation.
Feynmann diagrams for the $\chi_{1}^{0} \chi_{1}^{0} \rightarrow$
YZ are given in Fig. \ref{annchannel}
 If we calculate the relic density by eq(\ref{approxsig}) and put the value of the thermally
averaged annihilation cross section for the dominating channel
($<\sigma v>$) and compare it with the relic density calculated by
DarkSusy, we find the results comparable.

 For Soln.2, first we
calculate the relic density when there is no coannihilation. If we
take $<\sigma v> = <\sigma v>_{\chi_{1}^{0}
\widetilde{\mu}_{R}\rightarrow \mu^{+}\mu^{-}}$ which is the
dominating channel and calculate $\Omega h^{2}$, we get $\Omega
h^{2}$ $\simeq$ 0.8 Similarly, we calculate the relic density
including coannihilation channels. Here, also if we take $<\sigma
v> =<\sigma v>_{\chi_{1}^{0}\widetilde{\mu}_{R}\rightarrow \gamma
\mu_{R}} $ then we get $\Omega h^{2}\simeq .09 $ from the
approximate formulae. In both the cases for soln.2 the values of
relic density are nearly equal to relic density calculated by Dark
Susy considering all channels. So we can say that it is the cross
section for dominant channel which gives us the final relic
density for neutralino.
 Relic density is less when coannihilation channels are
included, it is due to the reason that
$<\sigma_{\chi_{1}^{0}\chi_{1}^{0}}v>$ is much less than
$<\sigma_{\chi_{1}^{0}\widetilde{\mu}_{R}} v>$. For Soln. 1 the total
cross section is very small ($10^{-5}$ times smaller than the maximum cross-section of Soln. 2)
hence the relic density is 5 orders of magnitude higher. Although we did not perform searches of
NMSGUT parameter space including dark matter constraints, it is clear form the above examples that
such searches are perfectly feasible. When we have complete incorporation of loop effects on sparticle
 masses in our searches as well as tan$\beta$ variation we intend to interface the DarkSusy Library with
 our code and search for realistic solutions compatible with DM cosmology. The preliminary exploratory
 survey of this chapter has laid the basis for future work in this direction.

\begin{table}[h]
 \centering
\begin{tabular}{|c|r|r|r|r|r|}
\hline\hline
   \multicolumn{1}{|c|}{Process:$\chi_{1}^{0} \chi_{1}^{0}\rightarrow YZ$}

 &\multicolumn{2}{|c|}{Exchanged Particles(X)}
 &\multicolumn{2}{|c|}{ $<\sigma
 v>$($cm^{3}s^{-1}$)}\\
\hline YZ   &  s channel & t and u channel & Soln.1 &Soln.2\\
\hline
HH,hH,AA,HA,ZA & h,H & $\chi_{i}^{0}$ &0.00000                  &0.00000\\
hh             & h,H & $\chi_{i}^{0}$ &6.61712 $\times 10^{-34}$& 2.8868 $\times 10^{-34}$\\
hA,ZH          & A,Z &$\chi_{i}^{0}$  &0.00000 &0.000000\\
Zh             & A,Z &$\chi_{i}^{0}$ &2.5789 $\times 10^{-38}$
&3.0151$\times 10^{-38}$\\
$W^{-},H^{+}$  & h,H,A & $\chi_{k}^{\pm}$ & 0.00000&0.0000\\
$Z^{0},Z^{0}$  & h,H   & $\chi_{i}^{0}$    & 8.37763 $\times
10^{-39}$    & 9.70311 $\times 10^{-40}$ \\
$W^{+},W^{-}$ & h,H,Z & $\chi_{k}^{\pm}$  & 4.33936$\times
10^{-38}$ & 1.509256 $\times 10^{-38}$\\
$\nu_{e}\overline{\nu_{e}}$ & h,H,A,Z & $\widetilde{\nu_{e}}$ &
1.96804 $\times 10^{-34}$ & 3.9409 $\times 10^{-35}$\\
$e^{+} e^{-}$ & h,H,A,Z & $ \widetilde{e}_{L,R}$ & 3.67648 $\times 10^{-33}$ & 2.18089 $\times 10^{-31}$ \\

$\nu_{\mu}\overline{\nu_{\mu}}$ & h,H,A,Z &
$\widetilde{\nu}_{\mu}$ &
1.96723 $\times 10^{-34}$ & 3.99257 $\times 10^{-35}$ \\
$\mu^{+}\mu^{-}$ & h,H,A,Z & $\widetilde{\mu}_{L,R}$ & 3.67588 $\times 10^{-33} $   & {\bf3.97164 $\times 10^{-27}$}   \\

$\nu_{\tau}\overline{\nu_{\tau}}$ & h,H,A,Z
&$\widetilde{\nu}_{\tau} $
&1.32008 $\times 10^{-35}$ & 1.04916 $\times 10^{-35}$ \\

$\tau^{+}\tau^{-}$ & h,H,A,Z & $\widetilde{\tau}_{L,R}$ &4.73294 $\times 10^{-34}$ & 4.7614 $\times 10^{-34}$  \\
$u \overline{u}$    &h,H,A,Z &$\widetilde{u}_{L,R}$ & 3.16611 $\times 10^{-33}$  & 4.75398 $\times 10^{-34}$  \\

$d \overline{d}$    &h,H,A,Z & $\widetilde{d}_{L,R}$ &2.2194 $\times 10^{-34}$   &9.52563 $\times 10^{-35}$ \\

$c \overline{c}$     &h,H,A,Z& $\widetilde{c}_{L,R}$ &3.1661$\times 10^{-33}$  & 4.75927 $\times 10^{-34} $  \\

$s \overline{s}$      &h,H,A,Z& $\widetilde{s}_{L,R}$ &2.26255 $\times 10^{-34}$  & 9.87178 $\times 10^{-35}$ \\

$t \overline{t}$       &h,H,A,Z & $\widetilde{t}_{L,R} $ &3.21770 $\times 10^{-35}$  & 2.42824 $\times 10^{-34}$ \\

$b \overline{b}$       &h,H,A,Z&    $\widetilde{b}_{L,R}$ &3.47445$\times 10^{-33}$  &3.59457 $\times 10^{-33}$ \\
\hline

\end{tabular}
 \caption{Thermally averaged annihilation cross section $<\sigma
v>$ for Neutralino pair-annihilation into tree level two body
final states (YZ). The indices i=1,..4. and
k=1,2.}\label{table:sol1}
\end{table}

\begin{table}[h]
 \centering
\begin{tabular}{|c|r|r|r|r|r| }
\hline\hline
   \multicolumn{1}{|c|}{Process:$\chi_{1}^{0} \widetilde{\mu_{R}} \rightarrow
   YZ$}

 &\multicolumn{3}{|c|}{Exchanged Particles X}
 &\multicolumn{1}{|c|}{ $<\sigma
 v>$($cm^{3}s^{-1}$)}\\
\hline YZ   &  s channel & t  channel & u channel  &Soln.2  \\
\hline Z$\mu_{R}$ & $\mu_{R}$ & $\widetilde{\mu}_{R,L}$ &
$\chi_{i}^{0}$& 8.57919 $\times 10^{-27}$\\
$\gamma \mu_{R}$ & $\mu_{R}$ &$\widetilde{\mu}_{R,L}$& & {\bf3.26385$\times 10^{-26}$}  \\
$ h^{0} \mu_{R}$ & $\mu_{R}$
&$\widetilde{\mu}_{R,L}$&$\chi_{i}^{0}$& 6.95715 $\times 10^{-29}$\\
$H^{0}\mu_{R},A\mu_{R}$ & $\mu_{R}$ & $\widetilde{\mu}_{R,L}$ &
$\chi_{i}^{0}$ & 0.00000\\
$W^{-}\,\overline{\nu}_{\mu}$ & $\mu_{R,L}$&
$\widetilde{\nu}_{\mu}$ & $\chi_{k}^{\pm}$ &
7.44170 $\times 10^{-32}$\\
$H^{-}\, \overline{\nu_{\mu}}$ & $\mu_{R,L}$ &
$\widetilde{\nu}_{\mu}$
& $\chi_{k}^{\pm}$ &  0.00000\\
\hline
\end{tabular}
\caption{Thermally averaged annihilation cross section $<\sigma
v>$ for Neutralino Smuon coannihilation into tree level two body
final states (YZ). The indices i=1,..4. and
k=1,2.}\label{table:sol2}
\end{table}

\section{Conclusion and discussion}
 The NMSGUT can give Neutralino as a dark matter candidate and can account
 for right relic density i.e. $\Omega h^{2}=.11$,
 if mass spectrum of s-particles have at least one particle whose mass is comparable to that of
Neutralino and there is also coannihilation channel. In one case
(all s-particles $\sim$ O(100TeV)) we get large value of relic
density which is not acceptable. For the second case (light
$\tilde{\mu}_R$ mass equal to Neutralino mass within 10$\%$) we
get a relic density in the right ballpark. In this way relic
density calculations constrain the parameter space of NMSGUT.
However this is just a start and a preliminary check. The most
suitable way is to interface the DarkSusy library with the NMSGUT
search criteria. In this way relic density calculations will guide
the search engine to look into a region where it fulfills the DM
relic density constraints.
\newpage

\chapter{NMSGUT RGEs}
\section{Introduction}
Renormalization Group Equations (RGE) are a mathematical tool to
investigate the evolution of the scale dependent (``running'')
parameters (couplings and masses) of a quantum field theory at
different energy scales. It involves loop (1-loop, 2-loop and so
on) corrections to the parameters in terms of powers of logarithms
of the ratio of running scale (Q) to definitional scale ($Q_0$).
The best known example is the evolution of three gauge couplings
of SM with energy. They tend to meet roughly around energy
$10^{16.33}$ GeV, pointing to the possibility of ``Grand
Unification'', the motivating idea of this thesis. Coupling
unification suggests that the SM is an effective theory of some
larger theory below the GUT scale. The sensitivity of SM Higgs
mass to  the GUT scale due to corrections $\Delta m_{H}^2$ $\sim$
$\alpha M_{GUT}^2$ indicates a structural instability in the SM.
However the introduction of supersymmetry solves this problem of
naturalness. With supersymmetry and one pair of Higgs the three
MSSM gauge couplings meet almost exactly at a point
\cite{marcianoamaldi}. The two loop RGEs for minimal
supersymmetric standard model are available \cite{martinvaughn}
and are being used by many who work on grand unified theories. The
RGEs are important to check the viability of GUTs i.e. whether
they are able to produce low energy data by downward RG flow of
GUT predictions at unification scale. For example to do a fit at
scale $M_{Z}$  using NMSGUT superpotential parameters and Soft
Susy breaking parameters of Sugry-NUHM type ($m_0$, $m_{1/2}$,
$A_0$, $m^2_{H,\bar{H}}$), we need two loop MSSM RGEs given in
\cite{martinvaughn} to run the parameters down from $M_X^0$. To do
this we randomly throw these parameters at GUT scale. However just
like SM is an effective theory of GUT scale, NMSGUT can also be an
effective theory of some higher theory at Planck scale. After the
Planck scale the gravity becomes strong so we restrict ourselves
to the energy range  up to the Planck scale ($(8 \pi G_N)^{-1/2}$
= $10^{18.34}$ GeV). Now above the GUT scale MSSM RGEs will not
work so we need NMSGUT RGEs which had so far not been computed.
Although seemingly straightforward, automated calculation programs
like \cite{fonseca} fail because their brute force approach
generates too many terms to sum.  At Planck scale, with canonical
kinetic terms, gravitino mass parameter ($m_{3/2}$) and  the
universal trilinear scalar parameter $A_0$ ($\sim m_{3/2}$ )
 will be the only free
parameters. The soft Susy breaking parameters at GUT scale
$10^{16.33}$ GeV are determined by running down soft parameters of
NMSGUT with just these two   soft parameters in input. In this
chapter we present two loop RGEs for soft Susy parameters of
NMSGUT. The motive of this work is to produce the type of soft
Susy parameters via NMSGUT RGE running from Planck scale to GUT
scale $10^{16.33}$ GeV that our search program finds are necessary
for successful fermion data fitting. Especially we hope to find an
explanation for the negative Higgs mass squared parameters
$m^2_{H,\bar{H}}$ which are found in NMSGUT fits (at least at
large tan$\beta$) whereas supergravity predicts that all soft
scalar masses squared are positive and equal to $m_{3/2}^2$ (at
the scale where they are generated : presumably the Planck scale).
This final chapter presents our calculation of RGE's and a
preliminary exploration of the RGE flows. We derive the 2-loop
RGEs for NMSGUT parameters using general RGE formulae given in
\cite{martinvaughn} and gauge invariance as guiding principle to
develop methods to overcome combinatorial complexity. The
calculation was done in collaboration and part of the results can
be found in \cite{charanthesis,AGK}. The sample tables we generate
are for distinct parameter sets.
\section{Generalized RGEs}
The general two loop running of a parameter X is defined as \be
\frac{dX}{dt}= \frac{1}{16 \pi^2} \beta^{(1)}_{X} + \frac{1}{(16
\pi^2)^2} \beta^{(2)}_X\ee Here $\beta^{1}_{X}$ and $\beta^{2}_X$
are the beta functions for couplings and masses. In Susy theories
the running of couplings arises entirely from anomalous
dimensions. The form of generic renormalizable Superpotential is
given by \be W= \frac{1}{6} Y_{ijk}
\Phi_{i}\Phi_{j}\Phi_{k}+\frac{1}{2}\mu_{ij}\Phi_i \Phi_j\ee Here
$\Phi$ is the chiral superfield which contains a complex scalar
$\phi_i$ and a Weyl fermion $\psi_i$. The generic lagrangian
corresponding to Soft Susy breaking terms is given by \be
L_{Susy}= -\frac{1}{6} h_{ijk}
\phi_{i}\phi_{j}\phi_{k}-\frac{1}{2} b_{ij}\phi_i
\phi_j-\frac{1}{2}(m^2)^i_j\phi^{*i}\phi_{j}-\frac{1}{2}
M\lambda\lambda +h.c. \ee
 $M$ is the gaugino mass parameter.
 We give here the general form of 2-loop beta function for the
 soft susy breaking parameters $h_{ijk},b_{ij},(m^2)^i_j, M $
 and gauge coupling $g$ from\cite{martinvaughn}.

  \be \frac{d g}{dt}= \frac{1}{16\pi^2}\beta^{(1)}_{g}+ \frac{1}{(16\pi^2)^2}\beta^{(2)}_{g}\ee
  Where

 \be \beta^{(1)}_{g}= g^3[ S(R)-3C(G)] \ee
 \bea \beta^{(2)}_{g}&=&g^5[-6 C(G)^2+2 C(G)S(R)+4 S(R)C(R)]\nonumber\\&&
 g^3Y^{ijk}Y_{ijk}C(k)/d(G)\eea

  \be \frac{d M}{dt}= \frac{1}{16\pi^2}\beta^{(1)}_{M}+ \frac{1}{(16\pi^2)^2}\beta^{(2)}_{ M}\ee
  Where

 \be \beta^{(1)}_{ M}= g^2[2 S(R)-6 C(G)] M\ee
 i.e. just double of $\beta_{g}^{(1)}$.
 \bea \beta^{(2)}_{ M}&=&g^4[-24 C(G)^2+8 C(G)S(R)+16 S(R)C(R)] M+\nonumber\\&&
 2g^2[h^{ijk}-M Y^{ijk}]Y_{ijk}C(k)/d(G)\eea
Here g is the gauge coupling of the group under consideration.
$Y_{ijk}$=$(Y^{ijk})^{*}$. S(R) is the Dynkin index for
representation R and C(G) is the Casimir invariant of group.
 S(R)C(R) is the weighted sum of Dynkin index over Casimir operator.
 d(G) is the dimensionality of the group.

\be \frac{d h^{ijk}}{dt}= \frac{1}{16\pi^2}[\beta^{(1)}_h]^{ijk}+
\frac{1}{(16\pi^2)^2}[\beta^{(2)}_h]^{ijk}\ee
  Where

 \bea [\beta^{(1)}_h]^{ijk}&=& \frac{1}{2}h^{ijl}Y_{lmn}Y^{mnk}+Y^{ijl}Y_{lmn}h^{mnk}-2(h^{ijk}-2M Y^{ijk})g^2 C(k) \nonumber\\&&
 +(k \longleftrightarrow i)+(k \longleftrightarrow j)\label{1looph}\eea
 \bea [\beta^{(2)}_h]^{ijk}&=& -\frac{1}{2}h^{ijl}Y_{lmn}Y^{npq}Y_{pqr}Y^{mrk}\\\nonumber&&
 -Y^{ijl}Y_{lmn}Y^{npq}Y_{pqr}h^{mrk}-Y^{ijl}Y_{lmn}h^{npq}Y_{pqr}Y^{mrk}\\\nonumber&&
+(h^{ijl} Y_{lpq}Y^{pqk}+2Y^{ijl} Y_{lpq}Y^{pqk}-2 M Y^{ijl}
Y_{lpq}Y^{pqk})g^2[2 C(p)- C(k)]\nonumber\\&& +(2h^{ijk}-8 M
Y^{ijk}) g^4[C(k)S(R)+2C(k)^2-C(G)C(k)]\nonumber\\&&+(k
\longleftrightarrow i)+(k \longleftrightarrow j)\eea
 Here C(k) represents the Casimir invariant of the representation carried by the chiral superfield.

\be \frac{d b^{ij}}{dt}= \frac{1}{16\pi^2}[\beta^{(1)}_b]^{ij}+
\frac{1}{(16\pi^2)^2}[\beta^{(2)}_b]^{ij}\ee
  Where

 \bea [\beta^{(1)}_b]^{ij}&=& \frac{1}{2}b^{il}Y_{lmn}Y^{mnj}+\frac{1}{2}Y^{ijl}Y_{lmn}b^{mn}+\mu^{il}Y_{lmn}h^{mnj}
\\\nonumber&& -2(b^{ij}-2 M \mu^{ij})g^2 C(i)+(i\longleftrightarrow j)\eea
 \bea [\beta^{(2)}_b]^{ij}&=& -\frac{1}{2}b^{ij}Y_{lmn}Y^{pqn}Y_{pqr}Y^{mrj}
 -\frac{1}{2}Y^{ijl}Y_{lmn}b^{mr}Y^{pqr}Y_{pqn}\\\nonumber&&
 -\frac{1}{2}Y^{ijl}Y_{lmn}\mu^{mr}Y_{pqr}h^{pqn}-\mu^{il}Y_{lmn}h^{npq}Y_{pqr}Y^{mrj}\\\nonumber&&
 -\mu^{il}Y_{lmn}Y^{npq}Y_{pqr}h^{mrj}+2Y^{ijl}Y_{ipq}(b^{pq}-\mu^{pq}M)g^2C(p)\\\nonumber&&
+(b^{il} Y_{lpq}Y^{pqj}+2\mu^{il} Y_{lpq}h^{pqj}-2M \mu^{il}
Y_{lpq}Y^{pqj})g^2[2 C(p)- C(i)]\nonumber\\&&
+(2b^{ij}-8M\mu^{ij})
g^4[C(i)S(R)+2C(i)^2-3C(G)C(i)]\nonumber\\&&+(i
\longleftrightarrow j)\eea

\be \frac{d (m^2)^{i}_j}{dt}=
\frac{1}{16\pi^2}[\beta^{(1)}_{(m^2)}]^{i}_j+
\frac{1}{(16\pi^2)^2}[\beta^{(2)}_{(m^2)}]^{i}_j\ee
  Where

 \bea [\beta^{(1)}_{(m^2)}]^{i}_j&=& \frac{1}{2}Y_{ipq}Y^{pqn} (m^2)^j_{n}+\frac{1}{2}Y^{jpq}Y_{pqn}(m^2)^n_{i}+
 2 Y_{ipq}Y^{jpr}(m^2)^q_r
\\\nonumber&& h_{ipq}h^{jpq}-8\delta^i_j M {\tilde M}^{\dag}g^2c(i)+2g^2t_i^{Aj}Tr[t^A m^2]\eea

 \bea [\beta^{(2)}_{(m^2)}]^{i}_j&=& -\frac{1}{2}(m^2)^l_i Y_{lmn}Y^{mrj}Y_{pqr}Y^{pqn}-\frac{1}{2}
 (m^2)^j_l Y^{lmn}Y_{mri}Y^{pqr}Y_{pqn}\\\nonumber&&
-Y_{ilm}Y^{jnm}(m^2)^l_r Y_{pqn}Y^{rpq}-Y^{ilm}Y_{jnm}(m^2)^r_n
Y^{rpq}Y_{lpq}\\\nonumber&& -Y_{ilm}Y^{jnr}(m^2)^l_n
Y_{pqr}Y^{pqm}-2Y^{ilm}Y_{jln}
Y^{npq}Y_{mpr}(m^2)^q_r\\\nonumber&&
-h_{ilm}Y^{jlm}Y_{npq}h^{mpq}-Y_{ilm}h^{jln}h_{npq}Y^{mpq}\\\nonumber&&
+[(m^2)^i_i Y_{lpq}Y^{jpq}+Y_{ipq}Y^{lpq}
(m^2)^j_l+4Y_{ipq}Y^{jpl}(m^2)^q_l+2h_{ipq}h^{jpq}\\\nonumber&&
-2h_{ipq}Y^{jpq}-2Y_{ipq}h^{jpq}M^{\dag}+4 Y_{ipq}Y^{jpq}M{\tilde
M}^{\dag} ]g^2[C(p)+C(q)-C(i)]\nonumber\eea \bea&& -2g^2t^{Aj}_i
(t^A m^2)^l_r Y_{lpq}Y^{rpq}+8g^4t^{Ai}_j Tr[t^A
C(r)m^2]\nonumber\\&& +\delta^i_j g^4
MM^{\dag}[24C(i)S(R)+48C(i)^2-72C(G)C(i)]\nonumber\\&&+
8\delta^i_j g^4C(i) (Tr[S(R)m^2]-C(G)M{\tilde M}^{\dag}) \eea
The
$\beta$ functions given above are in the $\overline{DR}$ scheme
\cite{DR}. The corresponding equations for hard parameters can be
found in \cite{martinvaughn} and the results for the NMSGUT RGE's
can be found in \cite{charanthesis,AGK}. In the next section we
present the NMSGUT RGE's for soft parameters and the way to
calculate the anomalous dimensions.

\section{SO(10) RGEs for soft parameters}
Superpotential of NMSGUT with terms bearing contraction on SO(10)
indices is : \bea W&=&\frac{1}{2}\mu_H H_i^2+
\frac{\mu_{\Theta}}{4!} \Phi_{ijkl} \Phi_{ijkl}
+\frac{\lambda}{4!} \Phi_{ijkl} \Phi_{klmn} \Phi_{mnij}+
\frac{\mu_{\Sigma,\bar\Sigma}}{5!}\Sigma_{ijklm}
\overline{\Sigma}_{ijklm}\nonumber\\&&+\frac{\eta}{4!}\Phi_{ijkl}\Sigma_{ijmno}
\overline{\Sigma}_{klmno}+\frac{1}{4!}H_i \Phi_{jklm}(\gamma
\Sigma_{ijklm}+ \overline{\gamma} \overline{\Sigma}_{ijklm}  )
\nonumber\\&& +\frac{\mu_{\Theta}}{2(3!)}\Theta_{ijk}\Theta_{ijk}
+ \frac{k}{3!}\Theta_{ijk}H_m \Phi_{mijk}
  +
 \frac{\rho}{4!}\Theta_{ijk}\Theta_{mnk}\Phi_{ijmn} \nonumber\\
  &&+  \frac{1}{2(3!)}
 \Theta_{ijk}\Phi_{klmn}(\zeta \Sigma_{lmnij}
  +  \bar\zeta \bar\Sigma_{lmnij})
 +h_{AB} \Psi^T_A C^{(5)}_2 \gamma_i \Psi_B H_i \nonumber \\&& +
\frac{1}{5!}f_{AB} \Psi^T_A C^{(5)}_2 \gamma_{i_1}....\gamma_{i_5}
\Psi_B\overline{\Sigma}_{i_1...i_5} +  \frac{1}{3!}g_{AB} \Psi_A^T
 C_2^{(5)}\gamma_{i_1}\gamma_{i_2}\gamma_{i_3}\Psi_B \Theta_{i_1 i_2
 i_3} \eea
 Here $i,j,k,l,m,n$ run over 1 to 10.
 Corresponding to each term in superpotential we have soft
 term (each soft trilinear coupling corresponding
 to hard coupling bears tilde on it).
 First we need to calculate the anomalous dimensions for each superfield.
 Anomalous dimensions are calculated from wave function renormalization only
 due to ``non-renormalization'' theorems of supersymmetry which prevent coupling constant
 renormalization in the superpotential \cite{WessZumino}.
 For example if we want to calculate the 1-loop $\beta$ function for $\tilde{\lambda}$
 (soft term corresponding to trilinear coupling $\lambda$) we
 need to follow steps given below:
 \begin{enumerate}
 \item Identify the superfields in the $\lambda$ coupling term (just 210-plet components $\Phi_{ijkl}$ in this case).
 \item Then identify the gauge invariant combinations of $\Phi$ with other superfields.
 \item  Then taking any independent component say $\Phi_{1234}$ as the
  (conserved by SO(10)) external line of the
 propagator, (the other two superfields
 from the trilinear combination will run inside the loop) calculate the number of ways
  it gets wave function corrections from the fields this
   chosen external component couples to in the $\lambda$ vertices.
   Since it must emerge with the same SO(10) quantum numbers as it
    entered with and the counting will apply equally
   to every such field component, a little practice suffices to get
    all 1-loop anomalous dimensions.
    It is clear from Eqn. (\ref{bargamphi}) that it gets corrections
    from $\eta, \kappa, \lambda,
  \rho,\gamma, \bar \gamma, \zeta, \bar \zeta$ trilinear coupling
  terms. The $\beta$ functions at one loop can be read off
  from Eqn.(\ref{1looph}).
 \end{enumerate}

 thus the one-loop beta function for the soft
parameter $\tilde{\lambda}$ is:

\be \beta_{\tilde{\lambda}}^{(1)}=3 \tilde{\lambda}
\bar{\gamma}_{\Phi}^{(1)}+6 \lambda \tilde{\gamma}_{\Phi}^{(1)}
-72g_{10}^2(\tilde{\lambda}-2 M \lambda) \ee

 where $\bar
\gamma_\Phi^{(1)}$ = $ \frac{1}{2} Y_{\Phi mn}Y^{mn\Phi}$ and
$\tilde{ \gamma}_\Phi^{(1)}$ = $\frac{1}{2} Y_{\Phi mn}h^{mn\Phi}$
are anomalous dimensions
 \be  \bar
\gamma_\Phi^{(1)}=4 |\kappa |^2+180 |\lambda|^2+2
|\rho |^2+240 |\eta |^2 +6 (|\gamma|^2+|\bar{\gamma}|^2)+60 (|\zeta
|^2+|\bar{\zeta}|^2)\label{bargamphi}\ee

\be   \tilde{ \gamma}_\Phi^{(1)}=  4 \tilde{\kappa} \kappa^*+180
   \tilde{\lambda} \lambda ^{*}+2
\tilde{\rho } \rho ^{*} +240
   \tilde{\eta } \eta ^{*} +6
   (\tilde{\gamma}  \gamma
   ^{*}+\tilde{\bar{\gamma}}
   \bar\gamma^{*})+60
   (\tilde{\zeta}\zeta
   ^{*}+\tilde{\bar{\zeta}}
   \bar \zeta ^{*})    \label{tilgamphi}   \ee
The numbers appearing in the above equations can be calculated as
follows. Say we want to find $\beta$ function for the
$\tilde{\lambda}^{1234,3456,5612}$. For this we need to calculate
the anomalous dimension for fields $\Phi_{1234}$,
$\Phi_{3456}$,$\Phi_{5612}$. Let's calculate for $\Phi_{1234}$.
The first term in $\bar \gamma_\Phi^{(1)}$ gives the number of
ways $\Phi_{1234}$ field line gets loop correction from $\kappa$
trilinear coupling term. It is calculated as follows: \bea
\frac{\kappa}{3!} H_{i} \Theta_{jkl} \Phi_{ijkl}&=& \kappa
\Phi_{1234} (H_1 \Theta_{234}-H_2
\Theta_{134}+H_{3}\Theta_{124}-H_{4}\Theta_{123})\nonumber\\&& +
{\rm terms ~without} ~\Phi_{1234}\eea So we have four terms which
will give 1-loop wave function
 correction to $\Phi_{1234}$ from $\kappa$ term.
 Similarly we can calculate the other factors.
  Care must be taken in the calculation of the factor corresponding
  to $\eta$ term as there will
   be doubling of counted terms due to (anti) self-duality  of
    $ (\bar\Sigma) \Sigma $.

1-loop $\beta$ functions for the SO(10) gauge coupling constant
and gaugino mass parameter are \be
 \beta_{g_{10}}^{(1)}=137g_{10}^3 \ee
\be \beta_{M}^{(1)}= 274 M g_{10}^2\ee
 Here S(R)-3C(G)=137 with
S(R)=161 and C(G)=C(45)=S(45)=8. The Table \ref{Dynkin} gives the
Dynkin index and Casimir invariant
 for different representations of NMSGUT.
 We get a total index S(R)=1+(3$\times$ 2)+28+35+35+56=161.

\begin{table}[htb]
 $$
 \begin{array}{|c|c|c|}
 \hline{\mbox{d}}&{\rm S(R)} &{\rm C(R)=45 S(R)/d(R)} \\
 \hline
 10 & 1 & 9/2 \\
 16 & 2 & 45/8 \\
 120 & 28 & 21/2\\
 126 & 35 & 25/2\\
 \overline{126} & 35 & 25/2\\
 210 & 56 & 12 \\
 \hline
  \end{array}
 $$
\caption{Dynkin index and Casimir invariant for different field
representations of NMSGUT.}
 \label{Dynkin} \end{table}

The anomalous dimensions for 2-loop beta functions can be
calculated once the anomalous dimensions for 1-loop are in hand by
extending the summation techniques explained above. For example,
the anomalous dimensions for 2-loop beta functions for
$\tilde{\lambda}$ (given below) contains anomalous dimensions for
all fields at 1-loop.

 \bea  \beta_{\tilde{\lambda}}^{(2)}&=&
3(-\tilde{\lambda}\bar{\gamma}_\Phi^{(2)}-\lambda
\tilde{\gamma}_\Phi^{(2)}-\lambda \hat{\gamma}_\Phi^{(2)} +2
g^2_{10} ((\tilde{\lambda}-2 M \lambda) (3120|\eta|^2+
12|\kappa|^2+2160|\lambda|^2\nonumber\\&&+18|\rho|^2+30|\gamma|^2+30|\bar\gamma|^2+660
|\zeta|^2+660 |\bar{\zeta}|^2)+ 2\lambda(3120\tilde{\eta}\eta^*+
12  \tilde{\kappa} \kappa^*+18 \tilde\rho  \rho^* \nonumber\\&&
+2160 \tilde{\lambda} \lambda^* +30 \tilde{\gamma} \gamma^*+30
\tilde{\bar\gamma}  \bar\gamma^* +660 \tilde{\zeta} \zeta^* +660
\tilde{\bar{\zeta}} \bar{\zeta}^*) )\nonumber\\&& + 1932g^4_{10}(2
\tilde{\lambda} -8 M \lambda  ))\eea
 Where the anomalous
dimensions $\bar{\gamma}_\Phi^{(2)}$ = -$\frac{1}{2}Y_{\Phi
 mn}Y^{npq}Y_{pqr}Y^{mr\Phi}$,
 $\tilde{\gamma}_\Phi^{(2)}$ = $Y_{\Phi mn}Y^{npq}Y_{pqr}h^{mr\Phi}$
 and $\hat{\gamma}_\Phi^{(2)}$ = $Y_{\Phi
 mn}h^{npq}Y_{pqr}Y^{mr\Phi}$ are given by

\bea  \bar{\gamma}_\Phi^{(2)}&=&240 |\eta|^2
(\bar{\gamma}_\Sigma^{(1)}+ \bar{\gamma}_{\bar{\Sigma}}^{(1)}  )+4
|\kappa|^2(\bar{\gamma}_H^{(1)}+ \bar{\gamma}_\Theta^{(1)}
)+360|\lambda|^2\bar{\gamma}_\Phi^{(1)} +
4|\rho|^2\bar{\gamma}_\Theta^{(1)} \nonumber\\&&+6
|\gamma|^2(\bar{\gamma}_H^{(1)}+ \bar{\gamma}_\Sigma^{(1)} )+6
|\bar{\gamma}|^2(\bar{\gamma}_H^{(1)}+
\bar{\gamma}_{\bar\Sigma}^{(1)} )+60
|\zeta|^2(\bar{\gamma}_\Theta^{(1)}+ \bar{\gamma}_\Sigma^{(1)}
)\nonumber\\&&+60 |\bar{\zeta}|^2(\bar{\gamma}_\Theta^{(1)}+
\bar{\gamma}_{\bar\Sigma}^{(1)} )\eea

\bea  \tilde{\gamma}_\Phi^{(2)}&=&480 \tilde{\eta}
\eta^*(\bar{\gamma}_\Sigma^{(1)}+
 \bar{\gamma}_{\bar{\Sigma}}^{(1)}  )+8 \tilde{\kappa} \kappa^*
(\bar{\gamma}_H^{(1)}+ \bar{\gamma}_\Theta^{(1)}
)+720\tilde{\lambda }\lambda^* \bar{\gamma}_\Phi^{(1)}
\nonumber\\&& +8\tilde{\rho} \rho^* \bar{\gamma}_\Theta^{(1)} +12
\tilde{\gamma} \gamma^* (\bar{\gamma}_H^{(1)}+
\bar{\gamma}_\Sigma^{(1)} )+12 \tilde{\bar{\gamma}}
 \bar{\gamma}^*(\bar{\gamma}_H^{(1)}+ \bar{\gamma}_{\bar\Sigma}^{(1)}
)\nonumber\\&&+120 \tilde{\zeta} \zeta^*
(\bar{\gamma}_\Theta^{(1)}+ \bar{\gamma}_\Sigma^{(1)} )+120
\tilde{\bar{\zeta}} \bar{\zeta}^* (\bar{\gamma}_\Theta^{(1)}+
\bar{\gamma}_{\bar\Sigma}^{(1)} )\eea

\bea  \hat{\gamma}_\Phi^{(2)}&=&480 |\eta|^2
(\tilde{\gamma}_\Sigma^{(1)}+ \tilde{\gamma}_{\bar{\Sigma}}^{(1)}
)+8 |\kappa|^2(\tilde{\gamma}_H^{(1)}+ \tilde{\gamma}_\Theta^{(1)}
)+720|\lambda|^2 \tilde{\gamma}_\Phi^{(1)} \nonumber\\&&
+8|\rho|^2 \tilde{\gamma}_\Theta^{(1)} +12
|\gamma|^2(\tilde{\gamma}_H^{(1)}+ \tilde{\gamma}_\Sigma^{(1)}
)+12 |\tilde{\gamma}|^2(\tilde{\gamma}_H^{(1)}+
\tilde{\gamma}_{\bar\Sigma}^{(1)} )\nonumber\\&&+120
|\zeta|^2(\tilde{\gamma}_\Theta^{(1)}+ \tilde{\gamma}_\Sigma^{(1)}
)+120 |\bar{\zeta}|^2(\tilde{\gamma}_\Theta^{(1)}+
\tilde{\gamma}_{\bar\Sigma}^{(1)} )\eea

 The rest of the anomalous
dimensions and $\beta$ functions for 1-loop and 2-loop are given
in Appendix A.

\section{Illustrative run down values of soft parameters at one loop}
\begin{figure}[h]
\centering
\includegraphics[scale=0.8]{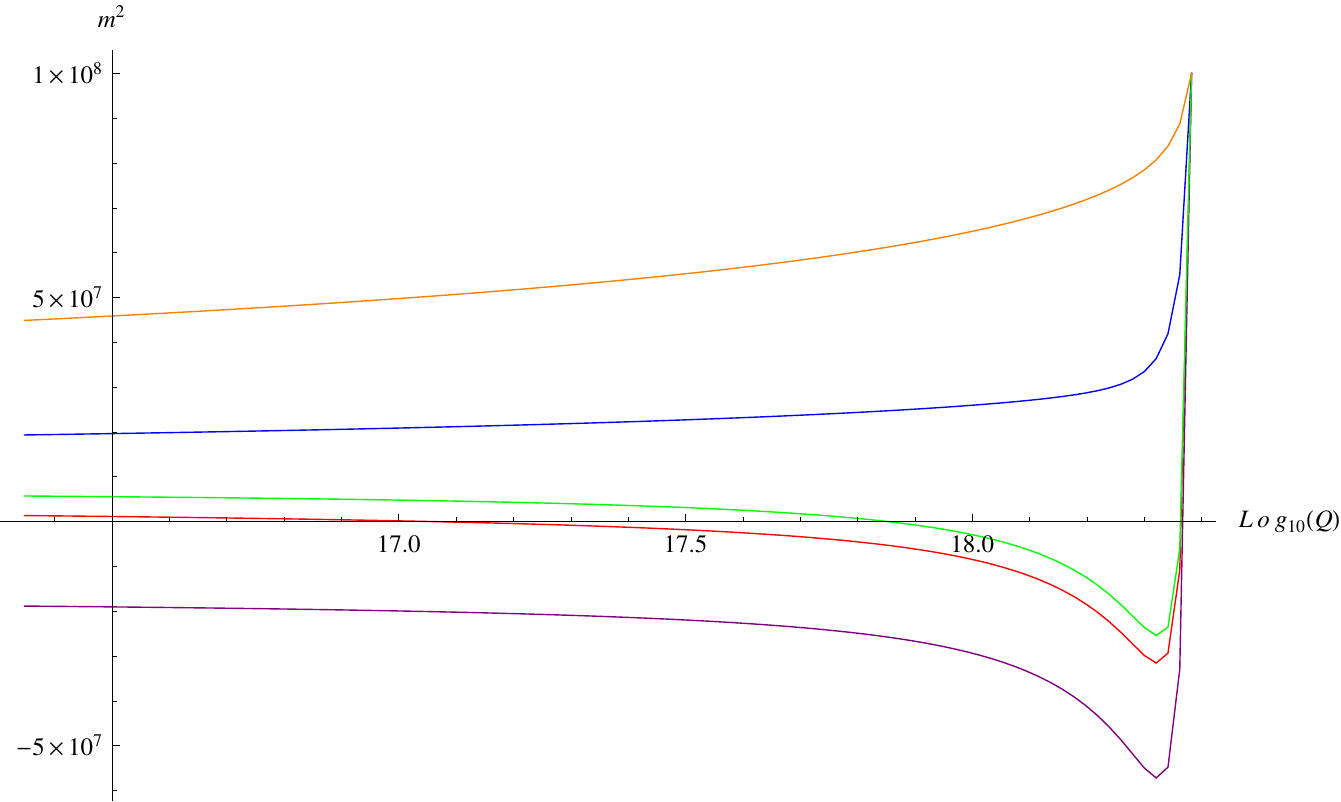}
\caption{Variation of Different Higgs masses squared
($m^2_{\Phi}$(purple), $m^2_{\bar{\Sigma}}$(red),
$m^2_{\Sigma}$(green), $m^2_{\Theta}$(blue), $m^2_{H}$(orange))
from t=$Log_{10}(10^{18.38})$ to t=$Log_{10}(10^{16.33})$.}
\label{sqmassflow}\end{figure}

In this section we present an example for run down soft Susy
breaking parameters of NMSGUT at 1-loop. The soft Susy parameters
chosen at $M_{Planck}$ scale in terms of $m_{3/2}$ are
\cite{godboleroy}: \bea A_0= 3m_{3/2};\quad
m_{0}^2=m_{3/2}^2;\quad B=A_0-m_{3/2};\quad M=0\eea and we take
$m_{3/2}$=10 TeV at $M_{Planck}$ scale. The choice of $M=0$ is
valid if we assume canonical gauge/gaugino kinetic terms
 in the Lagrangian. At one loop it comes out zero at
GUT scale too. However even $M(M_{X}^0)$=0 is acceptable in NMSGUT
\cite{BStabHedge} as we are able to get the acceptable gaugino
masses at $M_{S}$ with 2-loop MSSM RGE running from GUT scale. The
main motive of this preliminary study is to see the variation in
the soft sector parameters. In Figure \ref{sqmassflow} we present
the running of various squared Higgs mass parameters. This is the
main result as in NMSGUT the negative Higgs mass square are
required for fermion fitting and are the reason for the normal
hierarchy for the s-fermions \cite{nmsgutIII}. Table \ref{tabhard}
contains the sample values of hard parameters of NMSGUT
superpotential and Table \ref{softrge} contains the soft Susy
breaking parameters at $M_{Planck}$ and $M_{GUT}$ respectively.

 \begin{table}[!h]
 $$
 \small{\begin{array}{|c|c|c|}
 \hline{\mbox{Parameter}}&{\rm Value}
 &{\rm Value }\\
 & at~M_{Pl}= 10^{18.39} GeV & at ~M_{X}=10^{16.33} GeV\\
 \hline
 g_{10}&          1.252    &         0.336    \\
 h_{11}/10^{-5}&
            5.81
 &
            6.84
 \\
 h_{12}/10^{-7}&
            0.0
 &
           3.281-.088i
 \\
 h_{13}/10^{-7}&
            0.0
 &
           0.039 -1.095i
 \\
 h_{22}/10^{-3}&
           -5.86
 &
           -6.902\\
 h_{23}/10^{-7}&
            0.0
 &
            -.002+.12i
 \\
 h_{33}&
            0.0141
 &
             0.0166
 \\
 f_{11}/10^{-6}&
           0.236+ 1.62i
 &
            .158 + 1.083i
 \\
 f_{12}/10^{-6}& -2.50-1.78i

 &
           -1.668-1.191\\
 f_{13}/10^{-6}& 0.55+5.15i

 &
          36.745 + 3.437i
 \\
 f_{22}/10^{-6}&
            6.76+0.18i
 &
           4.51 + .12i
 \\
 f_{23}/10^{-4}&
           -1.26-.52i
 &
            -.842  + .349i
 \\
 f_{33}/10^{-4}& 3.89+.178i

 &
           2.599  + .119 i
 \\
 g_{12}/10^{-4}&
           -0.784+1.53i
 &
           -0.777  + 1.515i
 \\
 g_{13}/10^{-3}&
           -1.2+.163i
 &
          -1.188  + 0.161i
 \\
 g_{23}&
           -0.0116-0.014i
 &
           -.0115-0.004i
 \\
 \eta&
            1.21+            0.013i
 &
            0.16+            0.002i
 \\
 \lambda&
            0.22+            0.025i
 &
            0.02+            0.002i
 \\
 \gamma&
            0.241+            0.046i
 &
            0.058+            0.011i
 \\
 \bar\gamma&
            0.251+            0.053i
 &
            0.058+            0.012i
 \\
 \kappa&
            0.361+            0.02i
 &
            0.123+            0.007i
 \\
 \rho& 0.53+ .031i & 0.15+.009i\\
 \zeta&
            0.711+            0.065i
 &
            0.143+            0.013i
 \\
 \bar\zeta&
            0.801+            0.072i
 &
            0.155+            0.014i
 \\
  \hline
{\mbox{Parameter}}&{\rm Value}
 &{\rm Value}\\
 & {\rm( in ~GeV)} &{\rm( in ~GeV)}\\\hline
 \mu_{\Phi}&
10^{14}
 & 2.033 \times 10^{13}
 \\
 \mu_{\Theta}&
10^{14}
 &
6.381 \times 10^{13}
 \\
 \mu_{H}&
10^{14}
 &
8.997 \times 10^{13}
 \\
 \mu_{\Sigma,\bar \Sigma}&
10^{14}
 &
3.018 \times 10^{13}
 \\
 \hline
 \end{array}}
 $$
\caption{\small{Variation of hard parameters of NMSGUT
superpotential at 1-loop level.}}
  \label{tabhard}\end{table}

 \begin{table}
 $$
\small{ \begin{array}{|c|c|c|}
 \hline{\mbox{Parameter}}&{\rm Value~(in ~GeV)}
 &{\rm Value~(in~GeV) }\\
 & at~M_{Pl}= 10^{18.39} GeV & at ~M_{X}=10^{16.33} GeV\\
 \hline
\tilde{h}_{11}&
            1.74
 &
            1.84
 \\
 \tilde{h}_{12}&
          0.0
 &
            0.025-            0.007i
 \\
 \tilde{h}_{13}&
           0.0
 &
            0.003-            0.008i
 \\
 \tilde{h}_{22}&
         -176.07
 &
         -185.16
 \\
 \tilde{h}_{23}&
         0.0
 &
           -0.0002+            0.0009i
 \\
 \tilde{h}_{33}&
          423.0
 &
          444.64
 \\
 \tilde{f}_{11}&
            0.007+            0.049i
 &
            0.003+            0.022i
 \\
 \tilde{f}_{12}&
           -0.075-            0.054i
 &
           -0.034-            0.024i
 \\
 \tilde{f}_{13}&
            1.649+            0.154i
 &
            0.742+            0.069i
 \\
 \tilde{f}_{22}&
            0.203+            0.005i
 &
            0.091+            0.002i
 \\
 \tilde{f}_{23}&
           -3.78+            1.569i
 &
           -1.696+            0.704i
 \\
 \tilde{f}_{33}&
           11.67+            0.534i
 &
            5.236+            0.239i
 \\
 \tilde{g}_{12}&
           -2.352+            4.59i
 &
           -1.826+            3.564i
 \\
 \tilde{g}_{13}&
          -36.0+            4.89i
 &
          -27.95+            3.796i
 \\
 \tilde{g}_{23}&
         -348.0-           43.08i
 &
         -270.14-           33.44i
 \\
 \tilde{\eta}&
        36300.0+          397.5i
 &
         -140.48-            1.54i
 \\
 \tilde{\lambda}&
         6600.0+          757.5i
 &
         -104.24-           11.96i
 \\
 \tilde{\gamma}&
         7218.0+         1387.5i
 &
          335.37+           64.47i
 \\
 \tilde{\bar\gamma}&
         7518.0+         1597.5i
 &
          309.26+           65.72i
 \\
 \tilde{\kappa}&
        10818.0+          607.80i
 &
         1074.31+           60.36i
 \\
 \tilde{\rho}& 15900.0+930.0& 825.58+48.68i\\
 \tilde{\zeta}&
        21318.0+         1957.77i
 &
          360.24+           33.08i
 \\
 \tilde{\bar\zeta}&
        24018.0+         2167.77i
 &
          318.26+           28.72i
 \\
 M&
           0.0
 &
          0.0\\
 \hline
{\mbox{Parameter}}&{\rm Value}
 &{\rm Value}\\
 & {\rm( in ~GeV^2)} &{\rm( in ~GeV^2)}\\\hline
 b_{\Phi}&
2 \times 10^{18} & -7.01\times 10^{16}
 \\
 b_{\Theta}&
2 \times 10^{18}
 &
4.553 \times 10^{17}
 \\
 b_{H}&
2 \times 10^{18}
 &
1.237 \times 10^{18}
 \\ b_{\Sigma,\bar{\Sigma}}& 2 \times 10^{18}
 &
2.624 \times 10^{16}
 \\
 m^2_{\Phi}&
 10^8
 &
    -1.88 \times 10^7
 \\
m^2_{\Theta}&
    10^8
 &
     1.93 \times 10^7
 \\
 m^2_{H}&
   10^8
 &
     4.47 \times 10^7
 \\
 m^2_{\Sigma}&
    10^8
 &
      5.71 \times 10^6
 \\
 m^2_{\bar\Sigma}&
    10^8
 &
      1.36 \times 10^6\\
     {\rm Eval[m^2_{\tilde \Psi}]}& \{10^8,10^8,10^8\}&\{9.999,9.92,9.91\}\times 10^7
\\
 \hline
 \end{array}}
 $$
 \caption{\small{Variation of Soft Susy breaking parameters of NMSGUT at 1-loop level.}}
 \label{softrge}\end{table}

The soft masses ($b_H$, $b_{\Theta}$, $b_{\Phi}$,
 $b_{\Sigma, \bar{\Sigma}}$) for different NMSGUT Higgs
 representations ($b_{ij}$=B $\mu_{ij}$) comes out be very large
  O($10^{16}$) $GeV^2$ i.e. O($m_{3/2} M_X$) since the hard mass parameters $\mu_{ij}$ have GUT scale mass.
  So the fine tuning condition (discussed in Chapter 2) to
  determine a pair of light MSSM Higgs will need to be supplemented. The soft masses
   will also contribute to $m_{H,\bar{H}}^2 (M_S)$ that are O($m_{3/2}
   M_X$)$>>$ $M_{S}$.
   So we need to impose an additional condition
    that the contribution from the soft terms to $b_{H,\bar H}$ is O($m_{3/2}^2$).
     One important point to mention is that the value
    of $m_{H}^2$ given in Table \ref{softrge} is the soft squared mass for 10-plet Higgs of
    NMSGUT and not the $m_{H,\bar{H}}^2$ which is
     soft squared mass parameters for effective MSSM light Higgs doublets.
    The running of trilinear soft coupling and s-fermion mass squared parameters ($m_{\tilde\Psi}^2$) may give distinct values
     at GUT scale for the three generations: which was considered
    same in our studies of NMSGUT in previous chapters. This will be studied hereafter.

\section{Discussion}
The NMSGUT RGEs are important because they give us the soft Susy
breaking parameters at GUT scale derived from the Gravitino mass
parameter $m_{3/2}$ and the soft trilinear $A_0$ at the Planck
scale where they may have their origin. They give negative mass
squared for one NMSGUT Higgs multiplet which is required for the
explanation of the negative Higgs mass squared found by NMSGUT in
random searches for fits of the entire fermion spectra for large
tan$\beta$. We found the results at 1-loop only, so the
consistency checks at 2-loops are remaining.
 The incorporation of NMSGUT RGE's to our fitting code is
  a big project as it will
 impose an additional next to leading order fine tuning condition for
 having one pair of
 light Higgs doublets in effective MSSM from NMSGUT. We can have
  different soft trilinear coupling parameter
 $A_{0}$ and masses for different generations of sfermions, which were assumed same
 in the previous work \cite{nmsgut,nmsgutIII, BStabHedge}.
 Thus using our results one can calculate soft Susy
 breaking parameters by RGE runs of soft parameters
 from Planck scale and then use the resulting values at GUT scale for further
 fitting procedure. We note that the techniques we have used to actually
 evaluate the 2-loop RGEs have overcome the combinatorial
 complexity that prevented their calculation by automated means.
 They can be used for any Susy theory.

\clearpage
 \section{Appendix A}

 Anomalous dimension associated with different
superfields for 1-loop $\beta$ functions are:

 \be
\bar\gamma_\Sigma^{(1)}=  200 |\eta |^2 +10 |\gamma |^2+100 |\zeta
|^2\ee

\be    \bar\gamma_{\bar\Sigma}^{(1)}= 200 |\eta |^2+ 10
|\bar\gamma |^2+100 |\bar \zeta |^2+32 \text{Tr}[f^{\dag }.f] \ee

\be  \bar \gamma_H^{(1)}= 84 |\kappa |^2+126
(|\gamma|^2+|\bar{\gamma}|^2)+8 \text{Tr}[h^{\dag }.h] \ee

\be   \bar\gamma_\Theta^{(1)}= 7(|\kappa|^2+|\rho |^2)+ 105
(|\zeta |^2+|\bar\zeta |^2)+8\text{Tr}[g^{\dag }.g]         \ee

\be \bar\gamma_\Psi^{(1) }=252 f^{\dag}.f+120 g^{\dag}.g+10
h^{\dag}.h \ee

\be \tilde\gamma_i^{(1)j}=\frac{1}{2} Y_{ipq}h^{jpq} \ee

\be   \tilde{ \gamma}_\Sigma^{(1)}= 200 \tilde{\eta}  \eta ^{*}+10
\tilde{\gamma} \gamma ^{*}+100 \tilde{\zeta} \zeta ^{*} \ee

\be   \tilde{ \gamma}_{\bar\Sigma}^{(1)}= 200 \tilde{\eta} \eta
^{*}+10 \tilde{\bar{\gamma}} \bar\gamma ^{*}+100 \tilde{\bar\zeta}
\bar\zeta ^{*}+32\text{Tr}[f^{\dag}.\tilde{f}]  \ee

\be   \tilde{ \gamma}_H^{(1)}=  84\tilde{\kappa} \kappa^{*} +
      126(\tilde{\gamma} \gamma^{*} + \tilde{\bar{\gamma}}
      \bar\gamma^{*}) +8\text{Tr}[h^{\dag}.\tilde{h}]           \ee

\be   \tilde{ \gamma}_\Theta^{(1)}=7( \tilde{\kappa} \kappa^{*}+
\tilde{\rho}  \rho ^{*})+105 (\tilde{\zeta} \zeta
^{*}+{\tilde{\bar\zeta }} \bar\zeta^{*})+8
\text{Tr}[g^{\dag}.\tilde{g}] \ee

\be   \tilde{ \gamma}_{\psi}^{(1)}=252 f^{\dag}. \tilde{f}+120
g^{\dag}.\tilde{g}+10 h^{\dag}.\tilde{h} \ee

\be \hat\gamma_i^{(1)j}=\frac{1}{2} h_{ipq}h^{jpq} \ee

\be
\hat{\gamma}_{\Phi}^{(1)}=240|\tilde{\eta}|^2+4|\tilde{\kappa}|^2+180|\tilde{\lambda}|^2+2|\tilde{\rho}|^2+
6(|\tilde{\gamma}|^2+|\tilde{\bar{\gamma}}|^2)+60(|\tilde{\zeta}|^2+|\tilde{\bar{\zeta}}|^2)\ee

\be \hat{\gamma}_{\Sigma}^{(1)}=200|\tilde{\eta}|^2+10
|\tilde{\gamma}|^2+ 100|\tilde{\zeta}|^2 \ee

\be \hat{\gamma}_{\bar{\Sigma}}^{(1)}=200|\tilde{\eta}|^2+10
|\tilde{\bar{\gamma}}|^2+ 100|\tilde{\bar{\zeta}}|^2 +32 \text
Tr[\tilde{f}^{\dag}.\tilde{f}]\ee

\be \hat{\gamma}_{H}^{(1)}=84|\tilde{\kappa}|^2+
126(|\tilde{\gamma}|^2+|\tilde{\bar{\gamma}}|^2)+8 \text
Tr[\tilde{h}^{\dag}.\tilde{h}]\ee

\be
\hat{\gamma}_{\Theta}^{(1)}=7(|\tilde{\kappa}|^2+|\tilde{\rho}|^2)+
105(|\tilde{\zeta}|^2+|\tilde{\bar{\zeta}}|^2)+8 \text
Tr[\tilde{g}^{\dag}.\tilde{g}]\ee

\be \hat{\gamma}_{\Psi}^{(1)}=252
\tilde{f}^{\dag}.\tilde{f}+120\tilde{g}^{\dag}.\tilde{g}+10\tilde{h}^{\dag}.\tilde{h}
\ee

 One-loop beta functions for the soft
parameters:

\be \beta_{\tilde{\eta}}^{(1)}= \tilde{\eta}(
\bar{\gamma}_{\Phi}^{(1)}+ \bar{\gamma}_{\Sigma}^{(1)}+
\bar{\gamma}_{\bar{\Sigma}}^{(1)})+2\eta(
\tilde{\gamma}_{\Phi}^{(1)}+ \tilde{\gamma}_{\Sigma}^{(1)}+
\tilde{\gamma}_{\bar{\Sigma}}^{(1)}) -74g_{10}^2(\tilde{\eta}-2 M
\eta) \ee

\be \beta_{\tilde{\gamma}}^{(1)}= \tilde{\gamma}(
\bar{\gamma}_{\Phi}^{(1)}+ \bar{\gamma}_{\Sigma}^{(1)}+
\bar{\gamma}_{H}^{(1)})+2\gamma( \tilde{\gamma}_{\Phi}^{(1)}+
\tilde{\gamma}_{\Sigma}^{(1)}+ \tilde{\gamma}_{H}^{(1)}) -58
g_{10}^2(\tilde{\gamma}-2 M \gamma) \ee

\be \beta_{\tilde{\bar{\gamma}}}^{(1)}= \tilde{\bar{\gamma}}(
\bar{\gamma}_{\Phi}^{(1)}+ \bar{\gamma}_{\bar\Sigma}^{(1)}+
\bar{\gamma}_{H}^{(1)})+2\bar{\gamma}(
\tilde{\gamma}_{\Phi}^{(1)}+ \tilde{\gamma}_{\bar\Sigma}^{(1)}+
\tilde{\gamma}_{H}^{(1)}) -58g_{10}^2(\tilde{\bar{\gamma}}-2 M
\bar{\gamma}) \ee

\be \beta_{\tilde{\kappa}}^{(1)}= \tilde{\kappa}(
\bar{\gamma}_{\Phi}^{(1)}+ \bar{\gamma}_{\Theta}^{(1)}+
\bar{\gamma}_{H}^{(1)})+2\kappa ( \tilde{\gamma}_{\Phi}^{(1)}+
\tilde{\gamma}_{\Theta}^{(1)}+ \tilde{\gamma}_{H}^{(1)}) -54
g_{10}^2(\tilde{\kappa}-2 M \kappa) \ee

\be \beta_{\tilde{\rho}}^{(1)}= \tilde{\rho}(
\bar{\gamma}_{\Phi}^{(1)}+ 2 \bar{\gamma}_{\Theta}^{(1)})+2\rho(
\tilde{\gamma}_{\Phi}^{(1)}+ 2 \tilde{\gamma}_{\Theta}^{(1)}) -66
g_{10}^2(\tilde{\rho}-2 M \rho )\ee

\be \beta_{\tilde{\zeta}}^{(1)}= \tilde{\zeta}(
\bar{\gamma}_{\Phi}^{(1)}+ \bar{\gamma}_{\Sigma}^{(1)}+
\bar{\gamma}_{\Theta}^{(1)})+2\zeta ( \tilde{\gamma}_{\Phi}^{(1)}+
\tilde{\gamma}_{\Sigma}^{(1)}+ \tilde{\gamma}_{\Theta}^{(1)}) -70
g_{10}^2(\tilde{\zeta}-2 M \zeta) \ee

\be \beta_{\tilde{\bar{\zeta}}}^{(1)}= \tilde{\bar{\zeta}}(
\bar{\gamma}_{\Phi}^{(1)}+ \bar{\gamma}_{\bar{\Sigma}}^{(1)}+
\bar{\gamma}_{\Theta}^{(1)})+2\bar{\zeta} (
\tilde{\gamma}_{\Phi}^{(1)}+ \tilde{\gamma}_{\bar{\Sigma}}^{(1)}+
\tilde{\gamma}_{\Theta}^{(1)}) -70 g_{10}^2(\tilde{\bar{\zeta}}-2
M \bar{\zeta}) \ee

\be
\beta_{\tilde{h}}^{(1)}=\bar\gamma_{H}^{(1)}\tilde{h}+\tilde{h}.\bar\gamma_{\Psi}^{(1)}
+(\bar\gamma_{\Psi}^{(1)})^T.\tilde{h}+2\tilde{\gamma}_{H}^{(1)} h
+2 (h.\tilde{\gamma}_{\Psi}^{(1)} +
(\tilde{\gamma}_{\Psi}^{(1)})^T.h)-\frac{63}{2} g_{10}^2
(\tilde{h}-2 M h) \ee

\be \beta_{\tilde{g}}^{(1)}=\bar\gamma_{\Theta}^{(1)}\tilde{g}-
\tilde{g}.\bar\gamma_{\Psi}^{(1)} +(\bar\gamma_{\Psi}^{(1)})^T
.\tilde{g}+2\tilde{\gamma}_{\Theta}^{(1)}.g + 2(g.
\tilde{\gamma}_{\Psi}^{(1)} -(\tilde{\gamma}_{\Psi}^{(1)} )^T
.g)-\frac{87}{2} g_{10}^2 (\tilde{g}-2 M g) \ee

\be
\beta_{\tilde{f}}^{(1)}=\bar\gamma_{\bar{\Sigma}}^{(1)}\tilde{f}+\tilde{f}.\bar\gamma_{\Psi}^{(1)}
+ (\bar\gamma_{\Psi}^{(1)} )^T
.\tilde{f}+2\tilde{\gamma}_{\bar{\Sigma}}^{(1)}.f +2 (f
.\tilde{\gamma}_{\Psi}^{(1)} + (\tilde{\gamma}_{\Psi}^{(1)})^T
.f)-\frac{95}{2} g_{10}^2 (\tilde{f}-2 M f) \ee

\be \beta_{b_{\Phi}}^{(1)}=2 b_{\Phi} \bar{\gamma}_{\Phi}^{(1)}+4
\mu_{\Phi} \tilde{\gamma}_{\Phi}^{(1)}-48 g_{10}^2(b_{\Phi}-2 M
\mu_{\Phi})\ee

\be \beta_{b_{H}}^{(1)}=2 b_{H} \bar{\gamma}_{H}^{(1)}+4 \mu_{H}
\tilde{\gamma}_{H}^{(1)}-18 g_{10}^2(b_H-2 M \mu_H)\ee

\be \beta_{b_{\Theta}}^{(1)}=2 b_{\Theta}
\bar{\gamma}_{\Theta}^{(1)}+4 \mu_{\Theta}
\tilde{\gamma}_{\Theta}^{(1)}-42 g_{10}^2(b_{\Theta}-2 M
\mu_{\Theta})\ee

\be \beta_{b_{\Sigma}}^{(1)}= b_{\Sigma}(
\bar{\gamma}_{\Sigma}^{(1)}+\bar{\gamma}_{\bar{\Sigma}}^{(1)})+ 2
\mu_{\Sigma}(
\tilde{\gamma}_{\Sigma}^{(1)}+\tilde{\gamma}_{\bar{\Sigma}}^{(1)})-50
g_{10}^2(b_{\Sigma}-2 M \mu_{\Sigma})\ee

\bea \beta_{m^2_{\Phi}}^{(1)}&=&2 \bar{\gamma}_{\Phi}^{(1)}
m^2_{\Phi}+720 m^2_{\Phi}|\lambda|^2+
m^2_{H}(12|\gamma|^2+12|\bar{\gamma}|^2+8|\kappa|^2)\nonumber\\&&
+m^2_{\Theta}(8|\rho|^2+120(|\zeta|^2+|\bar{\zeta}|^2)+8|\kappa|^2)
+m^2_{\Sigma}(480|\eta|^2+12|\gamma|^2+120|\zeta|^2)\nonumber\\&&+m^2_{\bar{\Sigma}}(480|\eta|^2
+12|\bar{\gamma}|^2+120|\bar{\zeta}|^2)
+2\hat{\gamma}_{\Phi}^{(1)}-96|M|^2 g_{10}^2\eea

\bea \beta_{m^2_{H}}^{(1)}&=&2
\bar{\gamma}_{H}^{(1)}m^2_{H}+m^2_{\Phi}(252(|\gamma|^2+|\bar{\gamma}|^2)+168|\kappa|^2)
+168m^2_{\Theta}|\kappa|^2+252m^2_{\Sigma}|\gamma|^2\nonumber\\&&+252m^2_{\bar{\Sigma}}|\bar{\gamma}|^2
+2\hat{\gamma}_{H}^{(1)}-36|M|^2 g_{10}^2+32 \text
Tr[h^{\dag}.m^2_{\tilde{\Psi}}.h]\eea

\bea \beta_{m^2_{\Theta}}^{(1)}&=&2
\bar{\gamma}_{\Theta}^{(1)}m^2_{\Theta}+m^2_{\Phi}(14(|\kappa|^2+|\rho|^2)+210(|\zeta|^2+|\bar{\zeta}|^2))
+14m^2_{\Theta}|\rho|^2+14m^2_{H}|\kappa|^2\nonumber\\&&+210m^2_{\Sigma}|\zeta|^2
+210m^2_{\bar{\Sigma}}|\bar{\zeta}|^2
+2\hat{\gamma}_{\Theta}^{(1)}-84|M|^2 g_{10}^2+32 \text
Tr[g^{\dag}.m^2_{\tilde{\Psi}}.g]\eea

\bea \beta_{m^2_{\Sigma}}^{(1)}&=&2
\bar{\gamma}_{\Sigma}^{(1)}m^2_{\Sigma}+m^2_{\Phi}(400|\eta|^2+20|\gamma|^2+200|\zeta|^2)
+200m^2_{\Theta}|\zeta|^2+20m^2_{H}|\gamma|^2\nonumber\\&&+400m^2_{\bar{\Sigma}}|\eta|^2
+2\hat{\gamma}_{\Sigma}^{(1)}-100|M|^2 g_{10}^2\eea

\bea \beta_{m^2_{\bar{\Sigma}}}^{(1)}&=&2
\bar{\gamma}_{\bar{\Sigma}}^{(1)}m^2_{\bar{\Sigma}}+m^2_{\Phi}(400|\eta|^2+20|\bar{\gamma}|^2+200|\bar{\zeta}|^2)
+200m^2_{\Theta}|\bar{\zeta}|^2+20m^2_{H}|\bar{\gamma}|^2\nonumber\\&&+400m^2_{\Sigma}|\eta|^2
+2\hat{\gamma}_{\bar{\Sigma}}^{(1)}-100|M|^2 g^2_{10}+128 \text
Tr[f^{\dag}.m^2_{\tilde{\Psi}}.f]\eea

\bea \beta_{m^2_{\tilde{\Psi}}}^{(1)}&=&\bar{\gamma}_\Psi^{(1)}
.m^2_{\tilde\Psi}+ m^2_{\tilde\Psi}.\bar{\gamma}_\Psi^{(1)} +10
h^{\dag}.m^2_{\tilde{\Psi}}.h+120
g^{\dag}.m^2_{\tilde{\Psi}}.g+252 f^{\dag}.m^2_{\tilde{\Psi}}.f+10
m^2_{H} h^{\dag}.h\nonumber\\&&+120
m^2_{\Theta}g^{\dag}.g+252m^2_{\bar{\Sigma}}
f^{\dag}.f+2\hat{\gamma}_{\Psi}^{(1)}-45|M|^2 g^2_{10}\eea

The anomalous dimensions associated with the 2-loop beta functions
are given as:

\bea
\bar\gamma_i^{(2)j}&=&-\frac{1}{2}Y_{imn}Y^{npq}Y_{pqr}Y^{mrj}\eea

\be  \bar{\gamma}_\Sigma^{(2)}=200 |\eta|^2
(\bar{\gamma}_\Phi^{(1)}+ \bar{\gamma}_{\bar{\Sigma}}^{(1)}  ) +
10 |\gamma|^2(\bar{\gamma}_H^{(1)}+ \bar{\gamma}_\Phi^{(1)} ) +
100 |\zeta|^2(\bar{\gamma}_\Theta^{(1)}+ \bar{\gamma}_\Phi^{(1)}
)\ee

\bea  \bar{\gamma}_{\bar{\Sigma}}^{(2)}&=&200 |\eta|^2
(\bar{\gamma}_\Phi^{(1)}+ \bar{\gamma}_{\Sigma}^{(1)}  ) +  10
|\bar{\gamma}|^2(\bar{\gamma}_H^{(1)}+ \bar{\gamma}_\Phi^{(1)} ) +
100 |\bar{\zeta}|^2(\bar{\gamma}_\Theta^{(1)}+
\bar{\gamma}_\Phi^{(1)} )\nonumber\\&&+\text Tr[
f^{\dag}.\bar\gamma_{\Psi}^{(1)}.f]  \eea

\bea \bar{\gamma}_H^{(2)}&=& 84
|\kappa|^2(\bar{\gamma}_\Phi^{(1)}+ \bar{\gamma}_\Theta^{(1)} )
+126|\gamma|^2(\bar{\gamma}_\Phi^{(1)}+ \bar{\gamma}_\Sigma^{(1)}
)+126 |\bar{\gamma}|^2(\bar{\gamma}_\Phi^{(1)}+
\bar{\gamma}_{\bar\Sigma}^{(1)} )\nonumber\\&&+ \text Tr[
h^{\dag}.\bar\gamma_{\Psi}^{(1)}.h] \eea

\bea  \bar{\gamma}_{\Theta}^{(2)}&=&7 |\kappa|^2
(\bar{\gamma}_\Phi^{(1)}+ \bar{\gamma}_H^{(1)}  ) +7 |\rho|^2
(\bar{\gamma}_\Theta^{(1)}+ \bar{\gamma}_\Phi^{(1)} ) +105
|\zeta|^2(\bar{\gamma}_\Sigma^{(1)}+ \bar{\gamma}_\Phi^{(1)}
)\nonumber\\&& + 105
|\bar{\zeta}|^2(\bar{\gamma}_{\bar\Sigma}^{(1)}+
\bar{\gamma}_\Phi^{(1)} )+\text Tr[
g^{\dag}.\bar\gamma_{\Psi}^{(1)}.g] \eea

\bea \bar\gamma_{\Psi}^{(2)}&=&
h^\dag.\bar{\gamma}_{\Psi}^{(1)T}.h-g^\dag.\bar{\gamma}_{\Psi}^{(1)T}.g+f^\dag.\bar{\gamma}_{\Psi}^{(1)T}.f+h^\dag.h
\gamma_H^{(1)}- g^\dag.g \gamma_\Theta^{(1)}+f^\dag.f \gamma_{\bar
\Sigma}^{(1)} \eea

\be \tilde\gamma_i^{(2)j}=Y_{imn}Y^{npq}Y_{pqr}h^{mrj} \ee

\bea  \tilde{\gamma}_\Sigma^{(2)}&=&400 \tilde{\eta} \eta^*
(\bar{\gamma}_\Phi^{(1)}+ \bar{\gamma}_{\bar{\Sigma}}^{(1)}  ) +
20 \tilde{\gamma} \gamma^* (\bar{\gamma}_H^{(1)}+
\bar{\gamma}_\Phi^{(1)} ) + 200 \tilde{\zeta}
\zeta^*(\bar{\gamma}_\Theta^{(1)}+ \bar{\gamma}_\Phi^{(1)} )\eea

\bea  \tilde{\gamma}_{\bar{\Sigma}}^{(2)}&=&400 \tilde{\eta}
\eta^* (\bar{\gamma}_\Phi^{(1)}+ \bar{\gamma}_{\Sigma}^{(1)}  ) +
20 \tilde{\bar{\gamma}}
 \bar{\gamma}^*(\bar{\gamma}_H^{(1)}+ \bar{\gamma}_\Phi^{(1)} ) +  200
\tilde{\bar{\zeta}} \bar{\zeta}^*(\bar{\gamma}_\Theta^{(1)}+
\bar{\gamma}_\Phi^{(1)} )\nonumber\\&&+2 \text Tr[
f^{\dag}.\bar\gamma_{\Psi}^{(1)}.\tilde f]  \eea

\bea \tilde{\gamma}_H^{(2)}&=&168 \tilde{\kappa}
\kappa^*(\bar{\gamma}_\Phi^{(1)}+ \bar{\gamma}_\Theta^{(1)} )
+252\tilde{\gamma} \gamma^*(\bar{\gamma}_\Phi^{(1)}+
\bar{\gamma}_\Sigma^{(1)} )+252 \tilde{\bar{\gamma}}
 \bar{\gamma}^*(\bar{\gamma}_\Phi^{(1)}+ \bar{\gamma}_{\bar\Sigma}^{(1)} )\nonumber\\&&+
2 \text Tr[ h^{\dag}.\bar\gamma_{\Psi}^{(1)}.\tilde h]\eea

\bea  \tilde{\gamma}_{\Theta}^{(2)}&=&14 \tilde{\kappa} \kappa^*
(\bar{\gamma}_\Phi^{(1)}+ \bar{\gamma}_H^{(1)}  ) +14 \tilde{\rho}
\rho^* (\bar{\gamma}_\Theta^{(1)}+ \bar{\gamma}_\Phi^{(1)} ) +210
\tilde{\zeta} \zeta^*(\bar{\gamma}_\Sigma^{(1)}+
\bar{\gamma}_\Phi^{(1)} )\nonumber\\&& + 210 \tilde{\bar{\zeta}}
\bar{\zeta}^*(\bar{\gamma}_{\bar\Sigma}^{(1)}+
\bar{\gamma}_\Phi^{(1)} )+2 \text Tr[
g^{\dag}.\bar\gamma_{\Psi}^{(1)}.\tilde g]\eea

\be \tilde\gamma_\Psi^{(2)}= 2
h^\dag.\bar{\gamma}_{\Psi}^{(1)T}.\tilde h+2
g^\dag.\bar{\gamma}_{\psi}^{(1)T}.\tilde g+
2f^\dag.\bar{\gamma}_{\psi}^{(1)T}.\tilde f+2 h^\dag.\tilde h
\gamma_H^{(1)}+2 g^\dag.\tilde g \gamma_\Theta^{(1)}+2
f^\dag.\tilde f \gamma_{\bar \Sigma}^{(1)} \ee

\be \hat{\gamma}^{(2)}=Y_{imn}h^{npq}Y_{pqr}Y^{mrj} \ee

\bea  \hat{\gamma}_\Sigma^{(2)}&=& 400 |\eta|^2
(\tilde{\gamma}_\Phi^{(1)}+ \tilde{\gamma}_{\bar{\Sigma}}^{(1)}  )
+  20 |\gamma|^2(\tilde{\gamma}_H^{(1)}+ \tilde{\gamma}_\Phi^{(1)}
) + 200 |\zeta|^2(\tilde{\gamma}_\Theta^{(1)}+
\tilde{\gamma}_\Phi^{(1)} )\eea

\bea  \hat{\gamma}_{\bar{\Sigma}}^{(2)}&=&400 |\eta|^2
(\tilde{\gamma}_\Phi^{(1)}+ \tilde{\gamma}_{\Sigma}^{(1)}  ) +  20
|\bar{\gamma}|^2(\tilde{\gamma}_H^{(1)}+ \tilde{\gamma}_\Phi^{(1)}
) + 200 |\bar{\zeta}|^2(\tilde{\gamma}_\Theta^{(1)}+
\tilde{\gamma}_\Phi^{(1)} )\nonumber\\&&+2\text Tr[
f^{\dag}.\tilde\gamma_{\Psi}^{(1)}.f] \eea

\bea \hat{\gamma}_H^{(2)}&=&168
|\kappa|^2(\tilde{\gamma}_\Phi^{(1)}+ \tilde{\gamma}_\Theta^{(1)}
) +252|\gamma|^2(\tilde{\gamma}_\Phi^{(1)}+
\tilde{\gamma}_\Sigma^{(1)} )+252
|\bar{\gamma}|^2(\tilde{\gamma}_\Phi^{(1)}+
\tilde{\gamma}_{\bar\Sigma}^{(1)} )\nonumber\\&&+ 2\text Tr[
h^{\dag}.\tilde\gamma_{\Psi}^{(1)}.h] \eea

\bea  \hat{\gamma}_{\Theta}^{(2)}&=&14 |\kappa|^2
(\tilde{\gamma}_\Phi^{(1)}+ \tilde{\gamma}_H^{(1)}  ) +14 |\rho|^2
(\tilde{\gamma}_\Theta^{(1)}+ \tilde{\gamma}_\Phi^{(1)} ) +210
|\zeta|^2(\tilde{\gamma}_\Sigma^{(1)}+ \tilde{\gamma}_\Phi^{(1)} )
\nonumber\\&&+  210
|\bar{\zeta}|^2(\tilde{\gamma}_{\bar\Sigma}^{(1)}+
\tilde{\gamma}_\Phi^{(1)} )+2 \text Tr[
g^{\dag}.\tilde\gamma_{\Psi}^{(1)}.g]\eea

\be \hat\gamma_\Psi^{(2)}=  2 h^\dag.\tilde
{\gamma}_{\Psi}^{(1)T}.h+2 g^\dag.\tilde {\gamma}_{\Psi}^{(1)T}.g+
2f^\dag.\tilde {\gamma}_{\Psi}^{(1)T}.f+2 h^\dag.h
\tilde{\gamma}_H^{(1)}+2 g^\dag.g \tilde{\gamma}_\Theta^{(1)}+2
f^\dag.f \tilde{\gamma}_{\bar \Sigma}^{(1)}         \ee

\be \check{\gamma}=Y_{ilm}Y^{jln}h_{npq}h^{mpq} \ee

\bea  \check{\gamma}_\Phi^{(2)}&=&480 |\eta|^2
(\hat{\gamma}_\Sigma^{(1)}+ \hat{\gamma}_{\bar{\Sigma}}^{(1)} )+8
|\kappa|^2(\hat{\gamma}_H^{(1)}+ \hat{\gamma}_\Theta^{(1)}
)+720|\lambda|^2 \hat{\gamma}_\Phi^{(1)} \nonumber\\&& +8|\rho|^2
\hat{\gamma}_\Theta^{(1)} +12 |\gamma|^2(\hat{\gamma}_H^{(1)}+
\hat{\gamma}_\Sigma^{(1)} )+12
|\bar{\gamma}|^2(\hat{\gamma}_H^{(1)}+
\hat{\gamma}_{\bar\Sigma}^{(1)} )\nonumber\\&&+120
|\zeta|^2(\hat{\gamma}_\Theta^{(1)}+ \hat{\gamma}_\Sigma^{(1)}
)+120 |\bar{\zeta}|^2(\hat{\gamma}_\Theta^{(1)}+
\hat{\gamma}_{\bar\Sigma}^{(1)} )\eea

\bea  \check{\gamma}_\Sigma^{(2)}&=&400 |\eta|^2
(\hat{\gamma}_\Phi^{(1)}+ \hat{\gamma}_{\bar{\Sigma}}^{(1)}  ) +
20 |\gamma|^2(\hat{\gamma}_H^{(1)}+ \hat{\gamma}_\Phi^{(1)} ) +
200 |\zeta|^2(\hat{\gamma}_\Theta^{(1)}+ \hat{\gamma}_\Phi^{(1)}
)\eea

\bea  \check{\gamma}_{\bar{\Sigma}}^{(2)}&=&400 |\eta|^2
(\hat{\gamma}_\Phi^{(1)}+ \hat{\gamma}_{\Sigma}^{(1)}  ) +  20
|\bar{\gamma}|^2(\hat{\gamma}_H^{(1)}+ \hat{\gamma}_\Phi^{(1)} ) +
200 |\bar{\zeta}|^2(\hat{\gamma}_\Theta^{(1)}+
\hat{\gamma}_\Phi^{(1)} )\nonumber\\&&+2 \text Tr[
f^{\dag}.\hat\gamma_{\Psi}^{(1)}.f]  \eea

\bea \check{\gamma}_H^{(2)}&=&168
|\kappa|^2(\hat{\gamma}_\Phi^{(1)}+ \hat{\gamma}_\Theta^{(1)} )
+252|\gamma|^2(\hat{\gamma}_\Phi^{(1)}+ \hat{\gamma}_\Sigma^{(1)}
)+252 |\bar{\gamma}|^2(\hat{\gamma}_\Phi^{(1)}+
\hat{\gamma}_{\bar\Sigma}^{(1)} )\nonumber\\&&+ 2 \text Tr[
h^{\dag}.\hat\gamma_{\Psi}^{(1)}.h] \eea

\bea \check{\gamma}_{\Theta}^{(2)}&=&14 |\kappa|^2
(\hat{\gamma}_\Phi^{(1)}+ \hat{\gamma}_H^{(1)}  ) +14 |\rho|^2
(\hat{\gamma}_\Theta^{(1)}+ \hat{\gamma}_\Phi^{(1)} ) +210
|\zeta|^2(\hat{\gamma}_\Sigma^{(1)}+ \hat{\gamma}_\Phi^{(1)} )
\nonumber\\&&+ 210
|\bar{\zeta}|^2(\hat{\gamma}_{\bar\Sigma}^{(1)}+
\hat{\gamma}_\Phi^{(1)} )+2 \text Tr[
g^{\dag}.\hat\gamma_{\Psi}^{(1)}.g]\eea

\be \check\gamma_\Psi^{(2)}=2 h^\dag.\hat
{\gamma}_{\Psi}^{(1)T}.h+2 g^\dag.\hat {\gamma}_{\Psi}^{(1)T}.g+2
f^\dag.\hat {\gamma}_{\Psi}^{(1)T}.f+2 h^\dag .h
\hat{\gamma}_H^{(1)}+2 g^\dag.g \hat{\gamma}_\Theta^{(1)}+2f^\dag
.f \hat{\gamma}_{\bar \Sigma}^{(1)}          \ee

\be  \acute{\gamma}^{(2)}= h_{ilm } h^{jln  } Y_{npq  } Y^{mpq }
\ee

\bea  \acute{\gamma}_\Phi^{(2)}&=&480 |\tilde{\eta}|^2
(\bar{\gamma}_\Sigma^{(1)}+ \bar{\gamma}_{\bar{\Sigma}}^{(1)}  )+8
|\tilde{\kappa}|^2(\bar{\gamma}_H^{(1)}+ \bar{\gamma}_\Theta^{(1)}
)+720|\tilde{\lambda}|^2 \bar{\gamma}_\Phi^{(1)} \nonumber\\&&
+8|\tilde{\rho}|^2\bar{\gamma}_\Theta^{(1)} +12
|\tilde{\gamma}|^2(\bar{\gamma}_H^{(1)}+ \bar{\gamma}_\Sigma^{(1)}
)+12|\tilde{\bar{\gamma}}|^2(\bar{\gamma}_H^{(1)}+
\bar{\gamma}_{\bar\Sigma}^{(1)} )\nonumber\\&&+120
|\tilde{\zeta}|^2(\bar{\gamma}_\Theta^{(1)}+
\bar{\gamma}_\Sigma^{(1)} )+120
|\tilde{\bar{\zeta}}|^2(\bar{\gamma}_\Theta^{(1)}+
\bar{\gamma}_{\bar\Sigma}^{(1)} )\eea

\bea  \acute{\gamma}_\Sigma^{(2)}&=&400 |\tilde{\eta}|^2
(\bar{\gamma}_\Phi^{(1)}+ \bar{\gamma}_{\bar{\Sigma}}^{(1)}  ) +
20 |\tilde{\gamma}|^2(\bar{\gamma}_H^{(1)}+
\bar{\gamma}_\Phi^{(1)} ) + 200
|\tilde{\zeta}|^2(\bar{\gamma}_\Theta^{(1)}+
\bar{\gamma}_\Phi^{(1)} )\eea

\bea  \acute{\gamma}_{\bar{\Sigma}}^{(2)}&=&400 |\tilde{\eta}|^2
(\bar{\gamma}_\Phi^{(1)}+ \bar{\gamma}_{\Sigma}^{(1)}  ) +  20
|\tilde{\bar{\gamma}}|^2(\bar{\gamma}_H^{(1)}+
\bar{\gamma}_\Phi^{(1)} ) + 200
|\tilde{\bar{\zeta}}|^2(\bar{\gamma}_\Theta^{(1)}+
\bar{\gamma}_\Phi^{(1)} )\nonumber\\&&+2 \text Tr[ \tilde
f^{\dag}.\bar\gamma_{\Psi}^{(1)}.\tilde f] \eea

\bea \acute{\gamma}_H^{(2)}&=&168
|\tilde{\kappa}|^2(\bar{\gamma}_\Phi^{(1)}+
\bar{\gamma}_\Theta^{(1)} )
+252|\tilde{\gamma}|^2(\bar{\gamma}_\Phi^{(1)}+
\bar{\gamma}_\Sigma^{(1)} )+252
|\tilde{\bar{\gamma}}|^2(\bar{\gamma}_\Phi^{(1)}+
\bar{\gamma}_{\bar\Sigma}^{(1)} )\nonumber\\&&+ 2 \text Tr[ \tilde
h^{\dag}.\bar\gamma_{\Psi}^{(1)}.\tilde h] \eea

\bea  \acute{\gamma}_{\Theta}^{(2)}&=&14 |\tilde{\kappa}|^2
(\bar{\gamma}_\Phi^{(1)}+ \bar{\gamma}_H^{(1)}  ) +14
|\tilde{\rho}|^2 (\bar{\gamma}_\Theta^{(1)}+
\bar{\gamma}_\Phi^{(1)} ) +210
|\tilde{\zeta}|^2(\bar{\gamma}_\Sigma^{(1)}+
\bar{\gamma}_\Phi^{(1)} ) \nonumber\\&&+  210
|\tilde{\bar{\zeta}}|^2(\bar{\gamma}_{\bar\Sigma}^{(1)}+
\bar{\gamma}_\Phi^{(1)} )+2 \text Tr[ \tilde
g^{\dag}.\bar\gamma_{\Psi}^{(1)}.\tilde g] \eea

\be \acute\gamma_\Psi^{(2)}= 2 \tilde h^\dag.\bar
{\gamma}_{\Psi}^{(1)T}.\tilde h+2\tilde g^\dag.\bar
{\gamma}_{\Psi}^{(1)T}.\tilde g+2\tilde f^\dag.\bar
{\gamma}_{\Psi}^{(1)T}.\tilde f+2\tilde h^\dag.\tilde h
\bar{\gamma}_H^{(1)}+ 2\tilde g^\dag.\tilde g
\bar{\gamma}_\Theta^{(1)}+2\tilde f^\dag.\tilde f
\bar{\gamma}_{\bar \Sigma}^{(1)}  \ee

\be \grave{\gamma}^{(2)}= h_{ilm}Y^{jln}Y_{npq}h^{mpq}    \ee

\bea  \grave{\gamma}_\Phi^{(2)}&=&480 \tilde{\eta}^*
\eta(\tilde{\gamma}_\Sigma^{(1)}+
\tilde{\gamma}_{\bar{\Sigma}}^{(1)}  )+8 \tilde{\kappa}^* \kappa
(\tilde{\gamma}_H^{(1)}+ \tilde{\gamma}_\Theta^{(1)}
)+720\tilde{\lambda }^*\lambda \tilde{\gamma}_\Phi^{(1)}
\nonumber\\&& +8\tilde{\rho}^* \rho \tilde{\gamma}_\Theta^{(1)}
+12 \tilde{\gamma}^* \gamma (\tilde{\gamma}_H^{(1)}+
\tilde{\gamma}_\Sigma^{(1)} )+12 \tilde{\bar{\gamma}}^*
 \bar{\gamma}(\tilde{\gamma}_H^{(1)}+ \tilde{\gamma}_{\bar\Sigma}^{(1)}
)\nonumber\\&&+120 \tilde{\zeta}^* \zeta
(\tilde{\gamma}_\Theta^{(1)}+ \tilde{\gamma}_\Sigma^{(1)} )+120
\tilde{\bar{\zeta}}^* \bar{\zeta} (\tilde{\gamma}_\Theta^{(1)}+
\tilde{\gamma}_{\bar\Sigma}^{(1)} )\eea

\bea  \grave{\gamma}_\Sigma^{(2)}&=&400 \tilde{\eta}^* \eta
(\tilde{\gamma}_\Phi^{(1)}+ \tilde{\gamma}_{\bar{\Sigma}}^{(1)}  )
+ 20 \tilde{\gamma}^* \gamma (\tilde{\gamma}_H^{(1)}+
\tilde{\gamma}_\Phi^{(1)} ) + 200 \tilde{\zeta}^*
\zeta(\tilde{\gamma}_\Theta^{(1)}+ \tilde{\gamma}_\Phi^{(1)} )\eea

\bea  \grave{\gamma}_{\bar{\Sigma}}^{(2)}&=&400 \tilde{\eta}^*
\eta (\tilde{\gamma}_\Phi^{(1)}+ \tilde{\gamma}_{\Sigma}^{(1)} ) +
20 \tilde{\bar{\gamma}}^*
 \bar{\gamma}(\tilde{\gamma}_H^{(1)}+ \tilde{\gamma}_\Phi^{(1)} ) +  200
\tilde{\bar{\zeta}}^* \bar{\zeta}(\tilde{\gamma}_\Theta^{(1)}+
\tilde{\gamma}_\Phi^{(1)} )\nonumber\\&&+2\text Tr[ \tilde
f^{\dag}.\tilde\gamma_{\Psi}^{(1)}.f]  \eea

\bea \grave{\gamma}_H^{(2)}&=&168 \tilde{\kappa}^*
\kappa(\tilde{\gamma}_\Phi^{(1)}+ \tilde{\gamma}_\Theta^{(1)} )
+252\tilde{\gamma}^* \gamma(\tilde{\gamma}_\Phi^{(1)}+
\tilde{\gamma}_\Sigma^{(1)} )+252 \tilde{\bar{\gamma}}^*
 \bar{\gamma}(\tilde{\gamma}_\Phi^{(1)}+ \tilde{\gamma}_{\bar\Sigma}^{(1)} )\nonumber\\&&+
2 \text Tr[ \tilde h^{\dag}.\tilde\gamma_{\Psi}^{(1)}.h]\eea

\bea  \grave{\gamma}_{\Theta}^{(2)}&=&14 \tilde{\kappa}^* \kappa
(\tilde{\gamma}_\Phi^{(1)}+ \tilde{\gamma}_H^{(1)}  ) +14
\tilde{\rho}^* \rho (\tilde{\gamma}_\Theta^{(1)}+
\tilde{\gamma}_\Phi^{(1)} ) +210 \tilde{\zeta}^*
\zeta(\tilde{\gamma}_\Sigma^{(1)}+ \tilde{\gamma}_\Phi^{(1)}
)\nonumber\\&& + 210 \tilde{\bar{\zeta}}^*
\bar{\zeta}(\bar{\gamma}_{\tilde\Sigma}^{(1)}+
\tilde{\gamma}_\Phi^{(1)} )+2 \text Tr[ \tilde
g^{\dag}.\tilde\gamma_{\Psi}^{(1)}.g]\eea

\be \grave\gamma_\Psi^{(2)}= 2 \tilde h^\dag.\tilde
{\gamma}_{\Psi}^{(1)T}. h+2 \tilde g^\dag.\tilde
{\gamma}_{\Psi}^{(1)T}. g+ 2\tilde f^\dag.\tilde
{\gamma}_{\Psi}^{(1)T}. f+2 \tilde h^\dag.h
\tilde{\gamma}_H^{(1)}+ 2 \tilde g^\dag.g
\tilde{\gamma}_\Theta^{(1)}+2 \tilde f^\dag.f \tilde{\gamma}_{\bar
\Sigma}^{(1)} \ee

\be
\breve{\gamma}_i^{j(2)}=Y_{ilm}Y^{jln}Y_{npq}Y^{mpr}(m^2)^q_r\ee

\bea \breve{\gamma}_\Phi^{(2)}&=&
m^2_{H}(720|\kappa|^2|\lambda|^2+14|\kappa|^2|\rho|^2+28|\kappa|^4+
420|\kappa|^2|\zeta|^2+420|\kappa|^2|\bar\zeta|^2\nonumber\\&&+1080|\lambda|^2|\gamma|^2+1080|\lambda|^2|\bar\gamma|^2+2400|\eta|^2
|\gamma|^2+2400|\eta|^2|\bar\gamma|^2+60|\gamma|^4\nonumber\\&&+60|\bar\gamma|^4+
600|\zeta|^2|\gamma|^2+600|\bar\zeta|^2|\bar\gamma|^2)\nonumber\\&&+
m^2_{\Theta}(720|\lambda|^2|\kappa|^2+28|\rho|^2|\kappa|^2+360|\rho|^2|\lambda|^2
+14|\rho|^4+420|\zeta|^2|\rho|^2\nonumber\\&&+10800|\zeta|^2|\lambda|^2+
10800|\bar\zeta|^2|\lambda|^2+420|\bar\zeta|^2|\rho|^2+336|\kappa|^4+504|\kappa|^2|\gamma|^2\nonumber\\&&
+504|\kappa|^2|\bar\gamma|^2
+600|\zeta|^2|\gamma|^2+600|\bar\zeta|^2|\bar\gamma|^2+24000|\eta|^2|\zeta|^2+24000|\eta|^2|\bar\zeta|^2
\nonumber\\&&+600|\zeta|^4+600|\bar\zeta|^4
  )\nonumber\eea \bea&&+m^2_{\Phi}( 1008|\kappa|^2|\gamma|^2+1008|\kappa|^2|\bar\gamma|^2+364|\kappa|^4+
  816|\gamma|^4+1512|\gamma|^2|\bar\gamma|^2
\nonumber\\&&+816|\bar\gamma|^4
  +64800|\lambda|^4+42|\rho|^2|\kappa|^2+14|\rho|^4+630|\rho|^2|\zeta|^2\nonumber\\&&+630|\rho|^2|\bar\zeta|^2 +
 840|\kappa|^2|\zeta|^2+840|\kappa|^2|\bar\zeta|^2+
 12300|\zeta|^4+12600|\zeta|^2|\bar\zeta|^2\nonumber\\&&+12300|\bar\zeta|^4
+96000|\eta|^4+3600|\eta|^2|\gamma|^2+36000|\eta|^2|\zeta|^2
+1200|\gamma|^2|\zeta|^2
\nonumber\\&&+3600|\eta|^2|\bar\gamma|^2+36000|\eta|^2|\bar\zeta|^2+1200|\bar\gamma|^2|\bar\zeta|^2
)\nonumber\\&&+
m^2_{\Sigma}(1080|\gamma|^2|\lambda|^2+1200|\gamma|^2|\eta|^2+10800|\zeta|^2|\lambda|^2
+12000|\zeta|^2|\eta|^2\nonumber\\&&+43200|\eta|^2|\lambda|^2+48000|\eta|^4+210|\zeta|^2|\rho|^2+420|\zeta|^2|\kappa|^2+
6300|\zeta|^4\nonumber\\&&+6300|\zeta|^2|\bar\zeta|^2+756|\gamma|^4+756|\gamma|^2|\bar\gamma|^2+
504|\gamma|^2|\kappa|^2)\nonumber\\&&+
m^2_{\bar\Sigma}(1080|\bar\gamma|^2|\lambda|^2+1200|\bar\gamma|^2|\eta|^2+10800|\bar\zeta|^2|\lambda|^2
+12000|\bar\zeta|^2|\eta|^2
\nonumber\\&&+43200|\eta|^2|\lambda|^2+48000|\eta|^4+210|\bar\zeta|^2|\rho|^2+420|\bar\zeta|^2|\kappa|^2+
6300|\zeta|^2|\bar\zeta|^2\nonumber\\&&+6300|\bar\zeta|^4+756|\bar\gamma|^2|\gamma|^2+
756|\bar\gamma|^4+504|\bar\gamma|^2|\kappa|^2) \eea

\bea \breve{\gamma}_H^{(2)}&=&
m^2_{H}(924|\kappa|^4+1008|\gamma|^2|\kappa|^2+1008|\bar\gamma|^2|\kappa|^2
+ 2016|\gamma|^4+1512|\bar\gamma|^2|\gamma|^2\nonumber\\&&+
2016|\bar\gamma|^4+10\text Tr[h^\dag.h.h^\dag.h])\nonumber\\&&+
m^2_{\theta}(
336|\kappa|^4+924|\kappa|^2|\rho|^2+5040|\kappa|^2|\zeta|^2+5040|\kappa|^2|\bar\zeta|^2
+
504|\kappa|^2|\gamma|^2\nonumber\\&&+504|\rho|^2|\gamma|^2+20160|\zeta|^2|\gamma|^2+7560|\bar\zeta|^2|\gamma|^2
+
504|\kappa|^2|\bar\gamma|^2+504|\rho|^2|\bar\gamma|^2\nonumber\\&&+7560|\zeta|^2|\bar\gamma|^2+20160|\bar\zeta|^4
  +120\text
Tr[h^\dag.h.g^\dag.g])\nonumber\\&&+
m^2_{\Phi}(30240|\kappa|^2|\lambda|^2+45360|\gamma|^2|\lambda|^2+45360|\bar
\gamma|^2|\lambda|^2+588|\kappa|^4\nonumber\\&&+588|\kappa|^2|\rho|^2+1260|\gamma|^4+1260|\bar\gamma|^4+
12600|\zeta|^2|\gamma|^2+12600|\bar\zeta|^2|\bar\gamma|^2
\nonumber\\&&+8820|\kappa|^2|\zeta|^2+8820|\kappa|^2||\bar\zeta|^2+25200|\eta|^2|\gamma|^2+25200|\eta|^2|\bar\gamma|^2
)\nonumber\\&&+
m^2_{\Sigma}(504|\kappa|^2|\gamma|^2+20160|\kappa|^2|\eta|^2+13860|\kappa|^2|\zeta|^2+
30240|
\bar\gamma|^2|\eta|^2\nonumber\\&&+55440|\gamma|^2|\eta|^2+7560|\gamma|^2|\zeta|^2
+ 756|\gamma|^4+7560|\bar\gamma|^2|\zeta|^2
+756|\gamma|^2|\bar\gamma|^2)\nonumber\\&&+
m^2_{\bar\Sigma}(504|\kappa|^2|\bar
\gamma|^2+20160|\kappa|^2|\eta|^2+13860|\kappa|^2|\bar\zeta|^2+
30240|\gamma|^2|\eta|^2+756|\gamma|^2|\bar\gamma|^2\nonumber\\&&+7560|\gamma|^2|\bar\zeta|^2
+ 756|\bar
\gamma|^4+55440|\bar\gamma|^2|\eta|^2+7560|\bar\gamma|^2|\bar\zeta|^2
+252\text Tr[h^\dag.h.f^\dag.f])\nonumber\\&&+ Tr[(h.h^\dag(10
h.m^2_{\tilde{\Psi}}.h^\dag+252
f.m^2_{\tilde{\Psi}}.f^\dag+120g.m^2_{\tilde{\Psi}}.g^\dag)+31752|\bar\gamma|^2
f^\dag.f.m^2_{\tilde{\Psi}}\nonumber\\&&+10080|\kappa|^2
g^\dag.g.m^2_{\tilde{\Psi}})]           \eea

\bea \breve{\gamma}_{\Theta}^{(2)}&=&
m^2_{H}(28|\kappa|^4+77|\kappa|^2|\rho|^2+420|\kappa|^2|\zeta|^2+420|\kappa|^2|\bar\zeta|^2
+
42|\kappa|^2|\gamma|^2\nonumber\\&&+42|\rho|^2|\gamma|^2+1680|\zeta|^2|\gamma|^2+630|\bar\zeta|^2|\gamma|^2
+
42|\kappa|^2|\bar\gamma|^2+42|\rho|^2|\bar\gamma|^2\nonumber\\&&+630|\zeta|^2|\bar\gamma|^2+1680|\bar\zeta|^2|\bar\gamma|^2
+10\text Tr[g^\dag.g.h^\dag.h])\nonumber\\&&+
m^2_{\Theta}(42|\kappa|^2|\rho|^2+63|\rho|^4+630|\zeta|^2|\rho|^2+630|\bar\zeta|^2|\rho|^2
+
616|\kappa|^4\nonumber\\&&+630|\zeta|^2|\kappa|^2+630|\bar\zeta|^2|\kappa|^2
+
16800|\zeta|^4+12600|\bar\zeta|^2|\zeta|^2\nonumber\\&&+16800|\bar\zeta|^4
+120\text Tr[g^\dag.g.g^\dag.g ])\nonumber\\&&
+m^2_{\Phi}(49|\kappa|^2|\rho|^2+10500(|\zeta|^2|\gamma|^2+|\bar\zeta|^2|\bar\gamma|^2)
+49|\rho|^4+588|\kappa|^4\nonumber\\&&+10500(|\zeta|^4+|\bar\zeta|^4)+1260|\lambda|^2|\kappa|^2
+1260|\lambda|^2|\rho|^2+18900|\lambda|^2|\zeta|^2\nonumber\\&&+18900|\lambda|^2|\bar
\zeta|^2+21000|\eta|^2|\zeta|^2+21000|\eta|^2|\bar
\zeta|^2+882|\kappa|^2|\gamma|^2+882|\kappa|^2|\bar
\gamma|^2\nonumber\\&&+735|\rho|^2 |\zeta|^2 +735|\rho|^2
|\bar\zeta|^2)\nonumber\\&&+
m^2_{\Sigma}(924|\kappa|^2|\gamma|^2+42|\rho|^2|\gamma|^2+630|\zeta|^2|\gamma|^2+630|\bar\zeta|^2|\gamma|^2
+
420|\kappa|^2|\zeta|^2\nonumber\\&&+1155|\rho|^2|\zeta|^2+6300|\zeta|^4+6300|\bar\zeta|^2|\zeta|^2
+
1680|\kappa|^2|\eta|^2+1680|\rho|^2|\eta|^2\nonumber\\&&+25200|\zeta|^2|\eta|^2+46200|\bar\zeta|^2|\eta|^2
\nonumber\\&&+
m^2_{\bar\Sigma}(924|\kappa|^2|\bar\gamma|^2+42|\rho|^2|\bar\gamma|^2+
630|\zeta|^2|\bar\gamma|^2+630|\bar\zeta|^2|\bar\gamma|^2 +
420|\kappa|^2|\bar\zeta|^2\nonumber\\&&+1155|\rho|^2|\bar\zeta|^2+6300|\zeta|^2|\bar\zeta|^2+6300|\bar\zeta|^4
+
1680|\kappa|^2|\eta|^2+1680|\rho|^2|\eta|^2\nonumber\\&&+46200|\zeta|^2|\eta|^2+25200|\bar\zeta|^2|\eta|^2
+252\text Tr[g^\dag.g.f^\dag.f ])+
Tr[(g.g^\dag(h.m^2_{\tilde{\Psi}}.h^\dag\nonumber\\&&+f.m^2_{\tilde{\Psi}}.f^\dag+g.m^2_{\tilde{\Psi}}.g^\dag)+70|\kappa|^2
h^\dag.h.m^2_{\tilde{\Psi}}\nonumber\\&&+840|\rho|^2
g^\dag.g.m^2_{\tilde{\Psi}}+26460|\bar\zeta|^2
f^\dag.f.m^2_{\tilde{\Psi}})]  \eea

\bea \breve{\gamma}_{\Sigma}^{(2)}&=&
m^2_{H}(800|\eta|^2|\kappa|^2+40|\gamma|^2|\kappa|^2+1100|\zeta|^2|\kappa|^2
+
   1200|\eta|^2|\gamma|^2+60|\gamma|^4\nonumber\\&&+600|\zeta|^2|\gamma|^2  +
3200|\eta|^2|\bar\gamma|^2+60|\gamma|^2|\bar\gamma|^2+600|\zeta|^2|\bar\gamma|^2
)\nonumber\\&&+
m^2_{\Theta}(400|\eta|^2|\rho|^2+800|\eta|^2|\kappa|^2+12000|\eta|^2|\zeta|^2+32000|\eta|^2|\bar\zeta|^2
\nonumber\\&&+
20|\gamma|^2|\rho|^2+880|\gamma|^2|\kappa|^2+600|\gamma|^2|\zeta|^2+600|\gamma|^2|\bar\zeta|^2
+
900|\zeta|^2|\rho|^2\nonumber\\&&+400|\zeta|^2|\kappa|^2+6000|\zeta|^4+6000|\bar\zeta|^2|\zeta|^2
)\nonumber\\&&+
m^2_{\Phi}(700|\kappa|^2|\zeta|^2+2000|\bar\gamma|^2|\eta|^2+840|\kappa|^2|\gamma^2|+20000|\eta|^2|\bar\zeta|^2
\nonumber\\&&+36000|\eta|^2|\lambda|^2+1800|\gamma|^2|\lambda|^2+18000|
\zeta|^2|\lambda|^2+1260|\gamma|^2(|\gamma|^2+|\bar\gamma|^2)\nonumber\\&&+10500|\zeta|^4
+10500|\zeta|^2|\bar\zeta|^2+40000|\eta|^4)+\nonumber\\&&
m^2_{\Sigma}(3600|\eta|^2|\gamma|^2+1320|\gamma|^4+1200|\zeta|^2|\gamma|^2+
36000|\eta|^2|\zeta|^2\nonumber\\&&+16500|\zeta|^4+
88000|\eta|^4)+
m^2_{\bar\Sigma}(1200|\eta|^2|\bar\gamma|^2+1320|\bar\gamma|^2|\gamma|^2\nonumber\\&&+
600|\zeta|^2|\bar\gamma|^2+
12000|\eta|^2|\bar\zeta|^2+600|\gamma|^2|\bar\zeta|^2+16500|\bar\zeta|^2|\zeta|^2\nonumber\\&&+
48000|\eta|^4+2400|\gamma|^2|\eta|^2+24000|\zeta|^2|\eta|^2) \eea

\bea \breve{\gamma}_{\bar\Sigma}^{(2)}&=&
m^2_{H}(800|\eta|^2|\kappa|^2+40|\bar\gamma|^2|\kappa|^2+1100|\bar\zeta|^2|\kappa|^2
+
   3200|\eta|^2|\gamma|^2+60|\gamma|^2|\bar\gamma|^2\nonumber\\&&+600|\bar\zeta|^2|\gamma|^2  +
1200|\eta|^2|\bar\gamma|^2+60|\bar\gamma|^4+600|\bar\zeta|^2|\bar\gamma|^2
+10\text Tr[f^\dag.f.h^\dag.h])\nonumber\\&&+
m^2_{\Theta}(400|\eta|^2|\rho|^2+800|\eta|^2|\kappa|^2+32000|\eta|^2|\zeta|^2+12000|\eta|^2|\bar\zeta|^2
+
20|\bar\gamma|^2|\rho|^2\nonumber\\&&+880|\bar\gamma|^2|\kappa|^2+600|\bar\gamma|^2|\zeta|^2+600|\bar\gamma|^2|\bar\zeta|^2
+
900|\bar\zeta|^2|\rho|^2+400|\bar\zeta|^2|\kappa|^2\nonumber\\&&+6000|\bar\zeta|^4+6000|\bar\zeta|^2|\zeta|^2
+120\text Tr[f ^\dag.f.g^\dag.g ])\nonumber\\&&+
m^2_{\Phi}(700|\kappa|^2|\bar\zeta|^2+2000|\gamma|^2|\eta|^2+840|\kappa|^2|\bar\gamma^2|+20000|\eta|^2|\zeta|^2
+36000|\eta|^2|\lambda|^2\nonumber\\&&+1800|\bar\gamma|^2|\lambda|^2+18000|\bar
\zeta|^2|\lambda|^2+1260|\bar\gamma|^2(|\gamma|^2+|\bar\gamma|^2)+10500|\bar\zeta|^2(|\zeta|^2+|\bar\zeta|^2)
\nonumber\\&&+40000|\eta|^4)+
m^2_{\Sigma}(1200|\eta|^2|\gamma|^2+1320|\bar\gamma|^2|\gamma|^2+600|\bar\zeta|^2|\gamma|^2+
12000|\eta|^2|\zeta|^2\nonumber\\&&+600|\bar\gamma|^2|\zeta|^2+16500|\bar\zeta|^2|\zeta|^2+
48000|\eta|^4+2400|\bar\gamma|^2|\eta|^2+24000|\bar\zeta|^2|\eta|^2
)\nonumber\\&&+
m^2_{\bar\Sigma}(3600|\eta|^2|\bar\gamma|^2+1320|\bar\gamma|^4+1200|\bar\zeta|^2|\bar\gamma|^2+
36000|\eta|^2|\bar\zeta|^2+16500|\bar\zeta|^4\nonumber\\&&+
88000|\eta|^4 +252 \text Tr[f^\dag.f.f^\dag.f]) +Tr[(f.f^\dag(10
h.m^2_{\tilde{\Psi}}.h^\dag+252 f.m^2_{\tilde{\Psi}}.f^\dag
\nonumber\\&&+120g.m^2_{\tilde{\Psi}}.g^\dag)+100|\bar\gamma|^2
h^\dag.h.m^2_{\tilde{\Psi}}+12000|\bar\zeta|^2
g^\dag.g.m^2_{\tilde{\Psi}})] \eea

\bea \breve{\gamma}_{\Psi}^{(2)}&=&  10 h^\dag.(10
h.m^2_{\tilde{\Psi}}.h^\dag+ 120 g.m^2_{\tilde{\Psi}}.g^\dag + 252
f.m^2_{\tilde{\Psi}}.f)).h+ 120 g^\dag.(10
h.m^2_{\tilde{\Psi}}.h^\dag \nonumber\\&&+ 120
g.m^2_{\tilde{\Psi}}.g^\dag + 252 f.m^2_{\tilde{\Psi}}.f).g + 252
f^\dag.(10 h.m^2_{\tilde{\Psi}}.h^\dag+ 120
g.m^2_{\tilde{\Psi}}.g^\dag + 252
f.m^2_{\tilde{\Psi}}.f).f\nonumber\\&&+10 h^\dag.(10
m^2_{H}h.h^\dag+120 m^2_{\Theta} g.g^\dag +252 m^2_{\bar \Sigma}
f.f^\dag).h+120 g^\dag.(10 m^2_{H}h.h^\dag\nonumber\\&&+120
m^2_{\Theta} g.g^\dag +252m^2_{\bar \Sigma} f.f^\dag).g+252
f^\dag.(10 m^2_{H}h.h^\dag+120 m^2_{\Theta} g.g^\dag +252m^2_{\bar
\Sigma} f.f^\dag).f\nonumber\\&&+10 \text Tr[h^\dag. h](84
m_{\Phi}^2|\kappa|^2+m_{\Theta}^2(84|\kappa|^2+126|\gamma|^2+126|\bar
\gamma|^2)+126(|\gamma|^2 m_{\Sigma}^2+|\bar \gamma|^2 m_{\bar
\Sigma}^2)\nonumber\\&&+Tr[h^\dag.m^2_{\tilde{\Psi}}.h])+120 \text
Tr[g^\dag.g](7 m_{\Theta}^2|\rho|^2+7 m_{H}^2|\kappa|^2+
m_{\Phi}^2(7|\rho|^2+7|\kappa|^2+105|\zeta|^2\nonumber\\&&+105|\bar
\zeta|^2)+105(m_{\Sigma}^2|\zeta|^2+\tilde{m}_{\bar
\Sigma}^2|\bar\zeta|^2)+Tr[g^\dag.m^2_{\tilde{\Psi}}.g])+252 \text
Tr[f^\dag.f](10 m_{H}^2|\bar\gamma|^2\nonumber\\&&+100
m_{\Theta}^2|\bar \zeta|^2+ m_{\Phi}^2(200|\eta|^2+100|\bar
\zeta|^2+10|\bar \gamma|^2)+200
m_{\Sigma}^2|\eta|^2+Tr[f^\dag.m^2_{\tilde{\Psi}}.f]) \eea

 Two-loop beta functions for the soft
parameters:

\bea  \beta_{\tilde{\eta}}^{(2)}&=&
-\tilde{\eta}(\bar{\gamma}_\Sigma^{(2)}+\bar{\gamma}_{\bar\Sigma
 }^{(2)}+\bar{\gamma}_\Phi^{(2)})-\eta
(\tilde{\gamma}_\Sigma^{(2)}+\tilde{\gamma}_{\bar\Sigma}^{(2)}+\tilde{\gamma}_\Phi^{(2)}
)-\eta (\hat{\gamma}_\Sigma^{(2)}+\hat{\gamma}_{\bar\Sigma}^{(2)}
+\hat{\gamma}_\Phi^{(2)} )\nonumber\\&& +2 g^2_{10}((
\tilde{\eta}-2M \eta)(7920|\eta|^2+70  |\gamma|^2
+1660|\zeta|^2+70  |\bar\gamma|^2 +1660|\bar\zeta|^2\nonumber\\&&
+ 12|\kappa|^2 +2160|\lambda|^2+18|\rho|^2-40 \text
Tr[f^\dag.f])+2\eta( 7920\tilde{\eta}\eta^*+70  \tilde{\gamma}
\gamma^* \nonumber\\&&+1660\tilde{\zeta} \zeta^*+70
\tilde{\bar\gamma} \bar\gamma^* +1660\tilde{\bar{\zeta}}
\bar{\zeta}^* + 12 \tilde{\kappa} \kappa^* +2160 \tilde{\lambda}
\lambda^* +18 \tilde\rho \rho^*-40 \text Tr[f^\dag.\tilde f]  )
)\nonumber\\&&+5982g^4_{10}(2 \tilde{\eta} -8 M \eta )\eea

\bea   \beta_{\tilde{\gamma}}^{(2)}&=&
-\tilde{\gamma}(\bar{\gamma}_H^{(2)}+\bar{\gamma}_{\Sigma
 }^{(2)}+\bar{\gamma}_\Phi^{(2)})-\gamma
(\tilde{\gamma}_H^{(2)}+\tilde{\gamma}_{\Sigma}^{(2)}+\tilde{\gamma}_\Phi^{(2)}
)-\gamma (\hat{\gamma}_H^{(2)}+\hat{\gamma}_{\Sigma}^{(2)}
+\hat{\gamma}_\Phi^{(2)} )\nonumber\\&& + 2g^2_{10}((
\tilde{\gamma}-2M \gamma)(
1524|\kappa|^2+2590|\gamma|^2+2550|\bar{\gamma}|^2 +54\text
Tr[h^{\dag}.h] \nonumber\\&&+5520|\eta|^2 +1660|\zeta|^2
+2160|\lambda|^2+18|\rho|^2 +660
|\bar{\zeta}|^2)\nonumber\\&&+2\gamma( 1524\tilde{\kappa} \kappa^*
+2590\tilde{\gamma} \gamma^*+2550\tilde{\bar\gamma} \bar\gamma^*
+54\text Tr[h^{\dag}.\tilde h] +
5520\tilde{\eta}\eta^*\nonumber\\&&+1660\tilde{\zeta} \zeta^*
+2160 \tilde{\lambda} \lambda^* +18 \tilde\rho \rho^* +660
\tilde{\bar{\zeta}} \bar{\zeta}^*) )+4614g^4_{10}(2 \tilde{\gamma}
-8 M \gamma )\eea

\bea  \beta_{\tilde{\bar{\gamma}}}^{(2)}&=&
-\tilde{\bar\gamma}(\bar{\gamma}_H^{(2)}+\bar{\gamma}_{\bar\Sigma
 }^{(2)}+\bar{\gamma}_\Phi^{(2)})-\bar\gamma
(\tilde{\gamma}_H^{(2)}+\tilde{\gamma}_{\bar\Sigma}^{(2)}+\tilde{\gamma}_\Phi^{(2)}
)-\bar\gamma (\hat{\gamma}_H^{(2)}+\hat{\gamma}_{\bar\Sigma}^{(2)}
+\hat{\gamma}_\Phi^{(2)} )\nonumber\\&& +2 g^2_{10}((
\tilde{\bar\gamma}-2M \bar\gamma)(
1524|\kappa|^2+2550|\gamma|^2+2590|\bar{\gamma}|^2 +54\text
Tr[h^\dag.h] \nonumber\\&&-40\text Tr[f^\dag.f]+1660|\bar\zeta|^2
+5520|\eta|^2 +2160|\lambda|^2+18|\rho|^2+660
|\zeta|^2)\nonumber\\&& +2\bar\gamma( 1524\tilde{\kappa}
\kappa^*+2550\tilde{\gamma} \gamma^*+2590\tilde{\bar\gamma}
\bar\gamma^* + 54\text Tr[h^\dag.\tilde h]-40\text
Tr[f^\dag.\tilde f] \nonumber\\&&+ 1660\tilde{\bar{\zeta}}
\bar{\zeta}^* +5520\tilde{\eta}\eta^* +2160 \tilde{\lambda}
\lambda^* +18 \tilde\rho \rho^* +660 \tilde{\zeta} \zeta^* )
)\nonumber\\&&+4614g^4_{10}(2 \tilde{\bar\gamma} -8 M \bar\gamma
)\eea

\bea  \beta_{\tilde{\kappa}}^{(2)}&=&
-\tilde{\kappa}(\bar{\gamma}_H^{(2)}+\bar{\gamma}_{\Theta
 }^{(2)}+\bar{\gamma}_\Phi^{(2)})-\kappa
(\tilde{\gamma}_H^{(2)}+\tilde{\gamma}_{\Theta}^{(2)}+\tilde{\gamma}_\Phi^{(2)}
)-\kappa (\hat{\gamma}_H^{(2)}+\hat{\gamma}_{\Theta}^{(2)}
+\hat{\gamma}_\Phi^{(2)} )\nonumber\\&& +2 g^2_{10}((
\tilde{\kappa}-2M \kappa)(
1566|\kappa|^2+2550|\gamma|^2+2550|\bar{\gamma}|^2 +54\text
Tr[h^{\dag}.h]
\nonumber\\&&+102|\rho|^2+2130|\zeta|^2+2130|\bar\zeta|^2+6
Tr[g^{\dag}.g] +3120|\eta|^2+2160|\lambda|^2)
\nonumber\\&&+2\kappa( 1566\tilde{\kappa}
\kappa^*+2550\tilde{\gamma} \gamma^*+2550\tilde{\bar\gamma}
\bar\gamma^* +54\text Tr[h^{\dag}.\tilde h]+102 \tilde\rho
\rho^*\nonumber\\&&+2130\tilde{\zeta}
\zeta^*+2130\tilde{\bar{\zeta}} \bar{\zeta}^*+6 Tr[g^{\dag}.\tilde
g]  +3120\tilde{\eta}\eta^* +2160 \tilde{\lambda} \lambda^*))
\nonumber\\&&+4248g^4_{10}(2 \tilde{\kappa} -8 M \kappa )\eea

\bea  \beta_{\tilde{\zeta}}^{(2)}&=&-
\tilde{\zeta}(\bar{\gamma}_\Theta^{(2)}+\bar{\gamma}_{\Sigma
 }^{(2)}+\bar{\gamma}_\Phi^{(2)})-\zeta
(\tilde{\gamma}_\Theta^{(2)}+\tilde{\gamma}_{\Sigma}^{(2)}+\tilde{\gamma}_\Phi^{(2)}
)-\zeta (\hat{\gamma}_\Theta^{(2)}+\hat{\gamma}_{\Sigma}^{(2)}
+\hat{\gamma}_\Phi^{(2)} )\nonumber\\&& +2 g^2_{10}((
\tilde{\zeta}-2M
\zeta)(54|\kappa|^2+102|\rho|^2+3130|\zeta|^2+2130|\bar\zeta|^2+6
Tr[g^{\dag}.g]\nonumber\\&&+5520|\eta|^2+70 |\gamma|^2
+2160|\lambda|^2+30|\bar\gamma|^2) +2\zeta( 54\tilde{\kappa}
\kappa^*+102\tilde\rho \rho^*+3130\tilde{\zeta}
\zeta^*\nonumber\\&&+2130\tilde{\bar{\zeta}} \bar{\zeta}^*+6
Tr[g^{\dag}.\tilde g] + 5520\tilde{\eta}\eta^*+70 \tilde{\gamma}
\gamma^* +2160 \tilde{\lambda} \lambda^*+30 \tilde{\bar\gamma}
\bar\gamma^* ))\nonumber\\&&+5616g^4_{10}(2 \tilde{\zeta} -8 M
\zeta )\eea

\bea  \beta_{\tilde{\bar{\zeta}}}^{(2)}&=&-
\tilde{\bar\zeta}(\bar{\gamma}_\Theta^{(2)}+\bar{\gamma}_{\bar\Sigma
 }^{(2)}+\bar{\gamma}_\Phi^{(2)})-\bar\zeta
(\tilde{\gamma}_\Theta^{(2)}+\tilde{\gamma}_{\bar\Sigma}^{(2)}+\tilde{\gamma}_\Phi^{(2)}
)-\bar\zeta
(\hat{\gamma}_\Theta^{(2)}+\hat{\gamma}_{\bar\Sigma}^{(2)}
+\hat{\gamma}_\Phi^{(2)} )\nonumber\\&& +2 g^2_{10}((
\tilde{\bar\zeta}-2M
\bar\zeta)(54|\kappa|^2+102|\rho|^2+2130|\zeta|^2+3130|\bar\zeta|^2+6
\text Tr[g^{\dag}.g]\nonumber\\&&-40 \text Tr[f^\dag.f]
+5520|\eta|^2+70 |\bar\gamma|^2 +2160|\lambda|^2+30|\gamma|^2)
+2\bar\zeta( 54\tilde{\kappa} \kappa^*+102\tilde\rho
\rho^*\nonumber\\&&+2130\tilde{\zeta}
\zeta^*+3130\tilde{\bar{\zeta}} \bar{\zeta}^*+6 \text
Tr[g^{\dag}.\tilde g]-40 \text Tr[f^\dag.\tilde f]+
5520\tilde{\eta}\eta^*+70 \tilde{\bar\gamma} \bar\gamma^*
\nonumber\\&&+2160 \tilde{\lambda} \lambda^* +30 \tilde{\gamma}
\gamma^* ))+5616g^4_{10}(2 \tilde{\bar\zeta} -8 M \bar\zeta )\eea

\bea  \beta_{\tilde{\rho}}^{(2)}&=&-
\tilde{\rho}(2\bar{\gamma}_{\Theta
 }^{(2)}+\bar{\gamma}_\Phi^{(2)})-\rho
(2\tilde{\gamma}_{\Theta}^{(2)}+\tilde{\gamma}_\Phi^{(2)} )-\rho
(2\hat{\gamma}_{\Theta}^{(2)} +\hat{\gamma}_\Phi^{(2)}
)\nonumber\\&& + 2 g^2_{10}(( \tilde{\rho}-2M \rho)(
96|\kappa|^2+186|\rho|^2+3600|\zeta|^2+3600|\bar\zeta|^2+6
Tr[g^{\dag}.g]\nonumber\\&&+3120|\eta|^2+2160|\lambda|^2+30|\gamma|^2+30|\bar\gamma|^2)
+2\rho( 96\tilde{\kappa} \kappa^*+186\tilde\rho
\rho^*\nonumber\\&&+3600\tilde{\zeta}
\zeta^*+3600\tilde{\bar{\zeta}} \bar{\zeta}^*+6 Tr[g^\dag.\tilde
g] +3120\tilde{\eta}\eta^* +2160 \tilde{\lambda} \lambda^* +30
\tilde{\gamma} \gamma^*\nonumber\\&&+30 \tilde{\bar\gamma}
\bar\gamma^* ))+5250g^4_{10}(2 \tilde{\rho} -8 M \rho )\eea

\bea  \beta_{\tilde {h}}^{(2)}&=&- \tilde h \bar\gamma_H^{(2)}- h
\tilde{\gamma}_H^{(2)}-h \hat{\gamma}_H^{(2)}\nonumber\\&&+2
g_{10}^2((\tilde h-2 M
h)(1512|\kappa|^2+2520(|\gamma|^2+|\bar{\gamma}|^2 )+54\text
Tr[h^\dag.h])\nonumber\\&&+2h(1512\tilde\kappa
\kappa^*+2520(\tilde\gamma \gamma^*+\tilde {\bar{\gamma}} \bar
\gamma^*) +54\text Tr[h^\dag.\tilde h]))+657g_{10}^4(2 \tilde h-8
M h)\nonumber\\&&-\tilde h.\bar\gamma_{\Psi}^{(2)}-h.
\tilde\gamma_\Psi^{(2)} -h.\hat\gamma_\Psi^{(2)}+2
g_{10}^2((\tilde h -2 M h )(45 h^\dag.h+1260 g^\dag.g+3150
f^\dag.f)\nonumber\\&&+2h(45 h^\dag.\tilde h+1260 g^\dag.\tilde
g+3150 f^\dag.\tilde f)) +(2\tilde h-8 M h) \frac{26685
g_{10}^4}{32} \nonumber\\&&+(-\tilde
h.\bar\gamma_{\Psi}^{(2)}-h.\tilde\gamma_\Psi^{(2)}
-h.\hat\gamma_\Psi^{(2)}+2 g_{10}^2((\tilde h -2 M h )(45
h^\dag.h+1260 g^\dag.g+3150 f^\dag.f) \nonumber\\&&+2h(45
h^\dag.\tilde h+1260 g^\dag.\tilde g+3150 f^\dag.\tilde
f))+(2\tilde h-8  M h) \frac{26685 g_{10}^4}{32})^T\eea

\bea  \beta_{\tilde {f}}^{(2)}&=& -\tilde f
\bar\gamma_{\bar\Sigma}^{(2)}-f \tilde{\gamma}_{\bar
\Sigma}^{(2)}-f \hat{\gamma}_{\bar
\Sigma}^{(2)}\nonumber\\&&+2g_{10}^2((\tilde f-2 M
f)(2400|\eta|^2+40 |\bar\gamma|^2 +1000|\bar\zeta|^2-40\text
Tr[f^\dag.f])\nonumber\\&&+2f(2400\tilde \eta \eta^*+40
\tilde{\bar\gamma}\bar \gamma^* +1000\tilde{\bar\zeta}\bar
\zeta^*-40\text Tr[f^\dag.\tilde f]))+2025g_{10}^4(2 \tilde f-8 M
f)\nonumber\\&&-\tilde
f.\bar\gamma_{\Psi}^{(2)}-f.\tilde\gamma_\Psi^{(2)}-f.\hat\gamma_\Psi^{(2)}+2
g_{10}^2((\tilde f -2 M f)(45 h^\dag.h+1260 g^\dag.g+3150
f^\dag.f)\nonumber\\&&+2f(45 h^\dag.\tilde h+1260 g^\dag.\tilde
g+3150 f^\dag.\tilde f))+(2\tilde f-8  M f) \frac{26685
g_{10}^4}{32} \nonumber\\&&+(-\tilde
f.\bar\gamma_{\Psi}^{(2)}-f.\tilde\gamma_\Psi^{(2)}-f.\hat\gamma_\Psi^{(2)}+2g_{10}^2((\tilde
f-2 M f)(45 h^\dag.h+1260 g^\dag.g+3150
f^\dag.f)\nonumber\\&&+2f(45 h^\dag.\tilde h+1260 g^\dag.\tilde
g+3150 f^\dag.\tilde f))+(2\tilde f-8  M f) \frac{26685
g_{10}^4}{32})^T\eea

\bea  \beta_{\tilde {g}}^{(2)}&=& -\tilde g
\bar\gamma_{\Theta}^{(2)}-g \tilde{\gamma}_{\Theta}^{(2)}-g
\hat{\gamma}_{\Theta}^{(2)}+2g_{10}^2((\tilde g-2 M
g)(42|\kappa|^2+84|\rho|^2+1470|\zeta|^2+1470|\bar\zeta|^2\nonumber\\&&+6
Tr[g^\dag.g])+2g(42\tilde \kappa \kappa^*+84\tilde\rho
\rho^*+1470\tilde\zeta \zeta^*+1470\tilde{\bar\zeta}\bar\zeta^*+6
Tr[g^\dag.\tilde g]))\nonumber\\&&+1659g_{10}^4(2 \tilde g-8 M
g)\nonumber\\&&-\tilde
g.\bar\gamma_{\Psi}^{(2)}-g.\tilde\gamma_\Psi^{(2)}-g.\hat\gamma_\Psi^{(2)}+2g_{10}^2((\tilde
g -2 M g)(45 h^\dag.h+1260 g^\dag.g+3150
f^\dag.f)\nonumber\\&&+2g(45 h^\dag.\tilde h+1260 g^\dag.\tilde
g+3150 f^\dag.\tilde f))+(2\tilde g-8  M g) \frac{26685
g_{10}^4}{32} \nonumber\\&&+(-\tilde
g.\bar\gamma_{\Psi}^{(2)}-g.\tilde\gamma_\Psi^{(2)}-g.\hat\gamma_\Psi^{(2)}+2g_{10}^2((\tilde
g -2 M g)(45 h^\dag.h+1260 g^\dag.g+3150
f^\dag.f)\nonumber\\&&+2g(45 h^\dag.\tilde h+1260 g^\dag.\tilde
g+3150 f^\dag.\tilde f))+(2\tilde g-8  M g) \frac{26685
g_{10}^4}{32})^T\eea

\bea  \beta_{b_{\Phi}}^{(2)}&=& - 2 b_{\Phi} \bar{\gamma}
_\Phi^{(2)}-2\mu_{\Phi}(\hat{\gamma}_\Phi^{(2)}+\tilde{\gamma}_\Phi^{(2)})+4
g^2_{10}((b_{\Phi}-2 \mu_{\Phi}M)(3120|\eta|^2+ 12|\kappa|^2
\nonumber\\&&+2160|\lambda|^2+18|\rho|^2+30|\gamma|^2+30|\bar\gamma|^2+660
|\zeta|^2+660 |\bar{\zeta}|^2)\nonumber\\&&+4\mu_{\Phi}(3120
\tilde\eta \eta^*+ 12 \tilde\kappa \kappa^* +2160 \tilde\lambda
\lambda^* +18 \tilde\rho \rho^*+30 \tilde\gamma \gamma^*+30
\tilde{\bar\gamma} \bar{\gamma}^*\nonumber\\&&+660 \tilde\zeta
\zeta^*+660 \tilde{\bar{\zeta}}\bar{\zeta}^*))+3864g^4_{10}(2
b_{\Phi}-8\mu_{\Phi}M)\eea

\bea  \beta_{b_H}^{(2)}&=&  - 2 b_H \bar{\gamma}
_H^{(2)}-2\mu_H(\hat{\gamma}_H^{(2)}+\tilde{\gamma}_H^{(2)})
+4g^2_{10}((b_{H}-2\mu_HM)(1512|\kappa|^2+2520|\gamma|^2\nonumber\\&&+2520|\bar{\gamma}|^2
+54\text Tr[h^{\dag}.h])+4\mu_H(1512 \tilde\kappa \kappa^*+2520
\tilde \gamma
\gamma^*+2520\tilde{\bar{\gamma}}\bar\gamma^*\nonumber\\&&
+54\text Tr[h^{\dag}.\tilde{h}]))+1314g^4_{10}(2 b_H-8\mu_H M)\eea

\bea  \beta_{b_{\Sigma}}^{(2)}&=& -b_{\Sigma}(\bar{\gamma}
_\Sigma^{(2)}+\bar{\gamma}
_{\bar{\Sigma}}^{(2)})-\mu_{\Sigma}(\hat{\gamma}_\Sigma^{(2)}+\tilde{\gamma}_\Sigma^{(2)}+
\hat{\gamma}_{\bar{\Sigma}}^{(2)}+\tilde{\gamma}_{\bar{\Sigma}}^{(2)})
+2g^2_{10}((b_{\Sigma}-2\mu_{\Sigma}M)\nonumber\\&&
(4800|\eta|^2+40 |\gamma|^2 +1000|\zeta|^2+40  |\bar\gamma|^2
+1000|\bar\zeta|^2-40\text
Tr[f^{\dag}.f])\nonumber\\&&+2\mu_{\Sigma}(4800 \tilde\eta
\eta^*+40 \tilde\gamma\gamma^* +1000 \tilde\zeta \zeta^*+40
\tilde{\bar\gamma}\bar\gamma^* +1000\tilde{\bar\zeta}
\bar\zeta^*-40\text Tr[f^{\dag}.\tilde
f]))\nonumber\\&&+8100g^4_{10}(2b_\Sigma-8\mu_\Sigma M)\eea

\bea  \beta_{b_\Theta}^{(2)}&=& - 2 b_\Theta \bar{\gamma}
_\Theta^{(2)}-2\mu_\Theta(\hat{\gamma}_\Theta^{(2)}+\tilde{\gamma}_\Theta^{(2)})
+4g^2_{10}((b_{\Theta}-2\mu_{\Theta}M)(42|\kappa|^2+84|\rho|^2\nonumber\\&&+1470|\zeta|^2+1470|\bar\zeta|^2+6
Tr[g^{\dag}.g])+4\mu_\theta(42 \tilde\kappa \kappa^*+84 \tilde
\rho \rho^*+1470\tilde{\bar{\zeta}}\bar\zeta^*
\nonumber\\&&+1470\tilde\zeta \zeta^*+6\text Tr[g^{\dag}.\tilde
g])+3318g^4_{10}(2b_{\Theta}-8 \mu_\Theta M))\eea

\bea \beta_{m^2_{\Phi}}^{(2)}&=&-2 \bar{\gamma}_{\Phi}^{(2)}
m^2_{\Phi}-4(360m^2_{\Phi}\bar\gamma_{\Phi}^{(1)}|\lambda|^2
+m^2_{H}\bar\gamma_{H}^{(1)}(4|\kappa|^2+6(|\gamma|^2+|\bar\gamma|^2))
\nonumber\\&&+6m^2_{\Theta}\bar\gamma_{\Theta}^{(1)}(4|\rho|^2+4|\kappa|^2+60(|\zeta|^2+|\bar\zeta|^2))
+m^2_{\Sigma}\bar\gamma_{\Sigma}^{(1)}
(6|\gamma|^2+240|\eta|^2+60|\zeta|^2)\nonumber\\&&+m^2_{\bar{\Sigma}}\bar\gamma_{\bar
\Sigma}^{(1)}(6|\bar{\gamma}|^2+240|\eta|^2+60|\bar\zeta|^2))
-2(\bar\gamma_{\bar\Sigma}^{(1)}(240|\eta|^2m^2_{\Sigma}
+6|\bar\gamma|^2 m^2_{H}+60|\bar\zeta|^2m^2_{\Theta}
)\nonumber\\&&+\bar\gamma_{\Sigma}^{(1)}(240|\eta|^2m^2_{\bar\Sigma}
+6|\gamma|^2 m^2_{H}+60|\zeta|^2m^2_{\Theta} )
+\bar\gamma_{H}^{(1)}(4|\kappa|^2m^2_{\Theta}
+6|\gamma|^2m^2_{\Sigma}+6|\bar\gamma|^2m^2_{\bar\Sigma}
)\nonumber\\&&+ \bar\gamma_{\Theta}^{(1)}(4|\kappa|^2m^2_{H}
+2|\rho|^2m^2_{\Theta}+60|\zeta|^2m^2_{\Sigma}+60|\bar\zeta|^2m^2_{\bar\Sigma}
)+360m^2_{\Phi}|\lambda|^2\bar\gamma_{\Phi}^{(1)})\nonumber\\&& -
2\breve{\gamma}_\Phi^{(2)}    -
\check{\gamma}_{\Phi}^{(2)}-\acute{\gamma}_{\Phi}^{(2)}-2 \text
Re[\grave{\gamma}_{\Phi}^{(2)}]\nonumber\\&&+2 g_{10}^2(2
m_{\Phi}^{2}(3120|\eta|^2+ 12|\kappa|^2
+2160|\lambda|^2+18|\rho|^2+30|\gamma|^2+30|\bar\gamma|^2+660
|\zeta|^2\nonumber\\&&+660 |\bar{\zeta}|^2)+
2(3120|\tilde{\eta}|^2+ 12|\tilde{\kappa}|^2
+2160|\tilde{\lambda}|^2+18|\tilde{\rho}|^2+30|\tilde{\gamma}|^2+30|\tilde{\bar\gamma}|^2\nonumber\\&&+660
|\tilde{\zeta}|^2+660 |\tilde{\bar{\zeta}}|^2)-2(3120\tilde\eta
\eta^*+ 12\tilde\kappa \kappa^* +2160\tilde\lambda
\lambda^*+18\tilde\rho \rho^*+30 \tilde\gamma \gamma^*+30
\tilde{\bar\gamma} \bar \gamma^*\nonumber\\&&+660 \tilde\zeta
\zeta^*+660 \tilde{\bar{\zeta}} \bar \zeta^*)
M^\dag-2(3120\tilde\eta^* \eta+ 12\tilde\kappa^* \kappa
+2160\tilde\lambda^* \lambda+18\tilde\rho^* \rho+30 \tilde\gamma^*
\gamma \nonumber\\&&+30 \tilde{\bar\gamma}^* \bar \gamma +660
\tilde\zeta^* \zeta+660 \tilde{\bar{\zeta}}^* \bar \zeta)
M+4(3120|\eta|^2+ 12|\kappa|^2
+2160|\lambda|^2+18|\rho|^2\nonumber\\&&+30|\gamma|^2+30|\bar\gamma|^2+660
|\zeta|^2+660 |\bar{\zeta}|^2) M
{M}^\dag+4(2160m^2_{\Phi}|\lambda|^2
\nonumber\\&&+m^2_{H}(12|\kappa|^2+30|\gamma|^2+30|\bar{\gamma}|^2)
+m^2_{\Theta}(18|\rho|^2+12|\kappa|^2+660(|\zeta|^2+|\bar\zeta|^2))\nonumber\\&&+m^2_{\Sigma}
(30|\gamma|^2+3120|\eta|^2+660|\zeta|^2)+m^2_{\bar{\Sigma}}(30|\bar{\gamma}|^2+3120|\eta|^2
+660|\bar{\zeta}|^2))\nonumber\\&&+46368 g_{10}^4|
M|^2+96g_{10}^4(56 m_{\Phi}^2 -8|M|^2)\eea

\bea \beta_{m^2_{H}}^{(2)}&=&-2 \bar{\gamma}_{H}^{(2)}
m^2_{H}-4(m^2_{\Phi}
\bar\gamma_{\Phi}^{(1)}(126(|\gamma|^2+|\bar{\gamma}|^2)+84|\kappa|^2)
+84m^2_{\Theta}\bar\gamma_{\Theta}^{(1)}|\kappa|^2+126(m^2_{\Sigma}\bar\gamma_{\Sigma}^{(1)}|\gamma|^2
\nonumber\\&&+m^2_{\bar{\Sigma}}\bar\gamma_{\bar
\Sigma}^{(1)}|\bar{\gamma}|^2)) - 2 \text
Tr[h.\bar\gamma_{\Psi}^{(1)}.\tilde{m}_{\Psi}^2 .h^\dag] - 2 \text
Tr[h.\tilde{m}_{\Psi}^2.\bar\gamma_{\Psi}^{(1)}.
h^\dag]\nonumber\\&&-2(
m^2_{\Phi}(126(|\gamma|^2\bar\gamma_{\Sigma}^{(1)}
+|\bar{\gamma}|^2\bar\gamma_{\bar\Sigma}^{(1)})+84|\kappa|^2\bar\gamma_{\Theta}^{(1)}
)\nonumber\\&&+\bar\gamma_{\Phi}^{(1)}(126(|\gamma|^2
m^2_{\Sigma}+ 126(|\bar\gamma|^2 m^2_{\bar\Sigma}+84|\kappa|^2
m^2_{\Theta} )+ \text Tr[ \tilde{m}_{\Psi}^2.h^\dag.
\bar\gamma_{\Psi}^{(1)}. h
 ] ) - 2\breve{\gamma}_H^{(2)}-
\check{\gamma}_{H}^{(2)}\nonumber\\&&-\acute{\gamma}_{H}^{(2)}-\text
2 Re[\grave{\gamma}_{H}^{(2)}] +2g_{10}^2(2
m_{H}^{2}(1512|\kappa|^2+2520|\gamma|^2+2520|\bar{\gamma}|^2
+54\text Tr[h^{\dag}.h])\nonumber\\&&+2
(1512|\tilde{\kappa}|^2+2520|\tilde{\gamma}|^2+2520|\tilde{\bar{\gamma}}|^2
+54\text Tr[h^{\dag}.h]) -2 (1512\tilde \kappa \kappa^*+2520\tilde
\gamma \gamma^*\nonumber\\&&+2520\tilde{\bar{\gamma}}\bar \gamma^*
+54\text Tr[\tilde h.h^{\dag}]) M^\dag-2(1512\tilde \kappa^*
\kappa+2520\tilde \gamma^* \gamma+2520\tilde{\bar{\gamma}}^* \bar
\gamma +54\text Tr[\tilde h^\dag.h])
M\nonumber\\&&+4(1512|\kappa|^2+2520|\gamma|^2+2520|\bar{\gamma}|^2
+54\text Tr[h.h^{\dag}]) M
M^\dag+4(m^2_{\Phi}(5040(|\gamma|^2+|\bar{\gamma}|^2)\nonumber\\&&+3024|\kappa|^2)
+3024m^2_{\Theta}|\kappa|^2+5040(m^2_{\Sigma}|\gamma|^2+m^2_{\bar{\Sigma}}|\bar{\gamma}|^2)
+108 \text Tr[h^\dag.h.\tilde{m}_{\Psi}^2]))\nonumber\\&&+15768
g_{10}^4| M|^2+36g_{10}^4( m_{H}^2   -8| M|^2)\eea

\bea \beta_{m^2_{\Theta}}^{(2)}&=&-2 \bar{\gamma}_{\Theta}^{(2)}
m^2_{\Theta}-4(m^2_{\Phi}
\bar\gamma_{\Phi}^{(1)}(105(|\zeta|^2+|\bar{\zeta}|^2)+7(|\kappa|^2+|\rho|^2))
+7m^2_{H}\bar\gamma_{H}^{(1)}|\kappa|^2
\nonumber\\&&+7m^2_{\Theta}\bar\gamma_{\Theta}^{(1)}|\rho|^2
+105(m^2_{\Sigma}\bar\gamma_{\Sigma}^{(1)}|\zeta|^2
+m^2_{\bar{\Sigma}}\bar\gamma_{\bar \Sigma}^{(1)}|\bar{\zeta}|^2))
\nonumber\\&&-2 \text
Tr[g.\bar\gamma_{\Psi}^{(1)}.\tilde{m}_{\Psi}^2 .g^\dag] - 2 \text
Tr[g.\tilde{m}_{\Psi}^2.\bar\gamma_{\Psi}^{(1)}. g^\dag]-
2(\bar\gamma_{\Phi}^{(1)}(
7|\kappa|^2m^2_{H}+7|\Theta|^2m^2_{\Theta}\nonumber\\&&
+105|\zeta|^2m^2_{\Sigma}+105|\bar{\zeta}|^2m^2_{\bar\Sigma}
)+7|\kappa|^2m^2_{\Phi}\bar\gamma_{H}^{(1)}+7|\rho|^2m^2_{\Phi}\bar\gamma_{\Theta}^{(1)}
+105|\zeta|^2m^2_{\Phi}\bar\gamma_{\Sigma}^{(1)}\nonumber\\&&+105|\bar\zeta|^2m^2_{\Phi}\bar\gamma_{\bar\Sigma}^{(1)}
+\text Tr[\tilde{m}_{\Psi}^2.g^\dag.\bar\gamma_{\Psi}^{(1)} .g]) -
2\breve{\gamma}_{\Theta}^{(2)} -
\check{\gamma}_{\Theta}^{(2)}-\acute{\gamma}_{\Theta}^{(2)}-\text
2 Re[\grave{\gamma}_{\Theta}^{(2)}] \nonumber\\&&+2g_{10}^2(2
\tilde{m}_{\Theta}^{2}(42|\kappa|^2+84|\rho|^2+1470|\zeta|^2+1470|\bar\zeta|^2+6
Tr[g^{\dag}.g])\nonumber\\&&+
2(42|\tilde{\kappa}|^2+84|\tilde{\rho}|^2+1470|\tilde{\zeta}|^2+1470|\tilde{\bar\zeta}|^2+6
Tr[g^{\dag}.g])\nonumber\\&&-2(42\tilde \kappa
\kappa^*+84\tilde\rho \rho^*+1470\tilde \zeta
\zeta^*+1470\tilde{\bar{\zeta}}\bar \zeta^* +6Tr[\tilde
g.g^{\dag}]) M^\dag\nonumber\\&&-2(42\tilde \kappa^*
\kappa+84\tilde\rho^* \rho+1470\tilde \zeta^*
\zeta+1470\tilde{\bar{\zeta}}^*\bar \zeta +6Tr[\tilde g^{\dag}.g])
M\nonumber\\&&+4(42|\kappa|^2+84|\rho|^2+1470|\zeta|^2+1470|\bar\zeta|^2+6
Tr[g.g^{\dag}]) M
M^\dag\nonumber\\&&+4(m^2_{\Phi}(1470(|\zeta|^2+|\bar{\zeta}|^2)+42|\kappa|^2+84|\rho|^2)
+42m^2_{H}|\kappa|^2
+84m^2_{\Theta}|\rho|^2\nonumber\\&&+1470(m^2_{\Sigma}|\zeta|^2
+m^2_{\bar{\Sigma}}|\bar{\zeta}|^2) +6 \text Tr[g^\dag
g.\tilde{m}_{\Psi}^2])+39816 g_{10}^4|
M|^2\nonumber\\&&+84g_{10}^4(28  m_{\Theta}^2 -8| M|^2)\eea

\bea \beta_{m^2_{\Sigma}}^{(2)}&=&-2 \bar{\gamma}_{\Sigma}^{(2)}
m^2_{\Sigma}-4(m^2_{\Phi}
\bar\gamma_{\Phi}^{(1)}(200|\eta|^2+100|\zeta|^2+10|\gamma|^2)
+10m^2_{H}\bar\gamma_{H}^{(1)}|\gamma|^2
\nonumber\\&&+100m^2_{\Theta}\bar\gamma_{\Theta}^{(1)}|\zeta|^2
+200m^2_{\bar{\Sigma}}\bar\gamma_{\bar \Sigma}^{(1)}|\eta|^2)-
2(\bar\gamma_{\Phi}^{(1)}(10|\gamma|^2m^2_{H}+100|\zeta|^2m^2_{\Theta})
\nonumber\\&&+\bar\gamma_{\Theta}^{(1)}(200|\eta|^2m^2_{\bar\Sigma}+
100|\zeta|^2m^2_{\Phi}
)+200|\eta|^2m^2_{\Theta}\bar\gamma_{\bar\Sigma}^{(1)}
+10|\gamma|^2m^2_{\Phi}\bar\gamma_{H}^{(1)})\nonumber\\&& -
2\breve{\gamma}_{\Sigma}^{(2)}  -
\check{\gamma}_{\Sigma}^{(2)}-\acute{\gamma}_{\Sigma}^{(2)}-2
\text Re[\grave{\gamma}_{\Sigma}^{(2)}] \nonumber\\&&+2 g_{10}^2(2
m_{\Sigma}^{2}(2400|\eta|^2+40  |\gamma|^2 +1000|\zeta|^2)+
2(2400|\tilde{\eta}|^2+40 |\tilde{\gamma}|^2
+1000|\tilde{\zeta}|^2)\nonumber\\&& -2 (2400\tilde\eta \eta^*+40
\tilde\gamma \gamma^* +1000 \tilde\zeta \zeta^*)
M^\dag-2(2400\tilde\eta^* \eta+40 \tilde\gamma^* \gamma +1000
\tilde\zeta^* \zeta) M\nonumber\\&&+4(2400|\eta|^2+40 |\gamma|^2
+1000|\zeta|^2) M
M^\dag+4(m^2_{\Phi}(2400|\eta|^2+1000|\zeta|^2+40|\gamma|^2)
\nonumber\\&&+40m^2_{H}\bar\gamma_{H}^{(1)}|\kappa|^2
+1000m^2_{\Theta}\bar\gamma_{\Theta}^{(1)}|\zeta|^2
+2400m^2_{\bar{\Sigma}}\bar\gamma_{\bar
\Sigma}^{(1)}|\eta|^2)+48600 g_{10}^4|
M|^2\nonumber\\&&+100g_{10}^4(35  m_{\Sigma}^2   -8| M|^2)\eea

\bea \beta_{m^2_{\bar{\Sigma}}}^{(2)}&=&-2
\bar{\gamma}_{\bar\Sigma}^{(2)} m^2_{\bar\Sigma}-4(m^2_{\Phi}
\bar\gamma_{\Phi}^{(1)}(200|\eta|^2+100|\bar\zeta|^2+10|\bar\gamma|^2
) +10m^2_{H}\bar\gamma_{H}^{(1)}|\bar\gamma|^2
\nonumber\\&&+100m^2_{\Theta}\bar\gamma_{\Theta}^{(1)}|\bar\zeta|^2
+200m^2_{\Sigma}\bar\gamma_{ \Sigma}^{(1)}|\eta|^2)- 2 \text
Tr[f.\bar\gamma_{\Psi}^{(1)}.\tilde{m}_{\Psi}^2 .f^\dag]- 2 \text
Tr[f.\tilde{m}_{\Psi}^2.\bar\gamma_{\Psi}^{(1)}.
f^\dag]\nonumber\\&&- 2(
\bar\gamma_{\Phi}^{(1)}(10|\bar\gamma|^2m^2_{H}+100|\bar\zeta|^2m^2_{\Theta})
+\bar\gamma_{\Theta}^{(1)}(200|\eta|^2m^2_{\Sigma}+
100|\bar\zeta|^2m^2_{\Phi}
])+200|\eta|^2m^2_{\Theta}\bar\gamma_{\Sigma}^{(1)}\nonumber\\&&+10|\bar\gamma|^2m^2_{\Phi}\bar\gamma_{H}^{(1)}
+\text Tr[\tilde{m}_{\Psi}^2.f^\dag.\bar\gamma_{\Psi}^{(1)}.f ]) -
2\breve{\gamma}_{\bar\Sigma}^{(2)}-
\check{\gamma}_{\bar\Sigma}^{(2)}-\acute{\bar\gamma}_{\bar\Sigma}^{(2)}-2
\text Re[\grave{\gamma}_{\bar\Sigma}^{(2)}]\nonumber\\&&
+2g_{10}^2(2 \tilde{m}_{\bar\Sigma}^{2}(2400|\eta|^2+40
|\bar\gamma|^2+1000|\bar\zeta|^2) +2(2400|\tilde{\eta}|^2+40
|\tilde{\bar\gamma}|^2+1000|\tilde{\bar\zeta}|^2)\nonumber\\&&-2
(2400\tilde\eta \eta^*+40  \tilde {\bar\gamma} \bar\gamma^* +1000
\tilde{\bar\zeta} \bar\zeta^*)M^\dag -2(2400\tilde\eta^* \eta+40
\tilde{\bar\gamma}^* \bar\gamma +1000 \tilde{\bar\zeta}^*
\bar\zeta) M\nonumber\\&&+4(2400|\eta|^2+40 |\bar\gamma|^2
+1000|\bar\zeta|^2) M
M^\dag+4(m^2_{\Phi}(2400|\eta|^2+1000|\bar\zeta|^2+40|\bar\gamma|^2)
\nonumber\\&&+40m^2_{H}|\bar\gamma|^2
+1000m^2_{\Theta}|\bar\zeta|^2 +2400m^2_{\Sigma}\bar\gamma_{
\Sigma}^{(1)}|\eta|^2-40 \text
Tr[f^\dag.f.\tilde{m}_{\Psi}^2])\nonumber\\&&+48600 g_{10}^4|M|^2
+100g_{10}^4(35  m_{\bar\Sigma}^2 -8| M|^2)\eea

\bea \beta_{m^2_{\tilde{\Psi}}}^{(2)}&=&
-\tilde{m}^2_{\Psi}.\bar{\gamma}_{\Psi}^{(2)}-\bar{\gamma}_{\Psi}^{(2)T}.m^2_{\tilde{\Psi}}-4(20
m_{H}^2 \bar\gamma_{H}^{(1)} h^\dag.h+240 m_{\Theta}^2
\bar\gamma_{\Theta}^{(1)} g^\dag.g+504\tilde{m}_{\bar\Sigma}^2
\bar\gamma_{\bar \Sigma}^{(1)} f^\dag.f\nonumber\\&&+10 h^\dag
.m^2_{\tilde{\Psi}}.\bar\gamma_{\Psi}^{(1)}.h +120  g^\dag
.m^2_{\tilde{\Psi}}.\bar\gamma_{\Psi}^{(1)}.g+ 252 f^\dag
.m^2_{\tilde{\Psi}}.\bar\gamma_{\Psi}^{(1)}.f+10 h^
\dag.\bar\gamma_{\Psi}^{(1)T}.m^2_{\tilde{\Psi}}.h
\nonumber\\&&+120
g^\dag.\bar\gamma_{\Psi}^{(1)T}.m^2_{\tilde{\Psi}}.g+ 252f^ \dag
\bar\gamma_{\Psi}^{(1)T}.m^2_{\tilde{\Psi}}.f) -2(10m^2_{H} h^
\dag
 \bar\gamma_{\Psi}^{(1)T}.h +120 m^2_{\Theta} g^ \dag
 \bar\gamma_{\Psi}^{(1)T}.g\nonumber\\&&+ 252 m^2_{\bar \Sigma}f^ \dag
\bar\gamma_{\Psi}^{(1)T} f\nonumber\\&&+
m^2_{\tilde{\Psi}}(10\bar\gamma_{H}^{(1)} h^\dag
h+120\bar\gamma_{\Theta}^{(1)} g^\dag g+252\bar\gamma_{\bar
\Sigma}^{(1)} f^\dag f))- 2\breve{\gamma}_{\Psi}^{(2)}-
\check\gamma_\Psi^{(2)}-\acute{\gamma}_{\Psi}^{(2)}-\grave{\gamma}_{\Psi}^{(2)}
-\grave{\gamma}_{\Psi}^{(2)\dag}\nonumber\\&&+2g_{10}^2(\tilde{m}_{\Psi}^2(45
h^\dag.h + 1260 g^\dag.g +3150 f^\dag.f)+(45 h^\dag.h + 1260
g^\dag.g +3150 f^\dag.f)\tilde{m}_{\Psi}^{2T}\nonumber\\&&+4(45
m_{H}^2 h^\dag h+1260m^2_{\Theta} g^\dag g +3150m^2_{\bar \Sigma}
f^\dag f+m^2_{\tilde{\Psi}}(45 h^\dag.h+1260 g^\dag.g +3150
f^\dag.f))\nonumber\\&&+2(45 \tilde h^\dag.\tilde h + 1260\tilde
g^\dag .\tilde g +3150 \tilde f^\dag.\tilde f)-2(45 \tilde
h^\dag.h + 1260\tilde g^\dag.g +3150 \tilde f^\dag. f)
M\nonumber\\&&+2( 45h^\dag.\tilde h + 1260 g^\dag.\tilde g +3150
f^\dag.\tilde f) M^\dag+4( 45 h^\dag.h + 1260 g^\dag.g +3150
f^\dag.f) M  M^\dag)\nonumber\\&&+\frac{80055}{4}g_{10}^4 M
M^\dag+45 g_{10}^4(2 m_{\tilde \Psi}^2-8 M M^\dag)\eea

%\newpage\thispagestyle{empty} \mbox{}\newpage
\chapter{Summary}
The new minimal supersymmetric SO(10) GUT (NMSGUT) based on Higgs
system 10+120+210+126+$\overline{126}$, 16 dimensional matter
fermion representation and 45 dimensional gauge field
representation is a promising candidate theory of particle physics
beyond standard model. The structural features of NMSGUT e.g.
symmetry breaking scheme of GUT (in terms of a single complex
parameter), calculable GUT scale spectra, GUT scale threshold
effects etc. make this theory very appealing. The successful
embedding of Seesaw mechanism to explain neutrino masses
compatible with the neutrino oscillations, t-b-$\tau$ Yukawa
unification in the large tan$\beta$ regime, facilitation of LSP
WIMP as a dark matter candidate due to preservation of R-parity at
weak scale etc are other features of NMSGUT.

 Present work is
basically a step towards the promotion of this structurally rich
theory. This dissertation comprises the study of various problems
like effect of GUT scale threshold corrections on SO(10) Yukawa
and consequently suppression of d=5, $\Delta$ B$\neq$0 operator,
Supersymmetric seesaw inflection point inflation (SSI), reheating
in context of SSI, embedding of SSI in NMSGUT and BICEP2
revolution, relic density calculation using DarkSusy for NMSGUT
generated
 soft spectra and finally the NMSGUT RGEs for the evolution between $M_{Planck}$ and $M_{GUT}$.

 In Chapter 2, we reviewed the NMSGUT, its complete superpotential, Higgs system, symmetry breaking
 scheme, GUT scale spectrum, importance of GUT scale threshold corrections to $\alpha_3,\alpha_s,M_{X}$
 and d=5, $\Delta$ B$\neq$0 operator responsible for proton decay. But with these operators the resulting
 proton decay rate comes out 6-7 orders of magnitude faster than experimental bound.

 In Chapter 3, we presented a generic mechanism for suppression of $\Delta$ B$\neq$0 operators.
 The GUT scale threshold corrections to fermion, anti-fermion and Higgs line leading to SO(10)
 Yukawa vertex play an important role in fermion fitting. The corrections to Higgs line are large
 which results in the lowering of SO(10) Yukawa required to match with MSSM Yukawa while fitting.
 These lowered  Yukawas then determine the $\Delta$ B$\neq$0 operators which gives
 proton decay rates O($10^{-34} yrs^{-1}$).

   After reviewing inflation and inflection point
 inflation in the Dirac-neutrino-inflation scenario in Chapter 4, we showed that the Majorana-neutrino-inflation scenario
  is not only possible but more plausible than the Dirac neutrino based inflection in Chapter 5.
  In Majorana-neutrino-inflation scenario, the inflation parameters
   and thus the inflation conditions are determined in terms of Seesaw superpotential parameters.
    Fine tuning conditions are less severe and more stable. Reheating in such scenario can occur via
     instant preheating. After the end of inflation the inflaton can decay non-perturbatively
     in the modes which get inflaton vev induced mass. These modes further decay into MSSM degrees of
     freedom which gets thermalized due to scattering leading to a temperature of $10^{11}-10^{13}$
     GeV.

     In Chapter 7, the embedding of SSI in NMSGUT is studied. The
      required number of e-folds for sufficient
      reheating requires small quartic coupling which seems hard to achieve in NMSGUT along with fermion
      fitting. An important outcome of this study was that the fermion fitting in NMSGUT is possible even
      with small value of tan$\beta$ $\approx$ 2.50. It implies that NMSGUT is viable on the entire range of
      tan$\beta$.

 In Chapter 8 we showed that the lightest neutralino (Bino in our case) relic density
 calculations for dark matter put constraints on NMSGUT generated s-particle spectrum.
 For a heavy spectrum with all s-fermion of O(10 TeV) relic density comes out to be very large. But with spectrum
 having one light s-fermion like smuon having mass comparable to Bino gives reasonable relic density for
 smuon since it provides a viable co-annihilation channel.

In the Final Chapter 9 we presented the importance of NMSGUT RGEs
is in the region between Planck and GUT scale. The soft parameters
for the Higgs mass squares comes out to be negative which is a
very peculiar behavior of NMSGUT soft parameters at GUT scale and
it needs at least one of the mass square of a Higgs representation
out of six to be negative.  We see from the RGE flow for soft
parameters from Planck scale (throwing value of $m_{3/2}$)
 to GUT scale gives negative mass squares for some Higgs representations. Thus negative NUHM proceeding from a SUGRY seed of
 universal soft scalar masses may be an
 inherent property of NMSGUT.

 Our work has further strengthened the viability of the NMSGUT by solving
 the longstanding d=5 fast B violation problem. It has
 widened the ambit of the theory even further by showing its
 relevance for inflationary and dark matter cosmology.

 The SO(10) RGE's for this complex model have been calculated to
 two loops for the first time and shown to justify the negative
 Higgs squared soft masses assumed in our NMSGUT fits.
 
  These explorations have defined a number of problems for
  continued research which we are pursuing.

\newpage\thispagestyle{empty} \mbox{}

%\newpage\thispagestyle{empty} \mbox{}
%\addcontentsline{toc}{chapter}{Reprints}
%\chapter*{\vspace*{2in}\begin{center}\Huge \emph{
%Reprints}\end{center}\thispagestyle{empty}}
%\newpage\thispagestyle{empty} \mbox{}
\end{document}